\documentclass[dvipsnames,twocolumn,english,rmp]{revtex4-1}
\usepackage[T1]{fontenc}
\usepackage[utf8]{inputenc}
\usepackage{babel}
\usepackage{graphicx}
\usepackage{amsfonts}
\usepackage{amsmath}
\usepackage{natbib}
\usepackage{hyperref}
\usepackage{bm}
\usepackage{tikz}
\usepackage{bbold}
\usepackage{mathrsfs}
\usepackage{subfigure}
\usepackage{booktabs}
\usepackage{siunitx}

\newcommand{\bra}[1]{\left \langle #1 \right |}
\newcommand{\ket}[1]{\left | #1 \right \rangle}
\newcommand{\aver}[1]{\left \langle #1 \right \rangle}

\begin{document}
\graphicspath{{Figs/}}
\title{Long-range interacting quantum systems}

\author{Nicol\`o Defenu}
\affiliation{Institute for Theoretical Physics, ETH Zürich,\\ Wolfgang-Pauli-Str.\,27, 8093 Zürich, Switzerland}
\email[Corresponding author: ]{ndefenu@phys.ethz.ch}

\author{Tobias Donner}
\affiliation{Institute for Quantum Electronics, ETH Zürich,\\ Otto-Stern-Weg\,1, 8093 Zürich, Switzerland}

\author{Tommaso Macr\`i}
\affiliation{Departamento de F\'isica T\'eorica e Experimental,
  Universidade Federal do Rio Grande do Norte, \&
  \\ International Institute of Physics,\\ 59072-970 Natal-RN, Brazil}

\author{Guido Pagano}
\affiliation{Department of Physics and Astronomy, Rice University,\\ 6100 Main Street, Houston, TX 77005, USA}

\author{Stefano Ruffo}
\affiliation{SISSA and INFN, Sezione di Trieste,\\
Via Bonomea 265, I-34136 Trieste, Italy}

\author{Andrea Trombettoni}
\affiliation{Department of Physics, University of Trieste,\\
 Strada Costiera 11, I-34151 Trieste, \&\\
 CNR-IOM DEMOCRITOS Simulation Centre and SISSA,\\ Via Bonomea 265, I-34136 Trieste, Italy}

\begin{abstract}
  
  The presence of non-local
  and long-range interactions in quantum systems induces several
  peculiar features
  in their equilibrium and out-of-equilibrium behavior.
  In current experimental platforms 
  control parameters such as interaction range,
  temperature, density and
  dimension 
  can be changed. The existence of universal scaling regimes,
  where diverse physical systems and observables 
  display quantitative agreement, generates a common framework, where the
  efforts of different research communities can be -- in some cases
  rigorously -- connected. Still, the application of this general
  framework to particular experimental realisations requires
  the identification of the regimes where the
  universality phenomenon is expected to appear.
  In the present review
  we summarise the recent investigations of
  many-body quantum systems with long-range interactions, which
  are currently realised in Rydberg atom arrays, dipolar
  systems, trapped ion setups and cold atoms in cavity experiments.
  Our main aim is to present and identify the common and (mostly)
  universal features induced by long-range interactions in
  the behaviour of quantum many-body systems. We will discuss
  both the case of very strong
  non-local couplings, i.e. the non-additive regime,
  and the one in which energy is extensive, but nevertheless low-energy,
  long wavelength properties are altered with respect to the short-range limit.
  Cases of competition with other local effects in the above mentioned setups
  are also reviewed.

\tableofcontents{}
\end{abstract}
\maketitle

\section{Introduction}
\label{sec:intro}  

The bedrock of the success of mathematical models in the theory of critical
phenomena lies in the universal behaviour of continuous phase transitions.
Thanks to universality, it is possible to describe apparently different
physical situations within the same theoretical framework.
The $O(N)$ symmetric models provided privileged tools to investigate
the universal behaviour occurring in the vicinity of a second-order phase
transition in a large class of physical systems ranging from magnets
and superconductors to biological systems and cold atom
ensembles\,\cite{chaikin1995principles,pelissetto2002critical}.
Intense investigations 
of the properties of $O(N)$ models
within the last century, dating back from the original Ising's
paper\,\cite{ising1925beitrag},
have granted the physical community with a
deep insight in the physics of homogeneous phase
transitions\,\cite{cardy1996scaling,mussardo2009statistical,nishimori2015elements}. 

For several decades such 
understanding has been mostly limited to the universal behaviour
of systems with local, short-range interactions, such as
lattice systems with nearest neighbours couplings or local $\phi^4$ field
theories.
Only in more recent times the overall picture of the universal phenomena
appearing in classical systems due to long-range interactions has been
clearly delineated. We refer to
reviews\,\cite{dauxois2002dynamics,campa2009statistical,campa2014physics}
for discussions and references on equilibrium and out-of-equilibrium of
classical systems, including $O(N)$ models, with long-range interactions.
At the same time, the study of the influence of non-local couplings,
and especially of the competition between local and long-range interactions,
in \emph{quantum} systems has seen an extraordinary surge in the wake of
several experimental realisations in atomic, molecular and optical (AMO)
systems. 

Despite these outnumbering investigations, the current literature
still lacks a comprehensive perspective on long-range interacting quantum
systems, making it difficult to place
novel results in the existing framework. Indeed, most current publications present their findings in comparison with the traditional results on
short-range systems, rather then with more recent, but established, results in the
quantum long-range 
realm. While this 
has often helped to raise the interest of the broad physics community
on these investigations, it is eventually hindering the drawing of a
comprehensive picture on long-range interacting quantum systems as well as the
admission of this knowledge in the 
domain of general-interest physics. 

The main aim of the present review is to build up an exhaustive account of
the unique phenomena arising due to long-range couplings in quantum systems,
with special focus on the universal common features that may be observed
in AMO experiments. Our description will encompass both equilibrium and
transient phenomena, 
whenever possible relating them to their common long-range origin and
to their
counterparts in classical physics. After having 
reminded basic
notions of classical long-range models and discussed the phase
structure of non-local systems, we will extend our understanding beyond the equilibrium world and 
clarify paradigmatic questions regarding relaxation and thermalisation dynamics.
Experimental realisations of long-range quantum systems are mostly isolated
and their dynamics is governed by unitary time evolution. In this context,
several open questions derive from the comparison with the conventional local
interacting case, as motivated by recent remarkable progress in the experimental
simulation of quantum long-range systems with tunable range.
At variance, the strong coupling to the environment is inevitable
for cold atom ensembles in cavities and the discussion about their
properties necessarily connects with non-additive classical systems.

The main motivation of this review, i.e. identifying
the universal features induced by long-range interactions in
quantum many-body systems,
directly points 
to the overwhelming amount of novel research, appearing almost every
day in the literature, featuring both theoretical results
and state-of-the-art experimental measurements of the dynamical
universal behaviour of highly non-local interacting systems, such as
trapped ions, cavity quantum electro-dynamics (CQED) experiments,
Rydberg atom arrays and cold atoms experiments. All these experimental
platforms present a high degree of complexity and the comprehensive
picture, which we 
aim to draw, will serve as a necessary chart to set in context
both novel experimental realisations and recent theoretical findings.

Our ambition is not only the derivation of an all-in-one picture to direct
curious outsiders in the 
realm of long-range-induced physical effects, but also to pinpoint the
most relevant and broad results in the field. This effort will
hopefully provide a step 
towards the inclusion of the physics of long-range many-body systems
into the inventory of university-taught physics. Given this purpose and
the growing amount of publications in the field, we are necessarily forced to a
selection of themes and the reader
should not be surprised if 
not all the expected references are to be found. 
Indeed, for each topic we have tried to include only the
references relevant for our main goal of the discussion of universal properties
of quantum long-range systems or the ones that are better suited to
summarise the previous literature on the argument. Whenever possible,
we will point to the reader the references containing 
accounts of previous efforts on the different arguments. 

The review is organised as follows:
in the remaining part of the present section, Sec.\,\ref{sec:intro},
we will start with a definition of what we refer to as a long-range interaction
and we present reminders on the behaviour of classical long-range systems, that will
be used in the subsequent presentation.
We then move to the classification of quantum systems
in different groups and a brief account of the most relevant properties of
each group will be presented, with a special focus on the well-known
classical case. In Sec.\,\ref{sec:exp_real}, we will discuss the most
relevant experimental realisations of each of the aforementioned groups.
Sec.\,\ref{sec:classical} will be devoted to the definition and identification of critical and universal behaviour in classical many body long-range systems,
both at equilibrium and in the dynamical regime. The content of
Sec.\,\ref{
sec:quantum} will mainly concern the equilibrium critical
properties of long-range interacting quantum many body systems, evidencing the
analogies and differences with respect to the classical case.
Sec.\,\ref{sec:regimes} will dive further into the classification of
long-range quantum systems, trying to characterise each of the aforementioned
class as 
clearly as possible based on their common properties.
Finally, Sec.\,\ref{sec:dyn} will focus on the rich mosaics of dynamical
critical scalings observed in long-range systems, when driven out of their equilibrium
state. The concluding remarks and outlook are reported in Sec.\,\ref{sec:concl}.

\subsection{Classification of long-range systems}

Since the concept of long-range interactions encompasses non-local terms,
beyond on-site or nearest-neighbour couplings, it is rather natural to
classify long-range systems based on the shape of the considered interactions.
This arrangement does not only reflect differences in the interaction shapes,
but indicates the radically different properties that appear in each class. 

The word \textbf{long-range} conventionally, but not universally,
refers to couplings that decay as a power-law of the distance between
the microscopic components, i.e.
\begin{equation}
  V(r)\sim \frac{1}{r^{\alpha}},
  \label{def_alpha}
\end{equation}
in the large
$r$ limit, $r\to\infty$. The exponent $\alpha$ will be one the main characters
of this review, together with the related one
\begin{equation}
  \sigma \equiv \alpha-d,
  \label{def_sigma}
\end{equation}  
where $d$ is the dimension of the system.

A preliminary disclaimer is due at this point.
The word "long-range" is sometimes used to denote generic
\emph{non-local} couplings, where the latter are beyond on-site
or nearest-neighbour couplings, so that within this convention an exponentially
decaying coupling would be called "long-range".
In this review, for the sake
of clarity we prefer to stick (and to a certain extent
promote) the use of the wording "\textbf{non-local}" for a generic coupling which
is not local -- exponential or finite-range or power-law {\it et cetera} -- 
and "long-range" for interactions that at large distances decay as a power-law
of the form $1/r^\alpha$ with a power exponent $\alpha$ "small enough",
in a sense that will be defined below.

A classical result on the critical properties of systems
with power-law interactions \cite{sak1973recursion,defenu2020criticality}
is that if $\alpha$ is larger than a critical value, $\alpha_*$, then
the critical behaviour is indistinguishable from the short-range limit of the model,
retrieved for $\alpha \to \infty$. So, for $\alpha>\alpha_*$, the behaviour
of the model is not "genuinely" long-range and its  universal behaviour is
the same as in the short-range limit. The specific value of $\alpha_*$ depends on the
system and  on the transition under study.
At the same time, when $\alpha$ is smaller than the dimension of the system,
$d$, then the energy is not extensive. Since $\alpha_*$ is
larger than $d$, then there is an interval of values of $\alpha$ for which
the energy is extensive, yet the long distance properties of the system
are altered by the long-range nature of the interactions.

Given this, for the sake of our presentation we will employ
the following classification:
\begin{itemize}
\item \textbf{weak long-range 
  interactions}: \emph{infinite-range interactions with power-law behaviour
and} $\alpha$ such that $d<\alpha < \alpha_*$.
\item \textbf{strong long-range 
  interactions}: \emph{infinite-range interactions with power law behaviour
and} $\alpha<d$.
\end{itemize}
Therefore, with \textbf{"short-range 
  interactions"} we will refer to the limit $\alpha \to \infty$
and by extension to $\alpha$ larger than $\alpha_*$, bearing in mind
that for $\alpha>\alpha_*$ it is the {\it critical} behaviour to be
of short-range type, but non-universal properties may of course be affected.

In both the above definitions for weak and strong long-range
interactions, with "infinite-range interactions with power-law
behaviour" we mean that the power-law decay is present
for large distances, i.e. for
the tails of the potential, irrespectively of the short-range structure of
the interactions \cite{mukamel2008statistical}. 
To appropriately cover the cases in which there
is competition between excitations on different length-scales, e.g.
between a certain long-range interaction and another one acting at short-range,
we will use the following additional notation:
\begin{itemize}  
\item \textbf{competing non-local interactions}:
  \emph{finite- and/or infinite-range interactions with sign changing couplings}. 
\end{itemize}

It is worth noting that this classification has been introduced to ease
the following discussion, but it does not pretend to be rigorous or perfect.
Indeed, it may happen that certain strong long-range systems exhibit critical
scaling analogous to the general weak long-range scale, or that in an infinite range
interacting system the dominant effect is the creation if non-homogeneous
patters, so that its physics is more similar to the case of finite range sign
changing interactions. Similary, it could happen that in
a system with finite-range interaction plus a power law interaction with power
decay $\alpha$, the long-range tail does not affect ground-state properties
so that according the classification the interaction could be "long-range" and
nevertheless the system would behave as a non-long-range system.
Given the variety of situations, when needed for the sake of the clarity of the presentation,
we will regroup the material according
the phenomena exhibited by the different systems. Nevertheless, when not
misleading we will stick to the previous convention, which has the merit
to classify different interactions indipendently of further considerations and
of the knowledge of the actual behaviour of the quantity of interest studied
in the particular models at hand.

\subsection{Reminders on classical systems with long-range interactions}\label{sec_reminder}

In the rest of the Section, we will make a brief account of the
most established phenomena occurring in each of the previously introduced
classes in the classical physics case to set the ground for the quantum case.
In Tab.\,\ref{tab1} we schematically summarize physical
systems governed by long-range interactions.
Results for some of them are summarized in the remaining part of this 
Section, and further discussed in the quantum case in the next Sections.

\subsubsection{Strong long-range interactions}

Among the unique effects produced by long-range interactions, 
remarkable features appear in the case of strong long-range couplings $\alpha<d$.
There,
the interaction energy of homogeneous systems becomes infinite,
due to the diverging long distance contribution of the integral
$\int r^{-\alpha}d^{d}r$. Therefore, the common definitions for
the internal energy or the entropy turn out to be non extensive and
traditional thermodynamics does not apply. 

These properties 
are actually shared by a wide range of physical systems,
ranging from gravity to plasma physics, see Tab.\,\ref{tab1}.
Apart from the cases 
summarised there, 
the general results of strong long-range systems often apply also to mesoscopic
systems, far from the thermodynamic limit, whose interaction range,
even if finite, is comparable with the systems size. In the perspective
of quantum systems, this situation is particularly relevant for Rydberg gases\,\cite{boettcher2020new}. 
\begin{table*}
\begin{tabular}{cccc} \toprule
    System  & {$\alpha$} & {$\alpha/d$} &  {Comments} \\ \midrule
    Gravitational systems & 1 & \,\, 1/3 &  Attractive forces, possibly non homogenous states\\
    Non-neutral plasmas  &  1 & \, \, 1/3 & Some LR effects are also present in the neutral case\\
    Dipolar magnets & 3 & \,\, 1 & Competition with local ferromagnetic effects\\
    Dipolar Gases & 3 & \,\, 1 & Anisotropic interactions\\
    Single-mode cavity QED systems & 0 & \,\, 0 & Interactions mediated by cavity photons\\
    
    Trapped ions systems & $\sim$ 0-3 & \,\,  $\sim$ 0-3 & Interactions mediated by crystal phonons\\    \bottomrule
\end{tabular}
\caption{\label{tab1} Table listing different applications
  where systems are governed by long-range interactions
  (LR stands for long-range).
  These systems present interactions which remain long-range
  up to the thermodynamics limit. In the table the ratio $\alpha/d$, signaling
  how strong the long-range is, refers to $d=3$ in the first four lines (see the text for discussion
  of different $d$). Notice that for multi-mode cavity QED systems
$\alpha$ is tunable.}
\end{table*}

Due to the lack of extensivity, theoretical investigations in the strong
long-range regime need suitable procedure to avoid encountering divergent quantities.
This scope has been obtained in the literature 
scaling the long-range interaction term by a volume pre-factor $1/V^{\alpha-d}$, which is the
so-called Kac's prescription\,\cite{kac1963van}.
Interestingly, it has been possible to show\,\cite{anteneodo1998breakdown}
that this prescription in simple strong long-range models yields
the same physical picture obtained introducing a different definition of the entropy
via the non-extensive $q$-statistics\,\cite{tsallis2004nonextensive}.

The salient feature of the Kac prescription is that
it allows a proper thermodynamic description of strong long-range systems,
without disrupting their key property, i.e. non-additivity.
Indeed, other possible regularisations, where the long-range tails of the interactions
are cutoff exponentially or at a finite range tend to disrupt the peculiar physics of
these systems. Similar cutoff regularisations are often employed in neutral Coulomb systems,
where the $1/r$ potential tails are naturally screened by the presence of oppositely charged
particles. However, even in the screened case, the long-range tails of the interaction potential
may give rise to finite corrections to thermodynamic quantities from the boundary conditions,
which also remain finite in the thermodynamic limit\,\cite{lewin2015improved}.

Similarly, the appearance of non-additivity in strong long-range systems is connected with
a finite contribution of the system boundaries to the thermodynamic quantity, as
in the prototypical case of fully connected systems where boundary and bulk contributions
have the same order. In fact, it is in fully connected systems that most of the
spectacular properties of strong long-range systems have been first identified, such as
\emph{ensemble in-equivalence}\,\cite{barre2001inequivalence}. The latter
is the property of
non-additive systems to produce different results when described with different thermodynamical
ensembles, which leads to apparently paradoxical predictions such as negative
specific heats or susceptibilities. These models also present
the so-called \emph{quasi-stationary states} (QSS) in the out-of-equilibrium dynamics, i.e.
metastable configurations whose lifetime scales super-linearly with the system size.
An extensive account of the peculiar properties of long-range systems in the classical case can
be found in Refs.\,\cite{dauxois2002dynamics, campa2014physics}, while in the following
we are going to explicitly focus on the quantum case. 

Based on the discussion above, one may be tempted to exclusively relate such peculiar properties
such as ensemble inequivalence, negative specific heat and QSS to the non-extensive
scaling of strong long-range systems in the thermodynamic limit. However,
similar effects appear also
in mesoscopic systems, where the interaction range is finite,
but of the same order as the system size or for attractive systems where most of the density
is localised within a finite radius with flat interactions\,\cite{thirring1970systems}. 
 
\subsubsection{Weak long-range interactions}

The focus on short-range interactions in the theory of critical
phenomena\,\cite{nishimori2015elements} is not only motivated by
simplicity reason, but rather by the resilience of the universal
behaviour upon the inclusion of non-local couplings, at least in
homogeneous systems. Indeed, the common wisdom states that universal
properties close to a critical point do not depend upon variations
of the couplings between the microscopic components, but only on the
symmetry of the order parameter and the dimension of the system under study.
However, this statement is not generally true, when long-range interactions are
introduced into the system.

Indeed, while universal properties are insensible to the intermediate range
details of the interactions, for critical systems with homogeneous order
parameters, they are sensible to the power-law decaying
tails of long-range couplings (and, to be explicit, not on
the strength of the interaction itself). For $\alpha<d$,
the interaction energy diverges and the universal behaviour typically
belongs to the mean-field universality class. On the contrary, as a function
of the parameter $\sigma\equiv\alpha-d>0$ three different regimes may be
found\,\cite{defenu2020criticality}:
\begin{itemize}
\item for $\sigma\leq \sigma_{\rm mf}$ the mean--field approximation
  correctly describes the universal behavior; 
\item for $\sigma>\sigma_*$, 
the model has the same critical exponents of its short-range version, i.e. the limit $\sigma \to \infty$; 
\item for $\sigma_{\rm mf}<\sigma
  \le \sigma_*$ the system exhibits 
peculiar long-range critical exponents,
\end{itemize}
where the notation $\sigma_* \equiv \alpha_*-d$ has been used.
Therefore, it exists a range of long-range decay exponents
$0<\sigma \le \sigma_{*}$, where thermodynamics remains well defined and
the critical behaviour is qualitatively similar to the one appearing in
the limit $\sigma\to\infty$. Nevertheless, the universal properties
become $\sigma$-dependent and, loosely, mimic the dependence of
the short-range universal properties as a function of the geometric dimension
$d$\,\cite{fisher1972critical}. In other words,
varying $\sigma$ at fixed dimension is, loosely, equivalent to change
the geometric dimension in short-range systems. Notice that this equivalence is expected to be
not exact in general, but it does at gaussian level, as one can explictly see for the spherical
model\,\cite{joyce1966spherical}.

While the boundary $\sigma_{\rm mf}$ can be exactly calculated by appropriate
mean-field arguments, the location of the $\sigma_{*}$ is the result
of a complex interplay between long-range and short-range contributions to critical fluctuations.
This fascinating interplay is at the root of several 
interesting phenomena,
which appear in a wide range of different critical systems upon
the inclusion of long-range interactions in the weak long-range regime. The appearance of
novel effects is not limited to the equilibrium universal properties, but
also extends to the out-of-equilibrium realm, whose plethora of intriguing
long-range phenomena has only been partially understood. Given these considerations,
most of the focus of the forthcoming discussion on
weak long-range interacting systems will concern universal properties both at and
out of equilibrium.

\subsubsection{Competing non-local interactions} 

Systems with non-local interactions whose tails are rapidly decaying,
with $\sigma>\sigma_{*}$ or exponential decaying, may still produce interesting
universal features, due to the interplay with other local couplings or
to the presence of frustration in the system. Indeed, when long-range repulsive interactions
compete with short-range attractive ones the pertinent order parameter of the system may
form spatial modulations in the form of lamellae, cylinders, or spheres.
These \emph{modulated phases} are ubiquitous in nature and emerge in a large
variety of physical systems ranging from binary polymer mixtures, cold atoms and magnetic systems,
to high-temperature superconductors\,\cite{seul1995domain}. Especially in two dimensions,
the presence of modulated phases leads to rich phase diagrams with peculiar features,
which are far to be fully comprehended. 
In particular, modulation effects caused by
competing non-local interactions are responsible for the appearance of a nematic to
smectic phase transition in liquid crystal films, whose universality class remains an
open physical problem. 

At finite temperatures, another striking effect of modulated phase is inverse melting,
which is a consequence of reentrant phases. Indeed, a modulated phase may be
"too hot to melt"\,\cite{greer2000melting}, when the system recovers the
disordered state at very low temperature after  being in a symmetry broken state in
an intermediate temperature regime. The extension of this reentrance becomes appreciable
for systems, where the homogeneous and modulated phases present similar energy cost and
the order parameter remains small, and it is thus strongly influenced by the form and
intensity of non-local interactions\,\cite{mendoza2019mechanism}. 

The study of the universal properties of modulated phases has been
initiated long ago\,\cite{brazovskii1975phase}, but 
comprehensive picture of their critical properties is yet lacking,
despite the large amount of investigations\,\cite{cross1993pattern}, due to the
difficulty to devise reliable approximation schemes. However,
the increasing number of experimental realisations featuring striped phases 
could lead to a renovated interest in such problems within the framework of the
physics of long-range interactions.

\section{Experimental realisations}
\label{sec:exp_real}

As mentioned above, the rising interest for long-range physics has been made pressing
by the current developments of the experimental techniques for the control and
manipulation of AMO systems. Indeed, long-range \emph{quantum}
systems are being currently realised in several experimental platforms
such as Rydberg atoms\,\cite{saffman2010quantum}, dipolar quantum gases\,\cite{lahaye2009physics}, polar molecules \,\cite{carr2009cold}, quantum gases coupled to optical cavities\,\cite{ritsch2013cold, mivehvar2021cavity} and trapped
ions\,\cite{schneider2012experimental,blatt2012quantum,monroe2021programmable}.
Long-range interactions with tunable exponent $\alpha$ can currently be
realised using trapped ions off-resonantly coupled to motional degrees of
freedom stored in a Paul trap\,\cite{islam2013emergence,richerme2014nonlocal,jurcevic2014quasiparticle}, in a Penning trap\,\cite{dubin1999trapped,britton2012engineered} or neutral atoms coupled to photonic modes of a cavity\,\cite{douglas2015quantum,vaidya2017tunable}\,. 

Based on the aforementioned classification, we are going to focus our attention on three different classes of experimental systems: \emph{trapped ions},
\emph{quantum gases in cavities}
and \emph{dipolar systems},
including in particular Rydberg states.
All of these systems are quantum in nature and
represent prototypical applications of recent investigations in long-range physics.
Trapped ions 
present the almost unique possibility to experimentally realise
long-range interactions with decay exponent which may be tuned in the range
$\alpha\in 0 \sim 3$ exploring both the strong and weak long-range regimes. Conversely, cavity mediated interactions between atoms are typically flat ($\alpha=0$) and constitute the experimental counterpart of the celebrated Dicke or Lipkin-Meshkov-Glick models\cite{dicke1954coherence,hepp1973superradiant,lipkin1965validity}, two real workhorses of long-range interactions.
Finally, Rydberg states and dipolar atoms in general present several common
features with thin magnetic films, which have been the traditional experimental setup for the study of modulated critical phenomena at finite temperatures\,\cite{selke1988annni}. 

Thus, each of these 
experimental platforms
represents a realisation of the peculiar physics in each of the long-range regimes.
However, this statement should not be considered strictly, but mostly a
general guideline to ease our presentation. The reason for such a discalimer
is that in the following we
will describe several examples violating such correspondence --
such as the observation of QSS in the strong long-range regime of trapped
ions\,\cite{neyenhuis2017observation}; the presence of pattern formation in cavity systems\,\cite{baumann2010dicke,landini2018formation}; and the realisation of the Lipkin-Meshkov-Glick model in the
fully-blockade limit of Rydberg atoms\,\cite{henkel2010three, zeiher2016many}.

\subsection{Trapped ions}
\begin{figure*}[t]
\centering
\includegraphics[width=2\columnwidth]{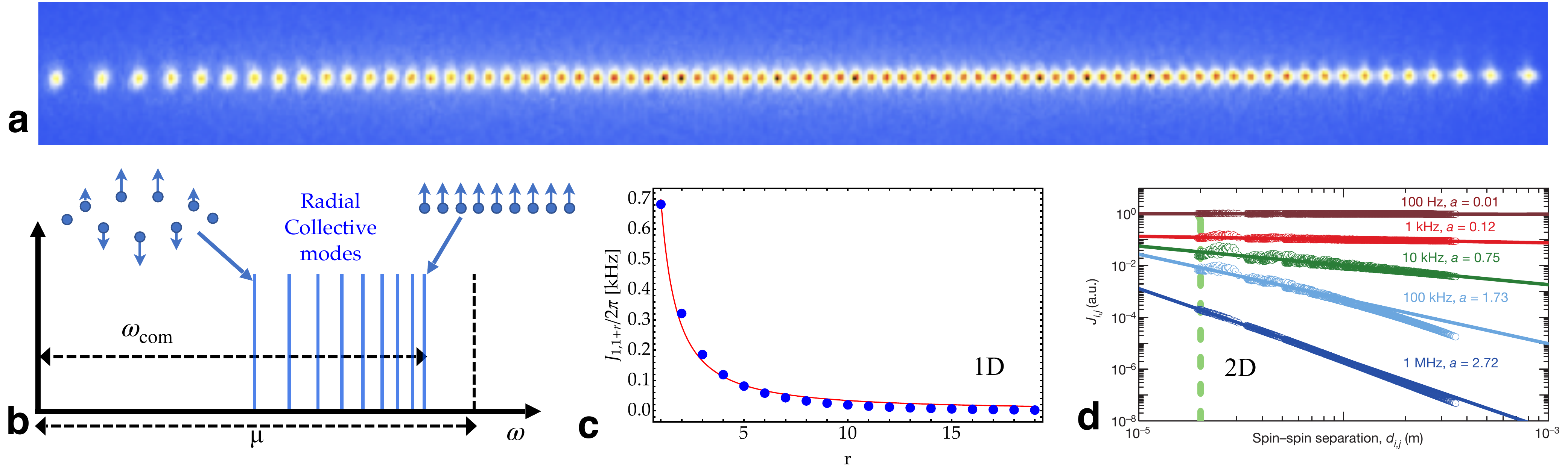}
\caption{{\bf Trapped ions systems.} 
{\bf (a)} A 77 linear chain of $^{171}$Yb$^+$ ions. The harmonic confinement and Coulomb interactions cause the spacing between ions to be inhomogenous, breaking translational invariance. {\bf (b)} A laser drive at frequency $\mu$ is detuned from the radial center of mass mode to create phonon-mediated spin-spin interacitons.
{\bf (c)} Calculated spin-spin interaction for a 1D chain of 20 ions versus distance from the edge ion. In this case $\delta=\mu-\omega_{\rm com}=2\pi \cdot100$ kHz and $J_{1,1+r}\sim 1/r^{1.3}$.
{\bf (d)}  Calculated Ising couplings in a 2D crystal of 217 ions versus a sampling of the distance $d_{ij}$ between ion pairs (circles). The lines are best-fit power law exponents $\alpha$ (lines) for various detunings from the center-of-mass (COM) mode of $795$ kHz. Adapted from Ref. \cite{britton2012engineered}.
}
\label{IonSpinInteraction}
\end{figure*}

Laser cooled ions confined in rf traps are one of the most advanced
platforms for both quantum computing \cite{ladd2010quantum} and
quantum simulation \cite{monroe2021programmable}. In these systems,
time-dependent electric fields create an effective harmonic,
eV-deep potential \cite{paul1990electromagnetic,dehmelt1967radiofrequency,Brown1986Geonium} allowing a long storage time of collections of charged
particles in vacuum systems \cite{Pagano2018cryogenic}.
When laser cooled \cite{leibfried2003quantum}, the atomic ions form
Wigner crystals with equilibrium positions and vibrational collective
modes well defined by the competition between Coulomb interactions and
harmonic confinement induced by the trap. The long-range character of Coulomb interactions present in these systems is directly translated into the effective spin models that can be engineered by applying a Floquet drive to the ion crystal.
In the following sections, we will first review the experimental techniques
used to realize spin models with tunable power law interactions and then
we will describe the experimental realizations of these models where the long-range
character of the interaction allowed the observations of new physical
phenomena in many-body quantum systems.

\subsubsection{Phonon-mediated interactions}\label{sec:TI_techniques}

In trapped ions systems the spin degree of freedom can be encoded in two
long-lived atomic states, either in the hyperfine ground state
manifold\,\cite{wineland2003quantum} or using a metastable
electronic state\,\cite{blatt2008entangled}. Both approaches
guarantee coherence time of the order of a few seconds,
near-perfect initialization via optical pumping\,\cite{happer1972optical}
and high-fidelity detection via state-dependent
fluorescence\,\cite{noek2013high,myerson2008high, Christensen2020high}.  

Without any spin-motion coupling, the ion crystal can be described as a
set of normal modes of motion (phonons) and an independent set of internal
(spin) degrees of freedom, with the Hamiltonian:
\begin{equation}
\label{eq_}
 H = \sum_m \hbar \omega_m a_m^\dagger a_m + \sum_{i} \vec{B_i}\cdot\vec{\sigma_i},
 \end{equation}
where $a^\dagger_m(a_m)$ is the creation(annihilation) operator of the $m$-th phonon mode with $[a_m,a^\dagger_n]=\delta_{mn}$, and
$\vec{\sigma}_i=\{\mathbb{1}_i,\sigma^x_i,\sigma^y_i,\sigma^z_i\}$ and
$\vec{B}_i$ are the Pauli matrix vector and effective magnetic fields
associated  with the $i$-th ion, respectively. The effective magnetic
fields are implemented experimentally with microwaves or one-photon and
two-photon laser-induced processes.

Laser cooling and sub-Doppler techniques, e.g. resolved
Raman sideband cooling \cite{Monroe1995resolved} and
Electromagnetic-Induced Transparency (EIT) cooling \cite{roos2000experimental,Lin2013sympathetic,Feng2020efficient,Jordan2019near},
can prepare all motional states near their ground states, which is
crucial for the simulation of spin models described below.

Quantum operations can be carried out by exerting a
spin-dependent optical force on the ion crystal, coherently coupling spin
and motional degrees of freedom. High-fidelity coherent spin-motion
coupling can be realized with one-photon optical transitions\,\cite{blatt2008entangled} in the case of optical qubits, two-photon stimulated
Raman transitions\,\cite{kim2009entanglement,harty2014high-fidelity,britton2012engineered} in the case of hyperfine qubits and near-field
microwaves\,\cite{Ospelkaus2011microwave,Harty2016high-fidelity,srinivas2021highfidelity}.

Considering the momentum $\hbar \Delta k$ imparted by the laser on the ions
confined in a harmonic potential well, the general light-atom Hamiltonian
in the rotating frame of the qubit is:
\begin{equation}
H =  \frac{\hbar \Omega}{2}\sum_{i} \left[ ( \vec{\theta} \cdot  \vec{\sigma}_i)\,\, e^{i(\Delta k X_i-\mu t-\phi)} + {\rm h.c.}  \right],
\label{eq_Hlaser}
\end{equation}
where $\Omega, \mu$ and $\phi$ are the Rabi frequency, the laser beatnote frequency and the laser phase, respectively. The spin Pauli operators $\vec{\sigma}=\{\mathbb{1},\sigma^x_i,\sigma^y_i,\sigma^z_i\}$ are multiplied by the complex coefficients $\vec{\theta}=\{\theta_0,\theta_1,\theta_2,\theta_3\}$ depending on the specific experimental configuration. The position operator can be written in terms of collective phononic modes as $$X_i=\sum_{m=1}^{N} \eta_{im}(a_m^\dag e^{i\omega_m t} + a_m e^{-i\omega_m t}),$$ with $\eta_{im}=\eta_m b_{im}$ where $b_{im}$\footnote{$\sum_i b_{im}b_{in} = \delta_{nm}$ and $\sum_m b_{im}b_{jm} = \delta_{ij}$} is the normal mode transformation matrix, $\eta_{m}=\Delta k\sqrt{\hbar/2m\omega_m}$ is the Lamb-Dicke parameter associated to the $m$-th normal mode at frequency $\omega_m$. 

In the Lamb-Dicke regime, $({\Delta k \langle{X_i}}\rangle\ll1)$, the first-order term of Hamiltonian (\ref{eq_Hlaser}) gives rise to spin-phonon couplings of the form $(\sigma^{\pm,z}_i a_m e^{i\omega_m t} + {\rm h. c.})$, where the spin operator depends on the experimental configuration. These terms generate an evolution operator under a time-dependent Hamiltonian that can be written in terms of Magnus expansions \cite{zhu2006trapped}. In the limit of $(\mu-\omega_m)\gg\eta_{im} \Omega$ for $\forall m$, the motional modes are only virtually excited meaning that only the second order term of the Magnus expansion is dominant and leads to the following pure spin-spin Hamiltonian:
\begin{equation}
H=\sum_{ij} J_{ij}\sigma^{\vec{\theta}}_i \sigma^{\vec{\theta}}_j,
\label{eq_Ising_ions}
\end{equation}
where the choice of the Pauli spin operator $\sigma^{\vec{\theta}}_i$ is controlled by the laser configuration\footnote{For a detailed derivation of Eq. (\ref{eq_Ising_ions}) we refer to \cite{monroe2021programmable}.}. One common configuration $\{\theta_1=1/2, \theta_2=i/2,\theta_0=\theta_3=0\}$ leads to the so-called M\o lmer-S\o rensen gate \cite{molmer1999quantum} where two laser beatnotes are tuned close to  the motional mode transitions with opposite detunings $\pm \mu$. In this configuration $\sigma^{\vec{\theta}}_i=\sigma^{\phi}_i=\sigma^x_i \cos(\phi) + \sigma^y_i \sin(\phi)$, where $\phi$ can be tuned by controlling the phases of the two laser beatnotes \cite{monroe2021programmable}. Another widely used laser configuration is $\{\theta_1=\theta_2=\theta_0=0, \theta_3=1\}$ \cite{leibfried2003experimental}, where $\sigma^\theta_i=\sigma^z_i$ and the ion motion is modulated by a spin-dependent light shift. 

The spin-spin interaction matrix $J_{ij}$  can be explicitly calculated given the normal modes frequencies $\omega_m$ and the detuning $\mu$ as follows:
\begin{equation}
\label{eq_}
J_{ij}=\Omega^2 \omega_{\rm rec} \sum_{m=1}^N \frac{b_{im}b_{jm}}{\mu^2-\omega_m^2}
\end{equation}
where $\omega_{\rm rec} = \hbar (\Delta k)^2/2M$ is the recoil frequency associated with the transfer of momentum $\hbar (\Delta k)$ (see Fig. \ref{IonSpinInteraction}). The spin-spin interaction can be approximated with a tunable power law:
\begin{equation}
\label{eq_}
J_{ij}=\frac{J_0}{|i-j|^\alpha}.
\end{equation}

The approximate power-law exponent can be tuned in the $0<\alpha<3$
range by tuning the detuning $\mu$ and the trap frequencies $\omega_m$.
In the limit of $\mu\gg{\Delta \omega}$, with ${\Delta\omega}$ being the
typical mode separation, all modes contribute equally and the spin-spin
interaction decays with a dipolar power law, e.g. $J_{ij}\sim1/|i-j|^3$.
On the other hand, when $\mu$ is tuned close to $\omega_{\rm com}$
(the center of mass, see Fig. \ref{IonSpinInteraction}), the exponent alpha decreases.

It is worth noting that in the quantum simulation regime, large transverse
fields ($\mu-\omega_{\rm com}\gg B_z \gg J_0$) have been used in the
Molmer-Sorensen configuration to tune Hamiltonian (\ref{eq_Ising_ions}) and experimentally realize a long-range XY model:
\begin{equation}
\label{eq_XY}
H=\sum_{ij}J_{ij}(\sigma^x_i\sigma^x_j+\sigma^y_i\sigma^y_j)=\sum_{ij}J_{ij}(\sigma^+_i\sigma^-_j+\sigma^-_i\sigma^+_j).
\end{equation}
Qualitatively, the large field $B_z$ transverse to the interaction direction suppresses energetically the processes involving two spin-flips ($\sim \sigma^+_i \sigma^+_j +\sigma^-_i \sigma^-_j$) of the Ising Hamiltonian (\ref{eq_Ising_ions}) and retains only the spin preserving part ($\sim \sigma^+_i \sigma^-_j+ \sigma^-_i \sigma^+_j $). Note that some works refer to Hamiltonian (\ref{eq_XY}) as XX Hamiltonian instead of XY. In the following we will use these two as synonyms, depending on the specific work that is being discussed.

In the past decade the possibility to have tunable power law interactions has stimulated a large body of theory work as well as ground-breaking experiments investigating both the equilibrium properties of the system as well as the non-equilibrium dynamics. In particular, it is challenging to calculate exactly the non-equilibrium dynamics of long-range interacting systems after a quantum quench for $N>25$ spins. In Sec.\,\ref{sec:dyn} we will address the experimental observations in trapped ions systems that are related to the long-range character of the underlying Hamiltonian.

\subsection{Quantum gases in cavities}
\label{q_cav_sec}

Dilute quantum gases of neutral atoms are a powerful platform to study many-body physics\,\cite{Bloch2008}. However, these gases typically only interact via collisional, short-range interactions. Long-range dipole-dipole interactions can nevertheless be implemented employing either particles with a large static dipole moment (such as heteronuclear molecules or atomic species with large magnetic dipole moments), or with an induced dipole moment, such as Rydberg atoms. These approaches will be discussed in section \ref{subsec:dipolar_gases}. A complementary route to exploit induced dipolar interactions is to couple the quantum gas to one or multiple modes of an optical cavity \cite{ritsch2013cold,mivehvar2021cavity}. In the following sections, we will first provide an introduction into the fundamental mechanism giving rise to cavity-mediated long-range interactions and then turn to experimental realizations of relevance for the current review.

\subsubsection{Cavity-mediated interactions}

The basic setting is shown in Fig. \ref{fig:CavityInteractions}(a).
A Bose-Einstein condensate (BEC) is trapped by an external confining potential at the position of the mode of an optical cavity. The quantum gas is exposed to a standing wave transverse pump laser field with wave vector $\mathbf{k}_p$ whose frequency $\omega_p$ is far detuned by $\Delta_a = \omega_p-\omega_a$ from the atomic resonance $\omega_a$. In this dispersive limit, the atoms are not electronically excited but form a dynamical dielectric medium, that scatters photons.
At the same time, the resonance frequency $\omega_c$ of a cavity mode with
wave vector $\mathbf{k}_c$ (where $|\mathbf{k}_c| \approx |\mathbf{k}_p| = k$)
is tuned close to the frequency of the transverse pump field, such that
photons scattered off the atoms are preferentially scattered into the cavity
mode. Compared to free space, such vacuum-stimulated scattering is greatly
enhanced by a factor proportional to the finesse of the optical cavity.

The scattering of a photon from the pump off a first atom into the cavity
and then back into the pump off a second atom is the microscopic process
mediating the interaction between two atoms. Such a photon scattering process
imparts each one recoil momentum along the cavity direction and the pump field
direction onto the atoms, such that atoms initially in the zero-momentum BEC
state $|\mathbf{p}_0\rangle = |p_x, p_y\rangle = |0,0\rangle$ are coupled to a
state $|\mathbf{p}_1\rangle$ which is the symmetric superposition of the four
momentum states $|\pm \hbar \mathbf{k}_c \pm \hbar \mathbf{k}_p\rangle$.
Since the photon is delocalized over the cavity mode this interaction is of global range. The strength of the interaction can be increased by either reducing
the absolute value of the detuning $\Delta_c= \omega_p - \omega_c$ between pump frequency and cavity resonance, or by increasing the power of the transverse pump field. The interaction inherits its shape from the interference of the
involved mode structures of transverse pump and cavity.

\begin{figure}[]
\centering
\includegraphics[width=1\columnwidth]{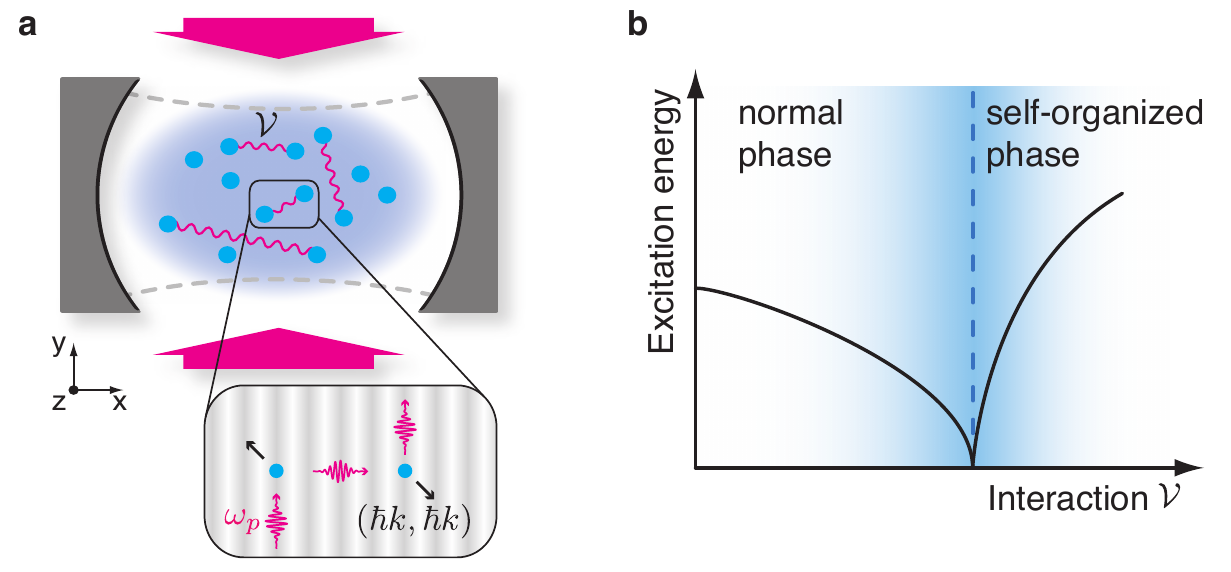}
\caption{{\bf Experimental scheme for realizing cavity-mediated interactions and mode softening at the superradiant phase transition.} 
{\bf (a)} A BEC (blue cloud) inside an optical cavity is transversally illuminated by a far red–detuned standing-wave laser field. In a quantized picture, atoms off-resonantly scatter photons from the pump field into a close-detuned cavity mode and back, creating and annihilating pairs of atoms in the superposition of momenta $(p_x, p_y)= (\pm \hbar k, \pm \hbar k)$ (one of four possible processes is shown schematically). This results in global interactions between all atoms. The interaction strength $\mathcal{V}$ is controlled via the power of the transverse laser field and the detuning $\Delta_c$. {\bf (b)} The cavity-mediated atom-atom interaction causes a softening of a collective excitation mode with energy $\hbar \omega_s$ at the momenta $(\pm \hbar k, \pm \hbar k)$ and a diverging susceptibility (blue shade) at a critical interaction strength (dashed line). Adapted from Ref. \cite{mottl2012roton}.
}
\label{fig:CavityInteractions}
\end{figure}

More formally, after adiabatically eliminating the electronically excited atomic states, a quantum gas driven by a standing wave transverse pump field with mode function $\chi(\mathbf{r})$ and coupled to a linear cavity with mode function $\xi(\mathbf{r})$ can be described by the many-body Hamiltonian  \cite{maschler2008ultracold} $H = H_c + H_a + H_{ac}$ with
\begin{equation}\label{eq:many_body_atom_field_Hamiltonian}
\begin{split}
   H_c &= -\hbar \Delta_c a^\dagger a \\
  H_a &= \int\mathrm{d}^3\mathbf{r} \Psi^\dag(\mathbf{r}) \left[\frac{\mathbf{p}^2}{2m} + V_p \chi^2(\mathbf{r}) + \frac{g}{2} \Psi^\dag(\mathbf{r}) \Psi(\mathbf{r}) \right] \Psi(\mathbf{r}) \\
  H_{ac} &= \int\mathrm{d}^3\mathbf{r} \Psi^\dag(\mathbf{r}) \hbar\left[\eta \chi(\mathbf{r})\xi(\mathbf{r}) (a+ a^\dag) + U_0 \xi^2(\mathbf{r}) a^\dag a \right]\Psi(\mathbf{r}),
\end{split}
\end{equation}
where $H_c$ describes the dynamics of a single cavity mode with photon creation (annihilation) operator $a^\dagger$($a$). The atomic evolution in the potential provided by the pump field with  depth $V_p$ is captured by the second-quantized term $H_a$, where $\mathbf{p}$ is atomic momentum, $m$ is atomic mass, $g$ describes the atomic contact interactions, and $\Psi(\mathbf{r})$ is the bosonic atomic field operator. The term $H_{ac}$ finally describes the interaction between atoms and light fields. Its first term captures the photon scattering between cavity and pump fields at a rate given by the two-photon Rabi frequency $\eta=\frac{g_0 \Omega_p}{\Delta_a}$, where $g_0$ is the maximum atom-cavity vacuum-Rabi coupling rate and $\Omega_p$ is the maximum pump Rabi rate. The second term describes the dynamic dispersive shift of the cavity resonance with $U_0=\frac{g_0^2}{\Delta_a}$ being the light-shift of a single maximally coupled atom.

The atomic system evolves on a time scale given by the energy $\sim\hbar \omega_r$ of the excited momentum state, where $\omega_r = \hbar k^2/(2m)$ is the recoil frequency of the photon scattering. If the cavity evolution is fast compared to this time scale, i.e. if the cavity decay rate $\kappa \gg \omega_r$, the cavity field can be adiabatically eliminated which yields
\begin{equation}\label{eq:steady_state_cavity_field}
  a=\frac{\eta \Theta}{\tilde{\Delta}_c+i\kappa},
\end{equation}
where $\tilde{\Delta}_c=\Delta_c - U_0\int \mathrm{d}^3\mathbf{r} \Psi^\dag(\mathbf{r}) \xi^2(\mathbf{r}) \Psi(\mathbf{r})$ is the dispersively shifted cavity detuning. Eq. (\ref{eq:steady_state_cavity_field})
shows that the cavity field is proportional to the order parameter operator
$\Theta=\int \mathrm{d}^3\mathbf{r} \Psi^\dag(\mathbf{r}) \chi(\mathbf{r})\xi(\mathbf{r}) \Psi(\mathbf{r})$ which measures the overlap between atomic density modulation and the mode structure of the interfering light fields. This relation is essential for the real-time observation of the atomic system via the light field leaking from the cavity.

Eliminating the steady-state cavity field of Eq. (\ref{eq:steady_state_cavity_field}) from Eqs. (\ref{eq:many_body_atom_field_Hamiltonian}), an effective Hamiltonian is obtained \cite{mottl2012roton},
\begin{equation}
  H_\mathrm{eff} = H_a + \int \mathrm{d}^3\mathbf{r}\mathrm{d}^3\mathbf{r'}
  \Psi^\dag(\mathbf{r}) \Psi^\dag(\mathbf{r'}) \mathcal{V}_{lr}(\mathbf{r}, \mathbf{r'}) \Psi(\mathbf{r}) \Psi(\mathbf{r'})\,,
\end{equation}
with the long-range interaction potential
\begin{equation}\label{eq:cavity_long_range_interactions}
 \mathcal{V}_{lr}(\mathbf{r}, \mathbf{r'}) = \mathcal{V} \chi(\mathbf{r})\xi(\mathbf{r}) \chi(\mathbf{r'})\xi(\mathbf{r'}).
\end{equation}
This periodic interaction potential with strength $\mathcal{V}=\hbar\frac{\eta^2 \tilde{\Delta}_c}{\tilde{\Delta}_c^2+\kappa^2}$ is of global range and favors a density modulation of the atomic system with a structure given by the interference of pump and cavity fields. For a standing wave transverse pump field impinging on the BEC perpendicular to the cavity mode, this interference has a checkerboard shape $\cos(kx)\cos(ky)$.

While integrating out the light field provides access to a simple description in terms of a long-range interacting quantum gas, it is important to keep in mind that the system actually is of driven-dissipative nature. The excitations of the system are polaritons that share the character of both the atomic and the photonic field. Furthermore as we detail below, in the sideband resolved regime $\kappa \lesssim \omega_r$ the cavity field cannot be integrated out anymore and the interaction becomes retarded \cite{klinder2015dynamical}.

The sign of the interaction $\mathcal{V}$ can be chosen by an according change in the detuning $\tilde{\Delta}_c$. For $\mathcal{V}<0$, this interaction leads to density correlations in the atomic cloud favouring a $\lambda$-periodic density structure, where $\lambda=2\pi/k$ is the wavelength of the pump laser field. This can also be understood inspecting the first term in $H_{ac}$ from Eqs. (\ref{eq:many_body_atom_field_Hamiltonian}). A $\lambda$-periodic density structure would act as a Bragg lattice, enhancing the coherent scattering of photons between pump and cavity. The emerging intra-cavity light field interferes with the pump lattice and builds an optical potential in which the atoms can lower their energy. However, the long-range interaction favoring the density modulation competes with the kinetic energy term. Only above a critical interaction strength, the system undergoes a quantum phase transition to a self-ordered state characterized by a density modulated cloud and a coherent field in the cavity mode, see Section \ref{subsec:selforganization}.

Also tunable-range interactions can be engineered by extending the scheme described above to multi-mode cavities \cite{gopalakrishnan2011frustration, gopalakrishnan2009emergent, gopalakrishnan2010atom}. In such cavities, a very large number of modes with orthogonal mode functions (in theory an infinite number, in practice several thousands) are energetically quasi-degenerate. An atom within the quantum gas will thus scatter the pump field into a superposition of modes, with the weights set by the position of the atom and a residual detuning between the modes. These modes interfere at large destructiveley, such that only a field wave packet localized around the scattering atom remains where constructive interference dominates. Accordingly, the effective atomic interaction acquires a finite range set by the number of contributing modes.

\begin{figure}[]
\centering
\includegraphics[width=1\columnwidth]{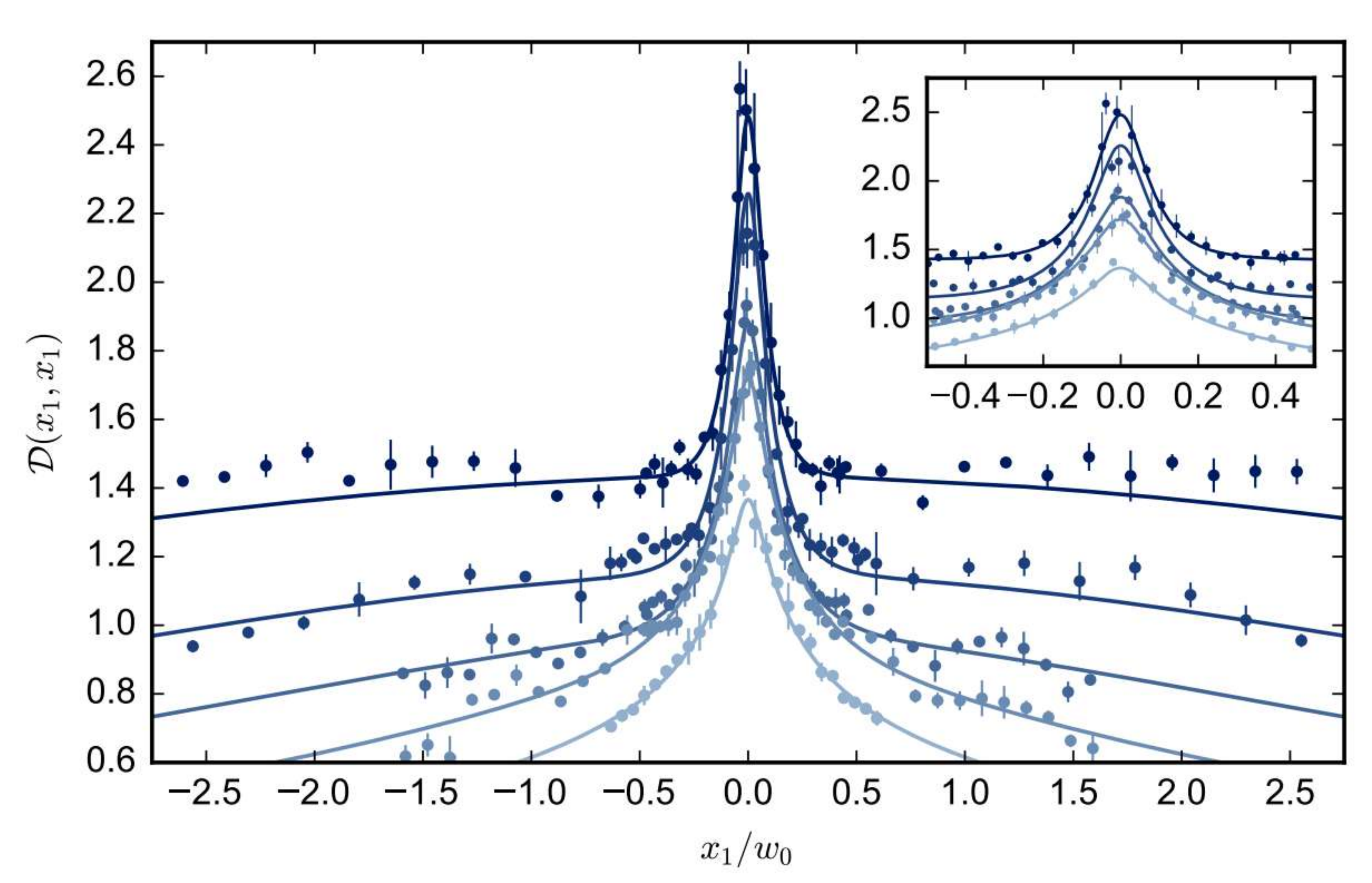}
\caption{{\bf Tunable-range cavity-mediated interaction in a multi-mode cavity.} Dimensionless interaction strength $\mathcal{D}(x_1,x_1)$ as a function of BEC position in a mode with waist $w_0$ for five different cavities, indicated by the saturation of the color. The darkest data corresponds to a confocal cavity at high degeneracy of modes, while the brighter colors correspond to fewer interacting modes. The inset shows a close-up near the cavity center, illustrating how larger number of interacting modes allows to engineer are more localized effective atomic interaction. Reproduced from\,\cite{vaidya2017tunable}.
}
\label{fig:CavityInteractions}
\end{figure}

Full degeneracy can only be reached in a multi-mode cavity that is either planar or concentric, both of which are marginally stable cavity configurations\,\cite{siegman1986lasers}. However, also the - experimentally stable - confocal cavity configuration supports a high degree of degeneracy, where either all even or all odd modes are degenerate. The resultant effective atomic interaction also features a tunable short-ranged peak. This interaction has been experimentally realized and mapped out\,\cite{vaidya2017tunable,kollar2017supermode}, and can be further employed to realize sign-changing effective atomic interactions\,\cite{guo2019signaling,guo2019emergent} Changing the range of the mediated interaction is expected to impact also the universality class of the self-ordering phase transition we describe in Section \ref{subsec:selforganization}. With increasing number of modes, the initially second-order phase transition is expected to develop into a weakly first-order phase transition\,\cite{vaidya2017tunable, gopalakrishnan2009emergent, gopalakrishnan2010atom}. 

Also thermal ensembles of cold atoms coupled to optical cavities  have proven to be a versatile platform for engineering long-range interactions. Nonlocal, tunable Heisenberg models and spin-exchange dynamics have been implemented using photon-mediated interactions in atomic ensembles, where the coupling between magnetic atomic sublevels is controlled via magnetic and optical fields \cite{norcia2018cavity,davis2019photonmediated,muniz2020exploring,davis2020protecting}. Furtheron, using multi-frequency drives in conjunction with a magnetic field gradient, interactions that are tailorable as a function of distance have been recently realized in arrays of atomic ensembles within an optical cavity \cite{periwal2021programmable} (see also \cite{hung2016quantum} for a theoretical proposal in crystal waveguides). With these tools, models that exhibits fast scrambling connecting spins separated  by distances that are powers of two, were proposed in \cite{bentsen2019treelike}, which neatly connects to 2-adic models.

\subsubsection{Mapping to spin models}\label{sec:SpinModelCavity}

One of the most fundamental models in quantum optics is the Dicke model, which describes the collective interaction between $N$ two-level atoms (captured as collective spin $\mathbf{J}$) with resonance frequency $\omega_0$ and a single electromagnetic field mode at frequency $\omega$\,\cite{dicke1954coherence,kirton2018introduction}. The Dicke model exhibits for sufficiently strong coupling $\Lambda$ between matter and light, $\Lambda>\Lambda_c=\sqrt{\omega \omega_0}/2$ a quantum phase transition to a superradiant ground state \cite{hepp1973superradiant, wang1973phase}, with a macroscopically populated field mode $\langle a \rangle$ and a macroscopic polarization $\langle J_- \rangle$ of the atoms. The observation of the Dicke phase transition employing a direct dipole transition was hindered due to the limited realizable dipole coupling strengths. However, it was theoretically proposed to make use of Raman transitions between different electronic ground states, allowing to reach the critical coupling in a rotating frame of the driven-dissipative Dicke model\,\cite{dimer2007proposed}.

\begin{figure}[]
\centering
\includegraphics[width=1\columnwidth]{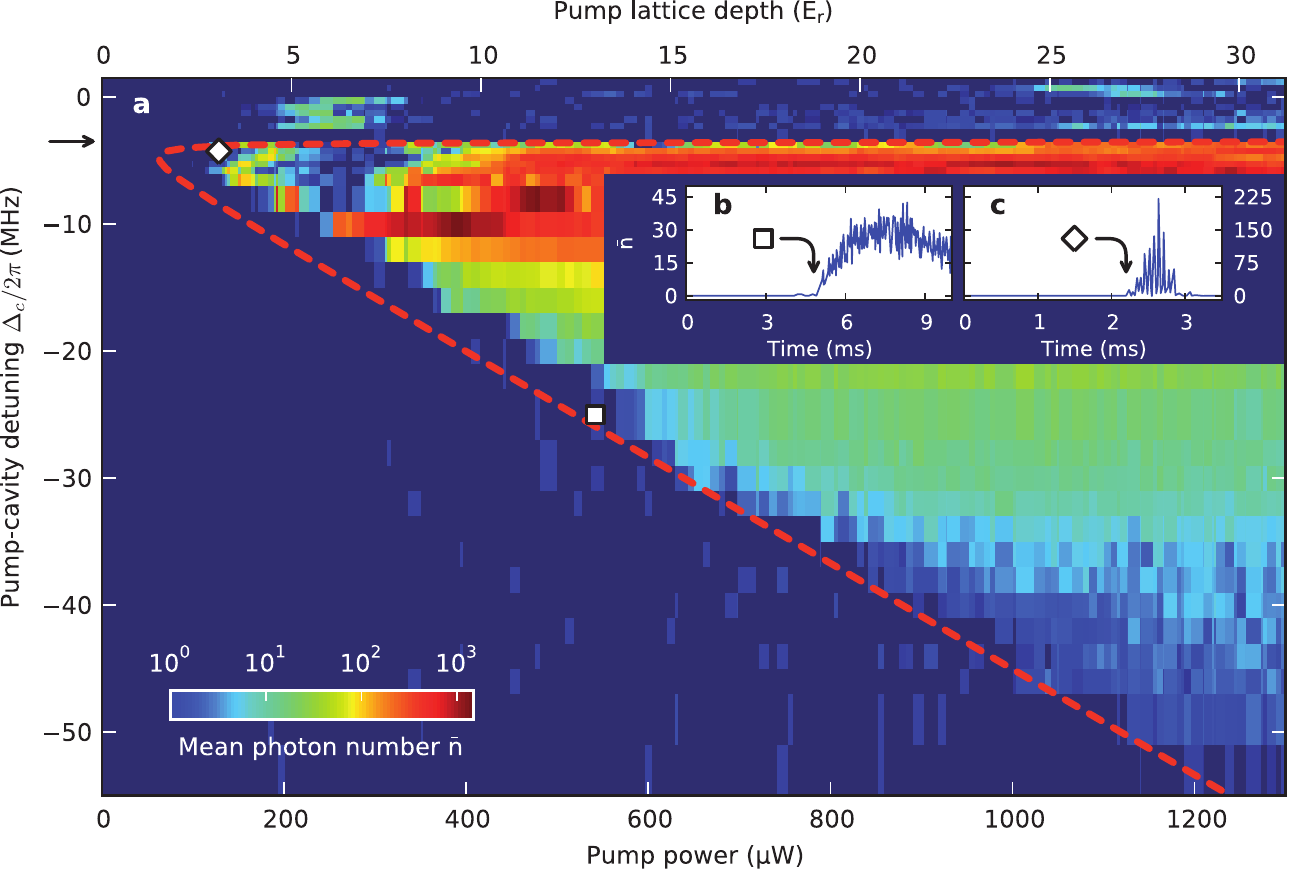}
\caption{{\bf Dicke model phase diagram.} \textbf{(a)} The power of the transverse pump is increased over 10 ms for different values of the pump-cavity detuning $\Delta_c$. The recorded mean intracavity photon number is displayed (colour scale) as a function of pump power (and corresponding pump lattice depth) and pump–cavity detuning, $\Delta_c$. A sharp phase boundary is observed over a wide range of values; this boundary is in good agreement with a theoretical mean-field model (dashed curve). The dispersively shifted cavity resonance for the non-organized atom cloud is marked by the arrow on the vertical axis.  \textbf{(b, c)} Typical traces showing the intracavity photon number for different pump–cavity detunings as indicated by the symbols. Reproduced from \cite{baumann2010dicke}.}
\label{fig:DickePhaseDiagram}
\end{figure}

Neglecting atomic collisonal interactions and the dispersive shift of the cavity, also the self-organization phase transition (see Section \ref{subsec:selforganization}) can be mapped to the superradiant quantum phase transition of the Dicke model \cite{baumann2010dicke,nagy2010dicke}. Exploiting the quantized atomic motion, the two-mode Ansatz $\Psi=\psi_0 c_0 + \psi_1 c_1 $ for the atomic wave function is inserted into the Hamiltonian Equations (\ref{eq:many_body_atom_field_Hamiltonian}). Here $c_0$ ($c_1$) are bosonic mode operators annihilating a particle in the flat BEC mode $\psi_0$, respectively in the excited motional mode $\psi_1 \propto \psi_0 \cos(kx)\cos(ky)$. Introducing the collective spin operators $J_+ = J^\dag_-=c_1^\dag c_0$ and $J_z=(c_1^\dag c_1 - c_0^\dag c_0)/2$, one arrives at the Dicke Hamiltonian
\begin{equation}
  \label{eq:DickeHamiltonian}
  H/\hbar = -\Delta_c a^\dag a + \omega_0 J_z + \frac{\Lambda}{\sqrt{N}}(a^\dag+
  a)(J_+ + J_-)\,,
\end{equation}
with bare energy $\hbar \omega_0$ of the motional excited state and coupling strength $\Lambda=\eta\sqrt{N}/2$. Compared to the original Dicke model, the mode frequency $\omega$ has been mapped to $-\Delta_c$ in the rotating frame of the pump field. The transversally pumped BEC in a cavity is the first realization of the Dicke phase transition\,\cite{baumann2010dicke}.
The phase diagram of the self-ordering phase transition is shown in
Fig. \ref{fig:DickePhaseDiagram} together with the well-matching
theoretical prediction for the open Dicke model phase transition. 

It is instructive to rewrite the long-range interaction Eq.~\ref{eq:cavity_long_range_interactions} in terms of  center-of-mass  and relative coordinates.
Focussing for simplicity on the 1D case, this results in
\begin{equation}
  \begin{split}
  \mathcal{V}_{lr}(x,x')=&\mathcal{V} \cos(kx)\cos(kx')\\
   =& \frac{\mathcal{V}}{2}\big[ \cos(2 k x_{\rm com}) + \cos(k x_{\rm rel}))\big]\,
  \end{split}
\end{equation}
with $x_{\rm com}=(x+x')/2$ and $x_{\rm rel}=x-x'$. The term $\cos[2k x_{\rm com}]$ originates from the cavity standig-wave mode structure and breaks continuous translational invariance, pinning the center of mass of the system at the phase transition onto the underlying mode structure with periodicity $\lambda/2$.
More interesting is the term $\cos[k x_{\rm rel}]$, which leads to the tendency of atoms to separate by a multiple of the wavelength $\lambda$. Due to the different periodicity of the two terms, a parity symmetry is broken at the self-ordering phase transition. The interaction term capturing the relative coordinate
allows to map this system to the Hamiltonian-Mean-Field
model\,\cite{ruffo1994hamiltonian,antoni1995clustering,dauxois2002hamiltonian,campa2014physics,schutz2014prethermalization}. This model is 
a paradigmatic model of the statistical mechanics of
non-additive long-range systems. By means of this mapping it was possible
to show that the transition to spatial self-organization is a second-order
phase transition of the same universality class as ferromagnetism, whose salient properties can be revealed by detecting the photons emitted by the cavity\,\cite{keller2017phases}. 

\subsubsection{Lattice models with cavity-mediated long-range interactions}
\begin{figure}[]
\centering
\includegraphics[width=1\columnwidth]{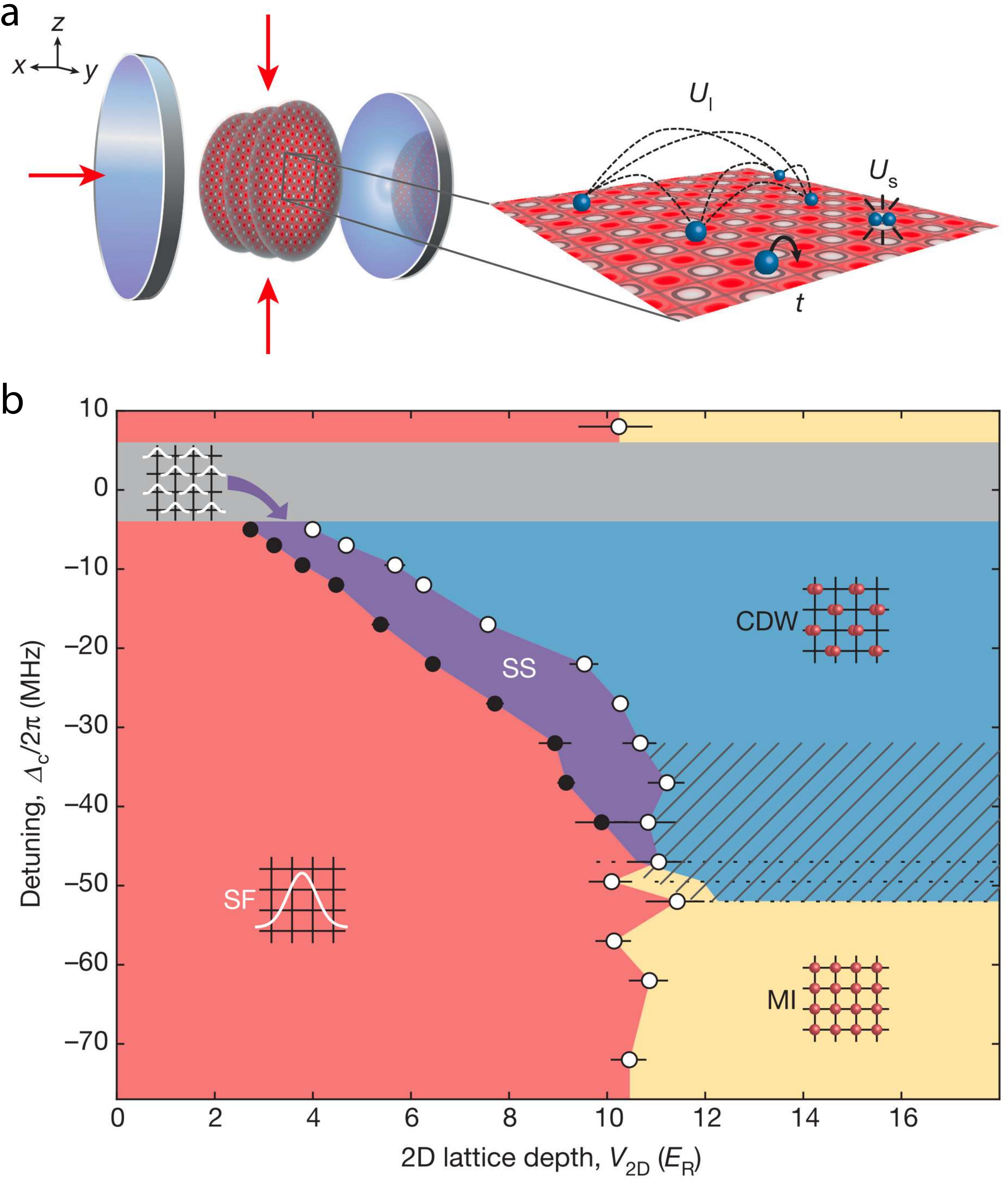}
\caption{{\bf Lattice models with cavity-mediated long-range interactions.} (\textbf{a}) Left, Experimental scheme. A stack of 2D systems along the $y$ axis is exposed to a 2D optical lattice in the $x-z$ plane (red arrows). Right, illustration of the three competing energy scales: tunneling $t$, short-range collisional interactions $U_s$ and global-range, cavity mediated interactions $U_l$. (\textbf{b}) Measured phase diagram as a function of detuning $\Delta_c$ between pump field and cavity, and 2D lattice depth $V_{2D}$, featuring superfluid (SF), lattice supersolid (SS), charge-density wave (CDW) and Mott insulating (MI) phases. Figure reproduced from~\cite{landig2015quantum}.
}
\label{fig:ExtendedHubbardCavity}
\end{figure} 
Ultracold atoms loaded into optical lattices are an unprecedented resource for the quantum simulation of condensed matter systems such as the Hubbard model~\cite{lewenstein2007ultracold,bloch2008many}. A prominent example is the experimental realization of the superfluid-to-Mott insulator quantum phase transition~\cite{greiner2002quantum}, caused by the competition of kinetic and interaction energy. However, since the dominant interaction in quantum gases is the collisional interaction, simulating models with long-range interactions poses a challenge. Adding cavity-mediated long-range interactions to this setting thus opens the path to access long-range interacting, extended Hubbard models. If this additional energy scale competes with the other two, the phase diagram will feature besides the superfluid and the Mott insulating phases also a density modulated superfluid phase -- the lattice supersolid -- and a density modulated insulating phase -- the charge density wave. Theoretical predictions discussed the resulting phases and phase diagrams in the case of commensurate and incommensurate lattices\,\cite{larson2008mott,fernandez-vidal2010quantum,habibian2013bose,li2013lattice,caballero-benitez2015quantum,bakhtiari2015nonequilibrium,chen2016quantum,dogra2016phase,lin2019superfluid,himbert2019mean}.

The system is captured in a wide parameter range by the extended Bose-Hubbard model:
\begin{equation}
  \begin{split}
  H=& -t\sum_{\langle e,o \rangle}(b^\dag_e b_o +\mathrm{h.c.}) + \frac{U_s}{2}\sum_{i\in e,o} n_i(n_i-1) \\ 
  & - \frac{U_l}{K}\left( \sum_e n_e  - \sum_o n_o \right)^2 - \sum_{i \in e,o} \mu_i n_i.
  \end{split}
\end{equation}
Here $e$ and $o$ refer to the even or odd lattice sites, $b_i$ is the bosonic annihilation operator at site $i$, $n_i=b^\dag_i b_i$ counts the number of atoms on site $i$, $K$ is the total number of lattice sites, and $\mu_i$ is the local chemical potential which depends on the external trapping potential. The first term captures the tunneling between neighboring sites at rate $t$. It supports superfluidity in the system since it favors delocalization of the atoms within each 2D layer. In contrast, the second term represents the on-site interaction with strength $U_s$, and leads to a minimzation of the energy if the atoms are localized on the individual lattice sites, favoring a balanced population of even and odd sites. The third term describes the effective global-range interactions of strength $U_l$, mediated by the cavity, and favors an imbalance between even and odd sites. The last term leads to an inhomogeneous distribution due to the trapping potential.

Self-organization in a cavity typically results in a 2D structuring of the atomic medium. If the cloud is additionally confined in a lattice along the third direction, it can be brought into an insulating, density modulated regime~\cite{klinder2015observation}.  An experimental scheme to implement a setting that in addition also features the above mentioned superfluid to Mott insulator phase transition, and thus also a transition between non-modulated and modulated insulating phases, is shown in  Fig. \ref{fig:ExtendedHubbardCavity}(a)~\cite{landig2015quantum}. A BEC is sliced into 2D systems which are subsequently exposed to a 2D optical lattice formed from one on-axis beam pumping the cavity and a standing wave lattice perpendicular to the cavity. The latter simultaneously acts as a transverse pump field inducing cavity-mediated global range interactions in the atomic system. The combined control over the lattice depth $V_{2D}$ and the detuning $\Delta_c$ allows to independently tune the ratios of collisional short-range interaction $U_s$, tunneling $t$, and global-range interaction $U_l$. The observables of this experiment are absorption images of the atomic cloud after ballistic expansion, indicating if the atomic system is insulating or superfluid, and the field leaking from the cavity, indicating a homogeneous or a density modulated system. Their combination allows to determine the phase diagram, as shown in Fig. \ref{fig:ExtendedHubbardCavity}(b), featuring the above mentioned phases.

Of special interest in the context of global-range interaction is the first-order phase transition between the non-modulated Mott insulating and the density modulated charge density wave phase. A system with only short-range interactions supports the formation of domain walls due to additivity: the reduction in energy scales with the volume of the domain, while the energy cost for the domain wall scales with its surface area. Fluctuations creating a domain will thus grow and lead to a decay of the metastable state~\cite{dauxois2002hamiltonian}. This is different in a global-range interacting system, where non-additivity makes  domain formation energetically costly: the energy of a domain wall here is proportional to the system size and not to the surface area. Accordingly, long-range interactions can stabilize metastable phases, whose lifetime then scale with system size and diverge in the thermodynmaic limit~\cite{antoni1995clustering,mukamel2005breaking,campa2009statistical,levin2014nonequilibrium}.

Quenching the system between these two insulating phases by changing the strength $U_l$ of the global-range interaction leads to hysteresis and metastability, which has been observed in the cavity field measuring the imbalance between even and odd sites~\cite{hruby2018metastability}. The quench eventually triggers a switching process that results in a rearranged atomic distribution and self-consistent potential. The timescale during which this process takes place is intrinsically determined by the many-body dynamics of the gas and is continuously monitored in the experiment. The Mott insulator, in which the system is initially prepared, forms a wedding-cake structure consisting of an insulating bulk surrounded by superfluid shells at the surface. Such an inhomogeneous finite-size system can exhibit a first-order phase transition of the bulk material (the Mott insulator), which is triggered by a second-order phase transition that took place previously on the system’s surface~\cite{lipowsky1983semi,lipowsky1987surface}, where the superfluid atoms possess a higher mobility than the insulating bulk~\cite{hung2010slow}.

\subsection{Dipolar systems and Rydberg atoms}
\label{subsec:dipolar_gases}

The study of modulated and incommensurate phases arising from the competition between short-range attractive interactions and long-range repulsive ones, has been a long standing topic in condensed matter physics\,\cite{blinc1986incommensurate,fisher1984multicritical}. Traditionally, several theoretical investigations have focused on simplified models, where the competition was limited to finite range interaction terms\,\cite{brazovskii1975phase,swift1977hydrodynamics,fisher1980infinitely}. However, natural occurrence of modulated phases is mostly due to repulsive interaction decaying as a power law of the usual form $1/r^{\alpha}$. The most relevant examples include dipolar ($\alpha=3$) and Coulomb
($\alpha=1$) interactions. 

In the framework of condensed matter experiments, dipolar interactions are known to produce modulated structures in monolayer of polar molecules\,\cite{andelman1987phase}, block co-polymers\,\cite{bates1990block}, ferrofluids\,\cite{dickstein1993science,cowley1967interfacial}, superconducting plates\,\cite{faber1958intermediate} and thin ferromagnetic films\,\cite{saratz2010experimental}. On the other hand, long-range Coulomb interactions are typical of low-dimensional electron systems, but experimental results are limited in this case. Evidences of stripe order have been found in 2D electron liquids\,\cite{borzi2007formation}, quantum Hall states\,\cite{lilly1999evidence, pan1999strongly}, doped Mott insulators\,\cite{kivelson1998electronic}. In this perspective, the appearance of stripe order is believed to be a crucial ingredient in high-temperature superconductivity\,\cite{parker2010fluctuating,tranquada1997coexistence}. 

The strong relation between traditional investigations in solid state systems and cold atomic 
platforms has clearly emerged, since the long-range nature of the forces between the atoms has 
begun to be exploited in experiments. Rydberg gases have been used to observe and study spatially
 ordered structures\,\cite{schauss2012observation,schauss2015crystallization} and correlated transport\,\cite{schempp2015correlated}. Dipolar spin-exchange interactions with lattice-confined polar molecules were as well observed\,\cite{yan2013observation}. Furthermore dipolar atoms\,\cite{lu2012quantum,park2015ultracold}
can open a new window in the physics of competing long-range and short-range interactions\,\cite{natale2019excitation}, clearing the path for the comprehension of modulated phases in strongly interacting quantum systems, as well as to higher-spin physics dynamics \cite{depaz2013nonequilibrium, lepoutre2019outofequilibrium,gabardos2020relaxation,patscheider2020controlling}.
 
In the remaining part of the Section we decided to focus on Rydberg atoms 
for their recent applications to the spin systems
with long-range and non-local interactions targeted by the present review, and 
therefore we are going to present the material needed
for the subsequent sections only in relation to Rydberg
systems.
We will not extend further the discussion on the interactions and platforms
on magnetic dipolar gases and polar molecules,
for which we refer to the
reviews\,\cite{lahaye2009physics,trefzger2011ultracold,baranov2012condensed,boettcher2020new}
for dipolar gases,
\,\cite{carr2009cold,gadway2016strongly,bohn2017cold,moses2017new} 
and recent developments 
\cite{valtolina2020dipolar,matsuda2020resonant,bause2021efficient}
for polar molecules,
even though we will anyway comment about these systems later in the text.

Highly excited Rydberg atoms display several fascinating properties
making them extremely appealing for diverse
applications in quantum information processing and quantum
simulation. The most relevant feature they show is the strong 
interaction between pairs of Rydberg atoms\,\cite{gallagher1994rydberg,saffman2010quantum,adams2019rydberg}. 

For the purposes of the present discussion, we briefly review
the main mechanisms leading to the simulation of
paradigmatic long-range spin Hamiltonians with Rydberg atoms
in the frozen-atom limit.
For two particles, denoted by $1$ and $2$, with dipole moments along the unit
vectors $\mathbf{e}_1$ and $\mathbf{e}_2$, and whose relative position is
$\mathbf{r}$, the energy due to their dipole-dipole interaction reads as
\begin{equation}
  U_{dd} = \frac{C_{dd}}{4\pi} \frac{(\mathbf{e}_1\cdot \mathbf{e}_2)r^2-
    (\mathbf{e}_1\cdot \mathbf{r})(\mathbf{e}_2\cdot \mathbf{r})}{r^5}.
\label{dip_dip}
\end{equation}
The coupling constant $C_{dd}$ is $\mu_0 \mu^2$ for particles having
a permanent magnetic dipole moment $\mu$ ($\mu_0$ 
is the permeability of vacuum) and $d^2/\epsilon_0$ for particles having a
permanent electric dipole moment $d$ ($\epsilon_0$
is the permittivity of vacuum) \cite{weber2017calculation}.
A relevant character of the dipolar interaction is its anisotropy.
In fact, the dipole-dipole interaction has the angular symmetry of
the Legendre polynomial of second order $P_2(\cos \theta)$, i.e. d-wave.

\begin{figure*}[]
\centering
\includegraphics[width=2\columnwidth]{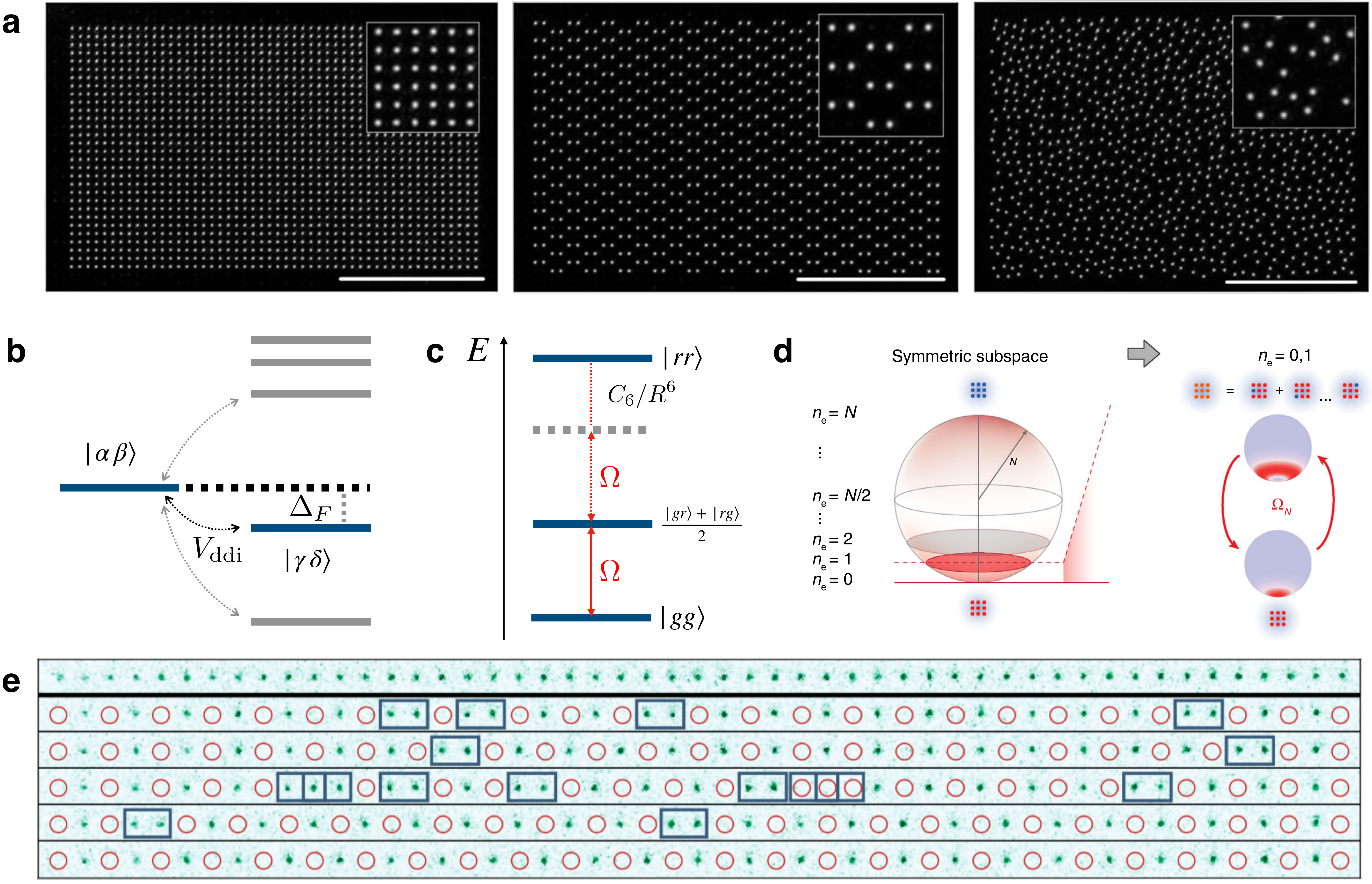}
\caption{\label{Fig11} 
  {\bf Long-range interactions in dipolar systems and Rydberg atoms
  for many-body dynamics.}
{\bf (a)} Realization of different, 
  possibly inhomogeneous, two dimensional lattices of Rydberg atoms
  with optical tweezers [courtesy of A. Omran].
{\bf (b)} Illustration of the interaction between pairs of atoms excited to Rydberg states.
Shown are the relevant dipole-coupled pair states labeled by quantum numbers 
$\alpha, \beta, \dots$ with the F\"orster defect $\Delta_F$ relative to the pair states
$\ket{\alpha \beta}$ and $\ket{\gamma \delta}$ [courtesy of A. Omran].
{\bf (c)} Principle of the Rydberg blockade. For two nearby atoms, the van der Waals interaction
$\propto C_6/R^6$ ($R$ is the interatomic distance) shifts the doubly-excited state $\ket{rr}$ preventing the double excitation of the atomic pair when 
$R<R_b=\left(C_6/\hbar\Omega \right)^{1/6}$.
{\bf (d)} Illustration of a superatom from the collective blockaded lattice of $N$ atoms.
Bloch sphere with its basis states (labeled by excitation numbers $n_e$) 
and coupled states highlighted [south pole ($n_e = 0$) and singly excited state ($n_e = 1$), 
represented by the red plane]. The small pictograms above and below the sphere depict the
lattice system with atoms in the ground (red) and Rydberg (blue) states. The dashed red line
indicates a zoom into the subspace spanned by the lowest two states. 
The Husimi distribution of these states and their enhanced coupling $\Omega_N$ 
is shown in the center.
Adapted from Ref. \cite{zeiher2015microscopic}.
{\bf (e)} The first row displays the experimental image of the initial state of a Rydberg atom
array. The following rows represent the atom array after a slow sweep across the phase
transition, showing larger average sizes of correlated domains for the slower sweep. 
Green spots (open circles) represent atoms in the ground (Rydberg) state. 
Blue rectangles mark the position of domain walls [courtesy of A. Omran].
}
\end{figure*}

Restricting to alkali atoms, denoting by
$\mathbf{d}_i$, $i=1,2$, the electric
dipole moments, when $r$ is much larger than the size of the electronic
wavefunction,
the dominant interaction term is the dipole-dipole interaction (\ref{dip_dip})
\begin{equation} 
\label{eq:dipole-dipole}
U_{dd}= \frac{1}{4\pi\epsilon_0} \frac{\mathbf{d_1} \cdot \mathbf{d_2}- 3(\mathbf{d_1\cdot n})(\mathbf{d_2\cdot n)}}{r^3},
\end{equation}
with $\mathbf{n}=\mathbf{r}/r$. Representing with $\ket{\alpha}$ and $E_\alpha$  the single eigenstates and eigenergies of
each atom one can
compute in perturbation theory the effect of the perturbation given by
Eq. (\ref{eq:dipole-dipole}).
The unperturbed eigenenergies of the two-atom states are given by
$E_{\alpha,\beta} = E_\alpha+E_\beta$, 
where for simplicity the Greek letters $\alpha$ describes the set of 
quantum numbers $(n,l,j,m_j)$.
Depending on the states involved, the relative energies and the 
dipole-dipole interaction strength, one identifies
two main regimes: the van der Waals regime and the resonant
dipole-dipole regime. 
To illustrate the main difference between the two, we assume that two atoms
that 
are in the state $\ket{\alpha \beta}$ are coupled to a single two-atom state 
$\ket{\gamma \delta}$, see Fig. \ref{Fig11}b. 
Then the reduced Hamiltonian in this two-state basis takes the form 
\begin{equation}
H_\text{red} = \left(
\begin{array}{cc}
0 & \tilde C_3/r^3 \\
\tilde C_3/r^3 & -\Delta_F \\
\end{array}
\right),
\end{equation}
where $\Delta_F=E_\gamma+E_\delta - E_\alpha-E_\beta$ 
is the F\"orster defect, $\tilde C_3$ is an effective
strength of the dipole-dipole interaction, and $r$ is the distance of the two atoms.
The eigenvalues of 
$H_\text{red}$ are then $ \Delta E = -\Delta_F/2 \pm  \sqrt{\Delta_F^2+4\left(\tilde C_3/r^3 \right)^2}$.
The van der Waals regime is recovered if $\tilde C_3/r^3\ll \Delta_F$,
then the state $\ket{\alpha \beta}$ is only weakly admixed to
$\ket{\gamma\delta}$. Its energy is perturbed to 
$\Delta E \approx \frac{1}{\Delta_F}\left(\frac{\tilde C_3}{r^3}\right)^2 \equiv 
\frac{\tilde C_6}{r^6} $. 
One obtains the scaling of the 
van der Waals coefficient with the principal quantum number
$n$ as $\tilde C_6\propto n^{11}$, as verified experimentally in a number of cases
\, \cite{beguin2013direct,weber2017calculation}.
More generally, to properly estimate the van der Waals coefficient, 
one has to formally include the contribution of all non-resonant states employing second-order perturbation theory to compute the two-atom energy shift
\begin{equation}
\Delta E_{\alpha \alpha} = \sum_{\beta, \gamma} 
\frac{|\bra{\alpha\alpha}
  U_{dd}\ket{\beta\gamma}|^2}{E_{\alpha \alpha}-E_{\beta \gamma}},
\end{equation}
where the sum extends to all the states that are dipole-coupled to 
$\ket{\alpha}$.

In the case where the $\ket{\alpha \beta}$ is resonant with $\ket{\gamma \delta}$, i.e.
$E_{\alpha \beta}\approx E_{\gamma \delta}$, or equivalently 
$\Delta_F\ll \tilde C_3/R^3$,
then the two eigenvalues of $H_\text{red}$ become $E_\pm \approx \pm \frac{C_3}{R^3}$
and the corresponding eigenstates are 
$\ket{\pm} = \frac{\ket{\alpha \beta} + \ket{\beta \alpha}}{\sqrt{2}}$. This is equivalent to a
resonant \textit{flip-flop} interaction $\ket{\alpha \beta}\bra{\gamma \delta}$ + h.c.
In this case the interaction energy scales as $1/R^3$ 
whatever the distance between the two atoms (F\"orster resonance). 
In the case of Rubidium it is easy to achieve resonance with very weak electric fields \cite{ravets2014coherent}.
The resonant dipole-dipole interaction is also naturally realised for two atoms in two dipole-coupled Rydberg states. Moreover, this interaction is anisotropic, varying as 
$V(\theta) = 1-3 \cos^2(\theta)$, with $\theta$ the angle between the internuclear axis and the quantization axis.

A central concept, essential for both many-body physics and applications, 
is the Rydberg blockade \cite{jaksch2000fast,lukin2001dipole,isenhower2010demonstration,gaetan2009observation,urban2009observation,wilk2010entanglement}, 
where the excitation of two or more atoms to a 
Rydberg state is prevented due to the interaction
\cite{browaeys2020many-body,morgado2020quantum}.
The blockade concept is illustrated in Fig. \ref{Fig11}c. 
The strong interactions between atoms excited to a Rydberg state
can be exploited to suppress the simultaneous excitation of two
atoms and to generate entangled states.
Consider a resonant laser field coherently coupling the ground state $\ket{g}$
and a given Rydberg state $\ket{e}$, with a Rabi frequency $\Omega$. 
In the case of two atoms separated by a distance $r$, the doubly excited state
$\left|ee\right>$
is shifted in energy by the quantity $C_6/r^6$ due to the van
der Waals 
interaction with $C_6$ being the interaction coefficient 
(all the other pair  states have an energy nearly independent of $r$). 
Assuming that the condition $\hbar\Omega\ll C_6/r^6$
is fulfilled, that is, $r\ll R_b=(C_6/\hbar\, \Omega)^{1/6}$ (blockade radius). 
Then, starting from the ground state $\ket{gg}$, 
the system performs collective Rabi oscillations
with the state $\ket{\psi} = \frac{\ket{eg}+\ket{ge}}{\sqrt{2}}$.
The above considerations can be extended to an ensemble
of $N$ atoms all included within a blockade volume. 
In this case, at most one Rydberg excitation is possible, leading to
collective Rabi oscillations with an enhanced frequency
$\Omega_\text{coll} = \sqrt{N}\Omega$, leading to the so-called superatom picture
illustrated in Fig. \ref{Fig11} (d). 
The system dynamics is confined to the symmetric subspace of zero ($n_e=0$) and one ($n_e=1$) excitations, whose basis are the Fock states $\ket{0}=\ket{g_1,\dots,g_N}$ and the entangled $W$-state
$\ket{1}=\frac{1}{\sqrt{N}}\sum_{i=1}^{N}\ket{g_1,\dots,r_i,\dots,g_N}$,
where $g_i$ and $r_i$ label the i-th atom in the ground or Rydberg state \cite{zeiher2015microscopic}.

An important objective is to implement interacting many-body systems
combining atomic motion with tunable long-range interaction via Rydberg atoms.
The main experimental challenge is to bridge the mismatch in energy and
timescales between the Rydberg excitation and the dynamics of ground state
atoms. A possible solution is the so-called Rydberg dressing where ground
state atoms are coupled off-resonantly to Rydberg states leading to
effectively weaker interaction with lower decay rates\,
\cite{jau2016entangling,pupillo2010strongly,johnson2010interactions,balewski2014rydberg,henkel2010three,macri2014rydberg}.
The main difficulty in this
approach is that decay and loss processes of Rydberg atoms have to be
controlled on these timescales that are much longer than for near-resonant
experiments such that also more exotic loss processes become relevant
\, \cite{zeiher2016many, zeiher2017coherent,guardado2021quench}.
One of the exotic states that might be realizable using Rydberg dressing
is a supersolid droplet crystal,
see Fig. \ref{Fig13} for details. Rydberg dressing also
allows to impose local constraints which are at the heart of the
implementation of models related to gauge theories, like the quantum spin ice\, \cite{glaetzle2014quantum}.
Other predictions include cluster Luttinger liquids in 1D
and glassy phases, see sec.\ref{sec:regimes}.
It might be even possible to implement a universal quantum simulator or
quantum annealer based on Rydberg dressing\, \cite{lechner2015quantum,glaetzle2017coherent}.

\begin{figure}[]
\centering
\includegraphics[width=1\columnwidth]{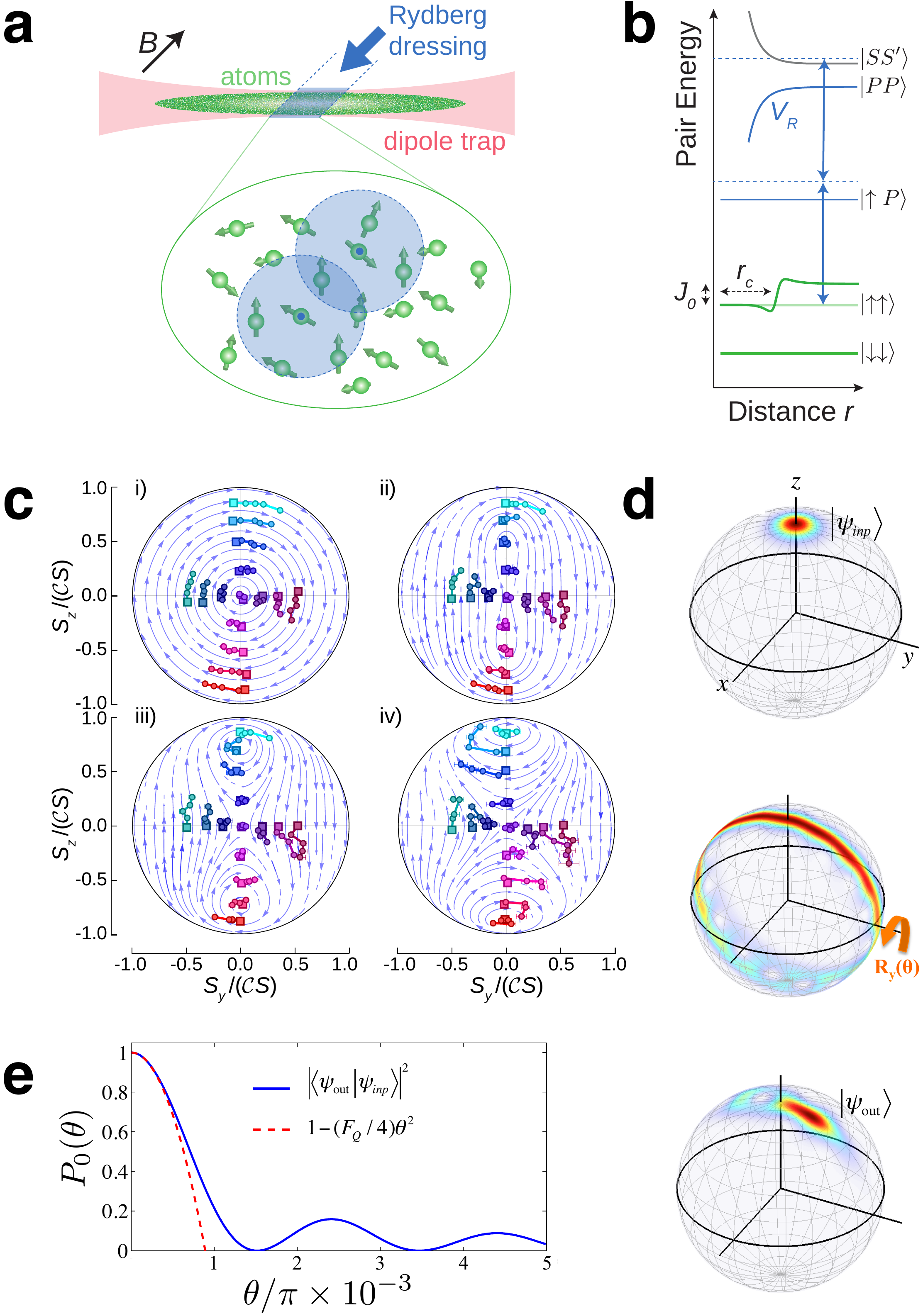}
\caption{\label{Fig12} 
{\bf Finite-range interactions in spin systems: dynamics and applications.}
{\bf (a)} Experimental setup and Rydberg dressing scheme for a 
cloud of cesium atoms is held in an optical dipole trap and locally illuminated 
with $319$ nm light to generate Ising interactions of characteristic range $r_c$ 
and strength $J_0$. The quantization axis is set by a $1$ G magnetic field $B$. 
{\bf (b)} Energy level diagrams for a pair of atoms.
{\bf (c)} Transverse-field Ising dynamics. Trajectories $S(k)$ for initial states $\ket{\theta,\phi}$
(square data points) and up to $k = 4$ Floquet cycles,
obtained with dressing parameters 
$(\Omega, \Delta) = 2\pi \times (2.8, 25)$ MHz. 
Plots (i-iv) are for $\Lambda_\text{eff} = 0, 1.2(2), 1.8(3), 2.7(4)$. 
Blue flow lines show mean-field theory for best fit $\Lambda = 0, 1.1, 1.5, 2.2$ (see main text). 
Figures (a)-(c) adapted from Ref. \cite{borish2020transverse}.
{\bf (d)} Loschmidt echo protocol applied to the Lipkin-Meshkov-Glick
(one-axis twisting) model.
Snapshot of the Husimi distribution.
(Top panel) A spin-polarized state is prepared at north pole of the Bloch sphere. (Central panel) Interaction is switched on for a time $t_1$ [transformation $U_1$].  
The state is then rotated of an angle $\theta$ [$R_y(\theta)$]. 
(Bottom panel) Interaction is switched on again for a time $t_2$ [transformation $U_2$] such that $U_1 U_2 = 1$.  
In these plots $\theta/\pi=0.01$ and $\tau/\pi=0.05$.
{\bf (e)} Probability $P_0(\theta)$ (solid line) as a function of phase shift.
as a function of $\theta$ for $\tau/\pi=0.05$.
The dashed line is the second-order expansion involving the 
quantum Fisher information $F_Q$. Here $N=101$.
Figures (d) and (e) adapted from Ref. \cite{macri2016loschmidt}.
}
\end{figure}

\subsubsection{Mapping to spin models}

The two-atom picture described in the previous section can be extended to
the many-body case. Including the coupling of single-atom states
to an external coherent laser drive, one obtains in the rotating frame of the laser the Ising Hamiltonian
\cite{schauss2012observation,schauss2015crystallization,labuhn2016tunable}
\begin{equation}
H_\text{Ising} = \frac{\hbar \Omega}{2} 
\sum_i \sigma_x^i
-\sum_i  \hbar\Delta n_i 
+\sum_{i<j} \frac{C_6}{r_{ij}^6} n_i n_j,
\end{equation}
where $n_i = \ket{e}_i\bra{e}=(1+\sigma_z^i)/2$ 
is the projector to the excited state $\ket{e}$, and $\Delta$ is the single-atom detuning from the Rydberg state $\ket{e}$. A discussion with references on the
simulation of quantum Ising models in a transverse field is in\,\cite{schauss2018quantum,morgado2020quantum}.

A relevant technical improvement to study the Ising model has been provided by the trapping and manipulation of Rydberg atoms in optical tweezers with defect-free configurations
\, \cite{barredo2016atom,endres2016atom,demello2016defect,covey2019times,wang2020preparation,festa2021motion,schymik2021single,anderegg2019optical,bohrdt2020multiparticle}.
Many interesting effects have recently been investigated, from the 
Kibble-Zurek mechanism 
and its related critical dynamics \, \cite{keesling2019quantum}, see Fig. \ref{Fig11}e,
to the realization of antiferromagnetic phases\, \cite{lienhard2018observing,guardado2018probing,scholl2020programmable}, and quantum spin liquids\, \cite{samajdare2021quantum,verresen2021prediction,semeghini2021probing}.

In addition to direct Rydberg excitation, 
Rydberg dressing provides an alternative way to implement quantum Ising models
with important implications beyond quantum simulation. 
In the dressing protocol, two internal ground states are used to encode spin-up and spin-down states. 
Coherent many-body dynamics of Ising quantum magnets built up by Rydberg dressing are experimentally studied both in an optical lattice
and in an atomic ensemble.
An illustration of the Ising dynamics in a finite-range model is
presented in Fig. \ref{Fig12}, where we show the trajectories of the collective spin $S(k)$ from \cite{borish2020transverse}. An important application of this Hamiltonian is for the study of Loschmidt echo protocol applied to the Lipkin-Meshkov-Glick (one-axis twisting) model for quantum metrology purposes\, \cite{gil2014spin}, e.g. for the preparation of non-gaussian states that can be detected via the quantum Fisher information
\, \cite{borish2020transverse,macri2016loschmidt}. Rydberg dressing of atoms in optical tweezers can also be employed for the realization of programmable quantum sensors based on variational quantum algorithms,  capable of producing entangled states on demand for precision metrology\cite{kaubruegger2019variational}.
This investigation is not limited to Rydberg atoms, but extends naturally
also to ion platforms \cite{davis2016approaching, morong2021observation}. 

A special case of the quantum Ising model arises when 
$a_{\text{latt}} < R_b < 2 a_{\text{latt}}$ with $a_{\text{latt}}$
the lattice
spacing (nearest-neighbor
blockade) and $V_{ij}\approx 0$ for everything beyond nearest
neighbors. 
Such a situation was experimentally realized 
in a 1D chain of Rydberg atoms in\,\cite{bernien2017probing,bluvstein2021controlling}.
In this case one can derive an effective Hamiltonian for
the low-energy subspace which amounts to neglecting
configurations with two adjacent excitations. In 1D the
resulting Hamiltonian takes the form of a PXP model
\begin{equation}
H = \sum_i \frac{\Omega_i}{2} P_{i-1} \sigma_x^i P_{i},
\end{equation}
where $P_i=\ket{g}\bra{g}$ is the projector onto the ground state.

Resonant dipole-dipole interactions between Rydberg atoms
are at the basis of several proposals to simulate the quantum dynamics
of many-body spin systems. As a major example,
it is possible to see that a system containing two dipole-coupled Rydberg states
can be mapped to a spin-$1/2$ XY model, see the
review\,\cite{wu2021coincise} and references therein.
Coherent excitation transfer between two types of Rydberg states of different atoms
has been observed in
a three-atom system\,\cite{barredo2015coherent}.
The resulting long-range XY interactions give rise to
many-body relaxation\,\cite{orioli2018relaxation}.

Given the well known mapping between the XY model and hard-core
bosons\,\cite{friedberg1993equivalence}, it is possible to
provide an experimental realization of the
bosonic Su-Schrieffer-Heeger model\,\cite{su1979solitons} and its
symmetry protected topological order
with a single-particle edge state\,\cite{deleseluc2019obervation,lienard2020realization}, see also \cite{kanungo2021realizing}.
Proposals to observe topological bands\,\cite{peter2015topological}
and topologically protected edge states\,\cite{weber2018topologically}
were presented. 
Moreover, a realization of a density-dependent Peierls phase in a
spin-orbit coupled Rydberg system has been recently
demostrated\,\cite{lienard2020realization}.

 We finally mention
that with Rydberg systems one could implement
digital simulation techniques \cite{georgescu2014quantum}.
The total unitary evolution operator $U(t)$ is decomposed
in discrete unitary
gates\,\cite{weimer2010rydberg,weimer2011digital} and one can
study a braod class of dynamical regimes of spin systems, such as
nonequilibrium phase transitions and non-unitary
conditional interactions in quantum cellular automata
\,\cite{lesanovsky2019nonequilibrium,gillman2020nonequilibrium,wintermantel2020unitary}. Kinetically constrained Rydberg spin systems, in which
a chain of several traps each loaded with a single Rydberg atom and coupled
with the bosonic operators expressing the deviation from the trap centers,
also referred to as facilitated Ryberg lattices,
were as well studied\,\cite{mazza2020facilitated}.

A further promising line of research is provided by Rydberg ions both for quantum 
simulation purposes\, \cite{muller2008trapped,gambetta2020longrange} as well as for the realization of fast quantum gates for quantum
information processing \, \cite{muller2008trapped,mokhberi2020trapped}.
Two-dimensional ion crystals for quantum simulation of spin-spin
interactions using interactions of Rydberg excited ions have been recently proposed in
\, \cite{nath2015hexagonal} to emulate topological quantum spin liquids using the spin-spin interactions between ions in hexagonal plaquettes in a 2D ion crystal. 
The role of a Rydberg ion is to modify the phonon mode spectrum such that
constrained dynamics required for realizing the specific Hamiltonian of the
Balents-Fisher-Girvin model using a Kagome lattice.
There, the effective spin-spin interaction for the hexagonal plaquette can be written as an extended XXZ model
\begin{equation}
H_{SS}=\sum_{i<j} J_{ij}^z S_{i}^z S_{j}^z + \sum_{i<j} J_{ij}^\perp (S_{i}^x S_{j}^x+ S_{i}^y S_{j}^y).
\end{equation}
Long-range XXZ Hamiltonians with tunable anisotropies can be Floquet-engineered 
using resonant dipole-dipole interaction between Rydberg atoms and a periodic external microwave
field coupling the internal spin states \cite{geier2021floquet, scholl2021microwaveengineering}.

We finally comment that in a realistic Rydberg atom system, coherent driving offered by external fields often competes with dissipation induced by coupling with the environment. Such a controllable driven-dissipative system with strong and nonlocal Rydberg-Rydberg interactions can be used to simulate many-body phenomena distinct from their fully coherent counterparts, e.g., dynamical phase transitions that are far from equilibrium.
Evolution of such an open many-body system is often
governed by the master equation 
$\partial_t \rho = -i [H,\rho ] + L \rho$, 
where $\rho$ is the state of the system, $H$ the system Hamiltonian 
and $L$ is the Liouvillian superoperator\,\cite{gardiner2004noise,benatti2005open,manzano2020short}. Correspondingy, several
aspects of driven-dissipative dynamics in Rydberg
systems and dissipative Rydberg media were addressed
\,\cite{lesanovsky2013kinetic,lee2015emergence,levi2016quantum,goldschmidt2016anomalous,letscher2017bistability,lee2019coherent,torlai2019integrating,bienais2020photon,pistorius2020quantum}.

\section{Thermal critical behaviour}
\label{sec:classical}

The critical properties of quantum long-range models at $T=0$ are related
to the corresponding critical features of long-range
systems at finite temperature in a way which
is different from the usual paradigm valid for short-range
systems\,\cite{sachdev1999quantum}. In the latter, the critical behaviour
of a model in dimension $d$ at $T=0$ is put in
correspondence with the critical
behaviour at a finite temperature $T$
but in a dimension $d+1$\,\cite{sondhi1997continuous,sachdev1999quantum},
a typical example being the
short-range quantum Ising model in a
transverse field (at $T=0$) and the short-range classical Ising model at finite temperature\,\cite{mussardo2009statistical}. The situation changes in the long-range regime,
and for this reason we are going to review in this section
the basics properties of equilibrium
critical long-range systems at finite temperature, and compare them
in Sec. \ref{sec:quantum} with the corresponding properties at zero temperature.

Phase transitions are among the most remarkable phenomena
occurring in many-body systems. 
Among various kinds of phase transitions, continuous phase transitions are
particularly fascinating since they are tightly bound with the concept of
universality.
Thanks to the universality phenomenon the same
formalism can be applied both to phase transitions occurring at a finite
temperature and at $T=0$.
The latter are usually denoted as quantum phase
transitions\,\cite{sachdev1999quantum}. Nowadays the intense efforts of the
scientific community have paid their rewards and the critical properties of
several physical systems have been characterised\,\cite{pelissetto2002critical}.

Usually, universality is defined as the insensitivity of the
critical scaling behaviour of thermodynamic functions with
respect to variations of certain microscopic details of the system under
study, such as the lattice configurations or the
precise shape of the couplings.
This definition alone cannot be considered rigorous unless one
specifies all the possible adjustments of the microscopic features,
which preserve universality. In the following, we will reserve the adjective
"universal" to all those phenomena which may be quantitatively
described by a suitable continuous 
formulation. Therefore, in our language, the concept of universality
is strictly tied to the existence of a continuous field theory formulation,
which, albeit ignoring the microscopic details of the lattice description, is
able to produce exact estimate for the universal quantities. 

It is convenient to discuss this definition directly on the
traditional problem of classical $O(N)$ spin systems, whose Hamiltonian reads
\begin{equation}
\label{h_o(n)}
H=-\frac{1}{2}\sum_{i \neq j} J_{ij}\mathbf{S}_{i}\cdot\mathbf{S}_{j}\,.
\end{equation}
where $\mathbf{S}_{i}$ is a $N$-component spin vector with unit modulus,
$J_{ij}>0$ are ferromagnetic translational invariant couplings and the indices
$i,j$ run over all sites on any $d$-dimensional regular lattice of $V$ sites.
The usual terminology is that $N=1$ is the Ising model, $N=2$ the XY model,
$N=3$ the Heisenberg model and $N\to\infty$
is the spherical model\,\cite{stanley1968spherical}.  
It is well known\,\cite{mussardo2009statistical,nishimori2015elements}
that for $N \ge 1$ and $d>2$ the Hamiltonian in Eq.\,\eqref{h_o(n)}
and fast enough decaying couplings (i.e., in the short-range limit)
presents a finite temperature phase
transition between a low temperature state $T<T_{c}$ with finite magnetisation
$m=|\langle\sum_{i}\mathbf{S}_{i}\rangle|/N\neq 0$ and an high temperature
phase with $m=0$. For $N=1$, the phase transition
occurs of course also for $d=2$\,\cite{mussardo2009statistical,nishimori2015elements}.

Close to the critical point the thermodynamic quantities display power law
behaviour as a function of the reduced temperature $\tau \equiv (T-T_{c})/T_{c}$,
with universal critical exponents which only depend on the symmetry index
$N$ and the dimension $d$ of the system. These critical exponents are known
to coincide with the ones of the $O(N)$-symmetric field theory with action
\begin{align}
\label{s_o(n)}
S[\varphi]=\int d^{d}x\left\{\partial_{\nu}\varphi_{i}\partial_{\nu}\varphi_{i} + \mu|\boldsymbol{\varphi}|^{2}+g|\boldsymbol{\varphi}|^{4}\right\}
\end{align}
where $\boldsymbol{\varphi}$ is an $N$-component
vector with unconstrained modulus, the lattice summation has been replaced by
a real space integration, $\nu=1,\cdots,d$
runs over the spatial dimensions, $i=1,\cdots,N$ refers to the
different components, 
the quadratic coupling controls the distance from the critical point
($\mu\propto\tau$), the value of the constant coupling is 
$g>0$ and the summation over repeated indexes is intended.

An extensive amount of theoretical investigations has been performed
on the critical properties of $O(N)$ symmetric models, both
in their continuous and lattice formulation, reaching an unmatched accuracy in the
determination of universal properties with a fair degree of consistency in
the whole dimension range $2\leq d \leq 4$\,\cite{holovatch1992critical,kleinert2001critical,pelissetto2002critical,codello2015critical,cappelli2019critical}.
Numerical simulations, which are limited to integer
dimensional cases $d\in\mathbb{N}$, are mostly consistent with theoretical
investigations\,\cite{pelissetto2002critical}, while the
recently emerged conformal bootstrap results confirmed and extended
the existing picture\,\cite{poland2019conformal}.

The action \eqref{s_o(n)} is the one reproducing
the behaviour of the mean-field propagator $G_{\mathrm{mf}}$ for the spin Hamiltonian
\eqref{h_o(n)} in the zero-momentum limit
$G_{\mathrm{mf}}^{-1}\approx \frac{\delta^{2} S}{\delta\varphi^{2}_{q}}\propto q^{2}+\mu$.
Within this framework, it clearly appears that any modification of the spin
Hamiltonian \eqref{h_o(n)} which does not alter the large scale mean-field
propagator should not modify the universal properties.  

\subsection{The weak long-range regime}
\label{wlr_classical}

Having introduced the formalism and notation for universality problems,
we can start with the case of interest of long-range $O(N)$ spin systems:
\begin{equation}
\label{NvectorSystemHamiltonian}
H=-\frac{1}{2}\sum_{i \neq j} 
J_{ij}\, \mathbf{S}_{i}\cdot\mathbf{S}_{j}\,,
\end{equation}
with $J_{ij} =  J/r_{ij}^{d+\sigma}$, where $r_{ij}$ is the distance
between sites $i$ and $j$, a coupling constant $J>0$, and a positive decay exponent $d+\sigma \geq 0$.
The Fourier transform of the matrix $J_{ij}$ produces a long-wavelength
mean-field propagator of the form
$G_{\mathrm{mf}}\sim J_{\sigma} q^{\sigma} +J_{2} q^{2}$,
setting the mean-field threshold for the relevance of long-range
interactions to $\sigma_{*}^{\mathrm{mf}}=2$\,\cite{fisher1972critical}.

The renormalisation group (RG) approach\,\cite{wegner1973renormalization,polchinski1984renormalization} delivers a comprehensive picture for the universal
properties of long-range $O(N)$.
In the so-called functional RG (FRG) one writes an -- in principle -- 
exact equation for the flow of the effective average action, $\Gamma_{k}$,
of the
model and then resort to various approximation schemes
\,\cite{wetterich1993exact,berges2002non,delamotte2011introduction}.
The $\Gamma_{k}$ is obtained by the introduction of a momentum space
regulator $R_{k}(q)$, which cutoffs the infra-red divergences caused
by slow modes $q\ll k$, while leaving the high momentum model $q\gg k$
almost untouched.
The problem of weak long-range interactions in the continuous space could be
then represented by the scale dependent action 
\begin{equation}
\label{lr_lpa_eaa}
\Gamma_{k}[\varphi]=\int d^{d}x \left\{Z_{k}\partial^{\frac{\sigma}{2}}_{\nu}
\varphi_{i}\partial^{\frac{\sigma}{2}}_{\nu}\varphi_{i}+U_{k}(\rho)\right\}\,,
\end{equation}
where $\rho=\frac{1}{2}\varphi_{i}\varphi_{i}$ and the index $i=1,\cdots,N$ being
summed over as in the previous section.

The ansatz in Eq.\,\eqref{lr_lpa_eaa} is already sufficient to
qualitatively clarify
the influence of long-range interactions on the universal properties. Indeed,
the difference between the bare action \eqref{s_o(n)} and
the effective action \eqref{lr_lpa_eaa} is limited to
the presence of the fractional derivative
$\partial^{\frac{\sigma}{2}}_{\mu}$ into the kinetic term instead of
the traditional $\nabla^{2}$ term. The definition
of the fractional derivative
in the infinite volume limit\,\cite{pozrikidis2016book,kawasnick2017ten}
leads to the
straigthforward result that
its Fourier transform yields a fractional momentum term
$q^{\sigma}$.
The renormalization of such anomalous kinetic term $q^{\sigma}$ is parametrised
in Eq.\,\eqref{lr_lpa_eaa} by a running wave-function renormalization $Z_{k}$
as it is customary done in the short-range case\,\cite{dupuis2020non}.

The actual subtlety 
of the weak long-range universality resides in the competition
between the analytic momentum term $q^{2}$ and the anomalous one $q^{\sigma}$ 
arising due to long-range interaction. Such effect cannot be properly
reproduced by the ansatz in Eq.\,\eqref{lr_lpa_eaa}, which only includes
the most relevant momentum term at the canonical level in the low energy behaviour of
long-range $O(N)$ models. Yet, Eq.\,\eqref{lr_lpa_eaa}
reveals to be a useful approximation to recover and
extend the mean-field description of the problem at least in the limit
$\sigma\ll 2$, where the non-analytic momentum term is certainly the
leading one.

Close to the transition, the correlation length of the system, which controls
the spatial extent of the correlations,
$\langle \varphi(x)\varphi(0)\rangle\approx \exp(|x|/\xi)/x^{d-2}$,
diverges as $\xi\propto \tau^{-\nu}$. Thus, the diverging critical
fluctuations produce an anomalous scaling of the correlation functions
via the presence of a finite anomalous dimension $\eta$. 
The standard definition
used for short-range models\,\cite{nishimori2015elements} is 
\begin{equation}
  \label{definition_eta}
  \langle \varphi(x)\varphi(0)\rangle\approx \frac{1}{|x|^{d-2+\eta}}
  .
\end{equation}
Conventionally, we refer
to a correlated universality when $\eta\neq 0$ and 
anomalous scaling
appears. If one refers to the definition
\eqref{definition_eta} of the decay of correlation functions in
short-range systems, then the anomalous dimension of long-range model
is already finite at mean-field level giving
$\eta_{\mathrm{lr}}=2-\sigma$, 
due to the contributions of the power-law couplings
to the scaling of the correlations (here and in the following
the indices $_{\mathrm{lr}}$ and  $_{\mathrm{sr}}$ stand for long-
and short-range, respectively) . However, to have a proper account of
correlation effects, it is convenient to re-define the anomalous dimension
$\eta_{\mathrm{lr}}$ of the long-range $O(N)$ models as follows
\begin{equation}
\label{eta_lr_def}
\eta_{\mathrm{lr}}(d,\sigma) \equiv 2-\sigma + \delta\eta\,,
\end{equation}
with respect to the canonical dimension of the long-range terms,
in agreement with the definition in the
classic paper\,\cite{fisher1972critical}.

Therefore the low-momentum scaling of the critical propagator
shall become $G(q)^{-1}\approx q^{\sigma-\delta\eta}$. Within the RG formalism
such correction $\delta\eta$ is expected to appear as a divergence of the
wave-function renormalization, which signals the rise of a modified scaling.
Yet, the $\beta$-function of the wave-function renormalization for the
fractional momentum term identically vanishes
($k \, \partial_{k}Z_{k}=0$)
for any $d$ and $\sigma$, at least in the approximation parameterised by
Eq.\,\eqref{lr_lpa_eaa}. Therefore, the correlated correction for long-range interactions vanishes $$\delta\eta=0,$$
a result first obtained in the paper\,\cite{sak1973recursion}
by J. Sak in 1973.
The flow of the effective potential remains the only
non-trivial RG evolution
for the ansatz in Eq.\,\eqref{lr_lpa_eaa}.

Similarly to the wave-function flow, the RG evolution of
the effective potential $U_{k}(\rho)$ has been obtained following the
traditional derivative expansion approach of the
FRG\,\cite{delamotte2011introduction} by introducing a suitable regulator
function $R_{k}(q)= Z_k (k^{\sigma}-q^{\sigma})\theta(k^{\sigma}-q^{\sigma})$.
The resulting $\beta$-function for the effective potential reads
\begin{equation}
\begin{split}
&\partial_t \bar{U}_{k}= -d\bar{U}_{k}(\bar{\rho})+(d-\sigma)\bar{\rho}\,\bar{U}'_{k}(\bar{\rho})+\frac{\sigma}{2} c_d (N-1)\frac{1}{1+\bar{U}'_{k}(\bar{\rho})}\\
&+\frac{\sigma}{2} c_d \frac{1}{1+\bar{U}'_{k}(\bar{\rho})+2\bar{\rho}\,\bar{U}''_{k}(\bar{\rho})}\,,
\label{u_flow_lpa}
\end{split}
\end{equation}
with $c_d^{-1}=(4\pi)^{d/2} \Gamma\left(d/2+1 \right)$ and
as usual in RG calculations we set
$t=\log(k/\Lambda_{uv})$ as the RG time,
with $\Lambda_{uv}$ the ultra-violet scale, typically
$\sim 1/a_{\rm{latt}}$.
In Eq.\,\eqref{u_flow_lpa} rescaled units are
as well used:
$\bar{\rho}=Z_k k^{\sigma-d} \rho$ 
and $U_k(\bar{\rho})=k^{-d} U_k(\rho)$.

We are now in position to discuss the emergence
of an effective fractional
dimension, the determination of the value of $\sigma_*$
and the interplay below $\sigma=2$ of short- and long-range
contributions. Then, the case of anisotropic long-range
interactions, important
for quantum long-range systems, will be discussed.

\subsubsection{Effective fractional dimension}
\label{eff_frac_dim}
Interestingly, the $\beta$-function in Eq.\,\eqref{u_flow_lpa} can be
exactly related to the one of the effective potential of a short-range $O(N)$
model in an \textit{effective} fractional dimension\,\cite{defenu2015fixed} 
\begin{equation}
\label{d_eff}
d_{{\rm eff}}=\frac{(2-\eta_{\mathrm{sr}})d}{\sigma},
\end{equation}
in agreement with the result of scaling arguments\,\cite{angelini2014relations}, see as well\,\cite{brezin2014crossover}.
This is an example of dimensional equivalence, as it could occur in disordered
systems, an example being the dimensional reduction in the
random field Ising\,\cite{parisi1979random,fytas2019evidence}.

Then, for each point $(d,\sigma)$ in the parameter space of long-range $O(N)$
models, the universal scaling exponents for thermodynamic functions are
effectively (see below) the same as in a short-range model in
dimension $d_{\mathrm{eff}}$. Conversely, the scaling exponent of the correlation
length obeys the relation
\begin{equation}
\label{exp_rel}
\nu=\frac{2-\eta_{\mathrm{sr}}}{\sigma}\nu_{\mathrm{sr}}\,.
\end{equation}
It is important noting that the effective dimension relation in Eq.\,\eqref{exp_rel} is obeyed exactly in the Gaussian $\sigma<d/2$\, \cite{aizenman1988critical}
(corresponding to $d_{\mathrm{eff}}>4$) and in the spherical model ($N\to\infty$) limits\,\cite{joyce1966spherical}. Moreover, the effective dimension approach implies that, at a fixed value of the dimension $d$, a sequence of lower critical decay exponents $\sigma_{c,i}$ appears, below which the model presents multi-critical universality, in analogy with the traditional short-range case\,\cite{codello2012scaling}.

\begin{figure}
\centering
\includegraphics[width=.45\textwidth]{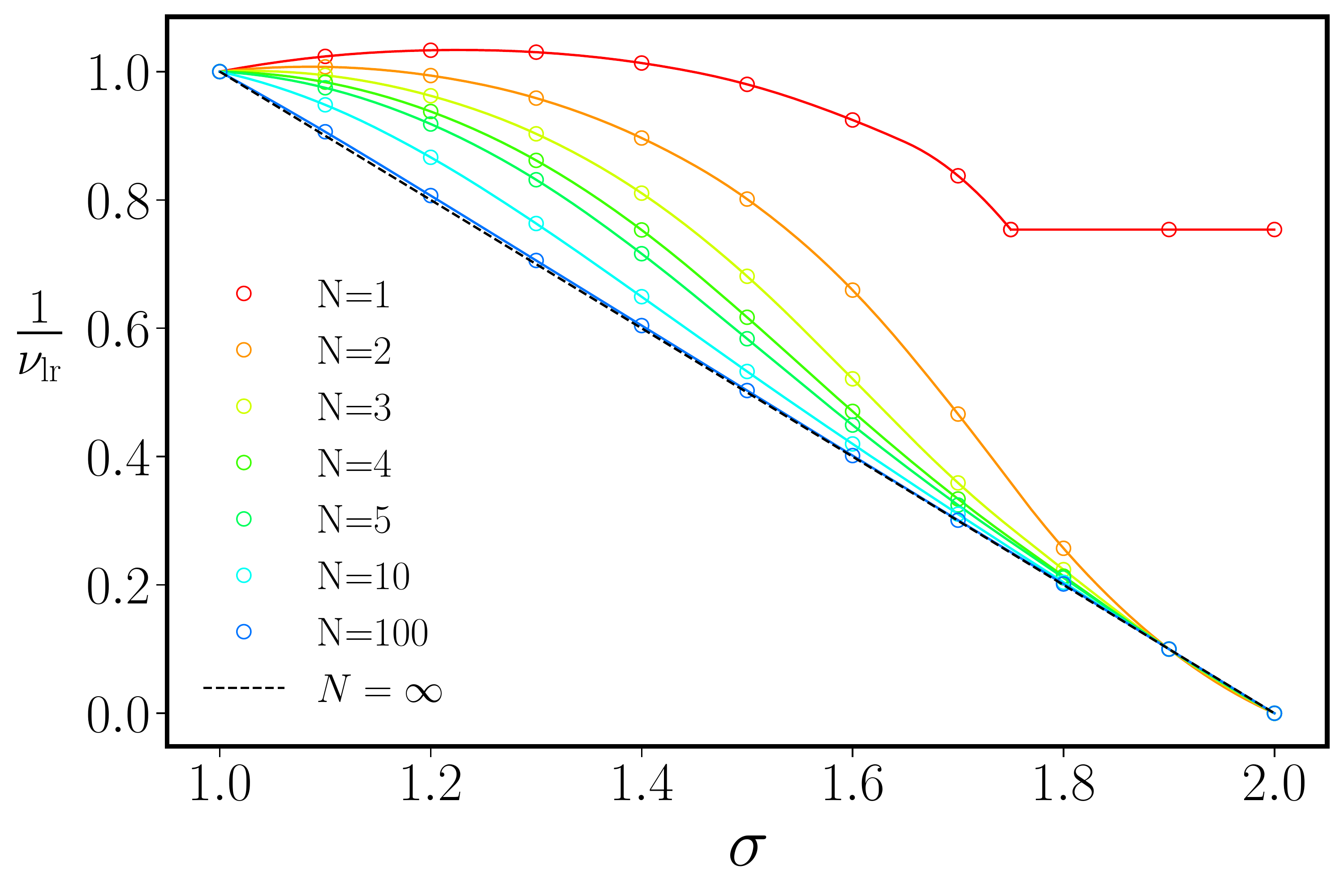}
\caption{\textbf{Classical correlation length exponent}.
Correlation length exponent
$1/\nu_{\rm lr}$ as a function of $\sigma$ in $d=2$ for several values of $N$ (from top: $N=1,2,3,4,5,10,100$). The discrepancy between the $N=1$ and the $N\geq2$ cases is in agreement with the expectations of the Mermin-Wangen theorem.
The black dashed line is the analytical result obtained for the spherical model $N=\infty$\,\cite{joyce1966spherical}.
}
\label{Fig3}
\end{figure}

The expression in Eq.\,\eqref{d_eff} is implicit since the anomalous dimension
exponent
$\eta_{\mathrm{sr}}$ depend in turn on the effective dimension $d_{\rm eff}$.
The critical exponent
$\eta_{\mathrm{sr}}$ as a function of the real parameter $d$ has to be computed
separately in order to obtain $d_{\rm eff}$. Several high precision results for this exponent exist in the literature\,\cite{holovatch1993critical,pelissetto2002critical,poland2019conformal,cappelli2019critical}.
For consistency sake, in the following we are going to base our analysis on the FRG estimates in Ref.\,\cite{codello2015critical}.
Applying the relation in Eq.\,\eqref{exp_rel}, the correlation length exponent $\nu$ as a function of $\sigma$ and $N$ in the long-range $O(N)$ model has been obtained and it is reported in Fig.\,\ref{Fig3}. 

The main result of this analysis is that qualitatively one can think
to the action of long-range coupling to the critical properties
of a system as an effective increasement of the dimension.
This qualitative fact helps to understand many properties of (classical and
quantum) long-range systems, but several evidences indicate that -- despite being exact at
one-loop level and giving very good estimates
of the critical exponents as a function of $\sigma$, with an estimated
error smaller than $1\%$ for classical Ising models\,\cite{defenu2015fixed} -- the mapping of a long-range model at criticality
on a short-range model with higher dimension does not hold
exactly [see the discussion of
  long-range critical exponents near the short-range crossover
  in Ref.\,\cite{behan2017long}].

\subsubsection{Competing momentum contributions}
\label{comp_mom_contr}

A crucial role in the critical behavior of long-range systems is played by
$\sigma_*$. For $\sigma=2-\eta_{\rm sr}$ the effective dimension result in Eq.\,\eqref{d_eff} reduces to $d_{\rm eff}=d$, implicitly suggesting the validity of the result in\,\cite{sak1973recursion} for $\sigma_*$. Actually, the $\sigma_{*}$ has represented, and still represents, one of the most fascinating questions in the study of weak long-range universality. Its value is the result of a subtle interplay between different momentum terms in the critical propagator and of their contribution to the universal behaviour. In particular, the question concerns the renormalization of the long-range ($p^\sigma$) term and its effect on the ($p^2$) one.  

A first answer to this question was given in Ref.\,\cite{fisher1972critical} 
by a second order $\varepsilon$-expansion approach. 
This analysis suggested that the mean-field result $\eta=2-\sigma$ 
holds at all orders in perturbation theory 
with respect to the parameter
$\varepsilon=2\sigma-d$, 
a result later extended by Ref.\,\cite{honkonen1990critical}. 
The conclusion of this study implied a
discontinuity of the anomalous dimension $\eta$ as a function of the parameter
$\sigma$, when $\sigma$ reaches $\sigma_*$, since $\sigma_*=2$\,\cite{fisher1972critical}.
The discontinuity issue was rather solved by the inclusion of both non-analytic $p^{\sigma}$ and analytic $p^{2}$ terms in the propagator, see Ref.\,\cite{sak1973recursion}, which confirmed the result $\eta=2-\sigma$, but found a different threshold value $$\sigma_*=2-\eta_{\rm sr}.$$

Most Monte Carlo (MC) investigations, featuring specific algorithms for long-range interactions\,\cite{luijten1997classical,fukui2009order,gori2017one}, 
appear to be in agreement with the so-called Sak's scenario ($\sigma_{*}=2-\eta$)\,\cite{luijten2002boundary,angelini2014relations,gori2017one,horita2016upper}.
Nevertheless, up to very recent times, several different theoretical pictures have been compatible with the $\sigma_*=2$ result\,\cite{suzuki1973critical, yamakazi1977critical, enter1982instability, picco2012critical,blanchard2013influence, grassberger2013two}.
Recently, conformal bootstrap results\,\cite{behan2017long} confirmed Sak's scenario and, albeit not giving numerical estimates for the long-range critical exponents,
furnished an exact framework for its understanding. A detailed study
of RG fixed points in a model of symplectic fermions with a nonlocal
long-range kinetic term is reported in\,\cite{giuliani2021gentle}

In the framework of the FRG approach, the absence of the analytic term in Eq.\,\eqref{lr_lpa_eaa} makes the aforementioned approximation not suitable to properly 
investigate the $\sigma\simeq\sigma_*$ regime, where the momentum terms interplay is crucial. 
A more complete parametrisation, which accounts for the leading and first sub-leading term in the expansion of the mean-field propagator, has been
introduced in\,\cite{defenu2015fixed}
\begin{equation}
\label{lr_lpa2_eaa}
\Gamma_{k}[\phi]=\!\int\! d^{d}x\! \left\{\!Z_{\sigma,k}\partial^{\frac{\sigma}{2}}_{\mu}
\phi_{i}\partial^{\frac{\sigma}{2}}_{\mu}\phi_{i}\!+\!Z_{2,k}\partial_{\mu}\phi_{i}\partial_{\mu}\phi_{i}\!+\!U_{k}(\rho)\!\right\}\,.
\end{equation}
In order to make the study as unbiased as possible, one needs also to introduce a proper regulator function
\begin{equation}
\label{lr_reg2}
R_{k}(q)=Z_{\sigma,k} (k^{\sigma}-q^{\sigma})\theta(k^{\sigma}-q^{\sigma})+
Z_{2,k} (k^{2}-q^{2})\theta(k^{2}-q^{2})\,.
\end{equation}
A close inspection of the ansatz in Eq.\,\eqref{lr_lpa2_eaa} already yields several results. 
Indeed, the effective action\,\eqref{lr_lpa_eaa} cannot describe the region $\sigma>\sigma_{*}$, as it cannot recover the standard short-range $O(N)$ form in Eq.\,\eqref{s_o(n)}. At variance, Eq.\,\eqref{lr_lpa2_eaa} simply reduces to the standard short-range form for vanishing non-analytic term. 

The first step in any fixed point calculation is rescaling the physical couplings with an appropriate power of the scale $k$. Since the RG flow displays scaling behaviour in the vicinity of a fixed point, this procedure, with appropriate choices for the scaling dimensions, shall represent the critical point as fixed points of the flow. Traditionally, one rescales the physical couplings based on the scaling dimension of the kinetic operator, but this choice is not unique for the effective action.\,\eqref{lr_lpa2_eaa}. Therefore, one has two possible definitions for dimensionless couplings summarised in the following table:
\begin{center}
\begin{tabular}{ccc}
\textit{Quantity} |& \textit{short-range\,dimensions} |& \textit{long-range dimensions}\\
$q$ & $k\bar{q}$ & $k\bar{q}$ \\
$\rho$ & $k^{d-2} Z_{2}^{-1} \bar{\rho}$ & $k^{d-\sigma} Z_{\sigma}^{-1} \bar{\rho}$ \\
$U(\rho)$ & $k^{d}\bar{U}(\bar{\rho})$ &  $k^{d}\bar{U}(\bar{\rho})$\\
$Z_{2}$ & $\bar{Z}_{2}$ &  $k^{\sigma-2}\bar{Z}_{2}$\\
$Z_{\sigma}$ & $k^{2-\sigma}\bar{Z}_{\sigma}$ &  $\bar{Z}_{\sigma}$\\
\end{tabular}
\end{center}

Clearly, as both definitions are arbitrary they shall yield the same physical results. Yet, for reason to be clarified in the following, a more consistent picture is found by employing short-range dimensions. Using the latter,
one has to consider three flow equations, respectively for the effective potential $\bar{U}_{k}$, the long-range coupling $J_{\sigma}=Z_{\sigma}/Z_{2}$ and the anomalous dimension $\eta=-\partial_t Z_{2}/Z_{2}$. The conventional short-range case is recovered in the limit $J_{\sigma}\rightarrow 0$. The resulting FRG equations can be obtained in analogy with the short-range case\,\cite{defenu2015fixed}, yielding
\begin{subequations}
\begin{equation}
\label{js_flow}
\partial_t \bar{J}_{\sigma} = (\sigma-2)\bar{J}_{\sigma}+\eta_{2}\bar{J}_{\sigma}\,,
\end{equation}
\begin{equation}
\label{eta}
\eta=\frac{(2+\sigma\bar{J}_{\sigma})^{2}\bar{\rho}_0\bar{U}''_{k}(\bar{\rho}_0)^{2}}{(1+\bar{J}_{\sigma})^{2}(1+\bar{J}_{\sigma}+2\bar{\rho}_0\bar{U}''_{k}(\bar{\rho}_0))^{2}}\,,
\end{equation}
\begin{equation}
\begin{split}
\label{pot_flow}
&\partial_t \bar{U}_{k}(\bar{\rho})=-d\bar{U}_{k}(\bar{\rho})+(d-2+\eta_{2})\bar{\rho}\,\bar{U}'_{k}(\bar{\rho})\\
+&(N-1)\frac{1-\frac{\eta_{2}}{d+2}+\frac{\sigma}{2}\bar{J}_{\sigma}}{1+\bar{J}_{\sigma}+\bar{U}'_{k}(\bar{\rho})}
+\frac{1-\frac{\eta_{2}}{d+2}+\frac{\sigma}{2}\bar{J}_{\sigma}}{1+\bar{J}_{\sigma}+\bar{U}'_{k}(\bar{\rho})+2\bar{\rho}\,\bar{U}''_{k}(\bar{\rho})}\,.
\end{split}
\end{equation}
\end{subequations}

The resulting picture for the universal behaviour is 
encoded in Eqs.\,\eqref{js_flow} and\,\eqref{eta}. A fixed point can emerge only if the r.h.s. of Eq.\,\eqref{js_flow} vanishes and, for non-vanishing long-range coupling $J_{\sigma}\neq 0$ this implies $\eta=2-\sigma$.  Therefore, the fixed point value for the long-range coupling $J_{\sigma}^{*}$ has to be such that Eq.\,\eqref{eta} is consistent with $\eta=2-\sigma$. Such solution is only possible for $d/2<\sigma<2-\eta_{\rm sr}$, consistently with Sak's scenario, where $\eta=2-\sigma$. Therefore, while at the short-range fixed point the long-range coupling vanishes $J_{\sigma}=0$, at the long-range one the short-range momentum term does not vanish, but its scaling dimension is increased to match the one of the long-range term.

This complex structures explains why short-range dimensions are more suited to describe the weak long-range criticality, at least as long as $\sigma>d/2$, but also demonstrates that the effective dimension approach described in previous section does not hold in the present case, as the critical propagator of the long-range universality class features a multiple power law structure, already noticed in MC simulations\,\cite{angelini2014relations}, which is absent in the short-range case.

The final summary for the universality picture for weak long-range ferromagnetic interactions is the following:
\begin{itemize}
\item for $\sigma\leq d/2$ the mean--field approximation
  correctly describes the universal behavior; 
\item for $\sigma$ greater than a threshold value, $\sigma_*=2-\eta_{\rm sr}$, 
the model has the same critical exponents of the short-range model
(the short-range model is strictly defined as the limit $\sigma \to \infty$); 
\item for $d/2<\sigma \le \sigma_*$ the system exhibits 
peculiar long-range critical exponents, which may be approximated by the one of the short-range model in the effective fractional dimension $d_{\rm eff}=(2-\eta_{\rm sr})d/\sigma$.
\end{itemize}

These results, albeit obtained in the approximated framework of ansatz in Eq.\,\eqref{lr_lpa2_eaa}, appear to 
hold also for the full theory and the result $\eta=2-\sigma$ seems now established\,\cite{defenu2015fixed, horita2016upper, gori2017one,  behan2017long}. In the FRG context, the validity of the Sak's scenario has been confirmed also for long-range disordered systems\,\cite{balog2014critical}.

\begin{figure*}
\subfigure[Long-Range Ising Model]{\label{Fig4a}\includegraphics[width=.45\linewidth]{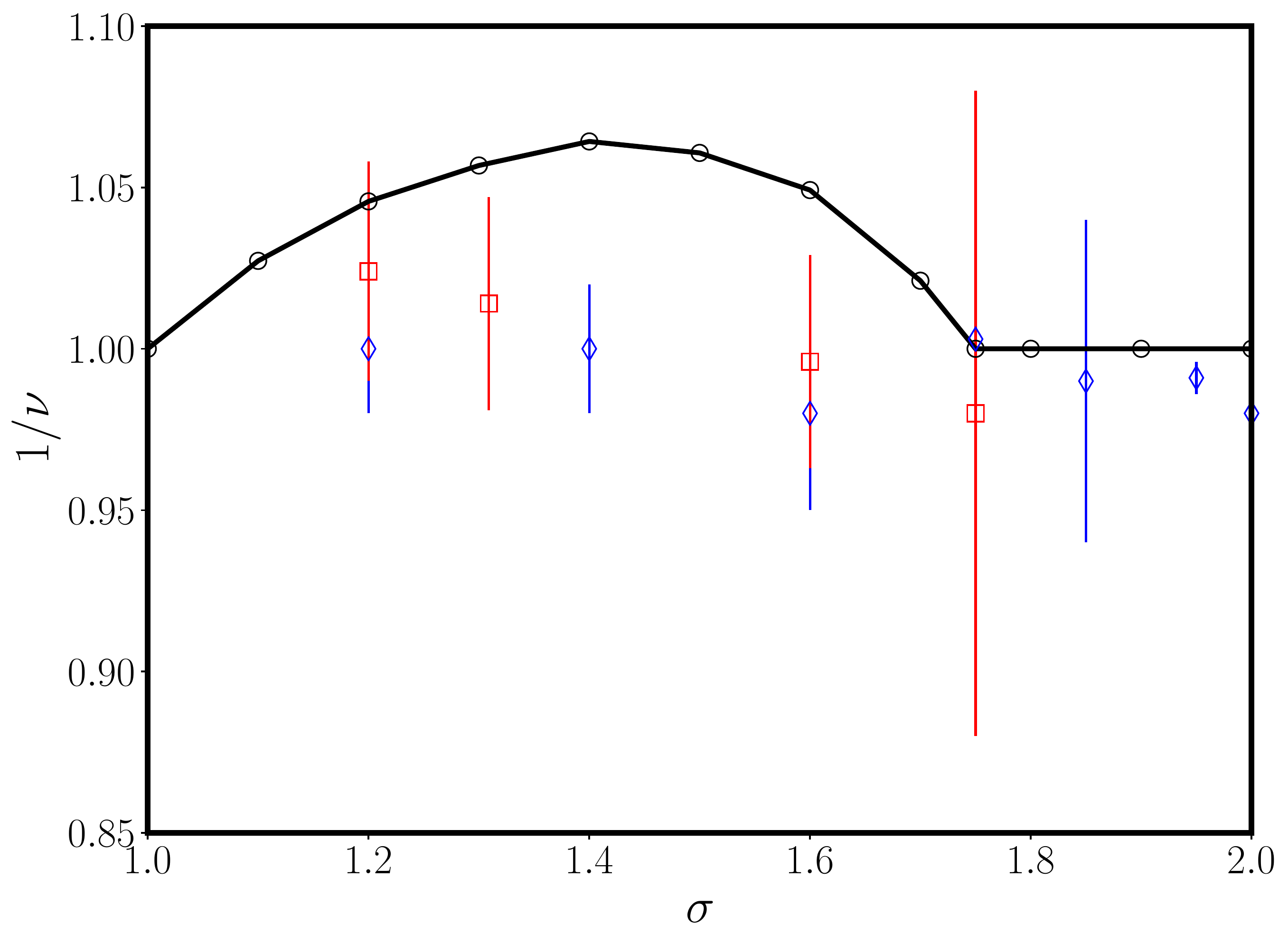}}
\hspace{1cm}
\subfigure[Long-Range Percolation]{\label{Fig4b}\includegraphics[width=.45\linewidth]{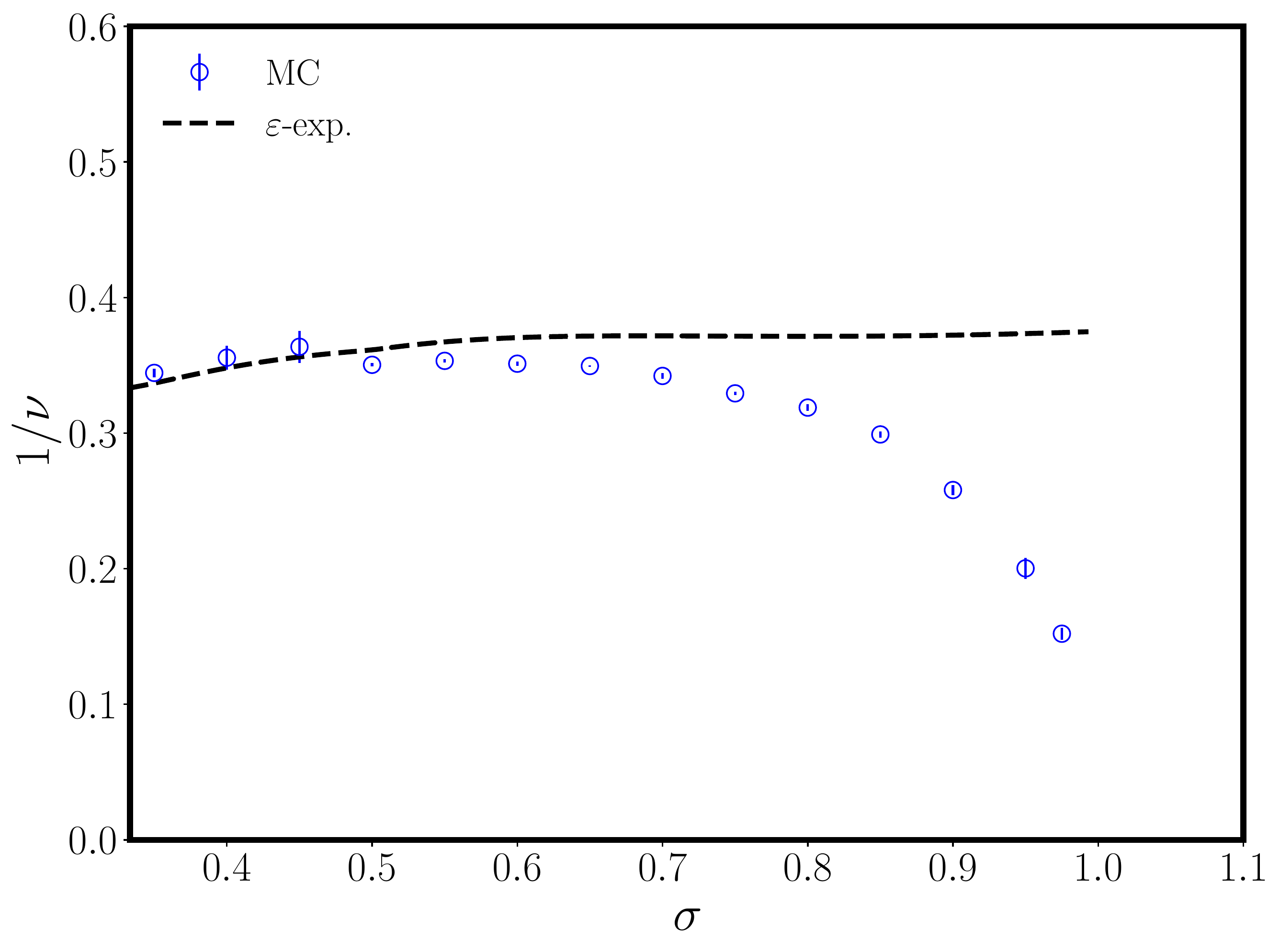}}
\caption{{\bf Inverse correlation length exponent of long-range interactions.}
  Reults obtained from MC simulations are compared with the results of the effective
  dimension approach. {\bf (a)} MC simulations for the long-range Ising model in $d=2$ in Refs.\,\cite{luijten1997classical, angelini2014relations} (blue and red respectively), the black points have been obtained by mapping the conformal bootstrap results for the short-range critical exponents\,\cite{elshowk2014conformal} via Eqs.\,\eqref{d_eff} and\,\eqref{exp_rel}. {\bf (b)} The MC data of Ref.\,\cite{gori2017one} are compared to the analytical curves of $\varepsilon$-expansion\,\cite{gracey2015four}. The low accuracy of the analytical result in the $\sigma\to 1$ limit is due to the appearance of BKT scaling\,\cite{cardy1981one}, which cannot be captured by $\varepsilon$-expansion.}
\end{figure*}

The approximate nature effective dimension formula in Eq.\,\eqref{d_eff}\,\cite{defenu2015fixed, behan2017long} shall not hinder its adoption to compute numerical estimates for the critical exponents. Indeed, the actual correction, rising from analytical contributions to the critical propagator, appears to be rather small and the application of the effective dimension approach produced rather accurate theoretical benchmarks for MC data, both in the long-range Ising and percolation models, see Fig.\,\ref{Fig4b}.

\subsubsection{Anisotropic Systems}

One of the main features of weak long-range interactions is the presence of non-analytic momentum terms in the critical propagator, which do not renormalise even in the correlated regimes. As a consequence, long-range critical models apparently display only a single non trivial critical exponent, namely the correlation length exponent $\nu$, and not two as in short-range critical models.
Indeed, the anomalous dimension of the model is not affected by long-range fluctuations and the correlation functions always display mean-field scaling $\eta=2-\sigma$. In general, this property remains unchanged in the case of anisotropic long-range interactions, where different directions of the system display different power-law interactions\,\cite{defenu2016anisotropic}.

A convenient ansatz for anisotropic $O(N)$ models has to account the presence of different non-analytic terms depending on the directions 
\begin{align}
\label{anis_eaa}
&\Gamma_{k}[\phi]=-\int d^{d}x \Bigl{(}Z_{\sigma_1}\varphi_{i}(x)\Delta_{\parallel}^{\frac{\sigma_1}{2}}\varphi_{i}(x)+Z_{\sigma_2}\varphi_{i}(x)\Delta_{\perp}^{\frac{\sigma_2}{2}}\varphi_{i}(x)\nonumber\\
&+\varphi_{i}(x)(\Delta_{\parallel}+\Delta_{\perp})\varphi_{i}(x)-U_{k}(\rho)\Bigr{)}\,,
\end{align}
where $\Delta_{\parallel}=\sum_{\mu=1}^{d_{1}}\partial_{\mu}^{2}$ and $\Delta_{\perp}=\sum_{\mu=d_{1}}^{d_{2}}\partial_{\mu}^{2}$ are the Laplacian in the two different subsystems of dimensions $d_{1}$ and $d_{2}$ respectively.
The action in Eq.\,\eqref{anis_eaa} is analogous to the case of anisotropic Lifshitz points\,\cite{hornreich1975critical}, where the critical correlations display anisotropic scaling with four critical exponents $\eta_{\parallel}$, $\eta_{\perp}$, $\nu_{\parallel}$ and $\nu_{\perp}$, two for each subsystem. However, these four critical exponents are not independent and the following relation may be obtained by means of scaling theory
\begin{align}
\theta=\frac{2-\eta_{\parallel}}{2-\eta_{\perp}}=\frac{\nu_{\perp}}{\nu_{\parallel}}
\end{align} 
which reduces the number of independent exponents to three, $\eta_{\parallel},\,\,\nu_{\parallel},\,\,\theta$.

In analogy with the long-range isotropic case the two anomalous dimensions satisfy the relation
\begin{align}
 \eta_{\parallel}&=2-\sigma_{1}\quad\mathrm{if}\,\,\sigma_{1}<\sigma_{*}\\
 \eta_{\perp}&=2-\sigma_{2}\quad\mathrm{if}\,\,\sigma_{2}<\sigma_{*}
 \end{align}
  yielding only a single non-trivial critical exponent also for anisotropic long-range interactions. Yet, anisotropic interactions may present a non-trivial anomalous dimension influenced by the long-range tails of the couplings. This occurs when only one of the decay exponents, say $\sigma_{1}$, overcomes the threshold value $\sigma_{*}$. Then, the leading analytic contribution in the critical propagator in the $d_{1}$ directions gets an anomalous contribution $\eta_{\parallel}=\eta(\sigma_{2})$, which depends on the decay exponent of long-range interactions in the other subsystem.
  
This also leads to the appearance of a non-trivial threshold value, which depends on the value of the decay exponent in the other subsystem, $\sigma_{*;1,2}=2-\eta(\sigma_{2,1})$. As $\sigma_{2}$ also approaches the short-range threshold $\sigma_{2}\to\sigma_{*}=2-\eta_{\rm sr}$ the anomalous dimension tends to its short-range value and isotropy is recovered.
This phenomenon shall play a crucial role in the study of quantum long-range $O(N)$ models.

\subsubsection{Berezinskii-Kosterlitz-Thouless scaling}
\label{wlr_classical_bkt}

For short-range interacting models with continuous symmetry, the occurrence of spontaneous symmetry breaking (SSB) in $d=2$ is forbidden by the Mermin-Wagner theorem\,\cite{mermin1966absence, hohenberg1967existence}. Yet, the inclusion of long-range interactions with $0<\sigma<\sigma_{*}$ modifies the scaling dimension of operators, allowing SSB also in low dimensions. The effect of such altered scaling is conveniently summarised by the effective dimension relation in Eq.\,\eqref{d_eff}, which proves the possibility for a long-range interacting system in $d$ dimensions to reproduce the scaling of any $d_{\rm eff}$-dimensional short-range system with $d_{\rm eff}\in [d,\infty]$.

Given these considerations, it is not difficult to generalise the results of the Mermin-Wagner theorem to long-range interactions\,\cite{bruno2001absence}, leading to the vanishing of the inverse correlation length exponent in the $\sigma\to2$ limit  for $N\geq 2$, see Fig.\,\ref{Fig3}. Then, for $d=N=\sigma=2$ the traditional picture for short-range models is recovered and Berezinsky-Kosterlitz-Thouless (BKT) shall occur\,\cite{kosterlitz1973ordering,kosterlitz1974critical,jorge201340}.

BKT scaling is 
a characteristic of two-dimensional systems, ranging from condensed matter\,\cite{nelson1977universal,yong2013robustness} and cold atoms\,\cite{hadzibabic2006berezinskii, murthy2015observation} to network theory\,\cite{dorogovtsev2008critical} and biology\,\cite{nisoli2014attractive}. Its most renown application is certainly the XY model, where its properties have been very well characterised\,\cite{gupta1988phase,gupta1992critical, hasenbusch1992high} and its relation with topological excitations first discovered\,\cite{kosterlitz2017nobel}.

First theoretical indications
of this topological phase transition have occured in long-range interacting classical systems\,\cite{thouless1969long}.
In particular, the Coulomb gas problem and the Ising model with $d=\sigma=1$ have been known to display such infinite order transition, well before its traditional formulation\,\cite{anderson1969exact,anderson1970exact}. This fact shall not surprise, since for $d=\sigma=1$ the scaling dimension of the operators are consistent with the one of short-range interactions in $d=2$. Understanding
in detail the difference between the number of degrees of freedom in the traditional short-range BKT scaling with $d=N=2$ and the long-range one occurring for $d=N=\sigma=1$ is a more complicated and possibly open task, but it is probably related with the irrelevance of amplitude fluctuations in $d=2$\,\cite{jakubczyk2017longitudinal,defenu2017non,krieg2017dual}. It is worth noting that long-range BKT scaling occurring in $d=\sigma=1$ does not only occur in $O(N)$ models, but also for long-range percolation and Potts models\,\cite{cardy1981one,gori2017one}.

Despite this long-lasting relation between BKT scaling and long-range interactions, the influence of power-law couplings on topological scaling
has been subject of a very limited amount of research so far.
Indeed, the applicabilty of the aforementioned threshold value $\sigma_{*}=2-\eta_{\mathrm{sr}}$ to BKT scaling seems questionable, since the anomalous dimension of two point correlations in $d=N=2$ does not originate from critical fluctuations, but from long wave-length phase fluctuations, which disrupt the zero-temperature magnetization. Interestingly, long-range interactions with  $\sigma<2$ can be mathematically proven to stabilise spontaneous magnetization in the 2D XY model\,\cite{kunz1976first}, implicitly suggesting that $\sigma_{*}=2$. On the other hand, early results concerning the XY model on diluted L\'evy graphs\,\cite{berganza2013critical}, which has been conjectured to lie in the same universality class of the long-range XY model, appeared to be consistent with $\sigma_{*}=2-\eta_{\mathrm{sr}}$. However, these results,
have been recently challenged\,\cite{cescatti2019analysis}. Moreover,
self-consistent harmonic approximation results give an upper bound for $\sigma_*$
equal to $2$\,\cite{giachetti2021self}. No MC results for the $2d$ XY with (non-disordered)
power-law long-range couplngs around $\sigma=2$ are avaialble, to the best of our knowledge. 

Extending the RG approach first employed by Kosterlitz\,\cite{kosterlitz1974critical},
in the very recent manuscript\,\cite{giachetti2021berezinskii}.
it has been possible to 
propose a scenario the complex phase diagram of the $d=2$ long-range XY model,
which features a novel transition between a low temperature magnetised state ($T<T_{*}$) and an
intermediate temperature state with topological scaling ($T_{*}<T<T_{c}$), which disappears
at higher temperatures ($T>T_{c}$). Interestingly, this unexpected transition only occurs for
$2-\eta_{\rm BKT}<\sigma<2$, placing the threshold value at $\sigma_{*}=2$ for the long-range XY model in $d=2$. 

The derivation of these findings follows from the study of the low-energy  euclidean action
\begin{equation} 
\label{bkt_a}
S[\theta] =  \int d^{2}x\left\{\frac{J}{2}   |\nabla \theta|^2 - \frac{g}{2} \int_{r>a} \frac{d^2 r}{r^{2+ \sigma}} \left(  1 - \cos  \Delta_{\mathbf{r}} \theta(\mathbf{x}) \right)\right\}
\end{equation}
where we introduced the notation  $\Delta_{\mathbf{r}}\theta=\theta(\mathbf{
  x}+\mathbf{r})-\theta(\mathbf{x})$ and the phase $\theta(x)$ is defined according to the polar representation of the field $\boldsymbol{\varphi}=\sqrt{\rho}e^{-i\theta}$.
The first and the second term account for the short- and long-range contributions respectively,
with $J\sim 1/T$ and $g\sim 1/T$ being the temperature dependent couplings of the continuum theory. As usual in the study of 2D topological phase transitions it is necessary to take into account the compact nature of the phase $\theta$, which is defined up to $2 \pi$ multiples\,\cite{bacso2015c}. Due to this compactness, the action in Eq.\,\eqref{bkt_a} supports low energy topological configurations in the form of vortexes. Upon the explicit inclusion of vortex configurations, one finds a low energy action parametrised by three couplings, the phase stiffness $J$ and the vortex fugacity $y$\,\cite{benfatto2013berezinskii}. 

With respect to the traditional case, the action in Eq.\,\eqref{bkt_a} also features the long-range coupling $g$. According to the RG approach, the relevance of the long-range perturbation depends on the scaling dimension $\Delta_{g}$ of the coupling $g$, which is defined according to the asymptotic behaviour $g_{k}\approx (k/\Lambda)^{-\Delta_{g}}$. Computing the scaling dimension of the long-range operator at the BKT fixed point line, one finds\,\cite{giachetti2021berezinskii}.
\begin{align}
\Delta_{g}=2-\sigma-\eta_{\mathrm{sr}}(T)
\end{align}
so that the long-range perturbation becomes relevant only if $ \sigma < 2-\eta_{\rm sr}(T)$, similarly to the traditional SSB case, but with a temperature dependent anomalous dimension.  Traditional BKT theory implies that the function $\eta_{\rm sr}(J)$ is maximal at the transition point $J_{\rm BKT}$, where it attains the universal value $\eta_{\rm BKT}=1/4$. Therefore, in the range of decay exponents $7/4 <\sigma < 2$ the relevance of the long-range perturbation is temperature dependent and the BKT fixed point line remains a stable solution to the RG flow in the temperature window $T_{*}<T<T_{\rm BKT}$.

The study of the extended BKT flow equations 
shows that long-range couplings with decay exponents in the range $7/4<\sigma<2$ disrupt \emph{quasi}-long-range order at low temperatures and produce a non-trivial finite temperature endpoint for the BKT line\,\cite{giachetti2021berezinskii}. The properties of this exotic low temperature phase are rather striking and can be depicted by the study of a pure quadratic phase action with non-analytic momentum terms. Due to this fractional kinetic, the Mermin-Wagner theorem does not apply and a finite magnetization is also allowed in 2D. Therefore, one finds that for $\sigma <2$ the system is ordered at small enough temperatures $T<T_{*}$.  Moreover,  approaching the transition point ($T \rightarrow T_*^{-}$), the magnetisation scales as
\begin{equation} \label{scaling}
\ln m \sim -e^{B(T_* - T)^{-1/2}}
\end{equation}
where $B$ is a non universal constant. It is interesting to see that all the derivatives of $m$ with respect to $T$ vanish at $T=T_*$, so that the phase transition between the ordered and disordered phase is actually of infinite order, but still presents an essential singularity. Furthermore, the connected correlation functions display  power-law decay for $T < T_*$ given by $\langle\boldsymbol{\varphi}(\mathbf{x}) \cdot \boldsymbol{\varphi}(0)\rangle_{c} \sim \frac{1}{r^{2 - \sigma}}$, as it is customary in the symmetry broken phase of critical models with $U(1)$ symmetry\,\cite{dupuis2011infrared}.

To summarize, the introduction of long-range interaction patterns in systems with $U(1)$ symmetry in $d=2$ generates exotic critical features, which have no counterpart in the traditional universality classification\,\cite{raju2019normal}. This is not surprising since the interplay between $U(1)$ systems and complex interaction patterns is known to generate peculiar critical behaviour
as in the anisotropic
3D XY model\,\cite{shenoy1995anisotropic},
coupled XY planes\,\cite{bighin2019berezinskii}, 2D systems with anisotropic
dipolar interactions\,\cite{maier2004ferromagnetic,vasiliev2014universality}
or four-body interactions\,\cite{antenucci2015statistical}, 
and high-dimensional systems with Lifshitz
criticality\,\cite{jacobs1983self,defenu2021topological}.

\subsection{Strong long-range regime}
\label{classcal_slr_criticality}

\subsubsection{Ensemble in-equivalence}
The traditional universality problem concerns the numerical characterisation of universal quantities, in the strongly correlated regime, where long-range collective correlations are relevant and mean-field, as well as other perturbative techniques, cannot be applied. Such questions have no actual application to the case of long-range interactions with $\sigma<0$, i.e. $\alpha<d$, since the divergent interaction strength stabilises the mean-field solution of the problem and the Gaussian theory reproduces the universal features also at the critical point.

Nevertheless, several interesting effects arise due to strong long-range interactions, in the thermodynamic behaviour of statistical mechanics models. These effects may be loosely regarded as universal, since they appear irrespectively of the particular model considered, as well as irrespectively of the introduction of any finite range couplings, and they may be ofter characterised starting from a continuous description\,\cite{antoniazzi2007exploring,bachelard2011vlasov}.

At equilibrium, the most striking feature of systems in the strong long-range regime is probably ensemble inequivalence, i.e. the appearance of substantial differences in the phase diagram of strong long-range systems depending on the application of the micro-canonical or the canonical thermodynamic descriptions\,\cite{barre2001inequivalence}. This property has been extensively revised in several review articles and books on the physics of classical long-range systems\,\cite{campa2009statistical, dauxois2002dynamics,campa2014physics} and there is no need to discuss it here, in details. For the sake of the following discussion we are only going to briefly mention the existence of two diverse issues of ensemble inequivalence.

The first, most common, example of ensemble in-equivalence is found in systems with long-range attractive or antiferromagnetic interactions, which feature a two-phase coexistence state. Such coexistence states are usually connected with a ‘dip' or a ‘convex intruder' in an otherwise concave entropy, possibly leading to a negative specific heat. The phase boundary associated with such coexistence states carries an infinite entropy cost, which makes them unstable in the canonical ensemble. On the other hand, in the micro-canonical description such entropy cost is not relevant and such equilibrium states may be actually realising by tuning the energy\,\cite{ispolatov2001first,lynden1999negative,dauxois2002hamiltonian}. Interestingly, the same phenomenon is observed on sparse random graphs, where the condition of a negligible surface in the thermodynamic limit is violated\,\cite{barre2007ensemble}.

The second example of ensemble inequivalence is conventionally found in long-range systems with a two parameter dependent free-energy $S(\varepsilon, \lambda)$, which present a line of second-order critical point along a line $\varepsilon_{c}(\lambda)$, terminating at a tricritical point at $\lambda_{c}$. The location of such tricritical point, as well as the structure of the first-order lines beyond it, strongly depend on the thermodynamic ensemble considered. In particular, the micro-canonical description as a function of the temperature $1/T=\partial S/\partial \varepsilon$ do not match the standard canonical description as it should be for short-range interacting systems\,\cite{barre2001inequivalence}.

It is worth noting that the "convex intruder" causing the first case of ensemble inequivalence is not exclusive of long-range interacting systems, but it is also present on short-range systems with finite sizes, where the boundary contribution is comparable to the one from the finite bulk\,\cite{ispolatov2001first}. This feature is then washed away in the thermodynamic limit for short-range systems, while it remains for strong long-range ones.
\subsubsection{Violation of hyperscaling}
Apart from ensemble in-equivalence, the relevance of boundaries in the scaling theory of strong long-range systems produces several anomalies, which influence the understanding of their critical behaviour. In particular,
let us comment on the usual finite size scaling theory, which relates the thermodynamic critical exponent of any quantity, e.g. the susceptibility
\begin{align}
\chi\propto |T-T_{c}|^{-\gamma}
\end{align}
with its finite size correction\,\cite{cardy1996scaling}
\begin{align}
\chi_{N}\propto N^{\gamma/\nu}
\end{align}
where the subscript $N$ indicates the corresponding quantity in a system of size $N$. In long-range systems the correspondence between thermodynamic exponents and finite size scaling ones is not obtained via  the correlation length exponent $\nu$, but via an exponent $\nu_{*}=\nu_{\rm mf}d_{\rm uc}$, where $\nu_{\rm mf}$ and $d_{\rm uc}$ are respectively the mean-field correlation length exponent and the upper critical dimension of the corresponding short-range system\,\cite{botet1982size}.

Such modification of finite size scaling theory has been related to the violation of hyperscaling  and, more in general, to a non-trivial power law scaling of the correlation length $\xi$ with the system size $N$\,\cite{flores2015finite}, leading to several anomalous differences between the actual finite-size scaling of strong long-range systems and the mean-field solution\,\cite{colonna2014anomalous}. Actually, these observations are not peculiar of strong long-range systems, but have been also found in the study of critical phenomena in short-range systems above the upper critical dimension\,\cite{flores2016role,luijten1996finite,binder1985critical}.

\subsection{Competing non-local systems}
\label{mod_phase}

Modulated phases, resulting from the competition of interactions at different length-scales, are ubiquitous in nature\,\cite{seul1995domain} and also display universal scaling close to their critical points. Despite this ubiquity, a comprehensive description of their universal behaviour not emerged yet and their understanding is apparently behind the one of homogeneous phase transition. A convenient effective action for modulated phases has been firstly introduced by Brazovskii\,\cite{brazovskii1975phase} and it reads
\begin{align}
\label{cbh}
S[\varphi]&=\frac{1}{2}\int \frac{d^{d}q}{(2\pi)^{d}}\vec{\varphi}(q)\left(\lambda+\frac{(q-q_{0})^{2}}{m}\right)\vec{\varphi}(q)\nonumber\\&+
u\int d^{d}x \frac{|\vec{\varphi}(x)|^{4}}{4!}
\end{align}
where $\varphi(q)$ is the Fourier transform of $\varphi(x)$, which is a $N$-components vector field, $q=|\vec{q}|$ is the momentum amplitude and $q_{0}$ a constant given by the nature of competing interactions. In writing Eq.\,\eqref{cbh} we assumed that the long-range tails of the interactions are not relevant ($\alpha>\alpha_{*}$).

The system described by the Hamiltonian in Eq.\,\eqref{cbh} represents a different paradigma with respect to the ordinary $N$-vector models.  Indeed, the Hamiltonian in Eq.\,\eqref{cbh} for $\lambda<0$ supports a condensate with any of the finite wave-vectors occurring on the $d-1$-dimensional sphere $|\vec{q}|=q_{0}$. Therefore, the condensed
phase of the model is somehow "doubly" symmetry broken, since the model does not only choose the  $i =1$ component of the field in which it condenses, but must also make a single choice for the wave-vector $\vec{q}= \vec{q}_{0}$, out of the infinite set of
equivalent order parameters with  $|\vec{q}|=q_{0}$. The diversity in the symmetry breaking procedure also reflects in a different phase space for fluctuations, since the d-dimensional phase
space around the $|\vec{q}|=q_{0}$ surface is anisotropic, with fluctuations parallel to the surface, which
are exactly degenerate, and  fluctuations away from it, which are only nearly degenerate. This discussion should have clarified that the Brazovskii model in Eq.\,\eqref{cbh}
does not belong to any of the usual universality classes of
isotropic models and presents its own set of universal properties as a function of the parameters
$N$ and $d$. 

Interesting applications of the physics described by the Brazovskii model occur in two dimensional or highly anisotropic systems, such as quantum hall platforms\,\cite{fradkin1999liquid}, high $T_{c}$ superconductors\,\cite{kivelson2003how,kivelson1998electronic} and ultra-thin magnetic films\,\cite{kashuba1993stripe,vaterlaus2000two,saratz2010experimental}. Nevertheless, the first efforts to apply the momentum shell renormalization group theory\,\cite{wilson1974renormalization} to the Hamiltonian in Eq.\,\eqref{cbh} with $d=2$ resulted in the impossibility to construct a reliable perturbative picture\,\cite{hohenberg1995metastability}. Applying the approach developed by Shankar on fermionic systems\,\cite{shankar1994renormalization}, Hohenberg and Swift\,\cite{hohenberg1995metastability} found out that momentum dependent corrections to the interacting coupling $u$ are relevant and no weak coupling expansion is possible in the treatment of modulated phases. Nevertheless, a symmetry analysis of these relevant corrections suggests the appearance of a second-order nematic-isotropic transition\,\cite{barci2007competing}. Similar difficulties have been encountered by more modern treatments\,\cite{shiwa2006exact} and the description of systems belonging to the Brazovskii universality has remained confined to mean field theory\,\cite{capati2015electronic, barci2007competing, barci2013nematic}, scaling arguments\,\cite{portmann2010scaling, mendoza2012coarse,barci2009orientational,barci2011microscopic} and numerical simulations\,\cite{poderoso2011new,cannas2006ising}. 

Recently, the study of the nematic-isotropic transitions in the Brazovskii model has been extended beyond the analytic momentum paradigm in Eq.\,\eqref{cbh} in order to include long-range repulsive interactions of the form $1/r^{\alpha'}$, with particular focus on the Coulomb ($\alpha'=1$) and dipolar ($\alpha'=3$) cases\,\cite{mendoza2015nature}. It is particularly interesting to note that, within the effective field theory approach of Ref.\,\cite{mendoza2015nature}, it is possible to show exact correspondence between the universality of the nematic-isotropic transition and the one of homogeneous rotor models at finite temperature with decay exponent $\alpha=\alpha'+2$\,\cite{mendoza2017quantum}. Therefore, for modulated phases in $d=2$, the relevant regime for long-range interactions is rigidly shifted in such a way that any power law decay $\alpha'>2$ is always irrelevant, while for $\alpha'<2$ the interaction energy remains finite also in absence of any rescaling, due to the modulation pattern of the order parameter.

Within this framework, the scalar $\varphi^{4}$-theory with long-range repulsive interactions lies in the same universality class of the long-range ferromagnetic $O(2)$ model with $\sigma=\alpha'$\,\cite{mendoza2015nature}, described in Sec.\,\ref{wlr_classical}. Therefore, for $\alpha'>2$ the isotropic nematic transition displays BKT scaling as in the short-range XY model, while for $\alpha'<2$ actual orientational order shall occur. Given this relation, one expects that for $\alpha'\in [1.75,2]$ the same phenomenology described in Sec.\,\ref{wlr_classical_bkt} shall occur.

\section{Quantum critical behaviour}
\label{sec:quantum}

Having summarised the main universal features of long-range models at equilibrium and at finite temperature, it is now time to dive into the quantum world.
Some of the previous results  for classical models are applicable to finite-temperature quantum long-range systems, albeit in a way sometime different from the usual
quantum-to-classical correspondence between quantum models at $T=0$ in dimension $d$ and classical ones at finite $T$ in dimension $d+1$ in the short-range, see the discussion below.

We start by observing that in the study of quantum critical phenomena the use of effective dimension relations is well established already for short-range models. Indeed, field theory approaches allow to relate the universal behaviour at a $T=0$ quantum critical point with the one of the corresponding $T\neq 0$ classical phase transition in dimension $d+z$\,\cite{sachdev1999quantum}. This correspondence is exact for continuous $O(N)$ field theories with dynamical
critical exponent $z$ given by $z=1$ and it can also be proven for the one-dimensional lattice Ising model in a transverse field\,\cite{mussardo2009statistical,dutta2015quantum}. Thus, it will be rather natural to connect, whenever possible, the universal behaviour in the quantum regime with the one of finite temperature phase transitions also for long-range models.

In the following, we are going to loosen our reference to the classification of 
long-range systems based on their long-distance tails, as for quantum systems this 
classification is naturally more blurred. 
Rather, we are going to divide our presentation according to the nature of variables at hand.

\subsection{Quantum rotor models}
\label{q_rot_mod}

Given the correspondence between quantum and classical universalities, $O(N)$ field theories constitute a paradigmatic model also for quantum critical behaviour. However, differently from the classical case, they do not describe the universality of ferromagnetic spin systems, since quantum spins possess $SU(N)$ rather than $O(N)$ symmetry. Nevertheless, their low energy behaviour describes the physics of several quantum models, such as antiferromagnetic quantum Heisenberg spin systems, which correspond to $N=3$, superfluid systems, $N=2$, and the Ising model, $N=1$. 

In this context, a convenient lattice representation of $O(N)$ field theories are quantum rotor models, whose Hamiltonian reads
\begin{align}
\label{q_rotor}
H_{{\rm R}}=-\sum_{ij}\frac{J_{ij}}{2} \boldsymbol{n}_{i}\cdot \boldsymbol{n}_{j}+\frac{\lambda}{2}\sum_{i}\mathcal{L}_{i}^{2}, 
\end{align}
where the $\hat{\boldsymbol{n}}_i$ are $n$ components unit length vector 
operators ($\hat{\boldsymbol{n}}_{i}^{2}=1$), $\lambda$ is a real constant 
and $\mathcal{L}$ is the invariant
operator formed from the asymmetric rotor space angular momentum tensor\,\cite{sachdev1999quantum}. As above, we are going to focus on power-law decaying ferromagnetic couplings $J_{ij}=\frac{J}{r_{ij}^{d+\sigma}}$ with $J>0$. 

In the short-range limit ($\sigma\to\infty$) the continuum formulation of quantum $O(N)$ rotor models would exactly correspond to a $d+1$-dimensional $O(N)$ field theory, with the extra dimension representing the temporal propagation of quantum fluctuations. However, in the long-range regime the field theory action is anisotropic as the spatial coordinates feature a leading non-analytic momentum term, at least for $\sigma<\sigma_{*}$. Following the same FRG approach as in Sec.\,\ref{wlr_classical}, one can introduce the following ansatz for the effective action of an $O(N)$ quantum rotor model 
\begin{align}
\label{qlr_eaa}
\Gamma_{k}=\int\,d\tau\,\int\,d^{d}x\{&K_{k}\partial_{\tau}\varphi_{i}\partial_{\tau}\varphi_{i}-Z_{k}\varphi_{i}\Delta^{\frac{\sigma}{2}}\varphi\nonumber\\
-&Z_{2,k}\varphi_{i}\Delta\varphi+U_{k}(\rho)\}
\end{align}
where $\Delta$ is the spatial Laplacian in $d$ dimensions, $\tau$ is the 
"Trotter"/imaginary time direction, $\varphi_{i}(x)$ is the $i$-th component 
($i \in \{1,\cdots,n\}$) of the system and $\rho\equiv \sum\frac{\varphi_{i}^{2}}{2}$ is the system order parameter. 
In Eq.\,\eqref{qlr_eaa} the summation over repeated indexes is intended.

It is worth reminding that the ansatz in Eq.\,\eqref{qlr_eaa} for the effective action, albeit sufficient to characterise the physics of long-range rotor models, it only approximately represents the exact critical action of correlated models. Indeed, it only contains  two kinetic terms in the $d$ spatial
directions, as necessary to represent the competition between long-range and short-range contributions to the critical propagator, but it does not contain momentum dependent corrections to the theory vertexes\,\cite{dupuis2020non}. As expected, the time direction $\tau$ does not contain any fractional derivative so that one obtains a non-unity value for the dynamical critical exponent $z$, 
defined by the relation $\omega\propto q^{z}$. %

\subsubsection{Effective dimension approach}

The characterisation of the critical properties of the action in Eq.\,\eqref{qlr_eaa} proceeds in full analogy with the case of classical anisotropic systems\,\cite{defenu2016anisotropic}, but it leads to a far more interesting picture. Scaling analysis allows to relate the universal properties of long-range quantum rotor models in $d$ dimensions with the ones of their short-range correspondents in an effective dimension 
\begin{align}
\label{d_eff_q}
d_{{\rm eff}}=\frac{2(d+z)}{\sigma},
\end{align}
where $d$ and $z$ are respectively the dimension and the dynamical critical exponent of the long-range model under study. Interestingly, the anisotropy between the time and spatial direction in the long-range model is already apparent in the mean-field estimations for the critical exponents\,\cite{dutta2001phase,monthus2015dyson}
\begin{align}
\eta&=2-\sigma,\label{eta_mf}\\
z&=\frac{\sigma}{2},\label{z_mf}\\
\nu &=1/\sigma.\label{nu_mf}
\end{align}
Upon inserting the result in Eq.\,\eqref{z_mf} into the effective dimension relation in Eq.\,\eqref{d_eff_q} one obtains the mean-field expression $d_{{\rm eff}}=\frac{2d}{\sigma}+1$, which proves that the effective dimension of quantum rotor models is increased by $1$ with respect to the classical case, as it occurs for traditional short-range systems.

\begin{figure}
\centering
\includegraphics[width=.45\textwidth]{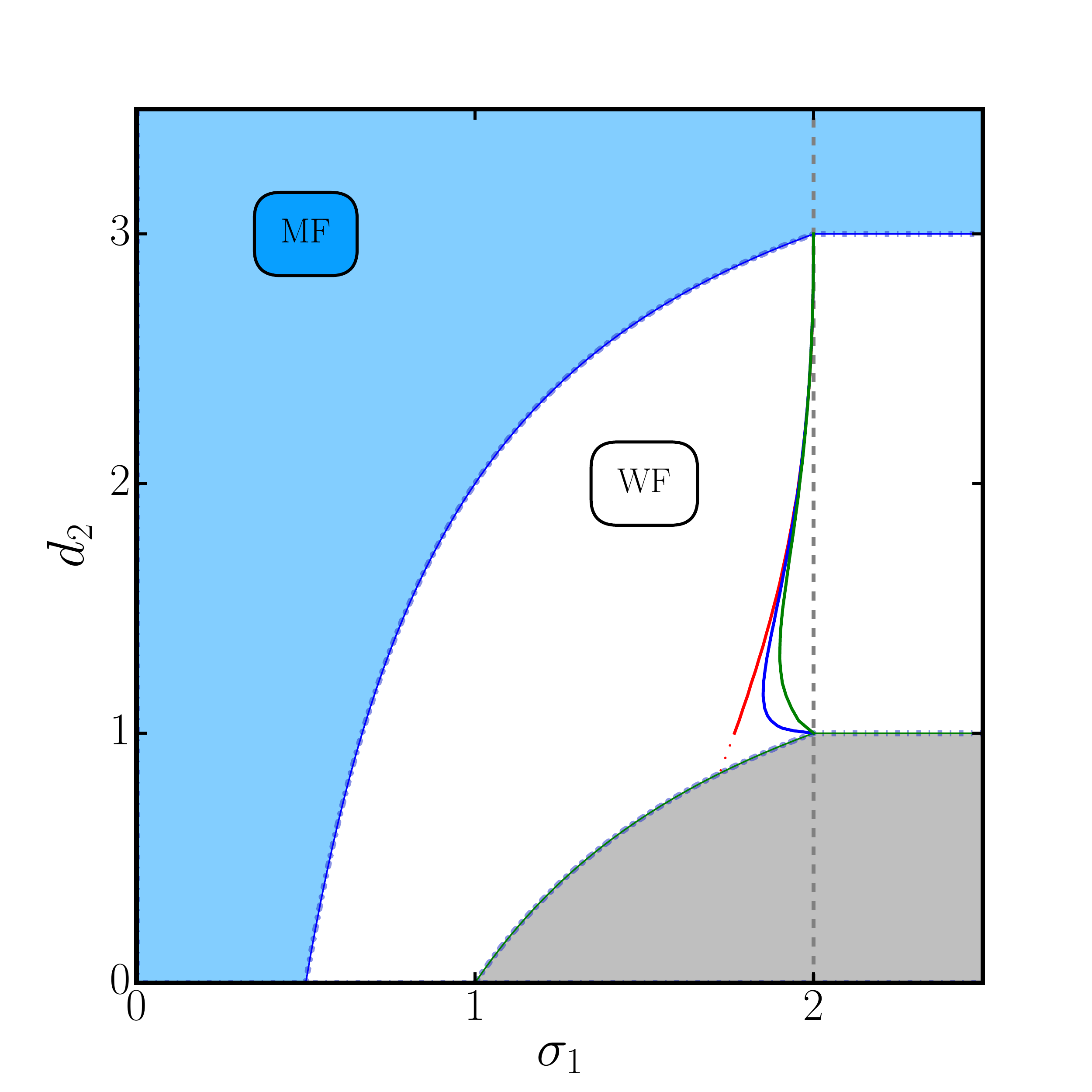}
\caption{{\bf Phase diagram of long-range quantum rotors models in the plane $d$, $\sigma$.} The universal behaviour features the mean-field critical exponents in Eqs.\,\eqref{eta_mf},\,\eqref{z_mf} and\,\eqref{nu_mf} in the cyan shaded region, while the universal properties are correlated in the white region. The color lines (red, blue, green) represent the boundary between long-range and short-range universality ($N=1,2,3$ respectively). Finally, the gray shaded region displays no phase transition at all.
}
\label{Fig6}
\end{figure}

The correspondence between quantum and classical $O(N)$ models based on the effective dimension approach in Eq.\,\eqref{d_eff_q} exactly applies close to the upper critical dimension and for quadratic models in general\,\cite{vojta1996quantum}. Then, we can employ the effective dimension approach to construct the phase diagram displayed in Fig.\,\ref{Fig6}.
Indeed, the upper critical dimension result can be derived by the condition $d_{{\rm}}\geq 4$, so that
\begin{align}
\label{duc_q}
d_{{\rm uc}}=\frac{3}{2}\sigma, 
\end{align}
as it follows also by standard scaling arguments\,\cite{Dutta2001}. Correspondingly, the lower critical dimension
for continuous symmetries $N\geq2$ follows from the condition $d_{\rm eff}\leq 2$, which yields
\begin{align}
\label{dlc_q}
d_{{\rm lc}}=\frac{\sigma}{2}.
\end{align}
It is worth stressing once again that relation \eqref{dlc_q} is only valid for continuous symmetries $N\geq 2$. As a result correlated universality shall be observed in the region $2\leq d_{\rm eff}<4$, i.e. the cyan shaded in Fig.\,\ref{Fig6}. There, the critical exponents will not coincide with mean-field result and we need to take into account the effective potential in Eq.\,\eqref{qlr_eaa}.

\subsubsection{Correlated critical exponents}

The study of the action in Eq.\,\eqref{qlr_eaa} closely follows the classical case, as the same mechanism is found for the transition between the long-range and the short-range universality which occurs at $\sigma_{*}=2-\eta_{\rm sr}$ as in the classical case. For $\sigma>\sigma_{*}$ the effective action of quantum rotors models is isotropic and analytic in the momentum sector, then its flow equations are identical to the ones in the classical $d+1$ case\,\cite{codello2015critical}. For $\sigma<\sigma_{*}$ however the anisotropy between spatial and imaginary time dimensions produce novel flow equations for the effective potential and the wave-function renormalization $K_{k}$:
\begin{align}
\label{ep_f_eq}
&\partial_t \bar{U}_{k}=(d+z)\bar{U}_{k}(\bar{\rho})-(d+z-\sigma)\bar{\rho}\,\bar{U}'_{k}(\bar{\rho})\nonumber\\
&- \frac{\sigma}{2}(N-1)\frac{1-\frac{\eta_{\tau} z}{3\sigma+2d}}{1+\bar{U}'_{k}(\bar{\rho})}
-\frac{\sigma}{2}\frac{1-\frac{\eta_{\tau} z}{3\sigma+2d}}{1+\bar{U}'_{k}(\bar{\rho})+2\bar{\rho}\,\bar{U}''_{k}(\bar{\rho})},\\
\label{eta_f_eq}
-&\frac{\partial_{t}K_{k}}{K_{k}}=\eta_{\tau}=\frac{f(\tilde{\rho}_{0},\tilde{U}^{(2)}(\tilde{\rho}_{0}))(3\sigma+2 d)}{d+(3\sigma+d)(1+f(\tilde{\rho}_{0},\tilde{U}^{(2)}(\tilde{\rho}_{0})))}.
\end{align} 
In the derivation of Eqs.\,\eqref{ep_f_eq} and\,\eqref{eta_f_eq}
analytic terms in the spatial direction are discarded\,\cite{defenu2017criticality}, setting $Z_{2,k}=0$ in Eq.\,\eqref{qlr_eaa}, as their contributions to the RG running of other quantities remain very small up to $\sigma\simeq\sigma_{*}$, see the discussion in Sec.\,\ref{comp_mom_contr}.

Interestingly, the numerical study of quantum long-range $O(N)$ models appears to be more extended with respect to the classical case, even if probably not so detailed. Numerical simulations have been performed both for the quantum long-range Ising and $O(2)$ rotor models have been performed, yielding numerical curves for both the critical exponents $z$ and $\nu$, while confirming the mean-field result $\eta=2-\sigma$ also in the correlated regime\,\cite{sperstad2012quantum}. 
\begin{figure*}
\subfigure[Dynamical Critical Exponent]{\label{Fig7a}\includegraphics[width=.45\linewidth]{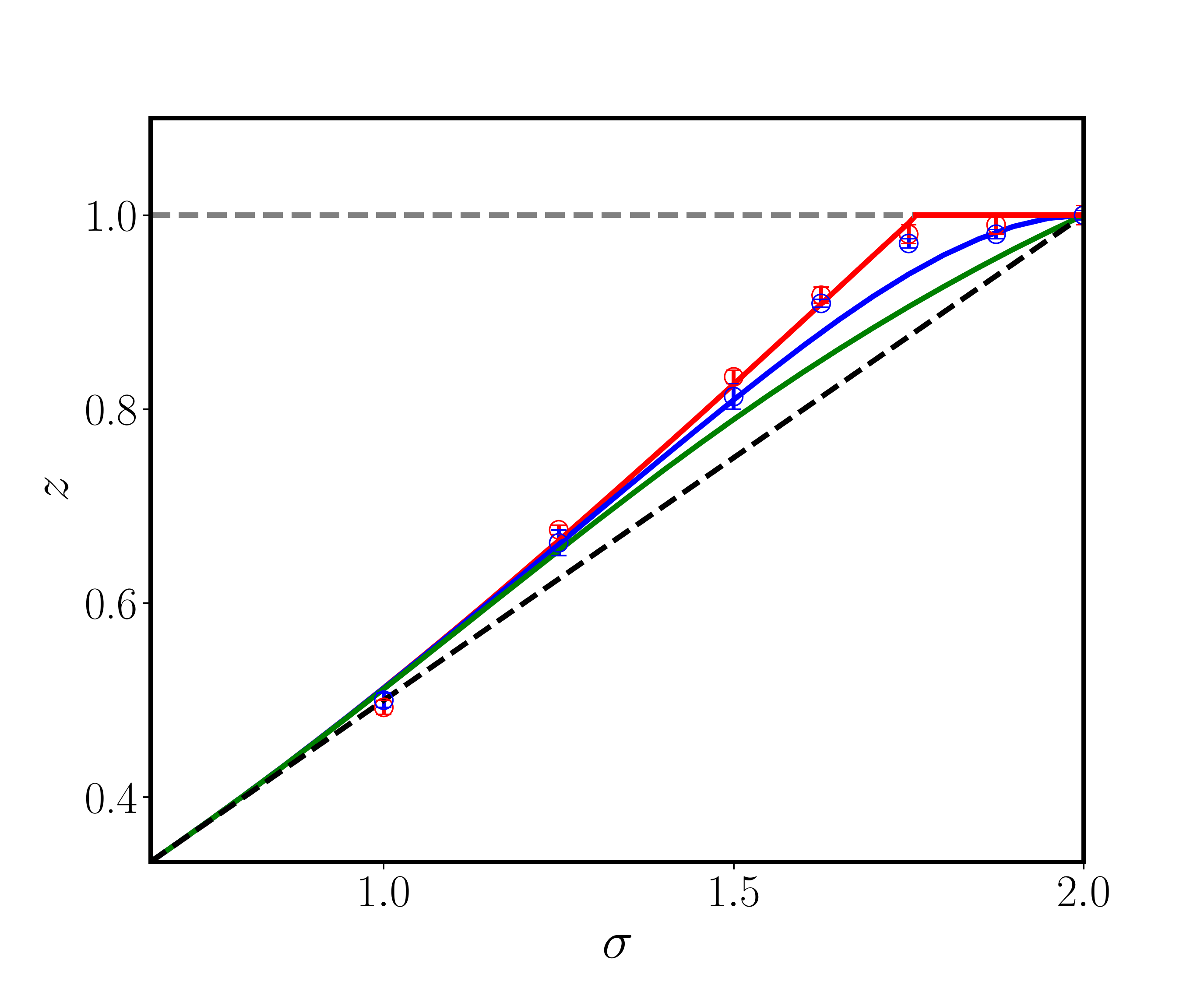}}
\hspace{1cm}
\subfigure[Correlation Length Exponent]{\label{Fig7b}\includegraphics[width=.45\linewidth]{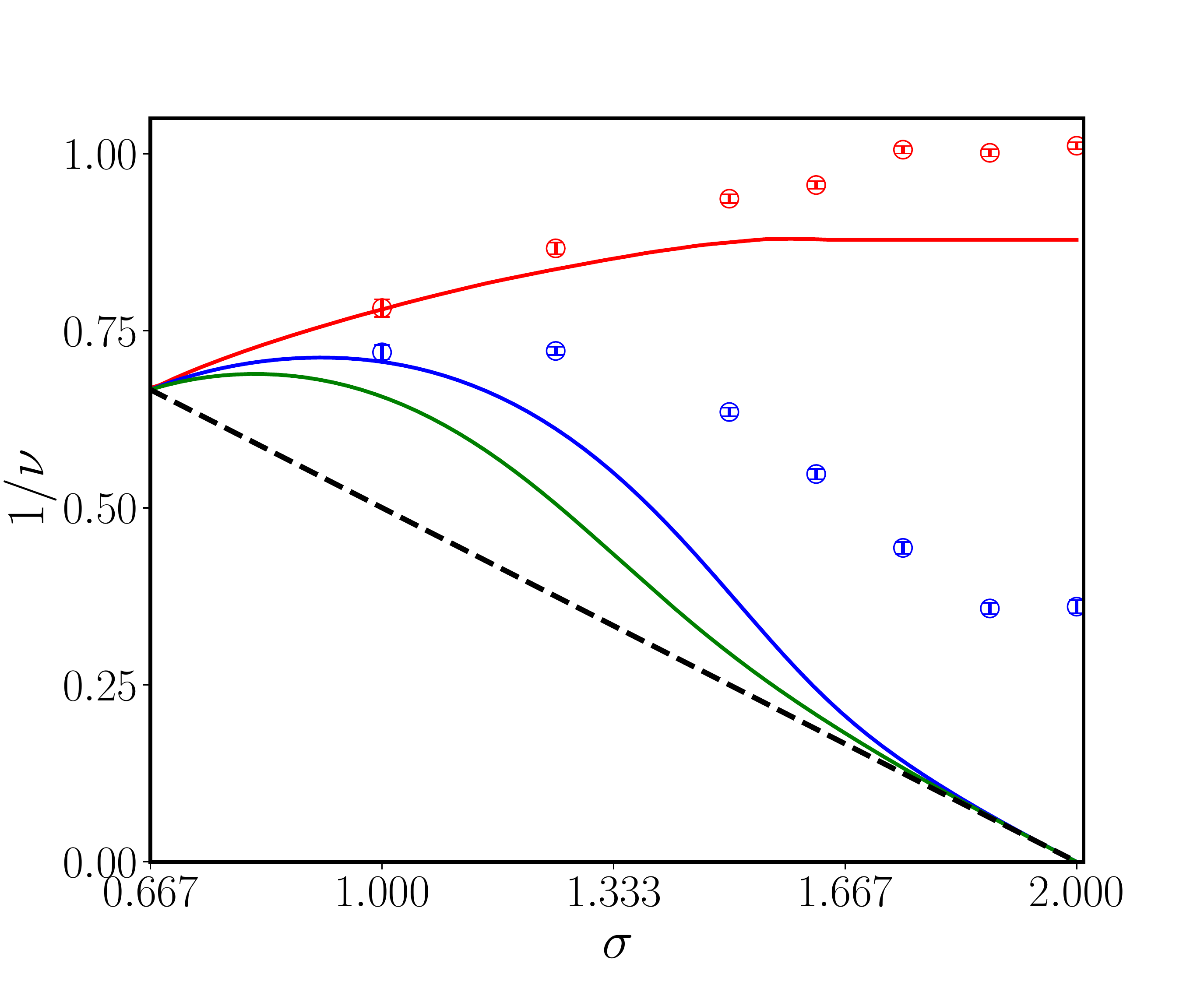}}
\caption{\textbf{Universal properties long-range quantum rotors.} \textbf{(a)} reports the estimates for the dynamical critical exponent $z=\sigma/(2-\eta_{\tau})$ obtained by the fixed point solution of the evolution Eqs.\,\eqref{ep_f_eq} and\,\eqref{eta_f_eq} in the cases $N=1,2,3$ in red, blue and green respectively. \textbf{(b)} reports the inverse correlation length exponent using the same color code. The MC simulations in Ref.\,\cite{sperstad2012quantum} are shown as empty circles in the $N=1,2$ cases in red and blue respectively.\label{Fig7}}
\end{figure*}
Fig.\,\ref{Fig7} compares the numerical estimates obtained by the flow Eqs.\,\eqref{ep_f_eq} and\,\eqref{eta_f_eq} using the solution approach described in Refs.\,\cite{defenu2015fixed,codello2015critical} with the results from MC simulations of Ref.\,\cite{sperstad2012quantum}.

In Fig.\,\ref{Fig7a} the dynamical critical exponent 
$z$ is reported as a function of $\sigma$ in $d=1$. These numerical results have been obtained solving Eqs.\,\eqref{ep_f_eq}, \,\eqref{eta_f_eq} at the fixed points and studying their stability matrix according as described in Refs.\,\cite{defenu2015fixed,codello2015critical}. The mean-field region $\sigma<\frac{2}{3}$ is not shown as it is exactly described by the analytical estimates in Eqs.\,\eqref{eta_mf},\,\eqref{z_mf} and\,\eqref{nu_mf}. Numerical results for $\sigma<1/2$ deviating from the mean-field expectation have not been reported\,\cite{fey2016critical}. 
The dynamical critical exponents of the transverse-field Ising model with long-range power-law interaction in the weak long-range regime has been derived in \cite{maghrebi2017continuous} up to the two-loop order within the renormalization group theory.
Recent QMC simulation have shown substantial agreement with the behaviours displayed in Fig.\,\ref{Fig7}\,\cite{koziol2021quantum}. It is worth noting that recent simulation have also targeted the finite temperature transition at $\alpha<d$\,\cite{gonzalezlazo2021finite}.

Out of the mean-field region, correlation effects 
tend to increase the value of the dynamical critical exponent, increasing the gap with the analytic prediction in Eq.\,\eqref{z_mf}. This effect is mitigated for continuous symmetries $N\geq2$ due to the vanishing of the anomalous dimension at the short-range threshold $\sigma_{*}=2$ . Accordingly, the agreement between the FRG curves and the numerical MC results (red solid line and circles in \ref{Fig7a})  remains consistent in the whole $\sigma$ range. On the other hand, the $N=2$ case displays overall poorer consistency, mostly due to the inaccuracy on MC estimates. Indeed, while the effective action parametrisation in Eq.\,\eqref{qlr_eaa} proved unable to properly describe the continuous BKT line\,\cite{graeter1995kosterlitz}, it consistently reproduces the scaling of critical exponents in the BKT limit\,\cite{codello2015critical}\footnote{It is worth noting that the power law scaling of BKT correlations originates from phase correlations and does not contradict the vanishing of the anomalous dimension defined according to Eq.\,\eqref{eta_f_eq}\,\cite{defenu2017non}.}. 

The lower accuracy found for the $N=2$ is confirmed by the comparison of MC simulation for the correlation length estimates with the FRG curve (blue circles and line in Fig.\,\ref{Fig7b}). Indeed, the MC data provide a finite correlation length exponent in the limit $\sigma\to 2$ for the $O(2)$ model, in contradiction with exact analytical predictions from $d_{\rm eff}-2$ expansion\,\cite{Brezin1976}. On the contrary, the FRG curve correctly reproduces the expected feature as it did in the classical case, see Fig.\,\ref{Fig3}. Therefore, the flow Eqs.\,\eqref{ep_f_eq} and\,\eqref{eta_f_eq} yield all the qualitative features and reach quantitative accuracy for all values $d$, $\sigma$ and $N$ in the phase diagram of quantum long-range $O(N)$ models, producing nice accuracy with exact numerical simulations.  The difficulties in the FRG characterisation of the BKT transition\,\cite{graeter1995kosterlitz,jakubczyk2014reexamination,jakubczyk2017longitudinal}  appear not to be problematic in this case, as MC simulations are as well plagued by severe finite size effects. 

An interesting fact is that the MC points in Fig.\,\ref{Fig7}appear provide $\sigma_{*}=2-\eta_{\rm sr}$ with $\eta=\frac{1}{4}$ also in the $N=2$, without any apparent distinction between the $N=1$ and $2$ cases. As already mentioned, this is in stark contradiction with the picture furnished by FRG, where one has $\sigma_{*}=2$ identically for all continuous symmetries. The correct picture is most likely in between, as suggested by the analysis pursued in Sec.\,\ref{wlr_classical_bkt}.

\subsection{Kitaev chain}
\label{subsec_kit_chain}
The introduction of long-range couplings in Fermi systems produces radically different results with respect to the bosonic case. The Kitaev
chain\,\cite{kitaev2001unpaired}
emerged
as on of the most studied playground in which effects of long-range terms
have been investigated. 
In the fermionic context we will first consider the generalized Kac-normalized\cite{kac1963van} long-range Kitaev chain\,\cite{maity2019one}. Its Hamiltonian reads
\begin{align}
\label{h_lrkc}
H=&\,-\sum_{j=1}^N\sum_{r=1}^{R}\big(J_rc_j^\dagger c_{j+r}+\Delta_rc_j^\dagger c_{j+r}^\dagger+\text{H.c.}\big)\\
&\,-h\sum_{j=1}^N\big(1-2c_j^\dagger c_j\big),
\end{align}
where
\begin{align}
\label{rs_ph}
&J_r=t\frac{d_{r}^{-\alpha}}{\mathcal{N}_\alpha},\,\,\,\,\,\,\,\Delta_r=g\frac{d_{r}^{-\beta}}{\mathcal{N}_\beta},
\end{align}
are the hopping and pairing profiles, respectively, with positive hopping $t>0$ and normalization satisfy one of the two relations
\begin{align}
\label{rescalings}
&\mathcal{N}_x=\begin{cases}
&\sum_{r=1}^{R}d_{r}^{-x},\,\,\mathrm{Kac\,\,rescaling}\\
&1\,\,\mathrm{otherwise}
\end{cases}
\end{align}
where $R$ denotes the range of the interactions, $d_{r}$ is the distance between the sites $i$ and $i+r$, $\alpha,\beta$  are the power-law exponents, $h$ the chemical-potential strength, and $c_j,c_j^\dagger$ the fermionic annihilation and creation operators, which obey the canonical anti-commutation relations $\{c_l,c_j^\dagger\}=\delta_{l,j}$ and $\{c_l,c_j\}=0$. 
In Eq.\,\eqref{rescalings} we allowed both the possibility to implement Kac rescaling or to leave the couplings unscaled as in the literature both conventions are employed.

The definition of distance depends on the choice of the boundary conditions. So, a ring structure, i.e. closed boundary conditions, lead to the definition $d_{r}=\min(r,L-r)$, while open boundary conditions simply produce $d_{r}=r$. Conventionally, closed boundary conditions allow straightforward analytical solution of the problem in the short-range limit. Yet, long-range couplings extending over the whole chain length will lead to the cancellation of the hopping (pairing) operators for anti-periodic (periodic) boundary conditions, due to the anti-commutation relations\,\cite{alecce2017extended}. This issue justifies the introduction of a finite interaction range $R$ into the Hamiltonian in Eq.\,\eqref{h_lrkc}. 

In the following we are going to mainly discuss the ring convention with $d_{r}=\min(r,L-r)$ and fix $R=N/2-1$. This choice allows us to adequately deal with closed boundary conditions, but still obtain a non-trivial thermodynamic limit $N\to\infty$, where the couplings display infinite range tails.
One can thus introduce the Fourier transform
\begin{align}
c_j=\frac{1}{\sqrt{N}}\sum_k^\text{B.z.}c_k\mathrm{e}^{\mathrm{i}kj}.
\end{align} 
On a finite ring the values of the momenta have to be chosen in order to comply with periodic ($k=\frac{2\pi n}{N}$) or anti-periodic ($k=\frac{2\pi (n+1/2)}{N}$) boundary conditions. The Hamiltonian in momentum space reads
\begin{align}
H=&\,\sum_k^\text{B.z.}\,\big[(c_k^\dagger c_k-c_{-k}c_{-k}^\dagger)(h-J_k)\nonumber\\\label{h_lrkc_ms}
&\,+(c_k^\dagger c_{-k}^\dagger-c_{k}c_{-k})\Delta_k\big],
\end{align}
where the momentum space couplings have been obtained by Fourier transforming $J_r$ and $\Delta_r$ 
\begin{align}\label{jk_fs}
J_k &=\frac{t}{\mathcal{N}_{\alpha}}\sum_{r=1}^{R}\frac{\cos(kr)}{r^\alpha},\\\label{dk_fs}
\Delta_k &=\frac{g}{\mathcal{N}_{\beta}}\sum_{r=1}^{R}\frac{\sin(kr)}{r^\beta},
\end{align}
The Hamiltonian in Eq.\,\eqref{h_lrkc_ms} is quadratic and it can be explicitly diagonalised via a Bogoliubov transformation
\begin{align}
\label{bg_trans}
c_k=\mathrm{i}\sin\frac{\theta_k}{2}\gamma_k+\cos\frac{\theta_k}{2}\gamma_{-k}^\dagger,
\end{align}
where $\gamma_k,\gamma_k^\dagger$ are fermionic  operators, which, respectively, annihilate and create Bogoliubov quasi-particles. They obey the conventional anti-commutation relations $\{\gamma_k,\gamma_p^\dagger\}=\delta_{k,p}$ and $\{\gamma_k,\gamma_p\}=0$.
The proper choice for the momentum dependent angle $\theta_{k}$ in oder to diagonalise Hamiltonian in Eq.\,\eqref{h_lrkc_ms} reads
\begin{align}
\label{bg_angle}
\theta_k=\arctan\frac{\Delta_k}{h-J_k}.
\end{align}
which leads to the diagonal Hamiltonian
\begin{align}
\label{h_lrkc_diag}
H=&\,\sum_k^\text{B.z.}\omega_k(\gamma_k^\dagger \gamma_k-\gamma_{-k}\gamma_{-k}^\dagger),
\end{align}
with the quasi-particle spectrum
\begin{align}
\label{lrkc_spec}
\omega_k=&\,\sqrt{(h-J_k)^2+\Delta_k^2}.
\end{align}

In the thermodynamic limit $N\to\infty$, the short-range model ($\alpha,\beta\to\infty$) features the familiar relations $J_{k}=t\cos(k)$ and $\Delta_{k}=g\sin(k)$. Accordingly, the minimal gap occurs at $k=0,\pi$, depending on the sign of $h$, and vanishes as the chemical potential approaches the critical values $h\to \pm t$. Interestingly, the two short-range critical points $h=\pm t$ feature a soft mode at respectively $k=0,\pi$, in correspondence with the appearance of ferromagnetic or antiferromagnetic order in the short-range Ising chain obtained by Jordan-Wigner transformation\,\cite{fradkin2013field}. Yet, the in terms of the fermionic operators of the Kitaev chain does no local order is found, but the quantum critical points divide different topological phases, where only non-local string orders are found\,\cite{chitov2018local}.

Without loss of generality, we can impose $t=g=1$ from now on, fixing the location of the short-range critical point. Upon crossing the critical point the system undergoes a quantum phase transition between a topologically trivial phase at $|h|>1$ and one featuring a finite winding number
\begin{align}
\label{w_def}
w=\frac{1}{2\pi}\oint d\theta_{k}
\end{align}
where the integral has to be taken along the periodic Brillouin zone.

In terms of topological properties the quantum phase transition occurs between the trivial phase $w=0$ at $|h|>1$ and the topologically non trivial phase at $|h|<1$. The existence of a non-trivial topological order in the bulk of the system is connected with the occurrence of zero energy Majorana modes at the boundaries with the normal phase. In particular, such zero energy Majorana modes are found at the edges of the finite chain with open boundaries\,\cite{kitaev2001unpaired}. The inclusion of interactions beyond the nearest-neighbours case radically modify and extend this traditional picture.

Before continuing such discussion, we observe that the use of open boundary conditions allow to
predict that the edge modes are exponentially localized
at the chain edges in the isotropic case, i.e. when
pairing and tunneling rates are equal, i.e. $\alpha=\beta$ \cite{jaeger2020edge}. 
Algebraic decay of the edge modes is found in the anisotropic case, when
either the exponent and/or the rates of tunneling and
pairing are different.
In this latter case, the smallest exponent causes the algebraic scaling of the tails, 
while at short distances the decay is exponential.

For power-law decaying superconducting pairings, 
the massless Majorana modes at the edges pair 
into a massive non-local Dirac fermion localized at both edges
of the chain dubbed topological massive Dirac fermion. 
with fractional topological numbers \cite{viyuela2016topological}.

It is worth noting that
signatures of Majorana edge modes have been 
studied in ferromagnetic atomic chains on top of superconducting leads\,\cite{nadj2014observation}. In this context, the realisation of power-law decaying couplings via Ruderman-Kittel-Kasuya-Yosida interactions has been proposed\,\cite{klinovaja2013topological}.

\subsubsection{Finite range couplings}

As usual, finite range interactions with $R<\infty$ cannot alter the universal critical scaling close to the quantum phase transition, but they may alter the topological phase diagram, leading to modifications in the number and properties of the edge modes. However, this is not the case if finite range interactions only appear in the hopping or the pairing sector separately, i.e. $\beta\to\infty$ or $\alpha\to\infty$ respectively. There the phase diagram remain almost unaltered with respect  to short-range case, apart for a modification of the critical boundaries, which become anisotropic, with the $k=0,\pi$ instabilities occurring at different values of $|h|$.

For a generic values of $\alpha$ and $\beta$, the topological phase diagram also contains regions with $w>\pm1$, with a maximum value equal to the range of the interactions $|w|_{\rm max}=R$. The range of parameters in which $w$ is maximum decreases with $\alpha$ and the phase diagram of the standard Kitaev chain model is recovered in the $\alpha\to\infty$ limit, independently of $\beta$. Interestingly, the winding number may also assume intermediate values between 1 and $R$ with steps of $2$. Therefore, for $R\in 2\mathbb{N} (2\mathbb{N}+1)$, the phase can be trivial, $w=0$, it can feature a pair of Majorana edge modes, $w=1$, or any even(odd) number of Majorana pairs smaller than the interaction range, $r\in 2\mathbb{N}(2\mathbb{N}+1)<R$.

The separation into even and odd numbers of Majorana modes depending on the range $R$ is justified by the possibility for Majorana modes on the same edge to annihilate each others one by one per edge, according to the mechanism described in Ref.\,\cite{alecce2017extended}. The topological phase with  $w = 1$, instead, persists for each interaction range $R\geq 1$, because the annihilation of a single Majorana pairs requires an overlap between the two
wave functions peaked the opposite edges of the chain.

In general, the influence of long-range interactions on topology has also been investigated for 
infinite-range couplings (see Sec.\,\ref{irp}) in antiferromagnetic spin-$1$ 
chains where the $\alpha_*$ for the survival of the topological phase 
strongly depends on the frustrated or unfrustrated nature of the long-range terms, i.e. 
$\alpha \simeq 0$ or $3$ \cite{gong2016topological, gong2016kaleidoscope}.
Moreover, the interplay between topology and long-range connectivity generates a wide range
of peculiar phenomena, including novel quantum phases \cite{gong2016kaleidoscope},
modifications of the area law \cite{gong2017entanglement}, and breaking of the Lieb-Robinson theorem \cite{maghrebi2016causality}.

\subsubsection{Infinite range pairing}
\label{irp}

First studies\,\cite{vodola2014kitaev} on the long-range Kitaev chain have been focusing on the case of infinite range long-range coupings $R=\infty$ only in the paring sector, leading to the thermodynamic limit expressions
\begin{align}
\label{jk_sr}
J_k&=\cos(kr),\\
\label{dk_lr}
\Delta_k&=\frac{1}{\mathcal{N}_{\beta}}\sum_{r=1}^\infty\frac{\sin(kr)}{r^\beta}=\frac{\mathrm{Im[\,Li}_\beta\left(e^{ik}\right)]}{2\zeta(\beta)},
\end{align}
where the case $\mathcal{N}_{\beta}=1$ is discussed first.
In absence of Kac rescaling, the critical line at $h=-1$ appearing in the short-range models persists independently of $\beta$, while the one at  $h=1$ disappears as soon as $\beta<1$. Notably, some references 
discuss
the persistence of the $h=-1$ critical line below $\alpha=1$ to prove that the long-range Kitaev chain does not require Kac rescaling\,\cite{lepori2015effective,lepori2017long}. 
Subsequent work clarified that the ground state energy of the system
\begin{align}
e_{\infty,\beta}=\int_{-\pi}^{\pi}\omega_{k}\,dk 
\end{align}
remains finite for all $\beta$ and $h$, due to the fermionic nature of the model and differently from the classical case. Yet, the zero momentum spectrum diverges $\lim_{k\to 0}\omega_{k}\to\infty$ for $\beta<1$ leading to the disappearance of the quantum critical point at $h=1$, which could be made stable by the introduction of Kac rescaling as in the classical case, see Eq.\,\eqref{rescalings}. This whole picture is in loose agreement with the discussion in Sec.\,\ref{mod_phase}, where we have shown that for modulated phases, characterised with instability at finite momentum, no internal energy divergence is detected for decay exponent $\alpha<d$, while ferromagnetic models with homogeneous order need Kac rescaling.

Then, the divergence in $k = 0$ is the cause for the disappearance of the $h=1$ quantum critical point for $\alpha<1$. At the same time, at every finite $\alpha$ divergences in some k-derivatives for $\omega_{k}$ occur both at $k=0$ and at $k=\pi$\,\cite{vodola2014kitaev,lepori2015effective}, giving rise to intersting effects both in the correlations decay and the dynamics\,\cite{lepori2017singular}.  In particular, these divergencies generate several novel features in the equilibrium behaviour of the Kitaev chain, which may be summarised in the following main effects:
\begin{itemize}
\item Hybrid decay of the static correlations with intermediate range exponential part and power law tails\,\cite{lepori2017singular}, which can be connected to the existence of a Lieb-Robinson bound peculiar to long-range systems\,\cite{foss2015nearly, regemortel2016information,hernandez2017correlation}.
\item Breakdown of conformal invariance for $\beta<2$ has been found\,\cite{lepori2015effective}. Nevertheless, the scaling of the von-Neumann entropy fulfils the area law up to $\alpha=1$, as in the short-range limit ($\beta\to\infty$)\,\cite{eisert2010area}. At the critical point, also the central charge defined by the logarithmic correction to the von-Neumann entropy remains $c=1/2$ as in the short-range limit as well\,\cite{lepori2015effective}.
\item Below the threshold $\beta=1$ logarithmic corrections to the area-law have been found out of criticality, modelled by the formula
\begin{align}
S(\ell)=\frac{c_{\rm eff}}{6}\log(\ell)
\end{align}
where $\ell$ is the size of the bipartition\,\cite{vodola2014kitaev, vodola2015long}. Notably, this correction, which is identical to the one of short-range systems at criticality\,\cite{holzhey1994geometric,calabrese2004entanglement}, has been also found in the Ising model\,\cite{koffel2012entanglement}.
\item Again below $\beta = 1$, the topological phase at $\mu < 1$ the Majorana edge modes, which remained well separated in the short-range limit, shall hybridise and produce a massive Dirac mode, effectively lifting the ground state degeneracy present for $\beta>1$. This mechanism is analogous to the one occurring in the short-range limit at finite size\,\cite{kitaev2001unpaired}. An
  explicit proof of this fact has been given in Ref.\,\cite{patrick2017topological} for $\alpha=\beta=0$.
\end{itemize}
 
All these striking features are also found in the general $\alpha,\beta<\infty$ case, almost independently from the value of $\alpha$\,\cite{alecce2017extended,vodola2015long,lepori2017long} and they can be straightforwardly reproduced by an continuous effective field theory description\,\cite{lepori2015effective}. Therefore, all the aforementioned properties can be classified as universal according to our definition. It is worth noting that the peculiar nature of the long-range Kitaev chain at $\beta<1$ is signalled by a non-integer value of the winding number defined in Eq.\,\eqref{w_def}, which is principle not admissible. This effect points towards a general breakdown of the traditional theory for topological phases  in short-range systems\,\cite{schnyder2008classification,kitaev2009periodic}, leading to modifications in the bulk-edge correspondance\,\cite{lepori2017long}.

\subsubsection{The $\alpha=\beta$ case and the relation with the long-range Ising model}
\label{lrkc_i_rel}

The topological features of the $\alpha<\infty$ are not substantially different from the $\alpha\to\infty$ case, as it is the paring term in Eq.\,\eqref{h_lrkc} that induces the topological behaviour. Yet, the presence of long-range hopping substantially alters both the critical and the dynamical properties of the long-range Kitaev chain. Before, discussing such properties, it is convenient to briefly discuss the case $\alpha=\beta$, which is strongly tied with the case of $1/2$-spins. In this perspective, it is convenient to first introduce the long-range Ising model Hamiltonian
\begin{align}
\label{h_lri}
H=-\sum_{l<j}V_{|l-j|} \sigma^x_l \sigma^x_j-h\sum_j \sigma^z_j,
\end{align}
where $\sigma^{\{x,y,z\}}_j$ are the Pauli spin matrices on site $j$, $h$ is the transverse-field strength, and $V_{r}$ is the spin coupling profile with power-law scaling
($\propto1/r^\alpha$, $\alpha\geq0$). As usual, in the limit $\alpha\to\infty$ one recovers the short-range Ising model, which is integrable and can be exactly solved with a Jordan-Wigner transformation\,\cite{fradkin2013field}. Another integrable limit is reached for $\alpha\to0$, where the Hamiltonian in Eq.\,\eqref{h_lri} represents the celebrated Lipkin-Meshkov-Glick model\,\cite{lipkin1965validity, meshkov1965validity, glick1965validity}. In this limit, the flat infinite-range interactions lead to permutation symmetry and allow to employ the Dicke basis\,\cite{nussenzveig1973introduction}, which scales linearly with the system size and yields a tractable description of the system amenable via exact diagonalization. 

The equilibrium phase diagram of the Hamiltonian in Eq.\,\eqref{h_lri} as well as its universal properties have been depicted in Sec.\,\ref{q_rot_mod} in the case $N=d=1$. In summary, the system displays a finite temperature phase transition for $\alpha<2$\,\cite{dyson1969existence,thouless1969long,dutta2001phase} within the same universality class of the classical long-range Ising model\,\cite{defenu2015fixed}. In the quantum limit $T\to0$, the system displays a quantum critical point at finite $h$, whose universal properties  depend on the value of $\sigma$ according to Fig.\,\ref{Fig6}\,\cite{defenu2017criticality}. In the nearest-neighbour limit $\alpha\to\infty$ the universal behaviour exactly corresponds with the ones of the Kitaev chain with $\alpha=\beta>2$, as a consequence of the Jordan-Wigner mapping.

Therefore, one may expect that a qualitative understanding of the Hamiltonian in Eq.\,\eqref{h_lri} shall result from the mapping of the spin operators  $\sigma^{\{x,y,z\}}_j$ onto fermions\,\cite{jaschke2017critical,vanderstraeten2018quasiparticles}
\begin{align}
\label{sz_op}
\sigma_j^z&=1-2 c_j^\dagger c_j,\\
\sigma_j^y&=-\mathrm{i}\Big[\prod_{m=1}^{j-1}\big(1-2 c_m^\dagger
  c_m\big)\Big]\big(c_j-c_j^\dagger\big),\\
\sigma_j^x&=-\Big[\prod_{m=1}^{j-1}\big(1-2c_m^\dagger c_m\big)\Big]\big(
c_j+c_j^\dagger\big),
\end{align}
where the fermionic annihilation and creation operators are represented, respectively, by $c_j,c_j^\dagger$ and, according to the canonical anticommutation relations, one has $\{c_l,c_j\}=0$ and $\{c_l,c_j^\dagger\}=\delta_{l,j}$. The fermionic Hamiltonian for the long-range Ising model reads
\begin{align}\nonumber
  H=&\,-\sum_{l<j}V_{|l-j|}\big(c_l^\dagger-c_l\big)\Big[\prod_{n=l+1}^{j-1}\big(1-2
    c_n^\dagger c_n\big)\Big]\big(c_j^\dagger+c_j\big)\\\label{h_lri_f}
&-h\sum_j\big(1-2 c_j^\dagger c_j\big).
\end{align}
An exact solution of the Hamiltonian in Eq.\,\eqref{h_lri_f} is not possible due to the inclusion of increasingly longer fermionic strings, due to the extended interaction range. In order to introduce a treatable model, we employ the approximation
\begin{align}\label{approx}
\prod_{n=l+1}^{j-1}\big(1-2c_n^\dagger c_n\big)=1,
\end{align}
for every $j\geq l+2$ and neglect all the non-quadratic string operators in the first line of Eq.\,\eqref{h_lri_f}. The resulting Hamiltonian reads
\begin{align}\nonumber
  H=&\,-\sum_{l<j}V_{|l-j|}\big(c_l^\dagger c_j+ c_l^\dagger c_j^\dagger-
  c_l c_j- c_l c_j^\dagger\big)\\
&-h\sum_j\big(1-2 c_j^\dagger c_j\big),
\label{lrk_h_app_i}
\end{align}
which corresponds to the Hamiltonian in Eq.\,\eqref{h_lrkc}  in the infinite range limit $R\to\infty$ with identical hopping and pairing functions, i.e. $g=t$ and $\alpha=\beta$. 

In the nearest neighbour limit, the fermions in the Hamiltonian\,\eqref{lrk_h_app_i} can be interpreted as domain-walls in the spin language. Consistently, long-range interactions introduce an effective non quadratic coupling between such domain walls, which we have discarded via the introduction of the approximation in Eq.\,\eqref{approx}\cite{fradkin2013field}. Since the relevance of the quartic terms in Hamiltonian in Eq.\,\eqref{h_lri_f} crucially depends on the interaction range, it is not surprising that the approximation in Eq.\,\eqref{approx} alters the universal properties of the model and, then, the Hamiltonian in Eq.\,\eqref{lrk_h_app_i} does not lie in the same universality as the long-range Ising model for $\sigma=\alpha-1<2$. The difficulty to reproduce the universal properties of the long-range Ising model at small $\alpha$ with the purely fermionic Hamiltonian can be also understood via an effective dimension argument. 

According to Eq.\,\eqref{d_eff_q}, the long-range Ising model displays the effective dimension $d_{\rm eff}=1$ for $\alpha>3$ and, there, it is not surprising that the universal properties of the fermionic theory in Eq.\,\eqref{lrk_h_app_i} correspond to the ones of the effective bosonic theory described by the EAA in Eq.\,\eqref{qlr_eaa}. Conversely, for $\alpha<5/3$ the effective dimension becomes large, $d_{\rm eff}>4$, and the universal features of the effective action in Eq.\,\eqref{qlr_eaa} are exactly captured by the mean-field approximation, which features bosonic excitations and cannot be reduced to the purely Fermionic theory in Eq.\,\eqref{lrk_h_app_i}. In the intermediate range $5/3<\alpha<3$ the model is not solvable and the low-energy excitations shall posses hybrid fermionic-bosonic character, which cannot be captured by the purely fermionic Hamiltonian in Eq.\,\eqref{lrk_h_app_i}.

\subsubsection{The general $\alpha\neq\beta$ case}
\label{kit_ch_sec}

In Sec.\,\ref{lrkc_i_rel}, we have discussed the relation between the universal properties of the Ising model and the ones of the Kitaev chain with $\alpha=\beta$ and $t=g$. Now, we will explicitly derive the critical exponents of the Kitaev chain in the general case. Long-range couplings may alter the equilibrium universality class only in the infinite-range case, then we shall consider Eqs.\,\eqref{jk_fs} and\,\eqref{dk_fs} in the thermodynamic limit at $R=\infty$
\begin{align}\label{jk_fs_tl}
J_k=\frac{1}{\zeta(\alpha)}\sum_{r=1}^\infty\frac{\cos(kr)}{r^\alpha}=\frac{\mathrm{Re[\,Li}_\alpha\left(e^{ik}\right)]}{2\zeta(\alpha)},\\\label{dk_fs_tl}
\Delta_k=\frac{1}{\zeta(\beta)}\sum_{r=1}^\infty\frac{\sin(kr)}{r^\beta}=\frac{\mathrm{Im[\,Li}_\beta\left(e^{ik}\right)]}{2\zeta(\beta)},
\end{align}
which are the momentum range couplings determining the single particle spectrum in Eq.\,\eqref{lrkc_spec}. In analogy with the nearest-neighbour case the long-range Kitaev chain features two quantum critical points, corresponding to the softening of the $k=0$ or $k=\pi$ modes. Employing the Kac normalised expressions in Eqs.\,\eqref{jk_fs_tl} and\,\eqref{dk_fs_tl} the location of the "homogeneous" critical point is fixed at $h^{h}_{c}=1$ independently on the choice of $\alpha$ or $\beta$. Conversely, the $k=\pi$ instability occurs at the $\alpha$ dependent critical point $h^{a}_{c}=1-2^{\alpha}$. The definition of critical exponents is given by the scaling of the excitation spectrum close to each of these quantum critical points
\begin{align}
\label{lrkc_exp_1}
\lim_{h\to h^{h,a}_{c}}\omega_{k}&\approx |h-h_{c}|^{z\nu}\quad k=0,\pi\\
\label{lrkc_exp_2}
\lim_{k\to 0,\pi}\omega_{k}&\approx k^{z}\quad h=h^{h,c}_{c}.
\end{align}
As in the case of rotor models, see Sec.\,\ref{q_rot_mod}, the two exponents $z$ and $\nu$ are sufficient to characterise the entire critical scaling.

Following the definitions in Eqs.\,\eqref{lrkc_exp_1} , it is straightforward to check that $\lim_{k\to0}\Delta_{k}=0$ and that the critical exponents combination is $z\nu=1$ for each of the two quantum critical points irrespectively on the values of $\alpha,\beta$. The determination of the dynamical scaling exponent $z$ close to $h^{h}_{c}$ quantum critical point requires the expansions of the Fourier couplings close at $k=0$
\begin{align}
J_{k}&=1+\sin(\alpha\pi/2)\frac{\Gamma(1-\alpha)}{\zeta(\alpha)}k^{\alpha-1}\nonumber
\\&-\frac{\zeta(\alpha-2)}{2\zeta(\alpha)}k^{2}+O(k^{3})\quad\mathrm{if}\,\,\alpha<3,\label{j_exp1}\\
J_{k}&=1+\frac{2\log(k)-3}{4\zeta(3)}k^{2}+O(k^{3})\quad\mathrm{if}\,\,\alpha=3,\label{j_exp2}\\
J_{k}&=1-\frac{\zeta(\alpha-2)}{2\zeta(\alpha)}k^{2}+O(k^{\alpha-1})\quad\mathrm{if}\,\,\alpha>3,\label{j_exp3}
\end{align}
and
\begin{align}
\Delta_{k}&=\cos(\beta\pi/2)\frac{\Gamma(1-\beta)}{\zeta(\beta)}k^{\beta-1}\nonumber\\
&+\frac{\zeta(\beta-1)}{\zeta(\beta)}k +O(k^{3})\quad\mathrm{if}\,\,\beta<2,\label{d_exp1}\,.\\
\Delta_{k}&=\frac{6(1-\log(k))}{\pi^{2}}k +O(k^{3})\quad\mathrm{if}\,\,\beta=2,\label{d_exp2}\,.\\
\Delta_{k}&=\frac{\zeta(\beta-1)}{\zeta(\beta)}k+O(k^{\beta-1})\quad\mathrm{if}\,\,\beta>2,\label{d_exp3}\,.
\end{align}
Apart from their relevance to the present case, the expansions above display the typical example of anomalous terms in the excitation spectrum generated by long-range interactions. A close inspection of the expressions above leads to the following result for the equilibrium dynamical critical exponent 
\begin{equation}
\label{dyn_exp_lrkc}
z = \begin{cases}\phi-1 \quad \textrm{if}\,\,\phi < 2,\\
 1\quad \textrm{if}\,\,\phi>2,
 \end{cases}  
 \end{equation} 
where $\phi = \min(\alpha,\beta)$. According to the result in Eq.\,\eqref{dyn_exp_lrkc} the relevant region for long-range couplings in the long-range Kitaev chain radically differs from the case of $O(N)$ rotors model described in Sec.\,\ref{q_rot_mod}. Indeed, long-range interactions in the Kitaev chain remain irrelevant also in the range $2<\alpha,\beta<3$, while long-range couplings in rotor models would be relevant in the whole $\alpha<3$ region. Yet, it is worth noting that even if long-range hopping couplings with $2<\alpha<3$ do not alter the critical behaviour, they still introduce relevant momentum terms in the hopping sector. Such discrepancy yields further proof that the approximation in Eq.\,\eqref{approx} crucially alters the universal behaviour at small $\alpha,\beta$.
 
For the sake of the forthcoming discussion it is crucial to notice that long-range interactions with different power law exponents $\alpha\neq\beta$ modify the influence of the hopping and pairing term on the critical scaling. Indeed, while for short-range interactions the dynamical critical scaling exponentsis determined by the low-momentum terms in the pairing coupling, for relevant long-range interactions with $\alpha<\beta$ it is the scaling of the hopping coupling which determines $z$. A similar scenario may also occur for finite range competing interactions and it is known to cause peculiar dynamical features\,\cite{deng2009anomalous,divakaran2009defect,defenu2019universal}, which will be discussed in the following sections.

Still, the altered relation between hopping and pairing couplings also generate interesting equilibrium effects, which, up to our knowledge, have never been discussed in the literature.
As an example, let us consider the low-energy expression for the Bogoliubov angle at the critical point
\begin{align}
\label{lrki_occ}
\tan(\theta_{k})\propto\begin{cases}
\frac{k^{\beta}}{k^{\alpha}}\quad\mathrm{if}\,\,\alpha<3\,\,\textrm{and}\,\,\beta<2\\
\frac{k^{\beta}}{k^{3}} \quad\mathrm{if}\,\,\alpha>3\,\,\textrm{and}\,\,\beta<2\\
\frac{k^{2}}{k^{\alpha}} \quad\mathrm{if}\,\,\alpha<3\,\,\textrm{and}\,\,\beta>2\\
\frac{1}{k} \quad\mathrm{if}\,\,\alpha>3 \quad\textrm{and}\,\,\beta>2\\
\end{cases}
\end{align}  
 Then, the traditional short-range system is characterised by the divergence of the argument of the $\arctan$ in Eq.\,\eqref{bg_angle}, which, as a consequence, leads the Bogoliubov angle to approach the limit $\theta_{k=0}=\pi/2$. This values is placed exactly in between the normal phase value $\theta_{k=0}=0$ and the topologically ordered one $\theta_{k=0}=\pi$. Thus, the occupation of low-momentum modes at the critical point stays exactly in between the ones of the two phases for short-range couplings. However, long-range interactions alter such behaviour according to Eqs.\,\eqref{lrki_occ} and the critical point does not anymore differ from each of the two phases, but may belong to the topologically trivial phase, according to such characterisation. 
\begin{figure*}
\subfigure[$g>g_{c}^{h}$]{\label{Fig8a}\includegraphics[width=.32\linewidth]{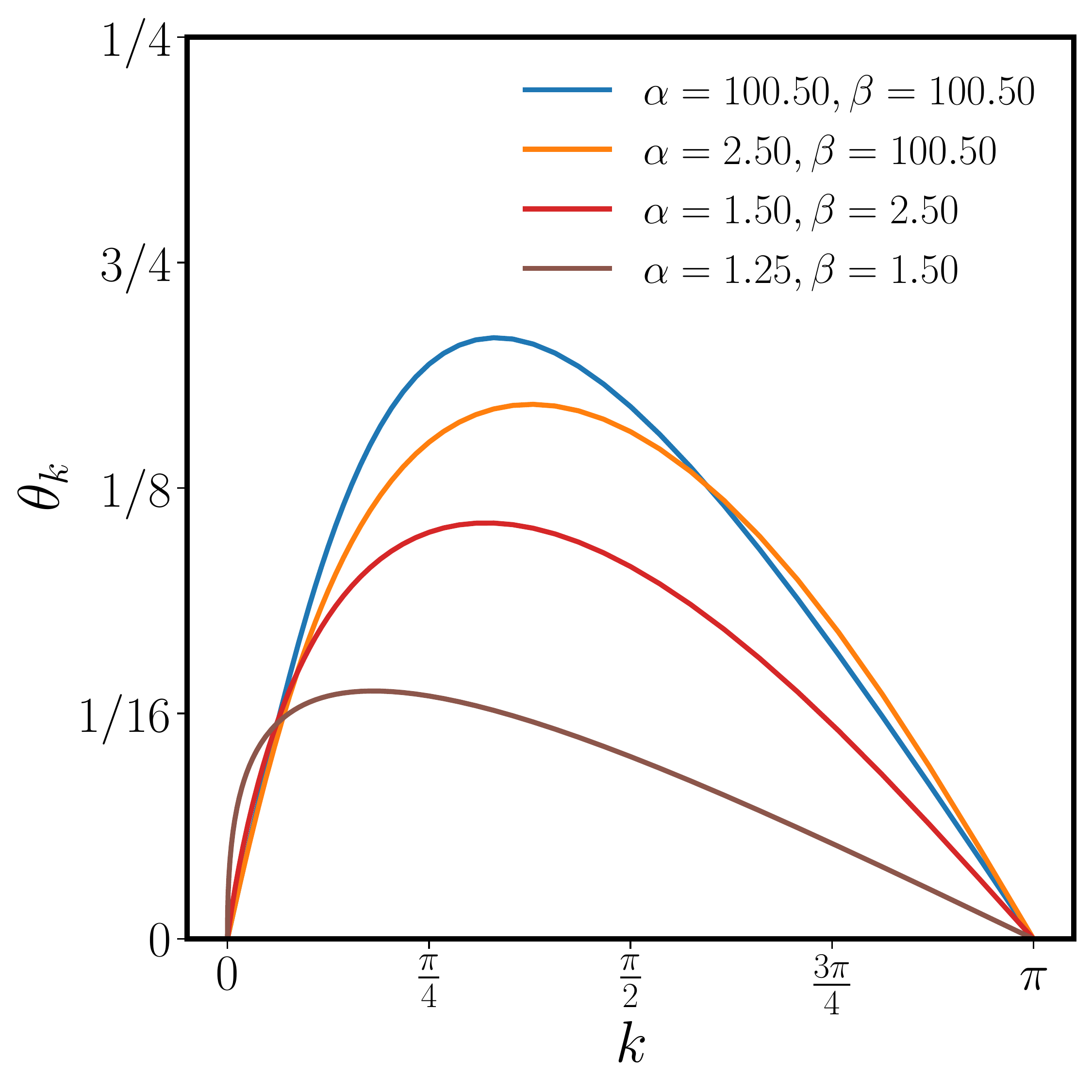}}
\subfigure[$g=g_{c}^{h}$]{\label{Fig8b}\includegraphics[width=.32\linewidth]{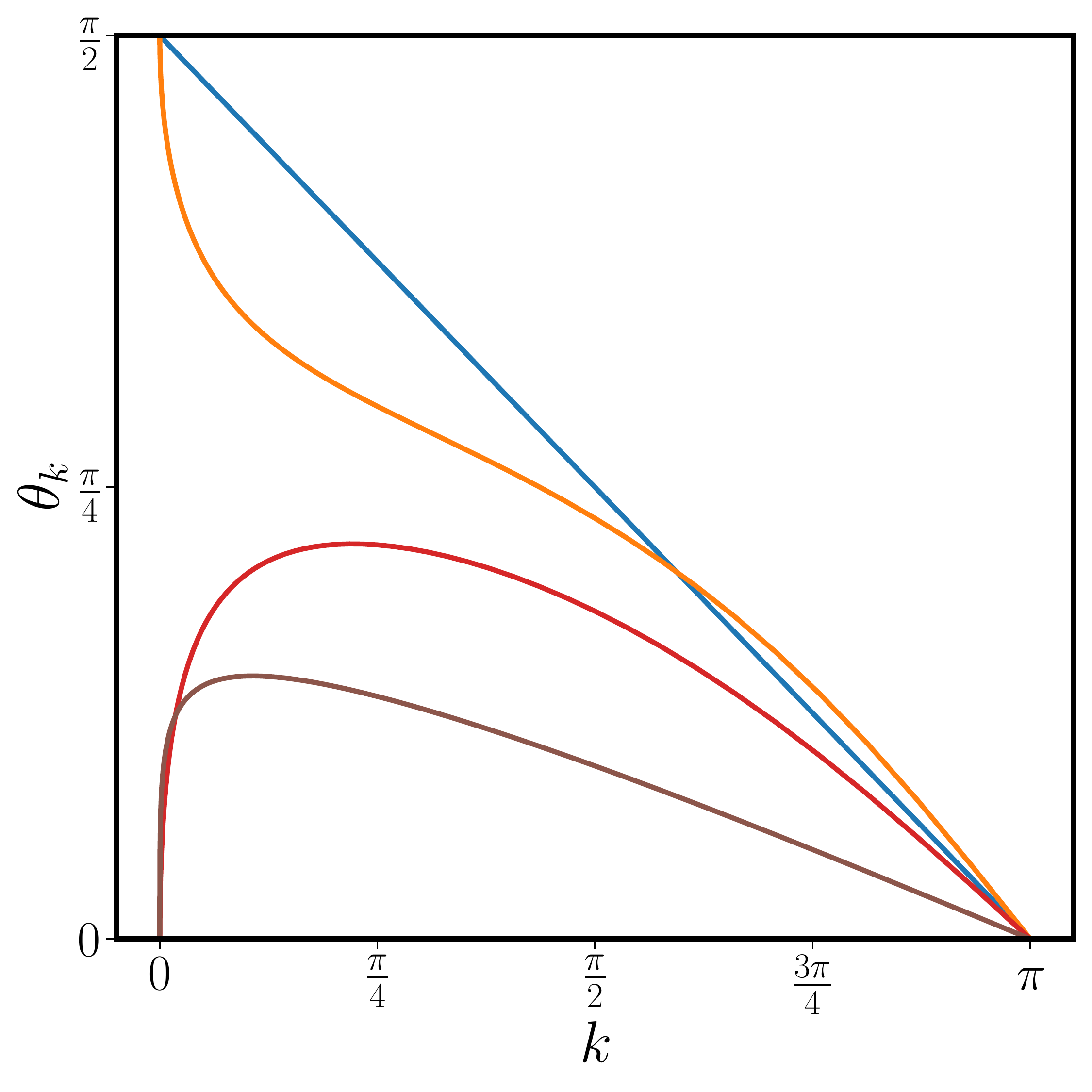}}
\subfigure[$g<g_{c}^{h}$]{\label{Fig8c}\includegraphics[width=.32\linewidth]{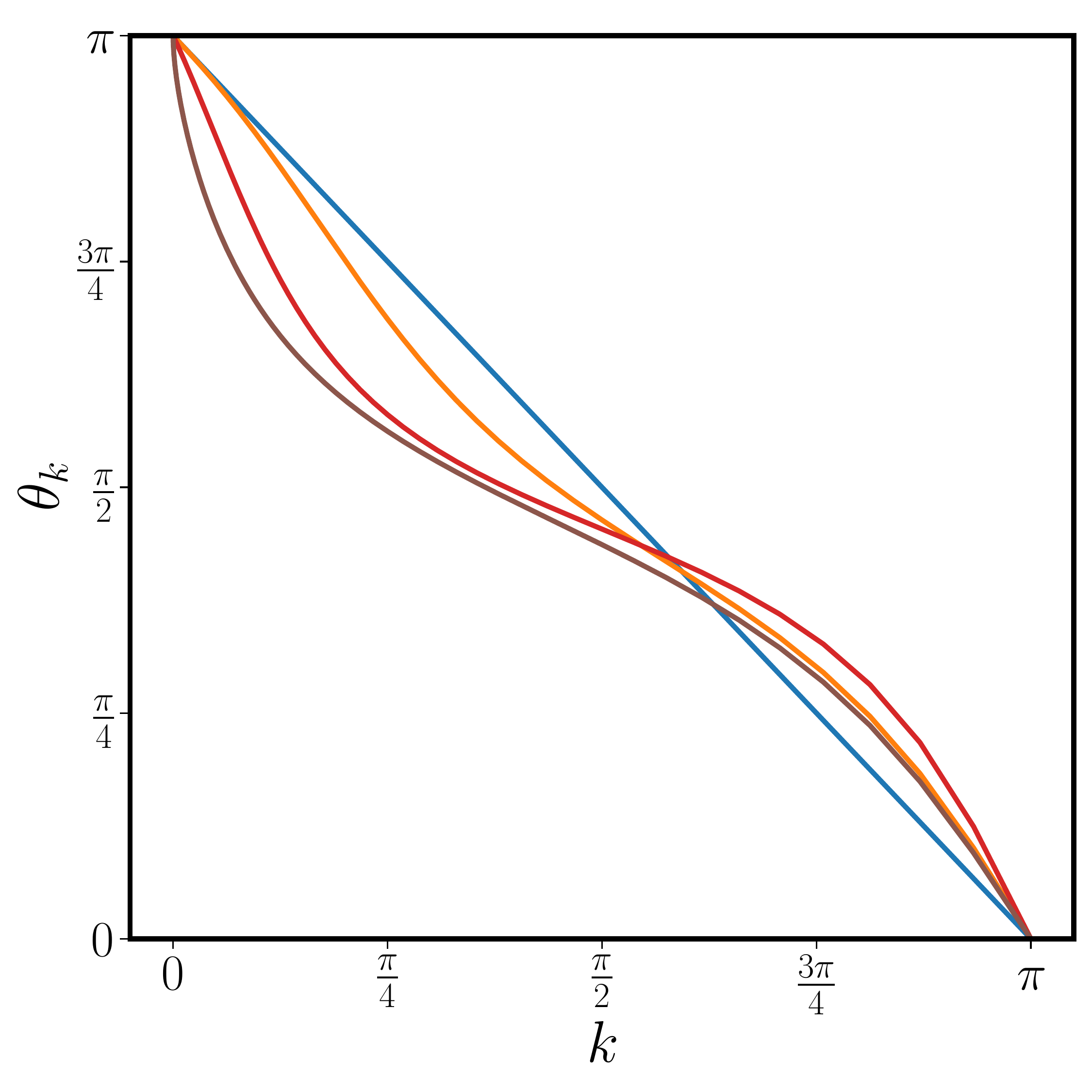}}
\caption{\textbf{The Bogoliubov angles of a long-range Kitaev chain.} They characterises the occupation number of the Bogoliubov quasi-particles via Eq.\,\eqref{bg_trans}, is displayed as a function of the quasi-particle momentum $k$. In the low- and high-temperature phases, in panel (a) and (c), the zero-momentum occupation is, respectively, $\pi$ or $0$ independently on the values of $\alpha$ and $\beta$. On the other hand, the critical point occupation, panel (b), in the $k\to 0$ limit attains the values $\pi/2$ or $0$ depending on the exponents ratio. \label{FigBogangles}}
\end{figure*}

The $\lim_{k\to 0}\theta_{k}=\pi/2$ is easily interpreted in terms of the Majorana edge modes of the finite chain. Indeed, in the thermodynamic limit, the spectral gap at $k=0$ vanishes and the Majorana modes at the two edges of the chain hybridise forming a massless Dirac mode. This picture looses its validity as soon as $\alpha<\beta<2$ or $\beta>2$ and $\alpha<2$, where the long-range hopping terms dominates with respect to the pairing coupling, see Eq.\,\eqref{lrki_occ}. Then, anisotropic long-range interactions may alter the low energy spectrum of the Kitaev chain, producing a purely electronic zero mode and remove the contribution from the Majorana edge modes from low-energy critical spectrum. This picture is consistent with the dependence of the Bogoliubov angle on the quasi-particle momentum displayed on Fig.\,\ref{FigBogangles}.

In summary, this section has delineated the equilibrium critical properties of quadratic fermion systems, with power-law decaying coupling of different decay rates. Yet, the same characterisation cannot be furnished in the case of fermionic systems with long-range non-quadratic interactions
\begin{align}
\label{if_h_dd}
H=\sum_{\langle ij\rangle,s}\left(c^{\dagger}_{i,s} c_{j,s}+\mathrm{h.c.}\right)+\sum_{i\neq j}V_{ij} n_{i} n_{j}
\end{align}
where the $c^{\dagger}_{i,s}$ operator and its conjugate create and annihilate a fermion with spin $s$ on the $i-th$ site of the lattice, while $n_{i}$ represents the total density operator on the same site. The comprehension of the influence of long-range density-density interactions on the critical behaviour of Fermi systems is still relatively obscure. One notable counter example is the one-dimensional case, where mapping of fermionic systems into bosonic or spin degrees of freedom is possible. 

In particular, the ground state of continuous 1d Fermions interacting via unscreened Coulomb repulsion was characterised by bosonization techniques, finding metallic features and a classical Wigner crystal phase with slow-decaying charge correlations\,\cite{schulz1993wigner, wang2001coulomb}. Numerical confirmation of such theoretical picture has been provided by density matrix renormalization group (DMRG)\,\cite{fano1999unscreened} and variational Monte Carlo methods\,\cite{casula2006ground,astrakharchik2011exact,lee2011ground}. The corresponding lattice systems with commensurate filling has been numerically shown to display an insulating ground-state, still with Wigner crystal character, in contradiction with the bosonization picture in the continuum\,\cite{poilblanc1997insulator, capponi2000effects}.

\subsection{XXZ models}
\label{sex_xxz}
The Hamiltonian of the long-range XXZ spin chain
 \begin{align}
 \label{h_xxz}
 H=\sum_{i>j}J_{ij}\left(-\sigma^{x}_{i} \sigma^{x}_{j}-\sigma^{y}_{i}\sigma^{y}_{j}+\sigma^{z}_{i} \sigma^{z}_{j}\right),
 \end{align}
 where $\sigma^{\mu}_{i}$ are quantum spin operators represented by the $\mu$-component of Pauli matrices and $J_{ij}\approx r_{ij}^{-\alpha}$ are the usual long-range couplings. Notice that in \eqref{h_xxz} all the couplings
 $x-x$, $y-y$ and $z-z$ are long-ranged. Putting the long-range
 couplings only in the $z-z$ directions corresponds actually to have hard-core
 bosons with long-range density-density interactions, see the next section
 for more details.
 
Conventionally, the solution in the $\alpha\to\infty$ limit is obtained through bosonization, which proves that the universal properties of the spin Hamiltonian in Eq.\,\eqref{h_xxz} are exactly described by the effective action of the quantum sine-Gordon model, which also describe the universality of $O(2)$ quantum rotors\,\cite{giamarchi2004quantum,sachdev1999quantum, fradkin2013field}.

However, such mapping is not possible in presence of long-range couplings. Nevertheless, one can split the Hamiltonian into long-range and short-range contributions and consider the long-range couplings only as a perturbation of the short-range action\,\cite{maghrebi2017continuous}, see also\,\cite{bermudez2016long}.
As a result one can consider the low energy action
\begin{align} 
\label{bkt_a_q}
S[\theta] &=  \frac{K}{2\pi u}\int d\tau\,dx\left\{  (\partial_{\tau}\theta)^2+ u^{2}(\partial_{x}\theta)^2\right\}\nonumber\\
&-g\int d\tau\int dx\,dy \frac{\cos(\theta(\tau,x)-\theta(\tau,y))}{|x-y|^{\alpha}}
\end{align} 
where $K$ is the so-called Luttinger parameter, $u$ is a velocity scale, and $g$ is the strength of long-range interactions term. Up to a rescaling of the spatial dimension $x$ the short-range part of the action in Eq.\,\eqref{bkt_a_q} corresponds to the one in Eq.\,\eqref{bkt_a}. This is not surprising since the two actions have been obtained under very similar premises. 

Despite this similarity, it worth noting that the long-range correction to the Hamiltonians in Eqs.\,\eqref{bkt_a_q} and\,\eqref{bkt_a} are not quite the same. Indeed, long-range interactions involve both spatial directions in the classical case. At variance, in the quantum case, part of the kinetic energy derives from fluctuations in the time direction, which is not influenced by long-range interactions.

Nevertheless, the final picture obtained for the critical behaviour of the action in Eq.\,\eqref{bkt_a_q} is analogous to the one discussed for the classical case. In fact, one can define the shifted decay exponent $\sigma=\alpha-d=\alpha-1$ and derive the flow equations
\begin{align}
\label{lr_bkt_fl_q}
\frac{dy_{k}}{dt}&=-(2-4K)y_{k}\nonumber\\
\frac{d\tilde{g}_{k}}{dt}&=-\left(2-\sigma-\frac{1}{2K}\right)\tilde{g}_{k}
\end{align}
where $y_{k}$ is the fugacity of topological excitations and $\tilde{g}_{k}$ the dimensionless long-range coupling, see the discussion in Sec.\,\ref{wlr_classical_bkt}. The phase diagram resulting from Eqs.\,\eqref{lr_bkt_fl_q} follows in close analogy the one obtained in the classical case, see Sec.\,\ref{wlr_classical_bkt}.

As long as $\sigma>2$ long-range interactions are irrelevant and the system displays universal BKT scaling. Conversely, for $\sigma<2$ a new phase emerges at large enough $K$, where the long-range couplings $\tilde{g}$ grow indefinitely. As a consequence a finite order parameter appears in the x-y plane $\langle \sigma^{+}\rangle\neq 0$ and the system undergoes spontaneous symmetry breaking. Evidence of this \emph{quasi-\,to true-} order transition have been also found in numerical density matrix renormalization group (DMRG) calculation. Indeed, computing the effective central charge of the model via an appropriate definition, Ref.\,\cite{maghrebi2017continuous} was able to show that this quantity changes from $c_{\rm eff}=c=1$, typical of the isotropic short-range sine-Gordon model\,\cite{mussardo2009statistical}, to $c_{\rm eff}>1$  at $\sigma<2$. Such change in the effective central charge is compatible with the appearance of a new phase with broken Lorentzian symmetry\,\cite{maghrebi2017continuous}. Correspondingly, also the dynamical critical exponents deviates from unity and acquires the expected value for anisotropic long-range field theories $z=\sigma/2$.Including the renormalization of the Luttinger parameter does not alter the aforementioned picture. 

By a thoughtful characterisation of the long-distance correlation functions, Ref.\,\cite{maghrebi2017continuous} could show that the ordered phase displays a finite correlation length $\xi$ that diverges exponentially as the critical point with the quasi-ordered phase is reached. Such exponential divergence is reminiscent of the behaviour of the correlation length at the BKT transition\,\cite{fradkin2013field}. Similar echoes of the essential BKT scaling in the ordered phase have been already evidenced in the classical case, see Eq.\,\eqref{scaling}.

\subsection{Hardcore bosons in 1d} 
\label{sec_hd_bosons}

In the section on the Kitaev chain, we already discussed the possibility to recover the homogeneous critical point of the Kitaev chain also for $\alpha,\beta<1$ by explicitly introducing Kac rescaling, contrary to existing studies\,\cite{vodola2014kitaev}. In the present section we are going to
review results on this matter by explicitly showing that the implementation (or not-implementation) of Kac rescaling, may significantly alter the equilibrium phase diagram of a long-range interacting quantum model.

Restricting our analysis to the one dimensional case, we can relate the findings discussed at the end of the above section with the study of hardcore bosons with arbitrary power-law interactions. The Hamiltonian under consideration reads
\begin{align}
H= -t \sum_{i=1}^{L} \left(c^{\dagger}_i  c_{i+1}  + {\rm h.c.} \right) + \sum_{i>j} V^{(\alpha)}_{ij} n_{i} n_{j},
\label{hcb_h}
\end{align}
with the usual power-law decaying potential 
\begin{align}
V^{(\alpha)}_{|i-j|}=\frac{1}{N_{\alpha}}\frac{V}{d_{i-j}^{\alpha}}  \quad V>0.
\label{in_pot_bh}
\end{align}
As in the Kitaev chain study presented in the above section, due to the quantum nature of the system, one can choose to implement or not Kac rescaling according to the physical situation, see Eq.\,\eqref{rescalings}. 
\begin{figure}[ht]
\includegraphics[width=\columnwidth]{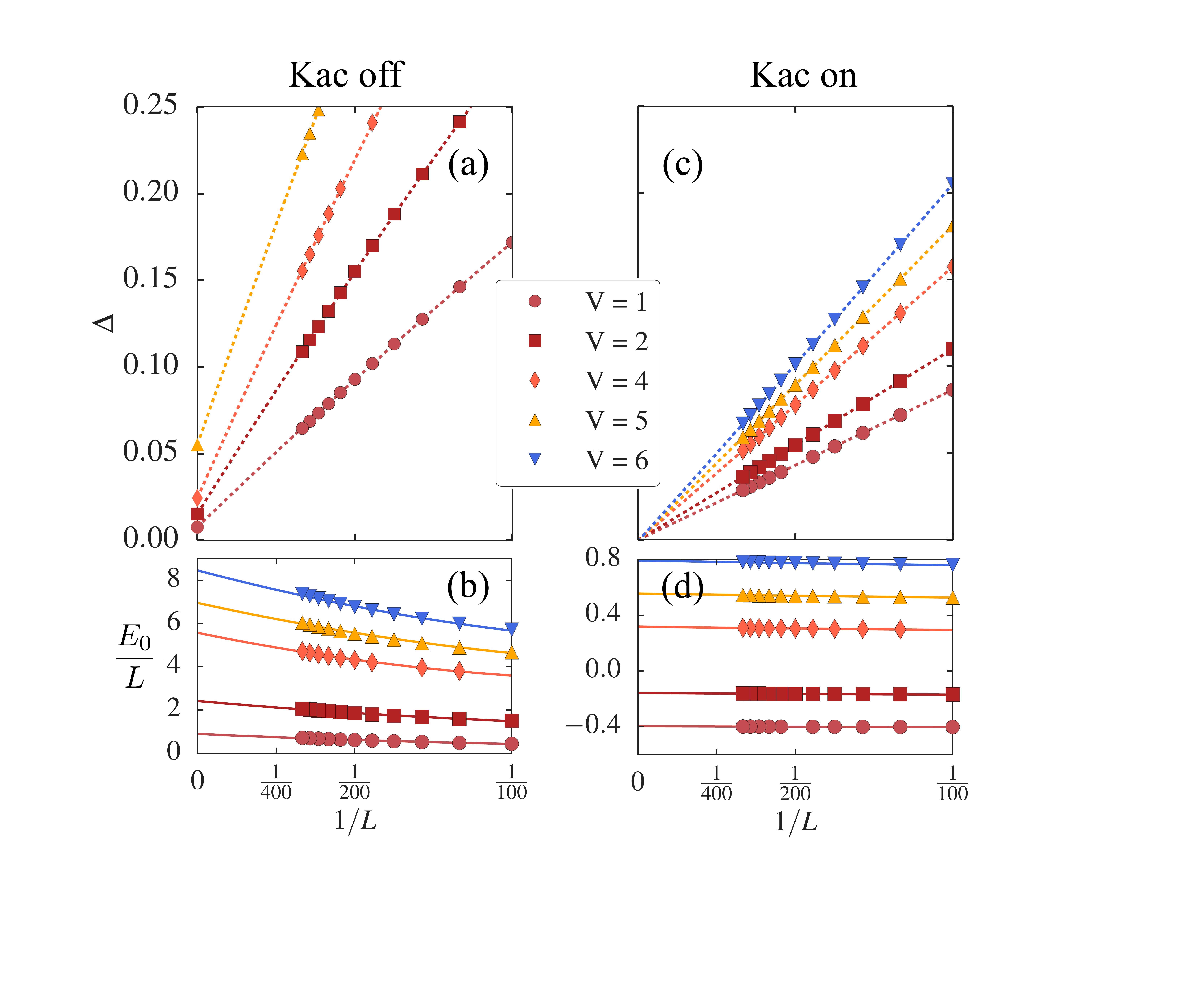}
\caption{\textbf{Comparison between Kac-on and Kac-off finite size scalings for hard-core bosons.} The scaling of the bosons single particle energy, as defined in Eq.\,\eqref{s_p_gap}, as a function of the system size. The results have been obtained via a DRMG computation at half-filling $\langle n_{i} \rangle =0.5$, for $\alpha = 1$ and different interaction strengths $V$ (in units of the hopping energy $t$).  The difference between the Kac scaled or unscaled scenario is rather evident, as in the first case the single particle energy always vanishes in the thermodynamic limit, while in the second case the system remains gapped up to the thermodynamic limit. }
\label{Fig9}
\end{figure}
DRMG simulations have been performed on the Hamiltonian in Eq.\,\eqref{hcb_h} in order to characterise the phase of the system. In particular, it has beed found that the single particle gap
\begin{align}
\label{s_p_gap}
\Delta(N)=E_{0}(N+1)+E_{0}(N-1)-2E_{0}(N)
\end{align}
displays radically different behaviours, depending on the implementation (or non-implementation) of the Kac rescaling in the interaction potential choice the situation changes radically, as it appears from the numerical results reported in Fig.\,\ref{Fig9}. 

In particular, the numerical simulations in absence of Kac rescaling predict a finite single-particle gap in the thermodynamic limit, which is consistent with an insulating phase for all values of the interaction coupling $V$ of the potential in Eq.\,\eqref{in_pot_bh}. This scenario has been first evidenced in Ref.\,\cite{capponi2000effects} for $\alpha=1$ and then confirmed, in the general $\alpha$ case, by the simulations discussed in Ref.\,\cite{botzung2021effects}. Conversely, the implementation of Kac rescaling induces metallic behaviour in the entire range $0\leq\alpha\leq1$ independently on the interaction coupling $V>0$. This proves that the restoration of an extensive interaction energy significantly alters the phase diagram of Hamiltonian \eqref{hcb_h}.

Theoretical understanding of the discrepancy between the scaled and unscaled theory can be obtained via the Luttinger liquid theory\,\cite{giamarchi2004quantum}, which reduces the universal behaviour of the Hamiltonian in Eq.\,\eqref{hcb_h} to the one of the continuous action
\begin{align}
\label{ll_h}
H_{\rm LL} &=  \frac{u}{2\pi}\int \,dx\left\{ K(\pi\Pi)^{2}+\frac{(\partial_{x}\varphi)^{2}}{K}-\frac{g}{\pi}\cos(4\varphi)\right\}
\end{align}
where the parameters $u$ and $K$ depend on the fermi velocity $v_{\rm F}$ and wave-vector $k_{\rm F}$ according to the relations
 \begin{align}
uK&=v_{\rm F},\\
\label{uK_eq2}
\frac{u}{K}&=v_{\rm F}+\frac{1}{\pi}\sum_{r=1}^{L}V_{r}^{(\alpha)}(1-\cos(k_{\rm F}r)).
\end{align} 
It is straightforward to show that the universal physics of the Luttinger Liquid Hamiltonian in Eq.\,\eqref{ll_h} is the same as in the sine-Gordon model\,\cite{malard2013sine} featuring a infinite-order transition between a line of free fixed points with power-law bosonic correlations $\langle a^{\dagger}_{i}a_{j}\rangle=|i-j|^{-1/2K}$. Therefore, the free field theory phase corresponds with the metallic phase of the Hamiltonian in Eq.\,\eqref{hcb_h}. For any finite um-klapp scattering term $g$
\begin{align}
\label{umklapp}
g=\sum_{r=1}^{L}V^{(\alpha)}_{r}\cos(2k_{\rm F}r)
\end{align}
the metallic phase breaks down beyond the critical coupling strength $K_{c}$, which at half-filling corresponds to $K_{c}=1/2$, neglecting multiple um-klapp processes. In the nearest-neighbour limit $\alpha\to\infty$ this scenario describe the metal-insulator transition appearing at $V_{c}=2t$. Such transition lies in the BKT universality and, indeed, the breakdown of the metallic phase is akin to vortex proliferation in the physics of the 2D XY model.

For $\alpha>1$ the introduction of the Kac rescaling does not influence the physics and the aforementioned picture does not change. Apart from obvious changes in the value of the critical interaction strength. On the other hand, the aforementioned universal picture is broken as soon as $\alpha=1$, since,  in absence of Kac rescaling, the first term in the summation of Eq.\,\eqref{uK_eq2} diverges in the thermodynamic limit, $\sum_{r}V_{r}^{1}\sim \log(L)$, leading to a vanishing $K$ coupling. At the same time the interaction coupling remains finite due to the alternating sign in Eq.\,\eqref{umklapp}  and, therefore, the system lingers in the insulating phase, as verified by numerical computations\,\cite{capponi2000effects,botzung2021effects}.

\begin{figure}[ht]
\includegraphics[width=\columnwidth]{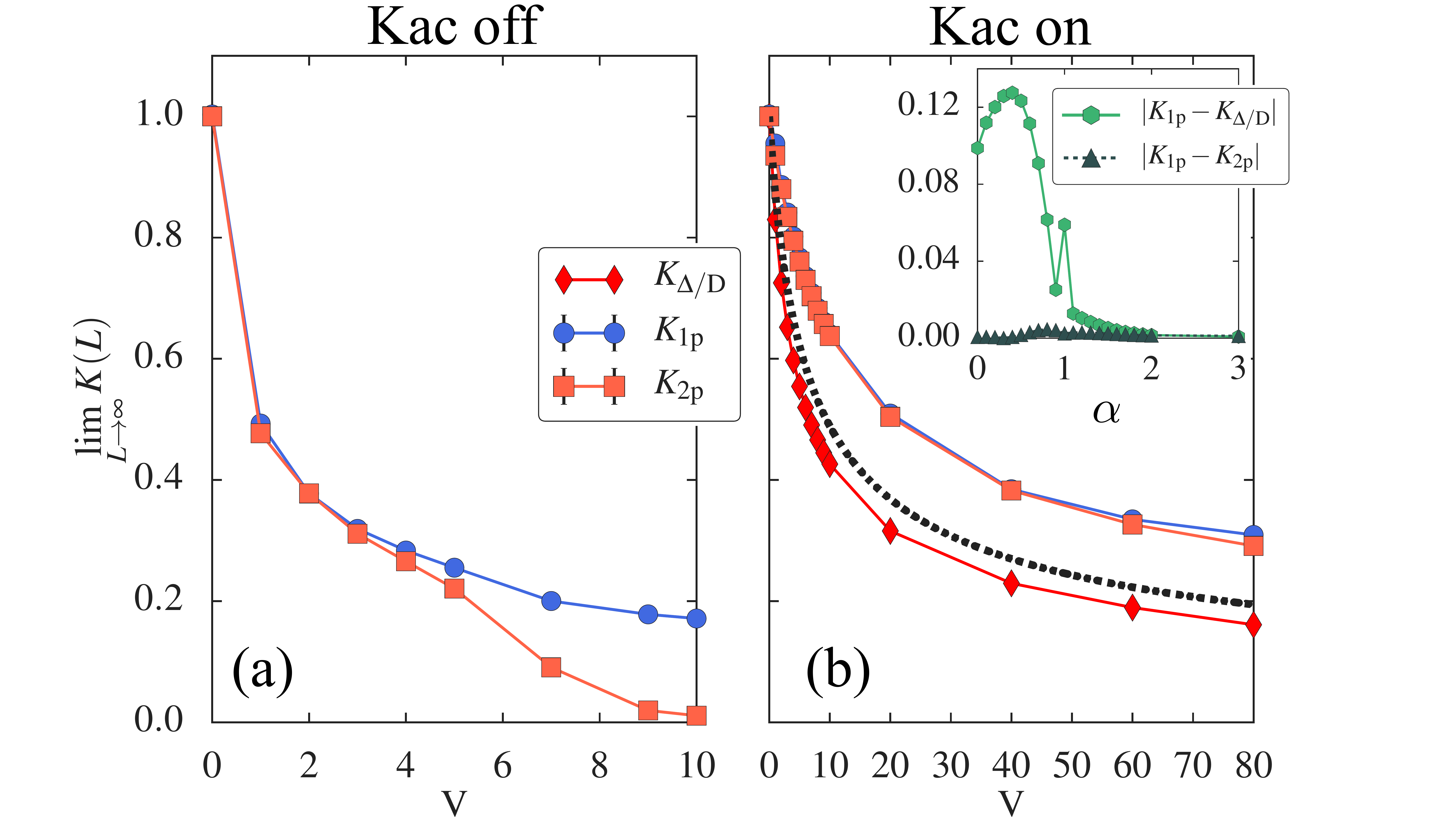}
\caption{\textbf{Luttinger parameter.} Thermodynamic limit extrapolation of the Luttinger parameter $K$ as a function of the long-range interaction strength V at half-filling, in the case $\alpha=0.5$ (the same scenario has been for several $\alpha$ values in the range $0<\alpha<1$). The three different definitions for the Luttinger parameter have been compared both in Kac unscaled, panel (a), and scaled, panel (b), cases\,\cite{botzung2021effects}. As a function of $\alpha$, given a fixed value of the interaction ($V = 1.5$ in the inset of panel (b)), there is no discrepancy between the Luttinger parameter obtained by correlation function $K_{1p} = K_{2p}$, confirming the metallic character of the system. Still the conventional Luttinger Liquid theory is not obeyed since the Luttinger parameter does not fit the gap scaling $K_{\Delta}$. As it is seen from the inset of panel (b), the traditional Luttinger Liquid picture is recovered for $\alpha>1$.}
\label{Fig10}
\end{figure}

The situation is reversed by the introduction of Kac rescaling, which imposes convergence on the first summation of Eq.\,\eqref{uK_eq2} irrespectively from the $\alpha$ value, while it makes the interaction coupling vanish identically.  It, then, does not come as a surprise that the Kac scaled systems always lies in the metallic phase. While the Luttinger-Liquid theory is able to reproduce the metallic(insulator) character in presence(absence) of Kac's rescaling, the actual features of the phase in both cases are not completely consistent with the continuous theory prediction. Indeed, the comparison between the numerical values for the $K$ coupling obtained by the single particle correlation functions ($K_{1p}$), the structure factor ($K_{2p}$) and the finite size scaling of the gap ($K_{\Delta}=\partial\Delta/\partial L^{-1}$)\,\cite{kohn1964theory} do not match each other and especially do not match the prediction of Luttinger Liquid theory in the Kac rescaled case, see Fig.\,\ref{Fig10}. Therefore, both the metallic and insulating character of the theory at $\alpha<1$ in the, respective, Kac scaled and unscaled case do not obey Luttinger liquid theory\,\cite{botzung2021effects}. It is worth noting that this picture does not apply to the flat interactions case $\alpha=0$, which is analytically solvable and may be treated separately.

\subsection{Flat interactions}
\label{lmg_model}

Systems with flat interactions ($\alpha=0$) constitute a unique in the realm of long-range interactions, since they often allow exact analytical solutions of their thermodynamic and critical properties ,at least at large scales. Yet, several of their qualitative features exactly reproduce the more complex physics of general strong long-range systems with $0<\alpha<d$. This special role makes such systems worthy of a special focus and in this section we are going to consider examples of fully connected quantum systems. 

\subsubsection{The Lipkin-Meshkov-Glick model}
\label{lmg_eq}

The Lipkin-Meshkov-Glick model, one the most famous example of strong long-range interacting model in the quantum realm, has been first introduced as a simple test for the validity of perturbative techniques in many-body theories\,\cite{lipkin1965validity,meshkov1965validity,glick1965validity}. Subsequently, the model has been applied to investigate many-body systems which allowed for a sensible descriptions in terms of mean-field interactions, such as coupled BECs\,\cite{cirac1998quantum} or BCS systems\,\cite{dusuel2005finite}. The Lipkin-Meshkov-Glick's Hamiltonian describes $N$ $1/2$-spins coupled by flat ferromagnetic interactions of strength $J/N$
\begin{align}
\label{h_lmg}
H_{\rm LMG}=-\frac{J}{N}\sum_{i<j}(\sigma^{x}_{i} \sigma^{x}_{j}+\gamma
\sigma^{y}_{i} \sigma^{y}_{j})-h\sum_{j=1}^{N}\sigma^{z}_{j}.
\end{align}
where $\gamma$ is the anisotropy parameter. At $\gamma=0$, the former Hamiltonian corresponds to the quantum extension of the HMF model the workhorse of classical strong long-range systems\,\cite{dauxois2002hamiltonian} and it also corresponds to the flat limit $\alpha=0$ of the long-range Ising Hamiltonian in Eq.\,\eqref{h_lri}.
The HMF model displays a rich dynamical behavior, including violent relaxation phenomena, common to several classical and quantum  long-range interacting systems \cite{barre2002outofequilibrium,plestid2018violent}.

The key property of any flat interaction problem is the possibility to rephrase it in terms of the a collective variable, which is the linear combination of all the microscopic variables. Indeed, in our case one can introduce the collective spin $S_{\mu}=\sum_{i=1}^{N}\sigma^{\mu}_{i}/2$, where $\mu\in\{x,y,z\}$. In terms of the new variables Eq.\,\eqref{h_lmg} reads
\begin{align}
\label{h_lmg_2}
H_{\rm LMG}=-\frac{2J}{N}(S_{x}^{2}+\gamma S_{y}^{2})-2h S_{z}+\frac{J}{2}(1+\gamma).
\end{align}
which describes a single self-coupled $N$-component spin immersed into a magnetic field. The Hamiltonian $H_{\rm LMG}$ preserves both the total spin and the total magentization values
\begin{align}
[H_{\rm LMG}, \boldsymbol{S}^{2}]=0\,\quad [H_{\rm LMG}, S_{z}]=0,
\end{align}
where $\boldsymbol{S}^{2}=S_{x}^{2}+S_{y}^{2}+S_{z}^{2}$. The highly symmetric nature of this model makes it particularly amenable also to numerical techniques, making it a prominent test-bed for novel algorithms\,\cite{bapst2012quantum,albash2018adiabatic}. Moreover, the model has proved the ideal tool to prove several generic properties of quantum critical points, such as finite size\,\cite{botet1982size} and entanglement scaling\,\cite{amico2008entanglement,wichterich2010universality}.

Nowadays, the Lipkin-Meshkov-Glick model is subject to renewed interest also due to its relation with the celebrated Dicke model, which is often used to describe driven-dissipative experimental setups, such as the cavity QED experiments outlined in Sec.\,\ref{q_cav_sec}. Its Hamiltonian contains $1/2$-spin operators coupled to the cavity electromagnetic field. In analogy with the long-range Ising model, the Dicke model displays a phase transition between a disordered ground state with $\langle \sigma_{x}\rangle=\langle a^{\dagger}a\rangle=0$ and a super-radiant one with polarised spins and finite photon density inside the cavity $\langle a^{\dagger}a\rangle\neq0$\,\cite{dicke1954coherence}. At equilibrium, it can be rigorously proven that the Hamiltonian of the Dicke and Lipkin-Meshkov-Glick models are equivalent in the thermodynamic limit and, then, produce the same critical behaviour\,\cite{gibberd1974equivalence,brankov1975asymptotically}.

One additional reason for the traditional interest in Hamiltonian\,\eqref{h_lmg}, and its extensions\cite{bapst2012quantum}, comes from its mean-field nature, which allows for an exact solution in the thermodynamic limit. In the form of Eq.\,\eqref{h_lmg_2}  it already evident that the problem is effectively a $0$-dimensional one. In fact, even if we started with a many-body system on the 1-dimensional chain, the flat nature of the interactions allowed the reduction to a single-body problem. The collective variable $\boldsymbol{S}$ can be seen as the zero-momentum Fourier component of the system and, then, differently from typical many body systems, the problem is completely characterised by such $k=0$ mode. Furthermore, the contribution from quantum fluctuations to the thermodynamic quantities vanishes in the thermodynamic limit, where the effective $N$-spin variable becomes classical. A rigorous proof of this fact can be obrained by  rewriting the problem into path integral formulation and by using the Suzuki-Trotter formula to disentangle the two non-commuting terms in the Hamiltonian\,\cite{chayes2008phase}.

Therefore, the control parameter for quantum fluctuations in the Lipkin-Meshkov-Glick model is $1/N$, which plays the same role of $\hbar$ in more traditional single-body problems. In the following we are going to restrict to $h>1$, as the spectrum of the model is symmetric under inversion $h\to -h$, and to ferromagnetic interactions $J>0$. A discussion on the physics of the antiferromagnetic problem $J<0$ and its relation to the super-symmetric formalism can be found in Ref.\,\cite{vidal2004entanglement}. For ferromagnetic interactions $J>0$, the ground state always belongs to the maximum spin $S=N/2$ sub-sector of the Hilbert space. 

Apart from the full isotropic limit $\gamma=1$, the Lipkin-Meshkov-Glick Hamiltonian cannot be analytically solved\,\cite{botet1983large}. Nevertheless, the Lipkin-Meshkov-Glick Hamiltonian is integrable and can be solved via algebraic Bethe ansatz\,\cite{pan1999analytical} or by mapping it to the Richardson-Gaudin Hamiltonian\,\cite{dukelsky2004colloquium}. Here, we are going to follow a simpler route by employing the $1/N$ expansion. First of all, we characterise the critical behaviour via mean-field approximation by using the non-interacting variational ansatz obtained via the external product of the single spin states 
\begin{align}
|\psi_{l}\rangle=\cos\left(\frac{\theta_{l}}{2}\right)e^{-i\frac{\varphi_{l}}{2}}|\uparrow\rangle+\sin\left(\frac{\theta_{l}}{2}\right)e^{-i\frac{\varphi_{l}}{2}}|\downarrow\rangle
\end{align}
since the system is translational invariant we can assume $(\theta_{l},\varphi_{l})=(\theta,\varphi)\,\,\forall\,l$, corresponding to the spin expectation values
\begin{align}
\boldsymbol{S}=\frac{N}{2}(\sin\theta\cos\varphi,\sin\theta\sin\varphi,\cos\theta)
\end{align}
which coincides with the a classical spin value. Due to the inversion symmetry of the model $S_{x}\to-S_{x}$ one can select $\varphi=0$ and $J=1$, without loss of generality. From the energy minimization within the mean-field ansatz, one obtains the explicit expression
\begin{align}
\label{mf_angle}
\theta=\begin{cases}
0\,\,\mathrm{if}\,\,h\geq1\\
\arccos(h)\,\,\mathrm{if}\,\,0\leq h\leq1
\end{cases}
\end{align}
for the angle $\theta$. The semiclassical equations of motion for the total spin operators yield the system gap in the thermodynamic limit\,\cite{botet1983large}
\begin{align}
\label{cl_gap}
\Delta=\begin{cases}
2\sqrt{(h-1)(h-\gamma)}\\
0
\end{cases}
\end{align}

A close inspection of the formulas above is all one needs to comprehend the quantum phase transition in the Lipkin-Meshkov-Glick problem. At $h\geq1$ only the solution $\varphi=\theta=0$  exists and the system is fully magnetised along the magnetic field direction, $\langle S_{z}\rangle=1$. As $h$ decreases below $h_{c}=1$ two state appears with $\theta\neq 0$ and $\varphi=\pm \pi$  and the in plane magnetisation continuously increases in the interval $[0,1]$, while the transverse magnetisation only vanishes at $h=0$. Accordingly, the gap $\Delta$ between the ground and the first excited state, which is finite at $h>1$, smoothly vanishes as $h\to 1^{+}$ with scaling behaviour characterised by the critical exponent $z\nu=1/2$. It is worth noting that the mean-field scenario can be only faithfully applied to the thermodynamic limit, while it cannot capture finite size fluctuations. Indeed, in the ordered phase $h<1$ the system gap $\Delta$ cannot vanish at finite size, since quantum fluctuations will lift the degeneracy and produce an exponentially vanishing gap $\Delta_{N}\propto \exp(-N)$\,\cite{newman1977metastability}.

In order to partially capture finite size fluctuations, it is convenient to perform an Holstein-Primakoff expansion\,\cite{holstein1940field} for the $N$-spin variable $\boldsymbol{S}$ around the mean field expectation value\,\cite{botet1983large,dusuel2005continuous}. First, one shall rotate the total spin in order to align it with the meand field magentization introducting the new variable $\bar{S}=R(\theta)S$, with the rotation matrix 
\begin{align}
R(\theta)=\begin{pmatrix}
\cos\theta & 0 & \sin\theta\\
0 & 1 & 0\\
-\sin\theta & 0 & \cos\theta
\end{pmatrix}
\end{align}
where $\theta$ is given by Eq.\,\eqref{mf_angle}. The re-aligned spin variables may be then expanded using the equivalence
\begin{align}
\label{hp_z}
\bar{S}_{z}&=\frac{N}{2}-a^{\dagger}a\\
\label{hp_+}
\bar{S}_{+}&=\bar{S}_{x}+i \bar{S}_{y}=\sqrt{N}\left(1-\frac{a^{\dagger}a}{N}\right)^{1/2}a\\
\label{hp_-}
\bar{S}_{-}&=\bar{S}_{x}-i \bar{S}_{y}=\sqrt{N}a^{\dagger}\left(1-\frac{a^{\dagger}a}{N}\right)^{1/2}
\end{align}
where the boson operators $[a,a^{\dagger}]=1$ have been introduced. This excitation characterises a small depletion of the mean-field spin expectation due to finite size quantum fluctuations. At leading order in $1/N$ only quantum corrections up to order $1/N$ have to be retained, yielding a quadratic Bosons Hamiltonian which can be subsequently diagonalised by a Bogoliubov transformation $a\to b$\,\cite{dusuel2005continuous}.
The net result is
\begin{align}
\label{lmg_large_n}
H_{\rm LMG}=N E_{0}+ e_{0}+\omega b^{\dagger}b+O\left(\frac{1}{N}\right)
\end{align} 
such that we have reduced the many-body problem in Eq.\,\eqref{h_lmg} to an effective $0$-dimensional one, described by a single harmonic oscillator mode. This is the peculiarity of several fully connected systems, the actual spectrum in the thermodynamic limit is not constituted by a continuum dispersion relation, but rather by a single quantum mode, whose contribution to the thermodynamic quantities is increasingly washed out approaching the thermodynamic limit.

The quantities appearing in Eq.\,\eqref{lmg_large_n} can be easily written in terms of the internal parameter and the average magnetization $m=2 \langle S_{z}\rangle/N$. The internal mean-field energy maintains the same form both in the symmetric and broken phases $E_{0}=(-1-2hm+m^{2})/2$, while the next-to-leading energy correction read
\begin{align}
e_{0}=\begin{cases}
-h +\frac{1+\gamma}{2}+\sqrt{(h-1)(h-\gamma)}\,\,\mathrm{for}\,\,h>1,\\
-\frac{1-\gamma}{2}+\sqrt{(1-h^{2})(1-\gamma)}\quad\mathrm{for}\,\,h<1,\\
\end{cases}
\end{align}
and the dynamical gap
\begin{align}
\label{dyn_gap}
\omega=\begin{cases}
2\sqrt{(h-1)(h-\gamma)}\,\,\mathrm{for}\,\,h>1,\\
\sqrt{(1-h^{2})(1-\gamma)}\,\,\mathrm{for}\,\,h<1.\\
\end{cases}
\end{align}
Notice that $\omega$ is not the actual gap $\Delta$ of the system, at least not in the ordered phase, where the minimal gap occurs between the two classical ground-states with different symmetry, but it rather represents the minimal gap between two states connected by the Hamiltonian dynamics.

As expected, the dynamical gap in Eq.\,\eqref{dyn_gap} vanishes approaching the transition with a dynamical critical exponent $z\nu=1/2$ in agreement with the semiclassical prediction for the disordered phase, see Eq.\,\eqref{cl_gap}. The exponent is symmetric on both sides of the transition and independent on the value of $\gamma\neq 1$ proving that the anisotropy plays no-role in the universal behaviour. The only exception is $\gamma=1$ where the system acquires continuous rotation symmetry, giving rise to a gapless ordered phase and a critical exponent $z\nu=1$; an analytical solution of the problem is available in this particular case\,\cite{dusuel2005continuous}.

The in-plane magnetisation $\langle S_{x}\rangle/N\propto  \sqrt{1-h^{2}}$ consistently with a critical exponent $\beta=1/2$. Similar arguments can be used to show that all the $\emph{thermodynamic}$ critical exponents, i.e. the ones associated with  global thermodynamic quantities, are in agreement with mean field theory. The question becomes, however, more complex if we consider the scaling of spatial dependent quantities such as the correlation length. Conventionally, the critical exponent $\nu$ is associated with the scaling of the correlation length $\xi$ at a (quantum) critical point $\xi\propto \lambda^{-\nu}$, where $\lambda$ is the control parameter which vanish at the critical point. Such critical exponent is particularly important since it relates the thermodynamic singularities of any critical quantity, with its finite size scaling close to the transition\,\cite{fisher1967theory,fisher1972scaling}. However, in a strong long-range system, and in particular in a fully connected one, no concept of length and, especially, of correlation length exists. 

However, even in absence of any definition of length, it is possible to define a correlation number, which diverges close to the critical point $N_{c}\propto |h-1|^{\nu_{*}}$. In general such correlation number will be proportional to the correlation volume $N_{c}\propto \xi^{d}$ and, assuming that such scaling has to remain the same for all systems in the mean-field regime, one obtains the estimate
\begin{align}
\label{fss_flat_int}
\nu_{*}=d_{\rm uc}\nu.
\end{align}
The quantity $d_{\rm uc}$ represents the upper critical dimension of the corresponding nearest neighbour model\,\cite{botet1982size}. Since the Lipkin-Meshkov-Glick Hamiltonian in Eq.\,\eqref{h_lmg} corresponds to the one of the quantum Ising model in a transverse field with $d_{\rm uc}=3$, the correlation number exponents shall read $\nu_{*}=3/2$.

Interestingly, this scaling theory, first introduced in Ref.\,\cite{botet1982size}, provides the exact value for the finite size scaling of the dynamical gap $\omega_{N}$ which can be obtained by incorporating higher order $1/N$ corrections into Eq.\,\eqref{dyn_gap} via the continuous unitary transformation approach, yielding $\omega_{N}\approx N^{-1/3}$\,\cite{dusuel2004finite,dusuel2005continuous} in perfect agreement with the generalised finite scaling theory $\omega_{N}\approx N^{-\frac{z\nu}{\nu_{*}}}$. In spite of this apparent simplicity, it has been shown that for large enough anisotropy parameters the spectrum of the Lipkin-Meshkov-Glick model may not actually converge to the prediction of Eq.\,\eqref{lmg_large_n}, due to the influence of two competing semiclassical trajectories\,\cite{ribeiro2007thermodynamic}.

More in general, the convergence to the "simple" thermodynamic limit solution in fully connected models has been shown to present several anomalous features\,\cite{colonna2014anomalous}. In particular, it has been shown that the actual picture for the finite size scaling of many-body systems above the upper critical dimension $d_{\rm uc}$ is actually more complicated than the one depicted in Ref.\,\cite{botet1982size}, since the zero and the fluctuations modes present different scaling behaviours and, therefore, different quantities may display different finite size corrections depending from the dominating contribution to that quantity\,\cite{sola2016role}.

\subsubsection{Self-organization phase transition in cavity QED}\label{subsec:selforganization}
The cavity-mediated long-range interaction, Eq. (\ref{eq:cavity_long_range_interactions}), favors for $\mathcal{V}<0$ a density modulation of the quantum gas and induces according density correlations with spatial periodicity $\lambda$ along pump and cavity directions. These density correlations are the collective elementary excitations of the system with energy $\hbar \omega_s$, and correspond to the creation and annihilation of correlated pairs of atoms in the momentum mode $|\mathbf{p}_1 \rangle$. However, the kinetic energy term in Eqs. (\ref{eq:many_body_atom_field_Hamiltonian}) stabilizes the gas against this modulation. 

Only if the long-range interaction becomes sufficiently strong, the gain in potential energy will overcome the cost in kinetic energy, and the system undergoes a quantum phase transition to a self-ordered state \cite{nagy2008self, piazza2013bose}. At this point, the energy $\hbar \omega_s$ of the collective excitation has softened such that the mode $|\mathbf{p}_1\rangle$ can be macroscopically populated without energetic cost. The atomic density acquires a checkerboard  modulation that efficiently scatters photons into the resonator, and the atoms can further lower their energy in the emerging optical interference lattice potential.

\begin{figure}[]
\centering
\includegraphics[width=0.9\columnwidth]{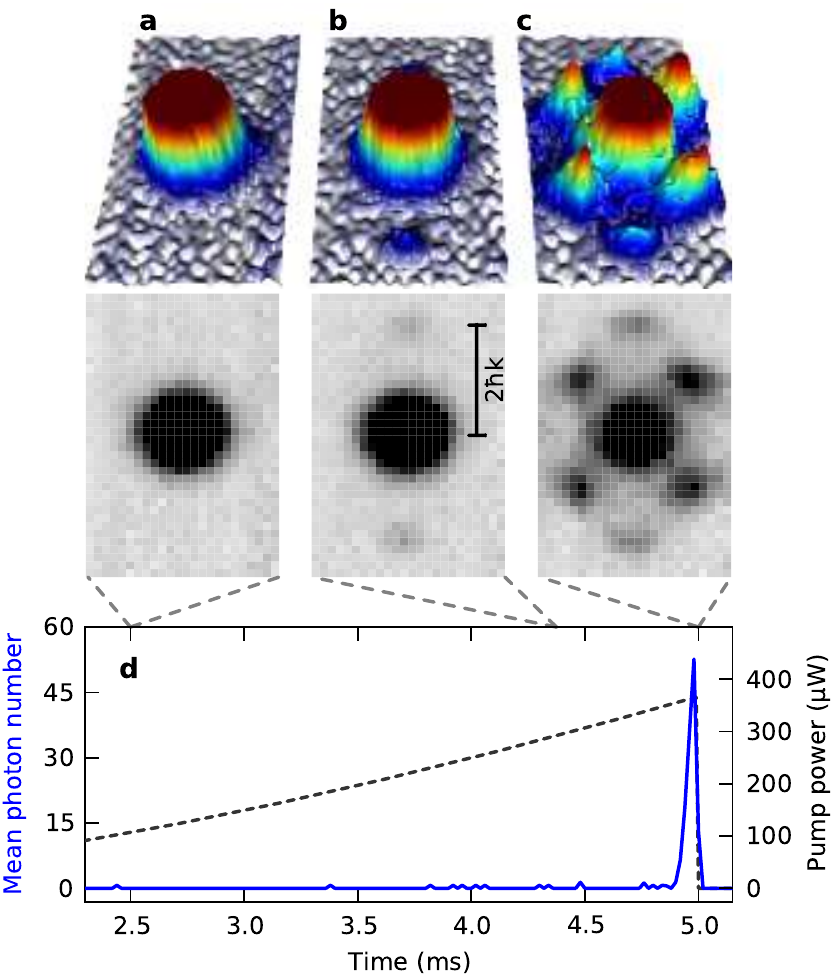}
\caption{{\bf Signatures of atomic self-organization in an optical cavity.} {\bf (a)} The transverse pump power (dashed) is gradually increased while the mean intracavity photon number (solid) is monitored. After the sudden release of the atomic cloud and its subsequent ballistic expansion, absorption images are made for pump powers corresponding to transverse pump lattice depths of 2.6~$\mathrm{E_r}$ (b), 7.0~$\mathrm{E_r}$ (c) and 8.8~$\mathrm{E_r}$ (d). Self-organization is manifested by an abrupt build-up of the cavity field accompanied by the formation of momentum components at $(p_x, p_y)=(\pm \hbar k, \pm \hbar k)$ (d). The weak momentum components at $(0,\pm 2 \hbar k)$ result from loading the atoms into the one-dimensional standing-wave potential of the transverse pump laser. Reproduced from \cite{baumann2010dicke}.}
\label{fig:SelfOrganization}
\end{figure}

A few years after self-organisation of a thermal gas coupled to an optical cavity had been observed \cite{black2003observation}, the phase transition to a self-ordered state of a bosonic quantum gas coupled to a cavity was realized \cite{baumann2010dicke}. While for a thermal gas the  threshold is set by thermal density fluctuations, for a quantum gas the critical point scales with the recoil energy. A BEC of $10^5$ $^{87}$Rb atoms is harmonically trapped at the location of a single mode of a high-finesse optical cavity. The transverse pump power is linearly increased over tens of milliseconds. The experimental signature for self-ordering of a BEC, where the motion is quantized, is two-fold as shown in
Fig. \ref{fig:SelfOrganization}: The cavity photon occupation rises abruptly when the critical interaction is reached, as can be observed via the light field leaking from the cavity. In addition, the momentum state distribution, as observed from absorption images after ballistic expansion, changes from occupying only the zero-momentum state $|\mathbf{p}_0\rangle$ below the critical point to a superposition of the momentum states $|\mathbf{p}_0\rangle$ and $|\mathbf{p}_1\rangle$ above the critical point. In real-space, this momentum state occupation corresponds to a chequerboard order of the atomic density. Ramping the transverse pump power down again, the normal phase with an empty cavity and macroscopic occupation of only the single momentum state $|\mathbf{p}_0\rangle$ is recovered.  As discussed in Section~\ref{sec:SpinModelCavity}, the self-organization phase transition can be mapped to the Dicke phase transition.

\begin{figure}[]
\centering
\includegraphics[width=1\columnwidth]{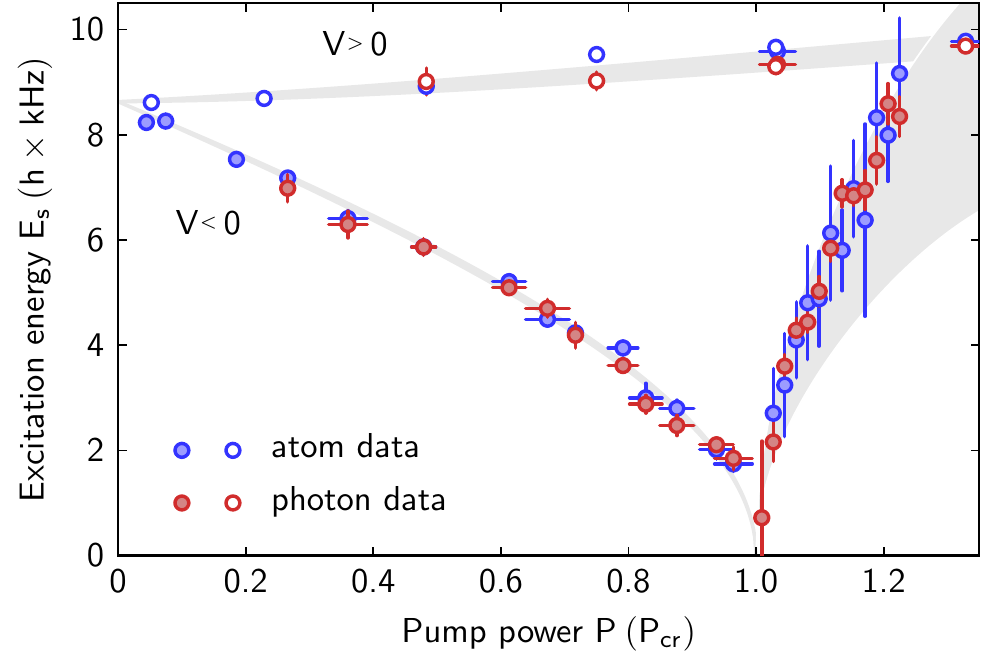}
\caption{{\bf Excitation spectrum across the self-organization phase transition.} Measured resonance frequencies $E_s=\hbar \omega_s$, obtained from atomic and photonic signals, are shown in blue and red, respectively, for positive (open circles) and negative (solid circles) interaction strength $\mathcal{V}$. Gray shading shows the theoretical prediction including experimental uncertainties. Reproduced from \cite{mottl2012roton}.}
\label{fig:ModeSoftening}
\end{figure}

The mode softening preceding the phase transition \cite{horak2001dissipative,nagy2008self,oztop2013collective} has been studied using a variant of Bragg spectroscopy \cite{mottl2012roton}. The cavity is seeded with a weak coherent field at a variable detuning with respect to the transverse pump frequency. If the detuning matches the soft mode frequency $\omega_s$, energy and momentum conservation are fulfilled, and the momentum mode $|\mathbf{p}_1\rangle$ becomes macroscopically occupied by the probing process. At the same time, photons from the transverse pump are scattered into the cavity. The measured mode frequency $\omega_s$ as a function of transverse pump power is displayed in Fig. \ref{fig:ModeSoftening}. For the case of negative long-range interaction $\mathcal{V}<0$, a clear mode softening towards the critical point of the self-organization phase transition is observed. In contrast, a positive long-range interaction $\mathcal{V}>0$ is leading to a  mode hardening without any phase transition. 

Also in the case of a sideband-resolving cavity, $\kappa < \omega_s$, a self-organization phase transition takes place. However, due to the increased photon lifetime, the intra-cavity field acquires a retardation with respect to the atomic evolution, and the effective cavity-mediated atom-atom interaction can not be captured any more in the simple form of Equation (\ref{eq:cavity_long_range_interactions}) \cite{klinder2015observation}. In this case it is more appropriate to stay with the coupled equation of motion. As we discuss below in Section \ref{sec:CavityKibbleZurek}, the sideband resolved regime allows to study quench experiments that can be interpreted with a Kibble-Zurek model.

The long-range interaction can not only be engineered to act on the atomic density. Instead, exploiting the atomic vector polarizablility or Raman schemes coupling different atomic ground states, an effective long-range interaction acting on the pseudo spin can be realized \cite{kroeze2018spinor, landini2018formation}.

\subsubsection{Discrete and continuous symmetry breaking}\label{sec:SymmetryBreaking}
The Dicke Hamiltonian (\ref{eq:DickeHamiltonian}) is invariant under the parity transformation $(a, J_\pm) \rightarrow (-a,-J_\pm)$. Accordingly, at the phase transition to the self-organized phase a discrete $\mathbf{Z}_2$-symmetry is broken, where the atomic density localizes either on the even or odd sites of the emergent checkerboard lattice and the cavity light field phase locks to either $0$ or $\pi$ with respect to the pump field phase. Site-resolving real-space imaging of the atomic system has not been achieved yet. However, this discrete symmetry breaking has been observed in the phase of the light field leaking from the cavity using a phase sensitive heterodyne detection system~\cite{baumann2011exploring}.

The discrete nature of this symmetry breaking is dictated by the boundary conditions of the single cavity mode. The symmetry can however be enhanced to a continuous $U(1)$-symmetry, as had been originally discussed for highly degenerate multimode cavities \cite{gopalakrishnan2009emergent}. Also the self-organization of a transversely driven BEC in the combined fields of two degenerate single mode cavities crossing under an angle of 60$^\circ$ allows to engineer an approximate continuous $U(1)$-symmetry, as was demonstrated experimentally~\cite{leonard2017supersolid}. Photons from the pump field were scattered into both cavities, and the atoms self-organized in the resulting interference potential. This system is invariant with respect to redistributing photons between the two modes, where the interference lattice potential breaks a continuous spatial symmetry depending on the relative photon occupation of the two cavities. 
The unique real-time access to the light field leaking from the optical cavities allowed to identify the fundamental collective excitations of the underlying $U(1)$-symmetry as a phase and an amplitude mode~\cite{leonard2017monitoring}. The continuous symmetry can be reduced to a $\mathbf{Z}_2 \otimes \mathbf{Z}_2$ symmetry if atom-mediated scattering between the two cavities is present~\cite{morales2017couplings,lang2017collective}.  Extending the scheme to multiple crossing cavities, also higher symmetries such as a continuous $SO(3)$ rotational symmetry might be realizable \cite{chiacchio2018emergence}. A continuous symmetry can furthermore be broken if instead two counterpropagating modes of a ring cavity are employed, as was proposed for a transversally driven BEC~\cite{mivehvar2018driven}, and realized for a BEC coupled to a ring cavity where two longitudinal modes were simulatenously driven~\cite{schuster2020supersolid}.

A self-organized BEC breaking a continuous $U(1)$-symmetry can be regarded as the minimal model of a supersolid state of matter. This paradoxical state combines the characteristics of crystalline and superfluid orders, and had been predicted to exist in solid helium-4~\cite{andreev1969quantum,thouless1969the,leggett1970can} but could never be unambigously observed~\cite{kim2004probable,kim2012absence}. Supersolidity was first demonstrated in a BEC coupled to two crossed cavities~\cite{leonard2017supersolid} and at the same time in spin-orbit coupled quantum gases~\cite{li2017stripe}. Lateron, supersolidity was also demonstrated in quantum gases with dipolar long-ranged interactions~\cite{tanzi2019observation,bottcher2019transient,chomaz2019long}.

\subsubsection{Criticality of the self-ordering phase transition}
\begin{figure*}[]
\centering
\includegraphics[width=2\columnwidth]{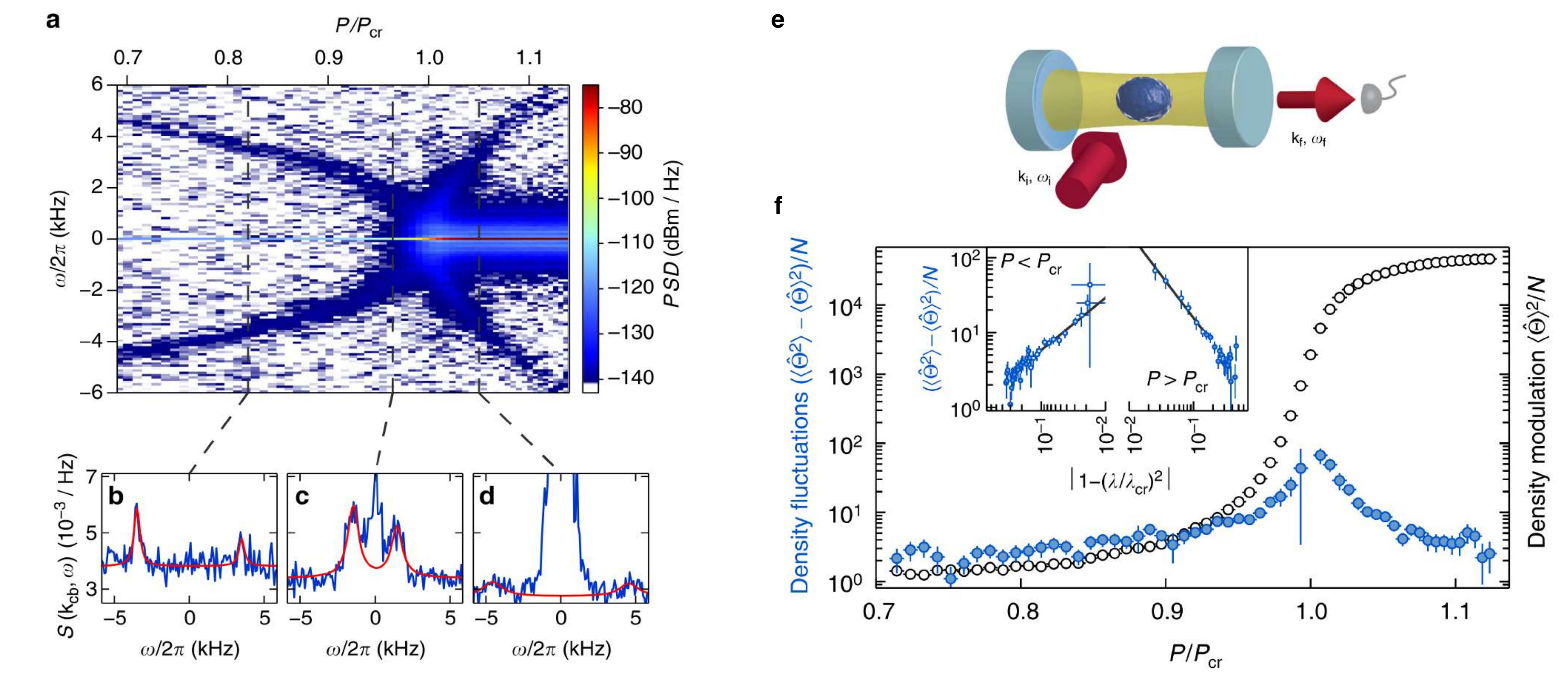}
\caption{{\bf Critical behavior of the self-organization phase transition.}  (\textbf{a}) The power spectral density PSD of the light field leaking out of the cavity is shown as a function of frequency shift $\omega$ with respect to the pump laser frequency and relative transverse pump power $P/P_\textrm{cr}$. Two sidebands are visible, corresponding to the incoherent creation ($\omega<0$) and annihilation ($\omega>0$) of quasi-particles. The energy of these quasi- particles vanishes towards the critical point. At the phase transition, a strong coherent field at the pump frequency appears ($\omega=0$). The panels (\textbf{b–d}) show the normalized dynamic structure factor for three different values of $P/P_\textrm{cr}$ (see dashed lines in upper panel). While the position and width of the sidebands give direct access to the energy and lifetime of the quasi-particles, the sideband asymmetry can be used to determine the occupation of the quasi-particle mode. Panel (\textbf{e}) is a sketch of the measurement setup: the atoms coupled to the cavity mode are illuminated by the transverse pump field at frequency $\omega_i$, while the frequency emitted from the cavity is $\omega_f$. A heterodyne detection system gives access to the PSD shown as a function of $\omega=\omega_i-\omega_f$ in  (\textbf{a}). The data can be used to extract the divergent density fluctuations and the emerging density modulation, shown in (\textbf{f}). The inset shows the density fluctuations on a double logarithmic scale, allowing to determine critical exponents of 0.7(1) and 1.1(1) on the normal and self-organized side, respectively. Figure reproduced from~\cite{landig2015measuring}.
 }\label{fig:CavityCriticalBehavior}
\end{figure*}
The critical behavior of the single-mode self-organization phase transition corresponds to that of the open Dicke model, falling into the universality class of the mean-field classical Ising model~\cite{nagy2010dicke,kirton2018introduction,emary2003chaos}.  The constant flow of energy from the pump laser to the cavity leakage causes additional fluctuations of the cavity field and accordingly larger  density fluctuations. The cavity dissipation thus makes the system leave its ground state and irreversibly evolve into a non-equilibrium steady state. The global range interaction turns the phase transition rather into a quantum bifurcation in a zero-dimensional system, such that there is no notion of a divergent correlation length. However, one can investigate the critical exponent of the fluctuations of the order parameter. While a mean-field exponent of 0.5 is expected for the closed system, the prediction for the open system is 1.0, given by the vanishing of the imaginary part of the spectrum at the critical point~\cite{nagy2011critical,oztop2012excitations}. The open system thus effectively behaves thermal. It is important to note that the actual steady state of the systen might not be reached in experiments, since close to the critical point the quasinormal modes vanish, leading to a critical slow down.
An analysis going beyond the mapping to the open Dicke model and considering also a finite temperature of the quantum gas allowed to study the interplay between the self-organization phase transition and Bose-Einstein condensation~\cite{piazza2013bose,piazza2014quantum}.

Monitoring the light field leaking from the cavity during self-organization  gives real-time access to the order parameter of the phase transition, see Equation~(\ref{eq:steady_state_cavity_field}). This allows not only to measure the mean density modulation of the atomic cloud, but also to detect the fluctuations of the system~\cite{brennecke2013real}. Heterodyne detection of the light field provides the low-energy spectrum of the system which can be directly converted into the dynamical structure factor of the gas at the wave vector of self organization~\cite{landig2015measuring}, see Fig. \ref{fig:CavityCriticalBehavior} (a-d). The observed spectrum features a carrier at zero frequency with respect to the pump laser frequency and sidebands at positive and negative frequencies. The sidebands are signatures of density fluctuations, indicating either the creation or annihilation of quasi-particles. Approaching the critical pump power $P_\textrm{cr}$, the mode softening is visible in the vanishing sideband frequency. At the critical point, a strong coherent field emerges at the carrier frequency, indicating the buildup of a static coherent density modulation. The amplitude of the carrier and of the integrated sidebands converted into density modulation and density fluctuations, respectively, is diplayed in Fig. \ref{fig:CavityCriticalBehavior}(f). While the density modulation changes by more than four orders of magnitude, the density fluctuations diverge towards the critical point. From this data, critical exponents of 0.7(1) and 1.1(1) for the fluctuations of the order parameter can be extracted on the normal and self-organized side, respectively. The sideband asymmetry visible in Fig. \ref{fig:CavityCriticalBehavior}(b-d) can be used to determine the occupation of the quasi-particle mode, but also to extract the irreversible entropy production rate~\cite{brunelli2018experimental} while the system crosses the phase transition.

\section{Non-local interactions regimes}
\label{sec:regimes}

In this section we are going to discuss the case of non-local interactions,
addressing also cases of competing interactions relevant for some of
the physical systems introduced in the previous sections. Given the
rich variety of physical behaviours in these systems,
including supersolid phases, we will not attempt
to cover all the phenomena associated to non-local interactions and
competing interactions and rather after a brief introduction 
we decided to focus on two main classes
of applications: the clustering phenomena induced by typical non-local
interactions, and the structural phase transitions occurring in mesoscopic
long-range interactions.

The phase behavior of systems whose constituent particles in free space or in the presence of an external confining potential interact with nonlocal potentials diverging at the origin is a problem that has been extensively studied in the last few decades both in the classical and more recently in the quantum regime \cite{likos2001effective}.
A major problem concerns the study of freezing transitions and the respective crystal structure, which 
depends on the steepness of the potential, the dimensionality, and the details of the external trapping.
At the classical level, power-law diverging potentials of the form $V(r)=\varepsilon (\sigma/r)^\alpha$, where $\varepsilon>0$ is an arbitrary energy scale, $\sigma$ has the dimension of a length, and $r$ is the interparticle distance, result into 
the formation of a crystalline state at arbitrarily high temperatures. 
Moreover, one can show that to ensure the stability against {\it explosion} 
(infinite thermodynamic observables, such as the energy per particle or pressure)
one has to impose $\alpha>d$\,\cite{weeks1981volume},
with $d$ the system dimensionality, i.e. to be in the weak long-range
or short-range regimes. 
If this condition is violated, i.e. $\alpha \le d$ a
neutralizing background could
be introduced to stabilize one-component systems, as e.g. the
one-component plasma
\cite{baus1980statistical}. Notice that both Kac rescaling
and the introduction of neutralizing background can be used
to perform calculations and regularize physical quantities,
but the reader should be alerted that while the Kac rescaling preserve
the functional power-law form of the interactions, a neutralizing background
may introduce screening effects for charged systems.
The study of quantum systems
with density-density power-law interactions without
any intrinsic lenght scale provides a quantum counterpart of these
results holding for classical systems and it 
has been subsequently
investigated\,\cite{buchler2007strongly,pupillo2010strongly,dalmonte2010onedimensional}.

Another interesting class of interactions are those which do not
diverge at the origin, i.e., they are bounded. 
In soft-matter physics, such {\it soft-core} potentials arise as effective interactions between the centers of mass of soft, 
flexible macromolecules such 
as polymer chains, dendrimers, polyelectrolytes, etc. Indeed, the centers of mass of two macromolecules can coincide 
without violation of the excluded volume conditions, hence bringing about a bounded interaction 
\cite{likos2007why}.
A relevant consequence of the removal of the onsite divergence is the possibility of overlapping of particles,
which under certain conditions can lead to clustering.
A rigorous criterion holding for a fluid at sufficiently high densities states that
given a nonattractive and bounded pair potential which satisfies the following requirements 
guaranteeing stability and the existence
of the thermodynamic limit which {\it i)} it is bounded, {\it ii)} it is
positive definite, {\it iii)} it decays fast enough to zero at large
separations, so that it is integrable and its Fourier transform
exists, and {\it iv)} it is free of attractive parts, it does not display clustering.
Otherwise, if the Fourier transform of the pair potential
has a negative value for a finite momentum $k_m$ then it
can freeze into clustered crystals with multiple occupied sites with an intercluster distance $\propto 1/k_m$ 
\cite{likos2001criterion}.
An intuitive way to understand such a criterion is via the high-density limit of the the structure factor $S(k)$ of a fluid, 
which is a measure of the susceptibility of the system to a spontaneous spatial modulation having wavenumber $k$.
Within the framework of the fluctuation-dissipation theorem, $S(k)$ appears as a proportionality factor between a weak external potential of wavenumber $k$ and the associated linear density response.
Employing the Ornstein-Zernike relation\,\cite{hansen2013theory,chaikin1995principles}
one finds that in the high-density limit, the structure factor can be well
approximated by
\begin{equation}
S(\mathbf{k}) = \frac{1}{1+ \rho\, \beta \, V(\mathbf{k})},
\end{equation}
where $V(\mathbf{k})$ is the Fourier transform of the potential and $\rho$ the system density.
Hence, a structure factor with a high peak at some wavenumber $k_m$ is
a signal of an incipient transition of the fluid to a spatially modulated system, i.e., a crystal.
Recently, \cite{mendoza-coto2021cluster}
presented a sufficient criterion for the emergence of cluster phases with low filling (up to two particles per cluster) 
in an ensemble of interacting classical particles with generic (also diverging at the origin)
repulsive two-body interactions in the {\it classical} zero-temperature limit
valid at intermediate densities. 
The basis of the criterion is a zero-temperature
comparison of the energy unbalance between the single particle lattice and the first cluster-crystal configuration
at small density obtained by the use of the Fourier transform of a regularized version of the potential. It determines the relevant characteristics of the interaction potential that make the energy of a two-particle
cluster-crystal become smaller than that of a simple triangular lattice in two dimensions. 
See also \cite{diaz-mendez2017glass} for an application to the formation of 
a vortex glass in clean systems of thin films of "type-$1.5$" superconductors.

In the quantum regime, it is possible to provide a connection between the emergence of 
a structural transition to the structure factor $S(k)$ 
via the analysis of the spectrum of elementary excitation
through the Feynman-Bijl relation \cite{feynman1954atomic}
$S(\mathbf{k}) = \hbar^2k^2/2m\varepsilon(\mathbf{k})$, 
where $\varepsilon(\mathbf{k})$ is the energy of 
excitations at momentum $\mathbf{k}$.
A peak at finite momentum $\mathbf{k}$ of $S(\mathbf{k})$ 
is associated to the presence of a roton minimum 
in the spectrum $\varepsilon(\mathbf{k})$.
Eventually, the upon softening of the roton minimum the
system to enter the roton instability.
This connection has recently worked out in detail in the context of long-range interacting quantum systems
in continuum space, see e.g. \cite{odell2000bose, odell2003rotons, santos2003roton, 
mottl2012roton, chomaz2018observation, hertkorn2021density}.
Dilute quantum gases can feature long-range interactions if the constituent particles have 
($i$) a strong magnetic dipole moment, or ($ii$) a strong permanent electric dipole moment as in polar molecules, or ($iii$) an induced electric dipole moment as in Rydberg atoms or in cavity-mediated systems.
Specifically, quantum gases of atoms with strong magnetic dipole moments have been extensively employed as experimental platform to detect the relation between the microscopic long-range interactions and the low-energy excitation spectra \cite{bismut2012anisotropic} and to study crystallization in a quantum many-body setting
\cite{lahaye2009physics,baranov2012condensed,trefzger2011ultracold,boettcher2020new}. 
The interplay between the collisional contact interactions, the magnetic dipolar interaction, 
and repulsive quantum fluctuations \cite{lima2011quantum}
can give rise to the stabilization of droplets \cite{chomaz2016quantum} or to 
the formation of a supersolid phase if the droplets share phase coherence 
 in the ground state
\cite{bottcher2021new,tanzi2019observation,tanzi2021evidence,norcia2021twodimensional,sohmen2021birth}, or to a rich set of patterns out of equilibrium \cite{parker2009structure}.
For sufficiently strong interactions dipolar systems display a 
roton instability which triggers the phase transition to a dipolar supersolid
and arrays of isolated quantum droplets\, \cite{baillie2016selfbound,baillie2018droplet}, or filaments in three dimensions\, \cite{cinti2017superfluid}.
A similar phenomenology of self-organized
ground-state density modulations was predicted  for a BEC illuminated
by a single, circularly polarized laser beam in the weak saturation limit in \cite{giovanazzi2002density}.
The appearance of a structural via the softening of roton minimum has been extesively studied also
in the context of Rydberg-dressed systems where an intrinsic soft-core potential can be engineered
via laser coupling to highly excited electronic states. In the following we focus on results
both in the continuum and on a lattice, leading to pattern formation in the presence of soft-core
pairwise interactions.

\subsection{Clustering in quantum soft-core systems}
We start by considering a system of $N$
bosons interacting via two-body soft-core potentials of the type
\begin{equation}
\label{eq:soft-core}
V(r) = \frac{V_0}{r^ \alpha + R_c^ \alpha},
\end{equation}
where $R_c$ is a characteristic length of the pair potential.
While the considered interactions do not straightforwardly 
occur in natural crystals, they can be designed in
ultracold atom experiments. 
As we commented in Sec.\ref{subsec:dipolar_gases} 
soft-core interactions of the type described by eq.(\ref{eq:soft-core})
can be realized with Rydberg-dressed atoms where  
$\alpha=6$, for which the
Hamiltonian provides a
prototype system for addressing the general physical picture. 
In general, this interaction approaches a constant value $V_0/R^\alpha$
as the inter-particle distance, $r$, 
decreases below the soft-core distance $R_c$, 
and drops to zero for $r \gg R_c$. The
limiting case $\alpha \rightarrow \infty$ 
yields the soft-disc model \cite{pomeau1994dynamics}, while
$\alpha = 3$ and $\alpha = 6$ correspond to soft-core dipole-dipole \cite{cinti2010supersolid}
and van der Waals \cite{henkel2010three,henkel2012supersolid} interactions that can be realized
with ultracold atoms \cite{maucher2011rydberg} or polar molecules \cite{buchler2007strongly,micheli2007cold}.

Following the classical case discussed above,
the analysis of the Fourier transform (which display negative values at finite momenta $k$) and the associated
structure factor $S(k)$,
we expect translational symmetry breaking in the form of a cluster crystal at sufficiently high densities for dimensions
larger than one at zero temperature.
Moreover, due to the bosonic symmetry of this single-component system, in a certain parameter interval 
of the phase diagram one might expect the system to display both crystalline and superfluid properties, i.e. the simultaneous breaking of continuous translational and global gauge symmetry, a supersolid state. 
The first mentioning of such a state goes back to Gross, who presented a theory for a density-modulated superfluid emerging from a mean-field model for solid Helium  \cite{gross1957unified}. 
A microscopic picture of supersolidity was proposed 
by Andreev, Lifshitz, and Chester \cite{andreev1969quantum}
and is based on two key assumptions: 
(i) that the ground state of a bosonic crystal contains defects such as vacancies and interstitials, 
and (ii) that these defects can delocalize, thereby giving rise to superfluidity. 
For a review on the subject and and the debate on the observation of such phase in solid Helium see
\cite{boninsegni2012supersolids}. For a more recent discussion of the 
observation of supersolid phases in dipolar systems both in quasi 
one- and two-dimensional setups see the review
\cite{bottcher2021new}.

\begin{figure*}[]
\centering
\includegraphics[width=2\columnwidth]{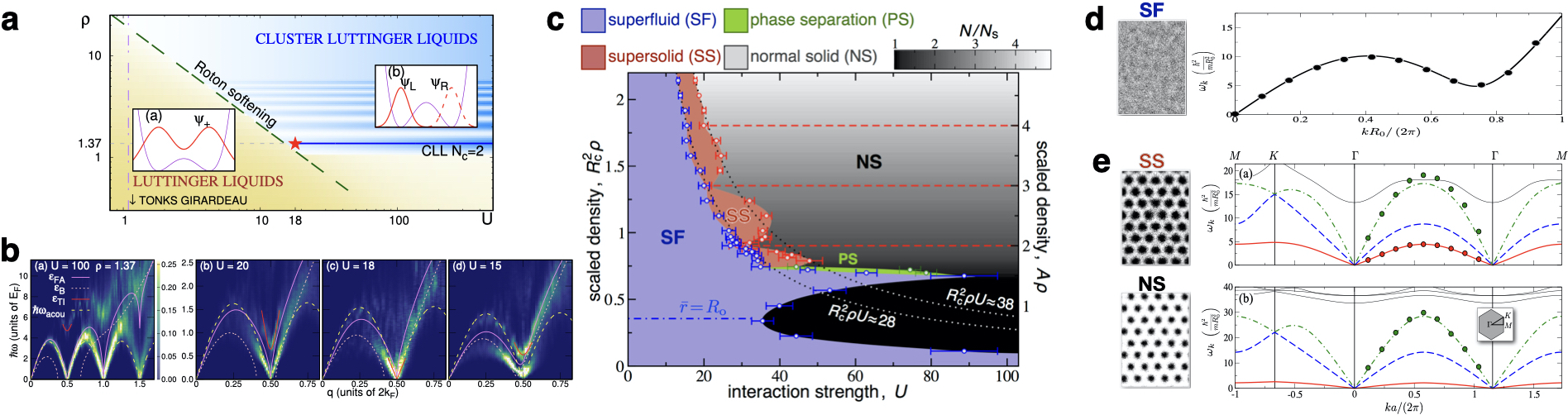}
\caption{\label{Fig13} {
\bf Zero-temperature phase diagram of one- and two-dimensional soft-core bosons in the continuum and their excitation spectrum.} 
(a) Phase diagram of one-dimensional soft-core bosons (log-log scale). 
A star marks the critical point between the LL and CLL phases for densities 
commensurate to $2$-particle clusters. The long-dashed line corresponds to the softening
of the Bogoliubov roton.
(b) Spectra at $\rho = 1.37$ with decreasing $U$, compared to Feynman $\epsilon_F$ and Bogoliubov $\epsilon_B$ approximations, and the harmonic chain
acoustic mode $\omega_\text{acou}$. At $q\approx q_c$, the secondary mode is fitted by the transverse Ising spectrum $\epsilon_{TI}$.
Figures (a) and (b) adapted from Ref. \cite{rossotti2017quantum}.
Two-dimensional system: (c) The phase diagram displays the emergence of superfluid (SF) and different solid (NS)
and supersolid (SS) phases for varying interaction strength $U$ and density $\rho$. The density
on the left $y$-axis has been scaled by the soft-core radius $R_{\rm c}$. The right axis gives the
density in units of the inverse area, 
$A=\sqrt{3}(1.6R_{\rm c})^2/2$, of the unit cell of the high-density solid phase, 
corresponding to the lattice site occupation $N/N_{\rm s}$ for a given number of 
particles and lattice sites, $N$ and $N_{\rm s}$, respectively. 
For $A\rho\gtrsim 1.5$, the grey region labeled as NS corresponds to a cluster crystal 
with $N/N_{\rm s}>1$, as indicated by the grey scale. Supersolid phases with
different occupation numbers are found between two hyperbolas, defined by 
$R_c^2 \rho U={\rm const.}$ (dotted lines). 
At high densities ($A\rho\gtrsim3.5$) they can be understood in terms of density 
modulated superfluids. In contrast, 
superfluidity within the low-density supersolid lobes emerges from delocalized 
zero-point defects according to the ALC scenario. 
Adapted from Ref. \cite{cinti2014defect}.
(d) (left) PIMC snapshot illustrating the particle density profile in the SF phase.
(right) Excitation spectrum in the superfluid phase compared to the PIMC data 
(circles) of \cite{saccani2012excitation}.
(e) (left) PIMC snapshot illustrating the particle density profile in the SS and NS phases.
(right) Mean field spectra (lines) at $\alpha = 16.93 $ (top) and $\alpha = 30.62$ (bottom) 
numerically computed along the three symmetry directions of the Brillouin zone 
[see inset of bottom panel]. 
The symbols represent the PIMC data of \cite{saccani2012excitation} 
for longitudinal excitations computed along the direction $\Gamma - M -\Gamma$ 
in the first two Brillouin zones.
Panels (d) and (e) adapted from Ref. \cite{macri2013elementary}.
}
\end{figure*}

\subsubsection{Quantum phases}
Soft-core potentials for hard-core bosons or spinless fermions 
on $1$D lattice systems described by the Hamiltonian
\begin{equation}
H = -t \sum_{\left<i,j \right>} b^ \dagger_i b_j + V \sum_{i<j; r_{ij}<r_c} n_i n_j,
\end{equation}
where $b_{i}$, ($b^\dagger_i$) are hard-core bosons annihilation (creation) localized on site $i$,
and $n_i=b^\dagger_i\, b_i$ is the density in $i$,
lead to correlated quantum liquid phases 
that do not fall into the conventional Tomonaga-Luttinger (LL) paradigm. Characteristic
features of these anomalous cluster Luttinger liquids (CLL)
include a deformation of the critical surface in momentum space and
are evident in correlation functions such as momentum
distributions and structure factors \cite{dalmonte2015cluster, mattioli2013cluster}
using DMRG and bosonization techniques.
A recent investigation of the spinful Fermi-Hubbard model 
with interspecies onsite interactions and density-density
soft-core interactions has been investigated \cite{botzung2019onedimensional},
which generalizes the extended Fermi-Hubbard model with soft-core radius equal to one lattice site studied in
\cite{nakamura2000tricritical}. It displays different types of CLL and a nontrivial supersymmetric critical line.  
The model in the continuum has been studied in 
\cite{rossotti2017quantum}, which showed evidence of the CLL via exact quantum Monte Carlo simulations.
The phase diagram of the system is shown in Fig.\ref{Fig13}a, together with the excitation spectrum in Fig.\ref{Fig13}b.  
The acoustic mode of the CLL phase (panels a-b) 
is gapless at $q = q_c$, corresponding to $k_F$, at this density.
After the transition, located at $U = U_c = 18$ (panel c),
this lowest excitation turns into the rotonic mode (panels d-e).
A weaker secondary mode appears also in the strongly correlated liquid
phase, in the form of a secondary roton. 
This secondary excitation in the LL phase can be linked to incipient cluster
formation, due to particles being preferentially localized close
to either the left or the right neighbor. The gap of both such LL excitations, and the
anharmonic optical modes of the CLL phase, vanishes
at the transition.

In the higher dimensional case in the continuum a good description is provided by a mean field treatment
\cite{pomeau1994dynamics,henkel2010three,macri2013elementary},
justified by the application of the first Born approximation to the two-body scattering problem, 
and the phases emerging from Eq.(\ref{eq:soft-core}) at zero temperature \cite{cinti2014defect}.
Using reduced units from now on till the end of the section, in mean field theory the system dynamics is described by a non-local Gross-Pitaevskii equation (GPE), which 
reads 
\begin{equation} 
\label{GPNH}
i \partial_t \psi({\bf r},t) = \left(-\frac{ \nabla^2}{2} + \alpha \int {\rm d}{\bf r}^{\prime} U({\bf r}-{\bf r}^{\prime}) 
|\psi({\bf r}^{\prime},t)|^2\right) \psi({\bf r},t) \;,
\end{equation}
where ${\bf r} \rightarrow {\bf r}/R_c$, $U({\bf r})=\frac{U_0}{1+r^6}$, and $\alpha =m\,n\, U_0 /\left(\hbar^2 R_c^2\right)$ 
is a dimensionless interaction strength that 
determines the ground state properties and the excitation dynamics.
The energy can be derived from the GP energy functional:
\begin{equation}
\label{GPenergy}
H = \int {\rm d}{\bf r}\; \frac{1}{2} \left| \nabla \psi_0 \right|^2 + 
\frac{\alpha}{2} \int {\rm d}{\bf r} \, {\rm d}{\bf r}^{\prime}\; |\psi_0({\bf r})|^2 U({\bf r}-{\bf r}^{\prime}) 
|\psi_0({\bf r}^{\prime})|^2 \;.
\end{equation} 
In order to numerically determine the location of the transition from a uniform to a modulated ground state, 
once can first expand the wavefunction 
$\psi_0(\mathbf{r})$ in Fourier series:
\begin{equation} \label{var}
\psi_0(\mathbf{r})=\sum_\mathbf{Q} C_\mathbf{\, Q}\; e^{i\, \mathbf{Q}\cdot \mathbf{r}}, 
\end{equation}
where $\mathbf{Q}=n\, \mathbf{b}_1+m\, \mathbf{b}_2$ with $n,m$ integers 
and $\mathbf{b}_1=\frac{2\pi}{a}\left(1,-\frac{1}{\sqrt{3}}\right)$,
$\mathbf{b}_2=\frac{2\pi}{a}\left(0,\frac{2}{\sqrt{3}}\right)$ are the reciprocal lattice basis vectors of a triangular lattice in two dimensions. 
One can then substitute Eq.(\ref{var}) into Eq.(\ref{GPNH})
and iteratively solve the non-linear equations for $C_\mathbf{\, Q}$
until convergence is reached \cite{PhysRevB.86.060510}. 
This procedure allows to determine the optimal lattice spacing,
the chemical potential and the coefficients $C_\mathbf{\, Q}$.
One finds that for low interaction strengths ($\alpha < 28$) the ground state
of the system is in a uniform superfluid phase. 
Upon increasing the interaction at $\alpha \approx 28$ one crosses
a first-order phase transition to a cluster supersolid phase 
characterized by a finite superfluid fraction
and broken translational invariance where particles arrange in clusters 
(each cluster contains an average
number of particles according to the density) in a triangular geometry. 
For even larger interactions $\alpha > 38$ the ground state preserves 
triangular symmetry but superfluidity vanishes resulting into an
uncorrelated cluster crystal.

The validity of the above mean field theory is limited to the regime of high densities, that is,
where the depletion of the condensate remains small for a wide range of interaction strengths. 
At lower densities one has to resort to {\it ab initio} methods to deal with the
development of nontrivial correlations. 
Numerical results were obtained
from path-integral Monte Carlo simulations \cite{ceperley1995path} based on the
continuous-space worm algorithm \cite{boninsegni2006worm} 
to determine the equilibrium properties of the system in the canonical ensemble,
that is, at a fixed temperature $T$ and a fixed
particle number (of the order of few hundreds).
The properties of the system ground state are obtained by extrapolating to
the limit of zero temperature, that is, by lowering the temperature $T$ 
until observables, such as the total energy, superfluid
fraction and pair-correlations did not change upon further decrease of $T$.

In Fig.\,\ref{Fig13}c it is presented the zero-temperature phase diagram of 
one- and two-dimensional soft-core bosons in continuum space.
At small densities $R_c^2 \rho \le 0.5$
one finds two phases: a superfluid and an insulating triangular 
crystal composed of singly occupied sites, that
is, where the number of lattice sites, $N_s$, equals the
particle number $N$.
A distinctive consequence
of the soft-core interaction is that the energy cost for
forming close particle pairs is bound by $V_0$. 
This potentially enables the formation of crystalline phases with
$N > N_s$ above a critical density where doubly occupied
lattice sites become energetically favorable on increasing
the lattice constant. 

The most interesting behavior takes place around the superfluid-solid quantum
phase transition at $N/N_s = 2$. 
Starting from the insulating solid with doubly occupied lattice sites, 
removing a small number of particles does not cause structural changes
of the ground state but rather creates a small fraction
$f_{def} = (2 N_s - N)/N_s>0$ of zero-point crystal defects in the
form of singly occupied sites. Such defects delocalize and give rise to a finite
superfluid fraction, in agreement with the ALC scenario.
It is interesting to notice that at the classical level one observes an intriguing scenario in which
the coexistence of a cluster crystalline structure, breaking translational symmetry in equilibrium, 
and of particle diffusion can be here explained in terms of
a thermally activated hopping mechanism, 
where particles delocalize without altering the underlying cluster
crystalline matrix \cite{diaz-mendez2015monodisperse}.

The $2$D extended Bose-Hubbard model with finite-range soft-core interactions
on the square lattice {\it with an hard-core constraint} 
displays an intriguing behavior \cite{pupillo2008cold, masella2019supersolid}
For intermediate interaction strengths $4\le V/t \le 4.45$ the stripes can turn superfluid, thus
leading to a self-assembled array of quasi one-dimensional superfluids. 
These bosonic superstripes turn into an isotropic supersolid with decreasing interaction strength. 
It is relevant to notice that the mechanism for stripe formation is based on cluster self-assemblying 
different from recently proposed mechanisms for dipolar magnetic atoms \cite{bottcher2021new}, spin-orbit coupled BECs \cite{li2017stripe}, or BECs with cavity-mediated interactions \cite{leonard2017supersolid}. 
A two-component version of this model in the square lattice
has also been recently proposed in \cite{li2018supersolidity}, where, among the several phases of the model, 
one can observe that the components that interact via a soft-core potential can 
induce a supersolid phase in the other component.
The out-of-equilibrium dynamics following a temperature quench to values well below the hopping amplitude $T/t \ll 1$
shows that together with classical solid phases and supersolids (for $3.8\le V/t \le 4.2$) 
also a normal glass is observed (for $V/t > 5.5$) without any remnant superfluidity
\cite{angelone2020nonequilibrium}. 
It is interesting to observe that in a triangular lattice, the same system after a temperature quench 
displays a superglass and a normal glass phase \cite{angelone2016superglass}.
For high enough temperature, the glass and superglass 
turn into a floating stripe solid and a supersolid, respectively.
Similar models of systems with nonlocal interactions 
diverging at the origin leading to glassy phases have been also recently investigated 
in the context of type-$1.5$ superconductors \cite{wang2020pinning} where the 
{\it particles} are point-like vortices in the presence of external disorder.

The three-dimensional soft-core model was investigated originally by \cite{henkel2010three,ancilotto2013supersolid}
for the repulsive case and by \cite{maucher2011rydberg} for the attractive one within a mean-field approach based 
on the solution of the $3$D GPE of eq.(\ref{GPNH}).
In the repulsive isotropic case, the ground-state phase diagram displays a transition from a superfluid phase at low density and interactions to an $fcc$ supersolid at intermediate densities, induced by a roton instability similar to the $2$D case. 
For attractive interactions one can prove the existence of (bright soliton) self-bound macroscopic states, stabilized purely by the competition of kinetic and negative mean-field energies.

\subsubsection{Elementary excitations}
The elementary excitations in the mean field approximation are found by expanding the GP energy functional around
the solution $\psi_0({\bf r})$, obtaining the so called Bogoliubov-de Gennes equations
\cite{macri2013elementary,macri2014ground}.
Denoting the change in $\psi(\mathbf{r},t)$ by $
\delta \psi(\mathbf{r},t)=e^{-i\mu t}\left[ u(\mathbf{r})
e^{-i\omega t}-v^*(\mathbf{r})e^{i\omega t}\right]$ 
and substituting this expression into the GPE eq.(\ref{GPNH})
one finds a set of two coupled linear differential equations:
for the Bogoliubov amplitudes $u(\mathbf{r})$ and $v(\mathbf{r})$.
The solution of the Bogoliubov equations in the uniform superfluid phase is analytical:
\begin{equation} \label{BdGuni}
\epsilon_q =\sqrt{\frac{q^2}{2}\left(\frac{q^2}{2}+2\alpha\,  U_q\right)},
\end{equation}
and depends only on the modulus of the excitation vector $\mathbf{q}$. 
Here $U_q$ is the Fourier
transform of the potential. 
Eq.(\ref{BdGuni}) can be extended to the case of multibody interactions \cite{laghi2017excitations}.
The spectrum is linear for small momenta and the slope defines the sound 
velocity of the system; for sufficiently large $\alpha$ (the specific value 
depends on the shape of the interaction)
one recovers the usual roton-maxon spectrum that is common
to other physical systems with non-local interactions as ultracold dipolar systems or superfluid $^4$He. 
In nonuniform phases one has to rely on a numerical solution of the Bogoliubov equations. One can use
a Fourier expansion of the Bogoliubov amplitudes followed by a 
diagonalization of the corresponding equations.
The results presented in Fig.\ref{Fig13} (d,e) are obtained using a grid based 
solution in real space for the lowest excitation bands and for $\mathbf{q}$ vectors
lying in the first Brillouin zone (FBZ) \cite{macri2013elementary} for a soft-shoulder potential.
The figure shows the excitation energies along the three
symmetry axes of the Brillouin zone corresponding to
the underlying triangular lattice. We find three gapless
bands, i.e. three Goldstone modes reflecting the symmetries that are broken in the supersolid phase \cite{watanabe2012unified, watanabe2013redundancies}. In
addition to the {\it superfluid band} due to the breaking of
global gauge symmetry, there are two bands corresponding to longitudinal and transverse phonon excitations of
the two-dimensional lattice. 
Even in the insulating phase, Bogoliubov-de Gennes equations yield excellent
agreement for the longitudinal phonon mode with quantum Monte Carlo calculations 
based on the so-called Genetic Inversion via falsification of the theories (GIFT) 
method for the inversion of the inverse Laplace transform 
$F(k, \tau)=\int d\omega e^{-\tau \omega} S(k, \omega)$
of the dynamic structure factor \cite{saccani2012excitation}, 
despite its evident inability to describe the break-down of global superfluidity. 
This indicates that each individual droplet
maintains a high condensate fraction despite the apparent lack of 
global phase coherence between the crystalline
ordered droplets. A proper identification of
each band can be done by computing local fluctuations
on top of the mean field solution $\psi_0(r)$.
One clearly distinguishes the transverse band
from the direction of the fluctuations, orthogonal to the
perturbing vector $k$. 
The contribution of this band to phase fluctuations is strongly suppressed. The first and
third band both contribute to density and phase fluctuations 
with different weight. The first band is
mostly responsible for phase whereas the third to density fluctuations. 
Therefore the lower band can be associated to the superfluid 
response of the system, whereas
the other two to the classical collective excitations of the
crystal. The results for a Rydberg-dressed potential of eq.(\ref{eq:soft-core}) are reported in \cite{macri2014ground}.
There the modes obtained by the solution of the BdG equations have been compared to quantum Monte Carlo calculations with the inclusion of the transverse excitation band. 
A good agreement between the two techniques has also been obtained for all three excitation bands.
We briefly comment that the calculation and the measurement 
of the excitation spectra received much attention also in the context of in dipolar systems,
both in trapped superfluid or droplet phases \cite{baillie2017collective,petter2019probing} and more recently also in supersolids \cite{tanzi2019supersolid,petter2020highenergy} 
in the ground states and in 
excited states, e.g. in vortices 
\cite{cidrim2018vortices,lee2018excitations,roccuzzo2020rotating}.

\subsection{Structural transitions in mesoscopic long-range systems.}
\label{sec_zig_zag}

The physics of structural transitions in power-law potentials has been deeply studied  in prototypical mesoscopic systems of ions and dipolar systems thanks to the  close connection to experimental realizations. 
The simplest one-dimensional case of 
a chain of singly-charged particles, confined by a harmonic potential, exhibits a sudden transition
to a zigzag configuration when the radial potential reaches a critical value, depending on the particle
number\,\cite{birkl1992multiple,bluemel1988phase}. For charged particle interacting via the Coulomb potential ($\alpha=1$) this structural change is a phase transition of second order, whose order parameter is the
crystal displacement from the chain axis \cite{schiffer1993phase, piacente2004generic, morigi2004eigenmodes, fishman2008structural} as was also experimentally observed\,\cite{enzer2000observation,kaufmann2012precise}. In the low temperature limit, this is quantum phase transition, whose universality lies in the same class as that of the ferromagnetic transition of an Ising chain in
a transverse field\,\cite{porras2004effective,shimshoni2011quantum,friednauer2008simulating}. The zig-zag transition also appears in strongly interacting one-dimensional electrons systems, i.e. quantum wires, whose Wigner-crystal phase corresponds to a splitting of the fermi gas into two chains\,\cite{meyer2007transition}. Interestingly, the zig-zag transition may be also related to the Peierls instability which occurs in antiferromagnetic chains coupled to phonon modes\,\cite{bermudez2012spin}.

As the range of the interactions decreases to $\alpha>2$ the nature of the transition is radically modified due to the coupling between  transverse and axial vibrations\,\cite{cartarius2014structural}, which leads to a weakly first order transition in analogy with the case of a ferromagnetic transitions in presence of phonon excitations\,\cite{larkin1969phase,imry1974tricritical}. This is particularly relevant to the study of self-organized phases in polar systems\,\cite{goral2002quantum,astrakharchik2007quantum,buchler2007strongly}. 

In higher dimensions crystals of repulsively interacting ions in planar traps form hexagonal lattices  undergo an instability towards a multilayer structure as the transverse trap frequency is reduced. 
The new structure is composed of three planes, with separation increasing
continuously from zero. A mapping to the six-state clock model can be performed, implying that 
fluctuations split the buckling instability into two thermal transitions, accompanied by the
appearance of an intermediate critical phase. A Berezinskii-Kosterlitz-Thouless phase is predicted interfacing the disordered and the ordered phase \cite{podolsky2016buckling}. 

Another important case is the generalization to the case of multi-scale potential which has
been recently studied in the quantum regimes
in\,\cite{cinti2019thermal,pupillo2020quantum,abreu2020superstripes} which can for specific
values of the parameters of the pairwise potential can
support quasicrystalline phases or stripe phases. 
The corresponding criteria to realize structural phase in these more 
complex potentials have been investigated \,\cite{mendoza2012coarse,mendoza2015nature,mendoza2015reply,mendoza2019mechanism,mendoza2017quantum}.

Finally we comment on the presence
of smectic, nematic, and hexatic phases in quantum systems with
competing non-local interactions, which
presents several analogies to the case of classical liquid-crystal systems\cite{abanov1995phase}. This parallel, which derives from the similarity between the anisotropic nature of the stripe order and the elongated shape of liquid-crystal molecules, allows the application of traditional results from liquid-crystal systems\,\cite{gennes1993physics,chaikin1995principles} to predict the qualitative, and to some extent also quantitative phase behavior of many systems with modulated order parameters.

In the context of dipolar Fermi gases theory has been, until now, ahead of experiments, with several preliminary theoretical calculation predicting exotic phases, such as p-wave superfluid
\cite{bruun2008quantum}, supersolid \cite{Lu:2015em}, hexatic
\cite{lechner2014role,bruun2014quantum}, and Wigner crystal phases \cite{matveeva2014fixed}.
In these systems stripe formation (in the form of charge density waves) and nematic phases should also occur with features analogous to the ones present in low temperature long-range solid state systems.


\section{Dynamical critical behaviour}
\label{sec:dyn}

In this section we review the multifaceted aspects of dynamical regimes in quantum
long-range aspects, emphasizing as much as possible universal behaviours. Given the vast amount
of literature on the subject, we decided to arrange the material presenting first
a discussion of metastability, a hallmark of long-range systems, followed by a presentation
of results on Lieb-Robinson bound, Kibble-Zurek mechanism, dynamical phase transitions and
confinement in quantum long-range systems. Miscellaneous material is presented in the last section.

\subsection{Metastability and diverging equilibration times}
\label{sec_5_1}

Diverging equilibration times in the thermodynamic limit are
a notorious characteristic of long-range interacting systems. Recently,
the absence of equilibration of strong long-range quantum systems has been
directly linked to their peculiar single particle spectrum, which leads
to a violation of Boltzmann's H-theorem and the appearance of finite Poincar\'e recurrence times
in the thermodynamic limit\,\cite{defenu2021metastability}. These observations are in agreement
with the aforementioned properties, see Sec.\,\ref{sec_reminder}, which are common to
thermodynamically large long-range systems and finite local ones, such as the impossibility to
fully disregard boundary over bulk phenomena\,\cite{barre2007ensemble, latella2015thermodynamics},
the existence of concave entropy regions\,\cite{ispolatov2001first} or the presence
of a macroscopic energy gap between the ground state and the first excited state\,\cite{gupta2012one,gupta2012overdamped}. 

The key point is that the excitation spectrum of non-interacting systems does not become continuous
in the thermodynamic limit, as the eigenvalues of a long-range coupling matrix
can be shown to remain discrete even in the infinite components limit,
forming a pure point spectrum\,\cite{last1996quantum} similar to the appearing in
strongly disordered systems\,\cite{thouless1972anderson,froehlich1983absence,simon1985some,scardicchio2017perturbation}. A discussion of the spectral discreteness of long-range couplings in the thermodynamic limit has been presented in Ref.\,\cite{defenu2021metastability} for few quadratic models and employed to justify the observation of diverging equilibration times in a long-range Ising model, quenched across its quantum critical point\,\cite{kastner2011diverging}.

\begin{figure*}[]
\centering
\subfigure[Ising model]{\label{Fig1a_sec5}\includegraphics[width=.32\linewidth]{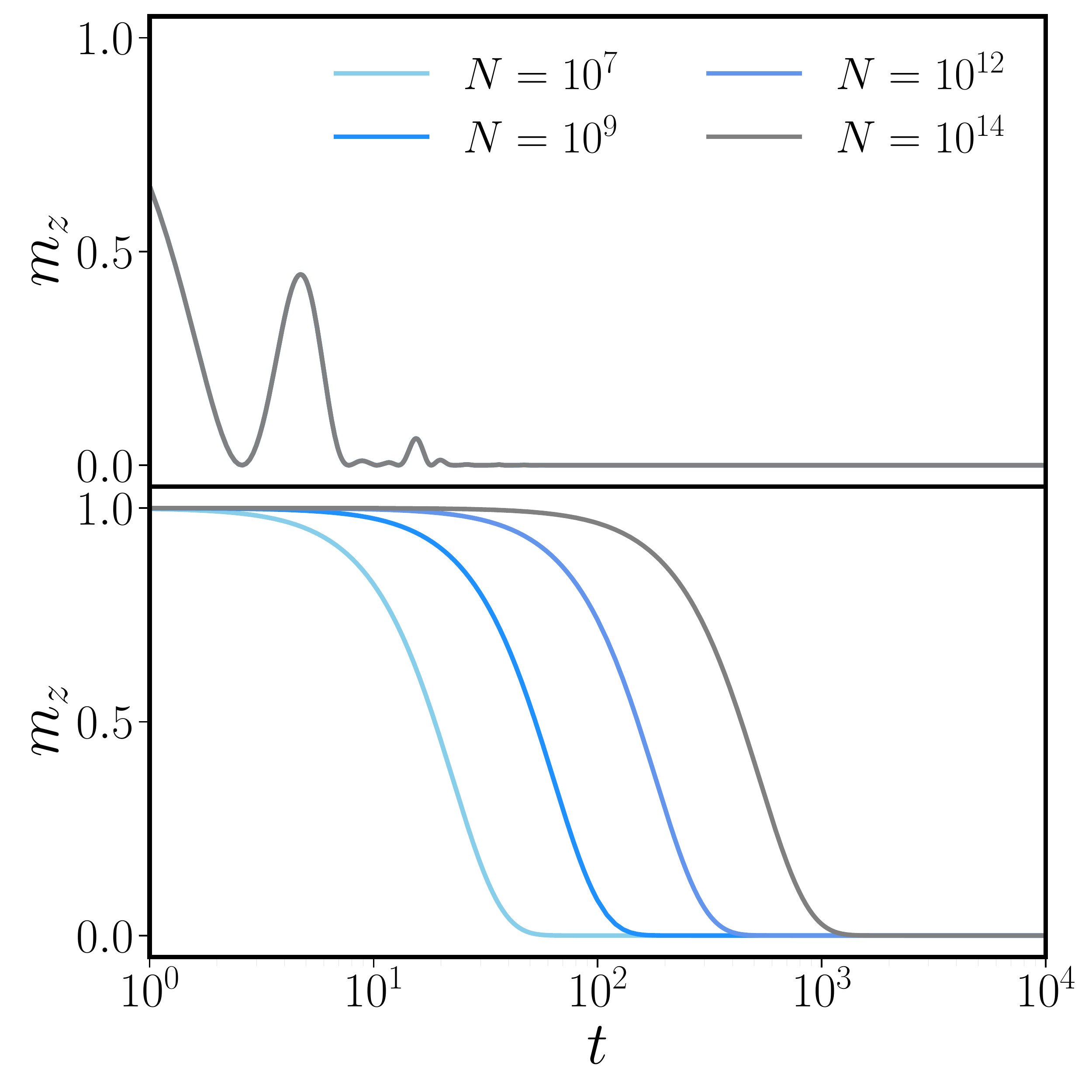}}
\subfigure[Kitaev chain]{\label{Fig1b_sec5}\includegraphics[width=.32\linewidth]{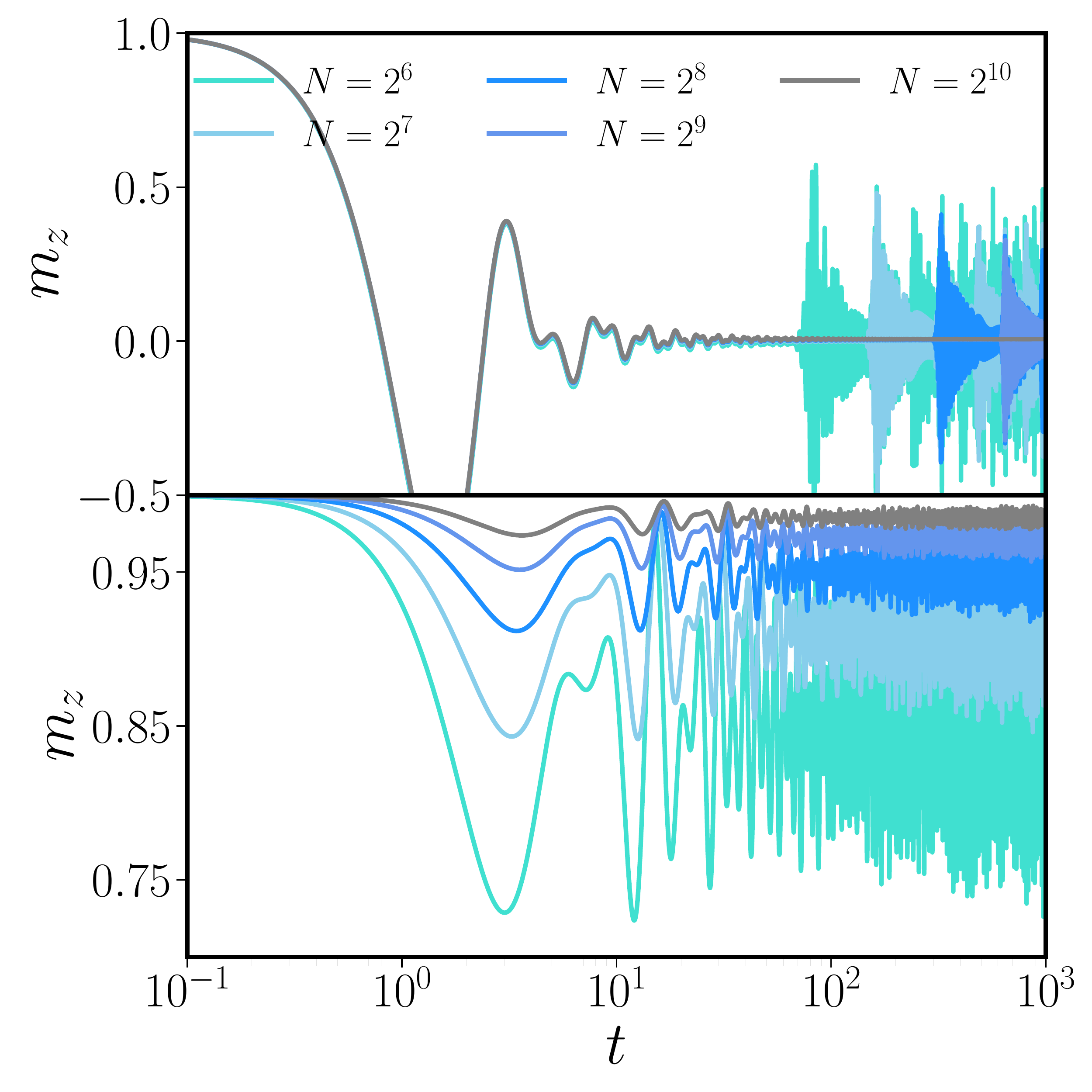}}
\subfigure[Spherical model]{\label{Fig1c_sec5}\includegraphics[width=.32\linewidth]{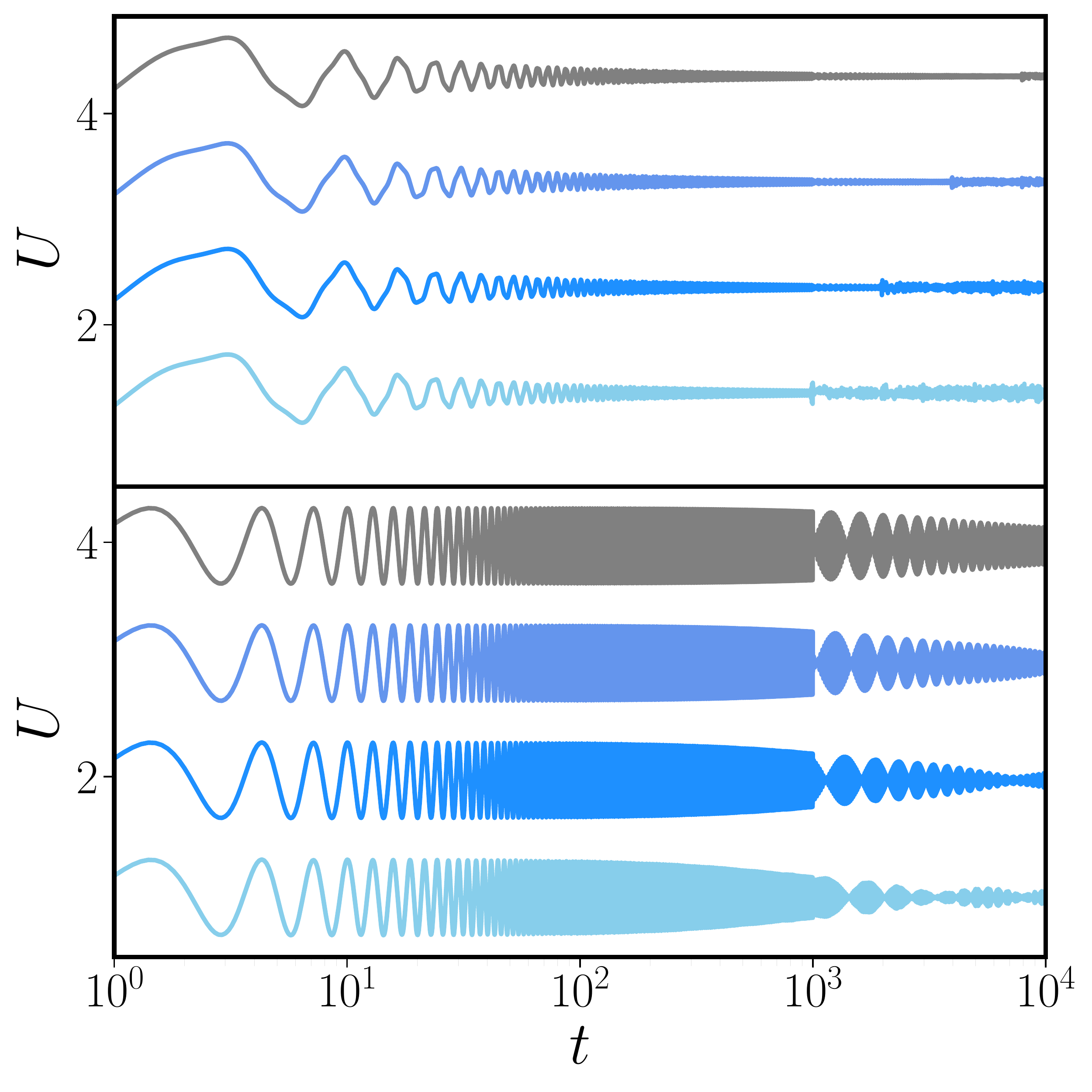}}
\caption{{\bf Evidences of QSS in the long-range Ising chain, the Kitaev chain and the one dimensional spherical model, from left to right respectively.} In all panels the top sub-panel displays the case of weak long-range interactions $\alpha>1$, where roughly the same equilibration properties of the nearest neighbour case are found. Conversely, the bottom sub-panel shows the case of strong long-range interactions $\alpha<d$, where dynamical fluctuations survive in the $t\to\infty$ limit. The leftmost panel displays the transverse magnetisation of the long-range Ising model, see the Hamiltonian in Eq.\,\eqref{h_lri}, after a quench from the fully paramagnetic state at $h\to\infty$ deep into the ordered phase at $h\to 0$. While the observable expectation equilibrates at long times for $\alpha=2.0$ (top sub-panel), it persists in its initial value for increasingly longer times as the system size increases for $\alpha=0.5$ (bottom sub-panel). See the discussion in Ref.\,\cite{kastner2011diverging}. A similar signature is noticed in the case of the Hamiltonian in Eq.\,\eqref{lrk_h_app_i}, i.e. the Kitaev chain representation of the Ising model, where the dynamics can be exactly solved for any quench across the phase boundary. The central panel shows the evolution of the spatial and quantum average of the $\sigma_{z}$ in Eq.\,\eqref{sz_op} for a long-range Kitaev chain with $\alpha\gtrsim 12$ (top sub-panel) and $\alpha=0.4$ bottom sub-panel for $h_{i}\gg 1$ to $h_{f}=0.4$. Lack of equilibration also appears for non-critical quenches, as it shown in the rightmost panel for the observable $A(t)$ of a quantum spherical model with long-range interactions. Dynamical fluctuations reduce as size increases for decay rates $\alpha>1$, see the upper sub-panel where the $\alpha\gtrsim12$ case is shown for increasing system sizes $N\in[2^{9},2^{10},2^{11},2^{12}]$ from bottom to top. Conversely dynamical fluctuations tend to increase for $\alpha<1$, as shown in the lower sub-panel for $\alpha=0.2$, see Ref.\,\cite{defenu2021metastability}. \label{fig_met}
 }
\end{figure*}

In fact, first evidences of QSS in quantum systems have been described in the prototypical example of the long-range Ising chain, see\,\eqref{h_lri}. The QSS have been shown to appear for quenches starting well inside the paramagnetic phase in the $h\to+\infty$ limit and terminating in deep in the ferromagnetic phase at $h=0$. Then, the system is prepared in the transversally polarised ground state and evolved according to the classical ferromagnetic Hamiltonian in absence of the transverse field. It follows that the expectation of the global operator $m_{z}=\langle\sum_{i} \sigma^z_i\rangle/N$ with the Hamiltonian in Eq.\,\eqref{h_lri} evolves from the initial value $\lim_{t\to0}m_{z}=1$ to the equilibrium expectation $\lim_{t\to\infty}m_{z}=0$, if the system actually equilibrates, see Fig.\,\ref{Fig1a_sec5}. These observations may be extended to any choice of the initial and final magnetic fields $h_{i},h_{f}$ using the Kitaev chain representation of the Ising model given in Eq.\,\eqref{approx}, see the discussion in Sec.\,\ref{lrkc_i_rel}.

It is worth recalling that for $1<\alpha<3$ the the correspondence between the fermion and spin Hamiltonians in respectively Eqs.\,\eqref{h_lri} and\,\eqref{approx} is not exact. Yet, the existence of the quantum critical points is preserved and the equilibration scenario for the two systems is analogous\,\cite{essler2012dynamical,regemortel2016information}. The analogy between the transition of the Ising and Kitaev chain has been discussed in Sec.\,\ref{lrkc_i_rel} and in the Refs.\,\cite{jaschke2017critical,defenu2019dynamical}. Within the Kitaev chain perspective, the critical point at $h=h_{c}=1$ is signalled by the property $\lim_{k\to 0^{\pm}}\theta_{k}=\pm\frac{\pi}{2}$, where the critical Bogoliubov quasi-particles are constituted by an equal superposition of electrons and holes ($|u_{k=0}|=|v_{k=0}|=1/\sqrt{2}$). This phenomenon is often interpreted as a Dirac mode resulting from the superposition of two Majorana edge states\,\cite{fradkin2013field}.

In the strong long-range regime ($0<\alpha<1$) and in presence of the Kac rescaling a full characterisation of the quantum phase transition in the Kitaev chain has not been attempted yet. Indeed, no clear continuous limit can emerge for this regime in thermodynamic limit due to the spectral discreteness evidenced in Ref.\,\cite{defenu2021metastability}.  Nevertheless, the existence of the quantum critical point can be also inferred in the strong long-range regime, analysing the $k\to0$ limit of the Bogoliubov angles.

The equilibration of a weak long-range Kitaev chain after a sudden quench of the chemical potential $h$ is summarised in the upper sub-panel of Fig.\,\ref{Fig1b_sec5}. The initial state of the system is the ground state at $h=h_{i}\gg 1$, deep in the normal phase, where $m_{z}\approx 1$. Then, this initial state is evolved according to the ferromagnetic Hamiltonian with $h=h_{f}<1$. The explicit description of the quench dynamics solution can be found in Ref.\,\cite{defenu2019dynamical}. In order to compare with the aforementioned investigations regarding QSS in the long-range Ising model the pictures displays the evolution of the observable 
\begin{align}
\label{k_mx}
m_{z}=1-\frac{2}{N}\sum_{i} \langle c^{\dagger}_{i} c_{i}\rangle,
\end{align}
which represents the transverse magnetisation in terms of the Fermi quasi-particles.
 
From the long-time dynamics of the observable in Eq.\,\eqref{k_mx} it is rather evident that the equilibration in the weak long-range Kitaev chain, see the upper panel Fig.\,\ref{Fig1b_sec5}, mimics the case of the long-range Ising with $\alpha=2$, see the upper panel Fig.\,\ref{Fig1b_sec5}. The initial value of the observable rapidly equilibrates to a long-time expectation which becomes time independent in the long time limit. In other words, any observable $A(t)$ relaxes to equilibrium if it approaches its Cesaro's average
 \begin{align}
 \label{ces_avg}
 \bar{A}=\lim_{T\to\infty}\langle A \rangle_{T}\quad\mathrm{with}\quad\langle\cdots\rangle_{T}=\frac{1}{T}\int_{0}^{T}\cdots dt.
 \end{align}
Moreover, the dynamical fluctuations, which are quantified by the parameter
\begin{align}
Q_{A}(T)=\langle\left|A(t)-\bar{A}\right|^{2}\rangle_{T}
\end{align}
must disappear in the long-time limit
\begin{align}
\label{eq_def}
\lim_{T\to\infty}Q_{A}(T)\approx 0.
\end{align}
Eq.\,\eqref{eq_def} is the conventional way to define equilibration in closed quantum systems\,\cite{short2011equilibration,reimann2008foundation,linden2009quantum,oliveira2018equilibration}.

In the weak long-range regime ($\alpha>d$) the result $\lim_{T\to\infty}Q_{m_{z}}(T)=0$ can be exactly proven for most quadratic models as well as for the Ising model for sudden quenches from $h_{i}=+\infty$ to $h_{f}=0$  thanks to the Riemann--Lebesgue lemma\,\cite{hughes2008calculus}. In other words, equilibration occurs in these systems as the Poincar\'e recurrence times diverge for $N\to\infty$. This phenomenon is evident in the numerical computation of the $m_{z}$ expectation value both for the Ising and the Kitaev chain with $\alpha>1$, see the upper panels of Fig.\,\ref{Fig1a_sec5} and\,\ref{Fig1b_sec5}.

This picture is radically altered in the $\alpha<1$ case, see the bottom panels in Fig.\,\ref{Fig1b_sec5}. Indeed, the dynamical evolution of the observable $m_{z}$ persists in the vicinity of its initial value for longer times as the system size is increased, in agreement with the $\tau_{\rm eq}\propto N^{\beta}$ expectation coming from classical systems\,\cite{campa2009statistical}. Interestingly, the $\beta=1/2$ scaling observed in the long-range Ising model appears to be related with the scaling of Poincar\'e recurrence time due to the discrete spectrum of long-range systems\,\cite{kastner2011diverging,defenu2021metastability}. It is also worth noting that the scaling of time-scales in long-range systems is crucially influenced by the Kac rescaling and, then, these observations may be altered modifying the regularization procedures\,\cite{bachelard2013universal}.

While the phenomenology of the Kitaev and Ising models are analogous, the quantitative features of the dynamical evolution display some peculiar differences. In particular, 
in the long-range Ising model no oscillatory fluctuations are present, while they occur in the Kitaev chain. 
These differences are probably due to the different quench boundaries between the two models. Despite these details, it is evident that the curves in the lower panels of Fig.\,\ref{Fig1a_sec5} and\,\ref{Fig1b_sec5} will both yield $\lim_{T,N\to\infty}Q_{m_{z}}(T)\neq 0$. This result do not depend on the order in which the two limits are taken.

The appearance of the QSS has been often connected to the scaling of equilibration times of critical observables such as the magnetization\,\cite{antoni1995clustering, mukamel2005breaking, campa2009statistical}. However, signatures of persistent time fluctuations in classical systems have been also found in generic thermodynamic observables, as for the evolution internal energy in systems of particle with attractive power-law pair interactions\,\cite{gabrielli2010quasistationary}. The same picture can be also found in many-body quantum systems, such as the spherical model, whose Hamiltonian reads\,\cite{vojta1996quantum,sachdev1999quantum}
\begin{align}\label{sph_ham}
H = \frac{g}{2} \sum_i p_i^2 + \frac{1}{2} \sum_{i,j} V_{|i-j|} s_i s_j
	+ \mu \left( \sum_i s_i^2 - \frac{N}{4} \right).
\end{align} 
The parameter $g$ controls the strength of quantum fluctuations and the coupling matrix $V_{|i-j|}\propto d_{ij}^{-\alpha}$ couples all pair of sites in the linear chain. The parameter $\mu$ plays the role of an effective chemical potential and it has to be chosen in order to enforce the constraint condition $\left\langle\frac{4 \sum_{i} s_{i}^{2}}{N}\right\rangle=1$. The connection between the equilibrium scaling of $O(N)$ field theories and the spherical model has been already discussed in Sec.\,\ref{q_rot_mod}. From the dynamical point of view, it can be shown that the spherical model corresponds to the time-dependent Hartee-Fock approximation of the Ising and $O(N)$ rotor models\,\cite{berges2007quantum}. In fact, the harmonic oscillator variables in Eq.\,\eqref{sph_ham} can be interpreted as the spin-wave exctiatations appearing in the Holstein-Primakoff description of the spin variables, see Eqs.\,\eqref{hp_z},\,\eqref{hp_+} and\,\eqref{hp_-}. 

Accordingly, several phenomena occurring in the out-of-equilibrium dynamics of critical systems may be approximated via the spherical model\,\cite{sotiriadis2010quantum}, including prethermalization\,\cite{chiocchetta2017dynamical, halimeh2021quantum}, defect formation\,\cite{degrandi2010adiabatic}, dynamical phase transitions\,\cite{syed2021dynamical}.
In particular, the dynamics induced by a sudden quench leads to the relaxation  of most observables according to the definition given in Eq.\,\eqref{eq_def}\,\cite{sotiriadis2010quantum,chandran2013equilibration, syed2021dynamical}.

However, equilibration does not occur in the non-additive regime  due to the discrete spectrum of the coupling matrix $V_{|i-j|}=-\frac{J_{0}}{d_{ij}^{\alpha}}$. This is clearly visible in the dynamical evolution displayed in Fig.\,\ref{Fig1c_sec5}. Each curve represents the evolution of the observable $U(t)=\left\langle\frac{4 \sum_{i} s_{i}^{2}}{N}\right\rangle$ after a sudden quench of the chemical potential from the initial value $\mu_{i}=2\mu_{c}$. The constraint is not imposed during the dynamics and the observable is let free to evolve according to the final Hamiltonian in Eq.\,\eqref{sph_ham} with $\mu=\mu_{f}=1.1\mu_{c}$.

Thus, the quench occurs within the normal phase $\mu_{f}>\mu_{c}$ and, apart for multiplicative factors, the observable $U(t)$ represents the potential energy of the system. In the weak long-range regime the amplitude of dynamical fluctuations decreases at long-times until the Poincar\'e recurrence time occurs. The value of such recurrence times grows increasing the system sizes $N\in[2^{9},2^{10},2^{11},2^{12}]$ from bottom to top in Fig.\,\ref{Fig1c_sec5}. The same does not occur for $\alpha<1$, where the width of dynamical fluctuations increases for growing system sizes, see the lower panel of Fig.\,\ref{Fig1c_sec5}, same size order as in the upper panel.

A simple explanation of this phenomenon is found into the fully connected limit ($\alpha\to 0$), where the spectrum of the coupling matrix $V_{|i-j|}$ separates between two distinct energy levels in the thermodynamic limit: a non-degenerate ground-state with energy $-J_{0}$ and a $N-1$ degenerate excited state with energy $0$. In presence of any given set of boundary conditions, full degeneracies does not occur and the system behaves at finite size as a set of harmonic oscillators with discrete energies. As the size increases, the spectrum accumulates at high energy where the eigenvalues $V_{q}$ of the coupling matrix become all identical, making the system equivalent to a single quenched harmonic oscillator. It follows that $\lim_{T,N\to\infty}Q_{\rm U}(T)\neq 0$ for any quench dynamics.

Interestingly, the metastability observed in the strong long-range regime appears to me more fundamental than the one observed in disordered systems. Indeed,  when flat interactions ($\alpha=0$) are perturbed by Gaussian distributed weak couplings $u_{ij}$, with
\begin{align}
P(u_{ij})\propto \exp\left(-N\,u_{ij}^{2}/2J^{2}\right)
\end{align}
whose width $2J$ represents the disorder strength. 
The disordered couplings lift the infinite ($\sim N-1$) degeneracy of the excited state at zero energy and the spectrum becomes continuous apart from the single non-degenerate ground-state at energy $J_{0}$, where $J_{0}>J$ is the strength of the flat homogeneous interactions\,\cite{kosterlitz1976spherical, edwards1976eigenvalue,dellanna2008critical}. The density of states of the continuous spectrum follows the celebrated Wigner semicircle law\,\cite{metha2004random}. In analogy with the non disordered case, the model lies in its equilibrium state at $\mu_{0}= 2\mu_{c}$ at $t=0$, when it is suddenly quenched at $\mu_{f}=1.1\mu_{c}$. The continuum nature of the spectrum leads the internal energy $U(t)$ to exponentially equilibrate according to the definition given in Eq.\,\eqref{eq_def}.
\begin{figure}[tbhp]
\centering
\includegraphics[width=1.\linewidth]{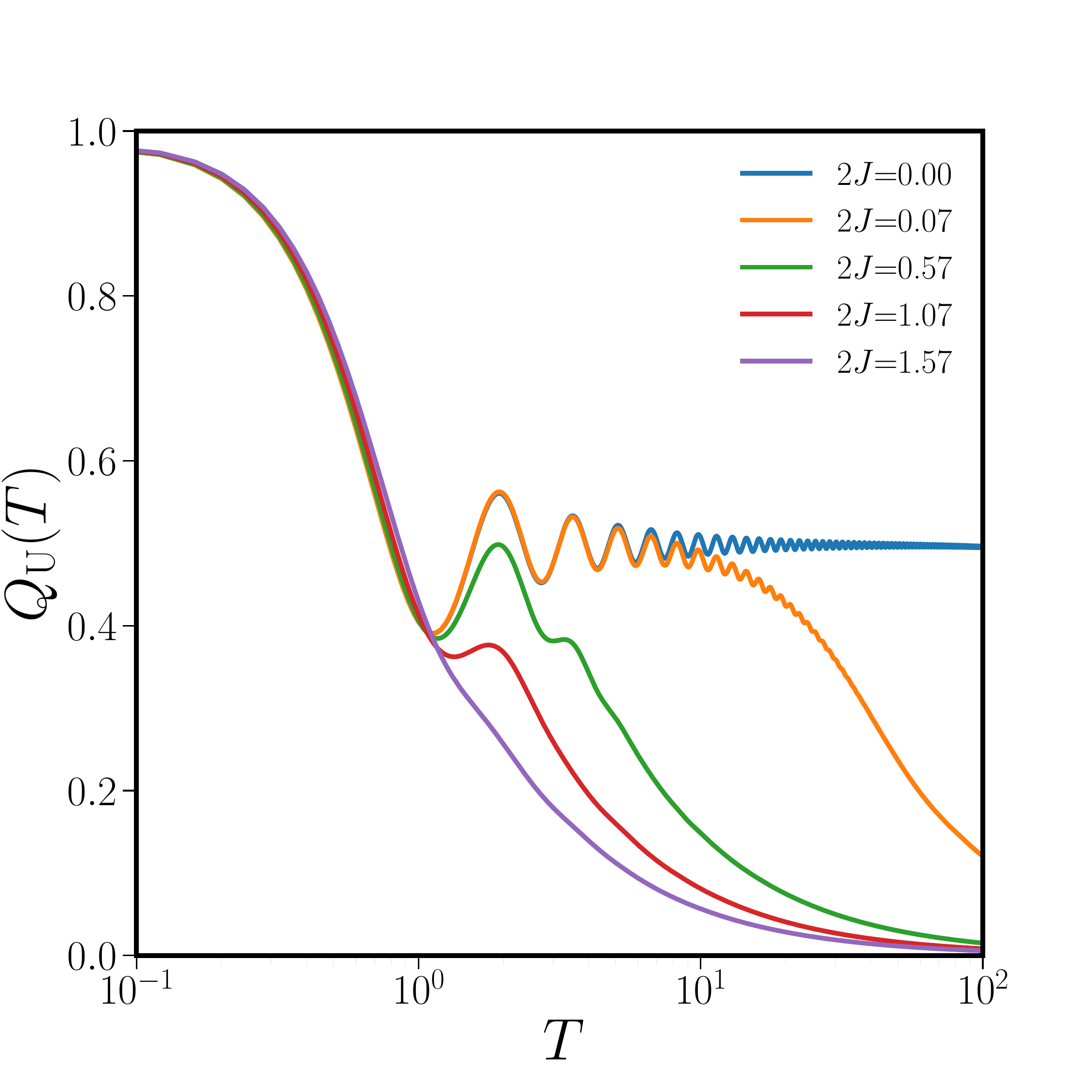}
\caption{\textbf{Equilibration of long-range spherical model.} Dynamical fluctuations decay as a function of the disorder strength for the potential energy observable $U$. As the disorder strength is decreased the decay rate also decreases until it vanishes in the zero disorder limit (upper blue curve), where dynamical fluctuations persist at all times and the equilibration condition in Eq.\,\eqref{eq_def} is never fulfilled.}
\label{Fig2_SecV}
\end{figure} 

The equilibration dynamics of the disordered quantum spherical model is accurately summarised by Fig.\,\ref{Fig2_SecV}. As long as the sudden quench occurs between Hamiltonian configurations within the normal phase, dynamical fluctuations tend to vanish in the long-time limit and the condition in Eq.\,\eqref{eq_def} is obeyed. The approach to the long-time limit is exponential with a finite equilibration time $\tau_{\rm eq}$, which strongly depends on the disorder strength $J$. As the disorder strength decreases the equilibration time grows, until it diverges for $2J=0$, where spectral discreteness is recovered and equilibration, as defined by Eq.\,\eqref{eq_def}, ceases to occur.

It shall be possible to connect most of the peculiar properties, which emerge in the dynamical behaviour of strong long-range systems, with the spectral discreteness and the finite size scaling properties, discussed in the present section. In the following we will try to evidence some of these connections while discussing few of the most celebrated dynamical properties of long-range systems.

\subsection{Lieb-Robinson bound}\label{sec:Lieb-Rob}

Understanding the maximum speed at which information propagates in many-body systems allows to put tight bounds on fundamental questions, such as how fast a quantum system can thermalize \cite{calabrese2006time} or the amount of quantum information that can be transmitted through a quantum channel \cite{Bose2007quantum}. 
In short-range interacting systems the Lieb-Robinson bound predicts a constant maximal velocity that confines the information to a linear effective light-cone \cite{lieb72finite}. 
Long-range interactions substantially alter this picture, since the traditional definition of group velocity does not apply to their case. Accordingly, the spreading of correlations, information, or entanglement speeds up dramatically, leading to a wide range of exotic dynamical properties, which may be exploited for fast information transmission, improved quantum state preparation, and similar applications.
Then, it is not surprising that a large body of theory work has emerged in recent years in order to find tighter propagation bounds for different values of the power-law exponent $\alpha$ \cite{hastings2006spectral,eisert2013breakdown,hauke2013spread,hazzard2013far,schachenmayer2013entanglement,lashkari2013towards,gong2014persistence, hazzard2014quantum,foss-feig2015nearly,rajabpour2015quantum,storch2015interplay,Matsuta2017Improving,Else2018Improved,guo2019signaling,Tran2019localitydigital,tran2019localityheating,Sweke2019Lieb,Chen2019finite,Kuwahara2020strictly,tran2020hierarchy,hermes2020dimensionality}

\begin{figure*}
{\center
\includegraphics[height=0.25\linewidth]{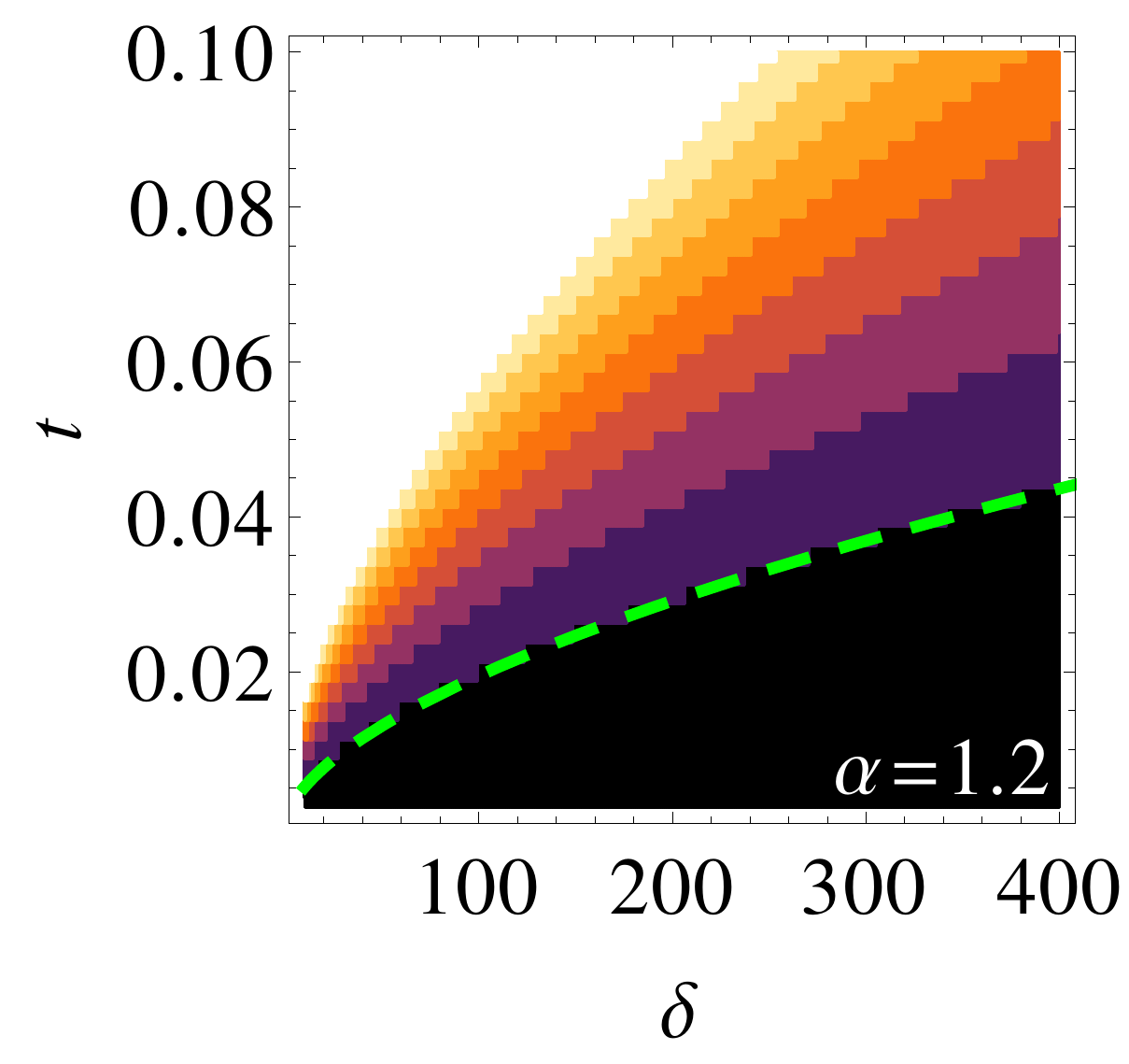}
\includegraphics[trim = 0mm -20mm 0mm -10mm, height=0.25\linewidth]{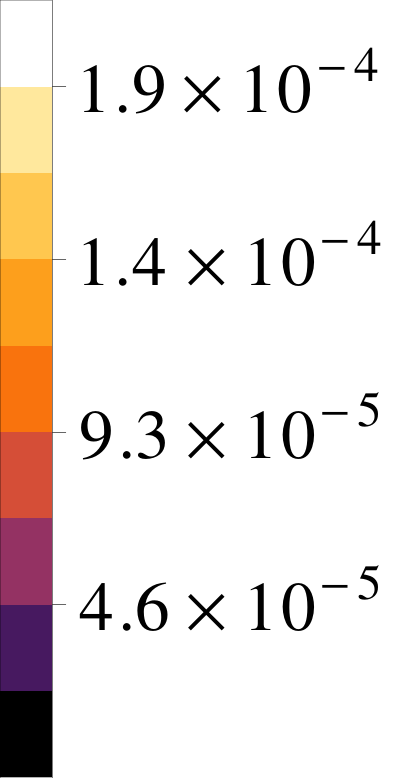}
\hspace{-4mm}
\includegraphics[height=0.25\linewidth]{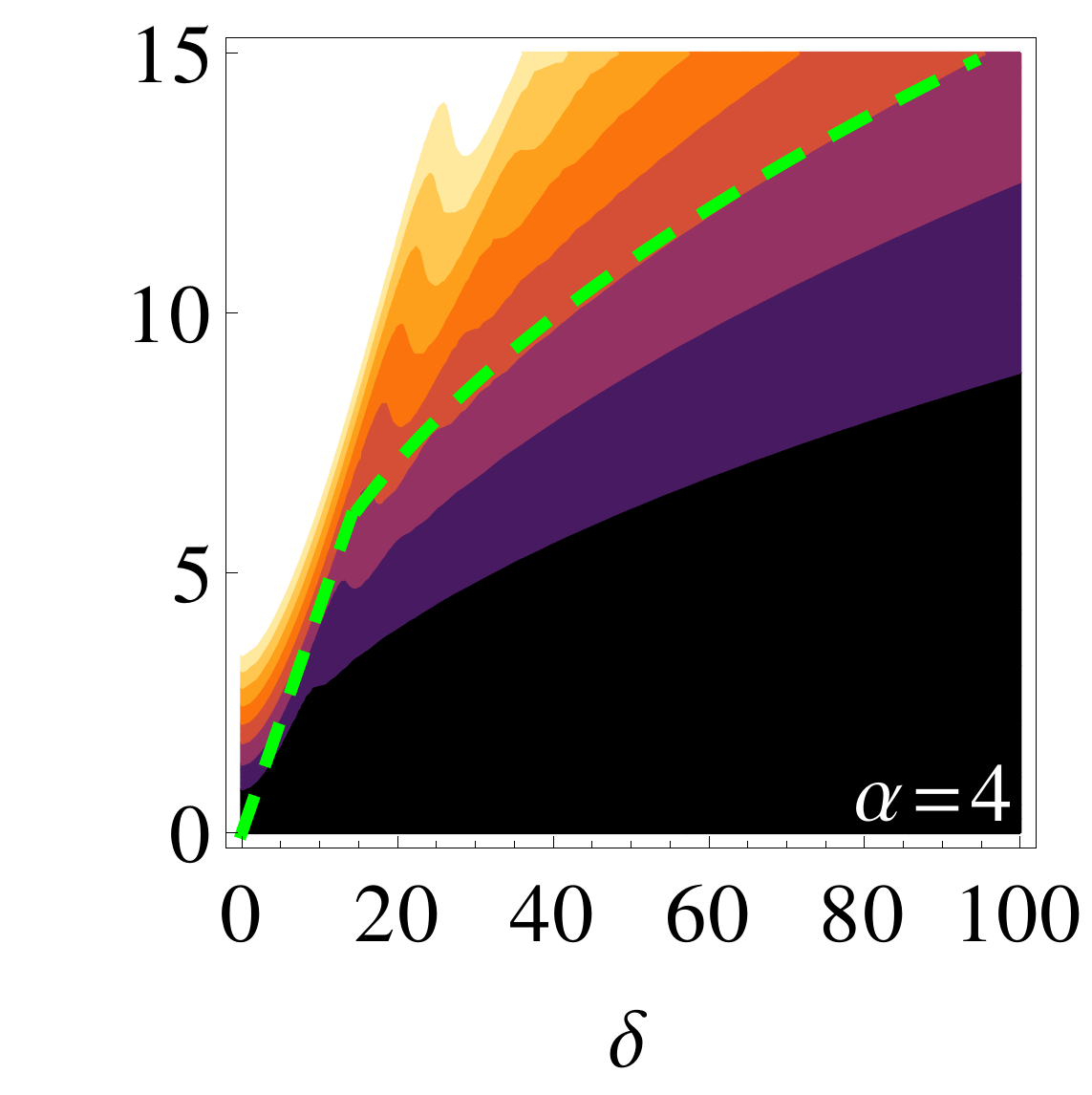}
\includegraphics[trim = 0mm -19mm 0mm -8.5mm, height=0.25\linewidth]{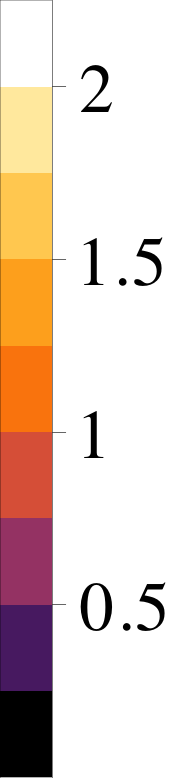}
\hspace{-4mm}
\includegraphics[height=0.25\linewidth]{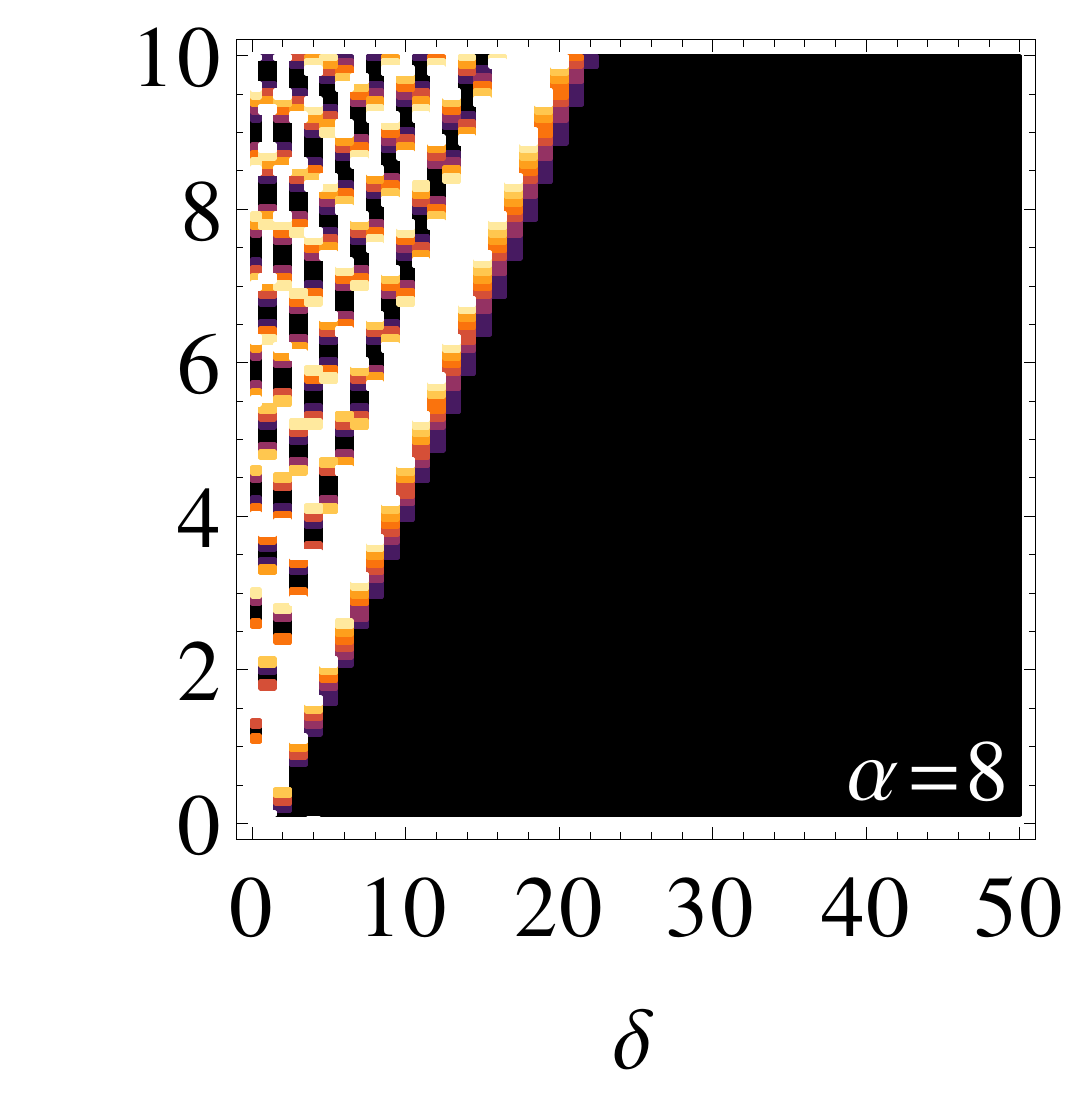}
\includegraphics[trim = 0mm -20mm 0mm -10mm, height=0.25\linewidth]{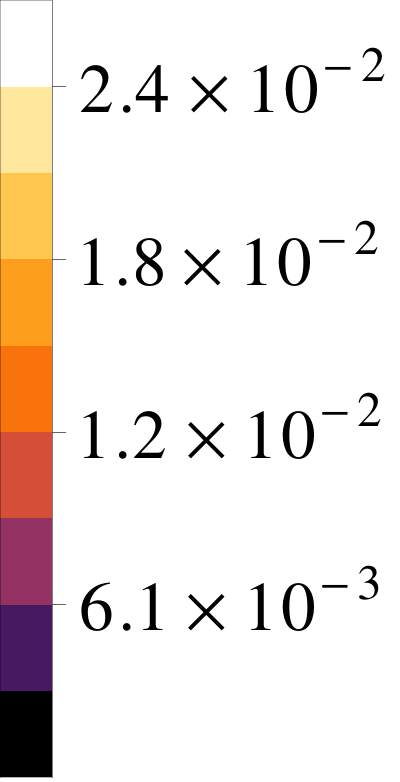}
}
\caption{\label{f:horizon}%
\textbf{Propagation patterns as a function of distance $\delta=r_{ij}$ and time $t$ for different long-range exponents $\alpha$.} Different models and physical quantities are shown in the different panels, but the overall picture remains the same. 
Left: The detection probability for a signal sent through a quantum channel between two sites at distance $\delta$ is shown for the long-range Ising chain\,\cite{eisert2013breakdown}. The green displays the  power law $\delta \propto t^{1.7}$. Center: Connected correlation functions between two sites at distance $\delta$ in a long-range field theory, see the effective action in Eq.\,\eqref{qlr_eaa} with $d=1$ and $\alpha=4$\,\cite{rajabpour2015quantum}. The short-distance spreading resembles the conventional light-cone observed with local interactions, while for larger distances long-range effects appear and power-law scaling is observed.  The green dashed curve is a guide to the eye. Right: The mutual information between two lattice sites at distance $\delta$ in the KItaev chain described by the Hamiltonian in Eq.\,\eqref{h_lrkc_ms} with vanishing pairing and long-range hopping ($\alpha=8$). The decay rate is large enough that only the light-cone is observed. Picture taken from Ref.\,\cite{storch2015interplay}.
}
\end{figure*}

Most of our understanding of correlations and entanglement in presence of long-range interactions has been based on prototypical systems. There, the synergy between analytical and numerical investigations has been particularly fruitful \cite{hauke2013spread,schachenmayer2013entanglement,hazzard2014quantum,nezhadhaghighi2014entanglement,rajabpour2015quantum,schachenmayer2015dynamics,schachenmayer2015manybody}. The general understanding of propagation in long-range systems is summarised in Fig.\,\ref{f:horizon}.
This qualitative picture applies almost independently on the particular model, the quantity or the decay range $\alpha$. In analogy for other universal results in the short-range regime $\alpha\gg 3$ (See Fig.\,\ref{f:horizon} on the right), entanglement scaling in long-range models reproduces the well-known light cone shape observed for local systems\,\cite{lieb1972finite}.
For intermediate values of $\alpha$ (see the central panel in Fig.\,\ref{f:horizon}) cone-light propagation is observed at short distances, while correlations between distant sites are heavily influenced by the presence of the long-range terms. 
Multi-speed prethermalization for lattice spin models with long-range interactions in the regime $d<\alpha<d+2$ was studied in \cite{frerot2018multispeed}.
The behaviour of correlations at intermediate decay is akin to the one found in the critical behaviour of the long-range Kitaev chain in Sec.\,\ref{kit_ch_sec}, where long-range hopping amplitudes with $2<\alpha<3$ do not modify the universal scaling behaviour, but they alter the overall shape of excitations. However, in the Kitaev chain long-range hopping only influence the subcritical behaviour for $\alpha<3$, while the light-cone bending is observed also for $\alpha=4$\,\cite{rajabpour2015quantum}.

Finally at smaller $\alpha$ (left panel in Fig.\,\ref{f:horizon}) the universal scaling is altered by long-range interactions and, accordingly, the correlations propagate faster than any possible group velocity, disrupting the linear light-cone shape.

Analytical insight into information propagation in long-range system may be also achieved by general Lieb-Robinson-type bounds. A first contribution in this direction has been given in Ref.\,\cite{hastings2006spectral}, yielding for $\alpha>d$
\begin{equation}
\label{first_bound}
\left|\left|\left[O_A(t),O_B(0)\right]\right|\right|
\leq C \left|\left| O_A\right|\right| \left|\left| O_B\right|\right|
\frac{\left| A\right|\left| B\right|  (e^{v\left| t\right|}-1)}{[\rho_{A,B}+1]^\alpha}.
\end{equation}
The regions $A,B$ are disjunct subset of the $d$ dimensional lattice. The generic operator expectations
$O_A$ and $O_B$ only receive contributions  from state in the Hilbert space, whose support lies in the spatial regions $A$ and $B$, respectively. In Eq.\,\eqref{first_bound} the symbol $\left|\left|\cdot \right|\right|$ denotes the operator norm, and $\rho_{A,B}$ is the topological distance between the regions $A$ and $B$. The topological distance is the minimum number of links connecting two nodes $i$ and $j$, which is sometimes referred to as graph distance, or chemical distance. The importance of the expression in Eq.\,\eqref{first_bound} derives from its generality, since it applies to a wide range of observables, while it is straightforwardly generalised also to other non-local quantities, such as the equal time correlators \cite{bravyi2006lieb,nachtergaele2006propagation}. In its regime of validity $\alpha>d$, the bound in Eq.\,\eqref{first_bound} qualitatively reproduces the shape in the left panel of Fig.\,\ref{f:horizon}. However, the
wave-front propagation obtained by Eq.\,\eqref{first_bound} is logarithmic rather than power-law and, then, does not faithfully describe larger $\alpha$ values. Further insight in this problem was obtained in Ref.\,\cite{gong2014persistence}, where a more general bound was derived, capable to reproduce both the Lieb-Robinson result in the local limit ($\alpha\to\infty$) and the expression in Eq.\,\eqref{first_bound}. Even this general bound does not appear to be tight on the entire $\alpha$ range, but rather to be more accurate at large $\alpha$. 

The extension of the previous picture to the strong long-range regime needs to account for the influence of diverging long-range interactions with $\alpha<d$ on the systems time-scales. In analogy with the equilibration rate of QSS, see Sec.\,\ref{sec_5_1}, also the fastest propagation scale in strong-long-range systems is found to vanish as a power-law approaching the thermodynamic limit $\tau_{\rm fastest}\propto N^{-q}$ with $q>0$\,\cite{bachelard2013universal}. Accordingly, signal propagation becomes increasingly faster as the system approaches the thermodynamic limit and hinders the traditional formulation of Lieb-Robinson bound. In order to circumvent such complication it is convenient to introduce rescaled time $\tau=t N^{q}$. In terms of this "proper" time variable the bound for $\alpha<d$ takes the same form as in the weak long-range regime, but with $\tau$ is spite of $t$ on the r.h.s. of Eq.\,\eqref{first_bound}\,\cite{storch2015interplay}.

The aforementioned results for $\alpha<d$ produce the shortest signalling time $t_{\rm ss}$ between the edges of a system of size $N$ to scale as $t_{\rm ss}\gtrsim N^{\frac{2\alpha}{d}-2}\log\,N$, which leads to the possibility of a vanishing time for transmitting information between linearly distant sites of a strong long-range system. However, such fast signals have never been observed nor described, rather a size independent signalling time was evidenced in several situations\,\cite{eisert2013breakdown,hauke2013spread,eldredge2017quantum}. Moreover, for specific initial states strong long-range interactions may be inconsequential to signal propagation, due to the so-called shielding effect\,\cite{santos2016cooperative}.

Focusing on quadratic Hamiltonians a much tighter bound can be obtained, $t_{\rm ss}\gtrsim N^{\frac{\alpha}{d}-1/2}$, which is saturated for $\alpha<d/2$ by the quantum state transfer protocol described in Ref.\,\cite{guo2019signaling}. The same reference also provides a stricter bound for general interacting spin systems. It is worth noting that the Lieb-Robinson bound can be also used to predict the velocity of quantum information scrambling, whose importance lies at the edge between high-energy and condensed matter physics\,\cite{sekino2008fast,maldacena2016bound,garttner2017measuring,bentsen2019fast}. In this context, the role of long-range interactions is particularly relevant due to their inclusion in most quantum mechanical models of black holes, possibly making these systems the fastest information scramblers in nature\,\cite{lashkari2013towards}.

Despite the fast propagation and scrambling of correlations due to long-range interactions, the growth of entanglement entropy after a sudden quench has been shown to be strongly reduced. In particular, in the strong long-range regime ($\alpha<d$) it can become as slow as logarithmic, even in the
absence of disorder\,\cite{schachenmayer2013entanglement, buyskikh2016entanglement, pappalardi2018scrambling}. This peculiar phenomenon is connected with a suppression of the quasi-particle contribution to the von Neumann entanglement entropy, which is known to be governed by collective spin-excitations related with spin-squeezing\,\cite{sorensen2001entanglement,toth2007optimal,pezze2009entanglement}. Extending to the dynamical case the bosonization procedure outlined in Sec.\,\ref{lmg_eq}\,\cite{ruckriegel2011time,lerose2019impact}, it has been possible to show that  the rate of divergence of semiclassical trajectories governs  the transient growth of entanglement. This provides a very transparent and quantitative relationship between entanglement propagation measures (such as entropy, quantum Fisher information, spin squeezing) and chaos quantifiers (such as Lyapunov exponents and out-of-time-order correlations) in the semiclassical regime\,\cite{lerose2020bridging,lerose2020origin}. Fast entanglement  growth is recovered only at criticality, corresponding to an unstable separatrix terminating onto a saddle point in phase space. Similarly, when the classical dynamics is chaotic (e.g. for kicked or multi-species models), the growth is fast, with a rate related to Lyapunov exponents. Interestingly, also long-but-finite-range interactions open up a finite layer of instability with fast entanglement growth, due to the presence of a chaotic dynamical phase\,\cite{lerose2017chaotic,lerose2019impact}
Correlation spreading with van der Walls interactions and the presence of positional disorder in $2$D was investigated in \cite{menu2020anomalous}.
Multifractality and localization of spin-wave excitations above a ferromagnetic ground state are observed. Also, the spreading of entanglement and correlations starting from a factorized state exhibits anomalous diffusion with variable dynamical exponent.

\subsubsection{Experimental observation}
\begin{figure*}[t!]
\centering
\includegraphics[width=2\columnwidth]{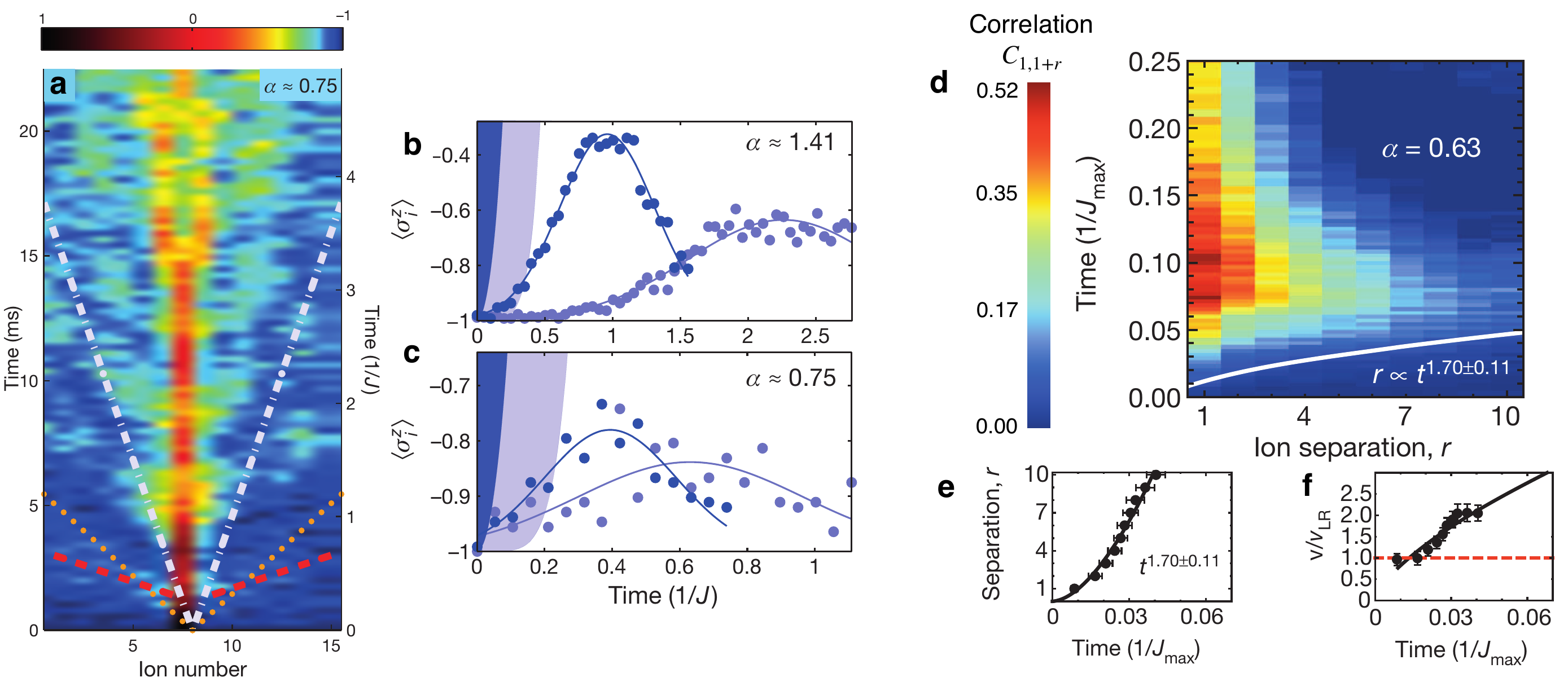}
\caption{\textbf{Propagation of quantum information in long-range trapped ions systems.}
{\bf (a)} Single-site magnetization $\aver{\sigma^z_i (t)}$ as a function of time, following a quantum quench of the long-range XY Hamiltonian (\ref{eq_XY}), with the central 8th ion initially flipped. Red lines are fits to the observed magnon arrival times (see in {\bf b}, bottom); white lines, light cone for averaged nearest-neighbour interactions; orange dots, after renormalization by the algebraic tail. The white lines are in clear disagreement with red lines.
{\bf (b-c)} Gaussian fits of magnon arrival time (red lines in {\bf a}) for ion 6 (dark blue) and 13 (light blue) with $\alpha=1.41$ (Top) and $\alpha=0.75$, a nearest neighbour Lieb–Robinson bound captures most of the signal (shaded region) in the $\alpha=1.41$ case and it does not for $\alpha=0.75$. Adapted from Ref. \cite{jurcevic2014quasiparticle}.
{\bf (d) } Spatial and time-dependent correlations following a global quench of a long-range Ising Hamiltonian (\ref{eq_Ising_ions}) with $\alpha=0.63$ . Correlation propagation velocity {\bf (e)} . The curvature of the boundary shows an increasing propagation velocity {\bf (f)}, quickly exceeding the short-range Lieb–Robinson velocity bound, $v$ (red dashed line) (c). Solid lines give a power-law fit to the data, which slightly depends on the choice of fixed contour $C_{i,j}$. Adapted from Ref. \cite{richerme2014nonlocal}.
} 
\label{fig_Lieb}
\end{figure*}
The propagation of correlations and the violation of the local Lieb-Robinson bound have been observed in trapped ions quantum simulators for $0.6\lesssim\alpha\lesssim1.2$ \cite{jurcevic2014quasiparticle, richerme2014nonlocal}. In Ref. \cite{jurcevic2014quasiparticle}, the authors have studied the dynamics following either a global or a local quench of a long-range XY Hamiltonian (see Eq.\,\ref{eq_XY}). The experimental system consists of a $15$ ions chain, prepared in a product state where only the central spin is flipped with respect to the rest of the system. In this system the global magnetization $S_z=\sum_i\sigma^z_i$ is a conserved quantity, therefore the excitation can be described as a magnon quasiparticle that propagates from the center throughout the system. After the local quench the authors observed that for $\alpha<1$ the light cone calculated considering only the nearest-neighbor couplings did not capture well the dynamics of the system (see Fig. \ref{fig_Lieb} a,b,c).

In Ref. \cite{richerme2014nonlocal}, a global quench was performed under both Ising (\ref{eq_Ising_ions}) and XY (\ref{eq_XY}) Hamiltonians, measuring the evolution of the connected two-body correlations 
$$C_{1,1+r}(t)=\aver{\sigma^z_1 (t) \sigma^z_{1+r} (t)}-\aver{\sigma^z_{1}(t)}\aver{ \sigma^z_{1+r}(t)}.$$ 
The light-cone boundary is extracted by measuring the time it takes a correlation
of fixed amplitude ($C_{i,j}\sim 0.1C^{\rm max}_{i,j}$, where $C^{\rm max}_{i,j}$ is the largest connected correlation between two ions) to travel an ion–ion
separation distance $r$. For strongly long-range interactions ($\alpha<1$),
accelerating information transfer is observed through the chain. This fast propagation of correlations is explained by the direct long-range coupling between distant spins. The increased propagation velocities quickly surpass the Lieb–Robinson
velocity for a system with equivalent nearest-neighbour-only interactions, $v=12 e J_{\rm max}$, where $e$ is Euler’s number and $J_{\rm max}$ is the maximum Ising coupling strength for a given spin–spin coupling matrix.

\subsection{Kibble-Zurek mechanism}
\label{kzm_gen}

The correlation length of a quantum system diverges approaching its quantum critical points, while the dynamical gap vanishes. As a result, the dynamical scaling of the observables when the system is driven across the transition is reminiscent of the thermodynamic scaling at equilibrium. Yet, in order for such scaling to be displayed, the drive has to be slow enough that the dynamical evolution actually occurs in the vicinity of the equilibrium critical point.

Let us consider a critical system with an internal control parameter $\lambda$ such that a (quantum) critical point occurs at $\lambda_{c}=0$ ($\lambda=|T-T_{c}|/T_{c}$ for finite temperature phase transitions). Conventionally, any slow enough drive of internal parameters $\lambda(t)=\delta t$ shall only produce adiabatic corrections $\sim\delta^{2}$ to the observables expectations with respect to the equilibrium value, as it can be deduced by simple thermodynamic arguments\,\cite{zwerger2008limited}. However, when crossing an equilibrium critical point, the traditional adiabatic picture breaks down and the residual energy  (heat) generated by the drive displays non-analytic behaviour $E_{\rm res}\approx\delta^{\theta}$ with $\theta<2$\,\cite{zurek1996cosmological}. In most local systems such non-analytic scaling emerges due to the formation of topological defects according to the celebrated Kibble-Zurek mechanism, as confirmed by several condensed matter experiments\,\cite{delcampo2014universality}.

In the quantum realm, the simplest example of defect production is furnished by the Landau-Zener problem, which describes a two level system driven through an avoided level crossing\,\cite{zener1932non,landau1965quantum,damski2005simplest}, but actual Kibble-Zurek scaling may be only observed in quantum many-body systems in the thermodynamic limit\,\cite{zurek2005dynamics,dziarmaga2010dynamics}.The heuristic scaling argument at the basis of the Kibble-Zurek mechanism can be proven to exactly apply to the nearest neighbour Ising model, i.e. the Hamiltoniam\,\eqref{h_lri} in the $\alpha\to\infty$ limit, since that problem can be mapped to an infinite ensemble of Landau-Zener transitions\,\cite{dziarmaga2005dynamics}.

In a general system, the Kibble-Zurek argument relies on the so called adiabatic-impulse approximation, where the dynamical evolution of a system starting in its ordered ground-state a $t=-\infty$ is assumed to adiabatically follow the drive until the so-called freezing time $-\hat{t}$. Beyond the ``freezing" time the equilibration rate of the system becomes too small with respect to the drive velocity and the system state cannot follow the Hamiltonian modification, as it is approaching the quantum critical point at $t=0$. Then, the dynamics is assumed to remain frozen at all times $t>-\hat{t}$ up to the crossing of the quantum critical point (at $t=0$)  and after; until the equilibration rate of the system grows back and the ``un-freezing" time $\hat{t}'$, where adiabaticity is restored, is reached. 

Once the system has unfrozen the state evolution will resume on the opposite site of the transition, where the Hamiltonian ground-state is supposed to break the Hamiltonian symmetry. Then, the dynamics will induce a transition between the symmetric and a symmetry-broken state. However, this transition will occur at finite correlation length $\hat{\xi}$, since the process only starts at $t\geq\hat{t}'=\hat{t}$, at least for a symmetric transition. The dynamics has thus modified the character of the continuous phase transition, making it rather similar to a
first-order one, and the system will likely form topological defects, whose size would be roughly proportional to the (finite) correlation volume $\hat{\xi}^{d}$. Therefore, the total defect density scales according to $n_{\rm exc}\approx \hat{\xi}^{-d}$.

During the adiabatic stage of the dynamics the system observables will acquire the equilibrium expectation of the instantaneous Hamiltonian and so does the minimal gap of the system $\Delta(t)=\Delta(\lambda(t))$. Then, a proper estimation of the drive strength on the system is $\dot{\Delta}/\Delta$ which has to be compared with the equilibration time $\Delta^{-1}$, leading to the adiabatic condition
\begin{align}
\label{ad_cond}
\dot{\Delta}\ll \Delta^{2}.
\end{align}
The freezing time $\hat{t}$ is defined by the breakdown of the adiabatic condition $\dot{\Delta}(\hat{t})\simeq \Delta(\hat{t})^{2}$. Appling the critical scaling of the minimal gap with $\lambda$, one obtains the scaling of the freezing time $\hat{t}\approx \delta^{-\frac{z\nu}{1+z\nu}}$ and, accordingly, the freezing length scaling $\hat{\xi}\approx\delta^{-\frac{\nu}{1+z\nu}}$, which lead to the defect density expression
\begin{align}
\label{def_sc_kz}
n_{\rm exc}\approx \hat{\xi}^{-d}\approx \delta^{\frac{d\nu}{1+z\nu}}.
\end{align}
The application of the traditional Kibble-Zurek picture is complicated by different effects depending on the strong or weak nature of long range interactions. In the first case, the additional relevance of boundaries with respect to local systems produces clear difficulties in the definition of the topological defects. While in the latter case, the presence of the competing scaling contributions discussed in Sec.\,\ref{kit_ch_sec} leads to novel scaling regimes, which are not encompassed by the Kibble-Zurek framework.

\subsubsection{Kitaev chain}

The appearance of multiple scaling contributions to the critical behaviour of long-range quantum systems has been already exemplified in the study of the Kitaev chain in Sec.\,\ref{kit_ch_sec}. In this sub-section we are going to consider the effect of such multiple scalings on the universal dynamics.

The study of exactly solvable toy-models is at the root of the current understanding of Kibble-Zurek scaling in general quantum systems. Indeed, first studies of defect formation in quantum systems have been pursued on the nearest neighbour Ising model, where finite size scaling arguments led to the prediction
\begin{align}
\label{fss_kz_arg}
n_{\rm exc}^{\rm fss}\approx \delta^{\frac{1}{2z}}
\end{align}
which produces $n_{\rm exc}\approx \sqrt{\delta}$ in agreement with the Kibble-Zurek prediction in Eq.\,\eqref{def_sc_kz} since $z=\nu=1$ in this case\,\cite{zurek2005dynamics}. Soon after this seminal investigation, an exact solution to the universal slow dynamics of the Ising model has been provided by mapping it to a infinite sum of Landau-Zener problems, each representing the dynamics of a single fermionic quasi-particle excitations\,\cite{dziarmaga2005dynamics}.

Indeed, the dynamical evolution of quadratic fermions can be described in terms of the Bogoliubov amplitudes via the equation
\begin{align}
\label{dyn_sys}
i\frac{d}{dt}\begin{pmatrix}u_{k}\\v_{k}\end{pmatrix}=\begin{pmatrix}\varepsilon_{\alpha}(k,t) & \Delta_{\beta}(k)\\
-\Delta_{\beta}(k) & \varepsilon_{\alpha}(k,t)
\end{pmatrix}\begin{pmatrix}u_{k}\\v_{k}\end{pmatrix},
\end{align}
which generically represent an ensemble of two level systems, whose energy and coupling are represented by the momentum space kinetic and pairing terms, respectively.  Thus, the Kibble-Zurek dynamics of the Kitaev chain can be studied exactly and this solution is not limited to the nearest neighbour case, which represents the Ising model, but it can be extended to any form of the long-range couplings.

Let us, then, consider a slow variation of the chemical potential $h$ in the Hamiltonian\,\eqref{h_lrkc} with the usual slow drive form $h(t)=h_{c}+\delta t$, with the time spanning in the interval $t\in[-h_{c}/\delta,h_{c}/\delta]$. Then, in the small $\delta$ limit, the system is adiabatically ramped from a point deep in the topological phase $h=0$ across the quantum phase transition and up into the trivial phase $h=2h_{c}$. In the following we are going to focus to a ramp across the quantum phase transition occurring at $h_{c}=1$.

Within this dynamical protocol the dynamical system in Eq.\,\eqref{dyn_sys} reduces to the $k$ dependent Landau-Zener problem\,\cite{landau1969statistical,damski2005simplest}. Thus, the excitation probability of each Bogoliubov quasi-particle can be computed according to the Landau-Zener formula 
\begin{align}
\label{exc_prob_lr_kit}
\langle \gamma_{k}^{\dagger}\gamma_{k}\rangle= n_{\rm exc}(k)= \exp\left(-\frac{\pi}{\delta^{2}}\Delta_{\beta}(k)^{2}\right)+O(\delta^{2}\Delta_{\beta}(k)^{4}).
\end{align}
The equation above only explicitly reports the leading term in the $k\to 0$ limit, which is the relevant one for universal behaviour. However, when considering a slow quench in a finite time interval  $t\in[-h_{c}/\delta,h_{c}/\delta]$, the discontinuity in the drive derivative at the borders of the interval induces $\delta^{2}$ corrections to the excitations probability\,\cite{dziarmaga2010dynamics,defenu2019universal}.

The crucial property of the excitation probability in  Eq.\,\eqref{exc_prob_lr_kit} only depends on the pairing term in Hamiltonian\,\eqref{h_lrkc}, so that the universal slow dynamics is fully determined by the low-momentum scaling of the pairing coupling. Accordingly, the excitation density can be obtained by integrating Eq.\,\eqref{exc_prob_lr_kit} over the Brillouin zone
\begin{align}
\label{kit_ex_kz}
\int n_{\rm exc}(k)dk\approx \delta^{\frac{1}{2z_{\Delta}}}
\end{align}
where we have defined $z_{\Delta}$ from the scaling of the pairing coupling $\lim_{k\to 0}\Delta_{\beta}(k)\approx k^{z_{\Delta}}$. The result in Eq.\,\eqref{kit_ex_kz} has been also employed to prove validity of the Kibble-Zurek argument in Kitaev chains with long-range pairing terms\,\cite{dutta2017probing} in addition to the purely local case\,\cite{dziarmaga2005dynamics}.

Apart for the aforementioned results, which explicitly refer to quadratic Fermi systems, the application of adiabatic perturbation theory to slow quenches close to quantum critical points predicts the scaling of the defect density to be in agreement with the Kibble-Zurek prediction $\theta=d\nu/(1+z\nu)$\,\cite{polkovnikov2005universal}. Such prediction comes from the assumption that the scaling form of the critical propagator reproduces the equilibrium critical exponents. Since for $1d$ Fermi systems, one has $z\nu=1$, the perturbative argument yields $d\nu/(z\nu+1)=1/2z$ in agreement with the finite size scaling argument in Eq.\,\eqref{fss_kz_arg}. 
However, it was realised long ago\,\cite{dziarmaga2010dynamics} that the correspondence between the exact scaling in Eq.\,\eqref{kit_ex_kz} and the perturbative prediction is tied to the relevance of the pairing term with respect to the momentum term in the scaling of the quasi-particle gap, see Eq.\,\eqref{lrkc_spec}.

As outlined in Sec.\,\ref{kit_ch_sec}, the presence of long-range (anisotropic) couplings in 1$d$ Fermi systems may produce equilibrium scaling exponents dominated by the kinetic term in the gap scaling, see Eq.\,\eqref{dyn_exp_lrkc}, differently from what occurs in short-range systems. Similarly, the introduction of non-local finite range couplings in the Kitaev model has been known to produce a modified equilibrium scaling with a kinetic dominated dynamical critical exponent.  The latter phenomenon is only found in proximity of multi-critical points, where finite range non-local couplings become relevant, and  is known to lead to a violation of the Kibble-Zurek result\,\cite{divakaran2009defect,deng2009anomalous,dziarmaga2010dynamics}.

\begin{figure*}
\subfigure[Dynamical phase diagram]{\label{FigSecV_3a}\includegraphics[width=.32\linewidth]{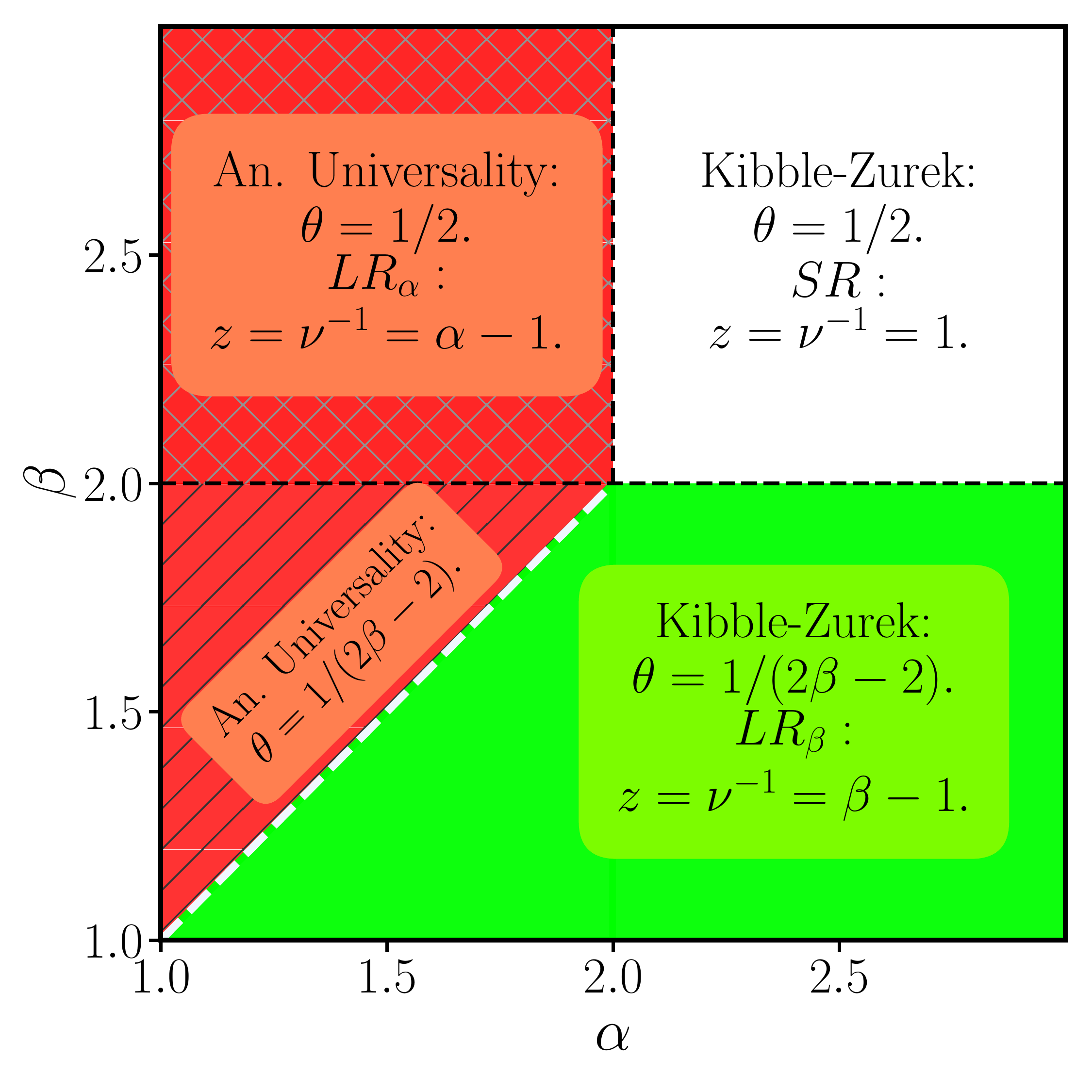}}
\subfigure[Excitations probability $\delta =0.5$]{\label{FigSecV_3b}\includegraphics[width=.32\linewidth]{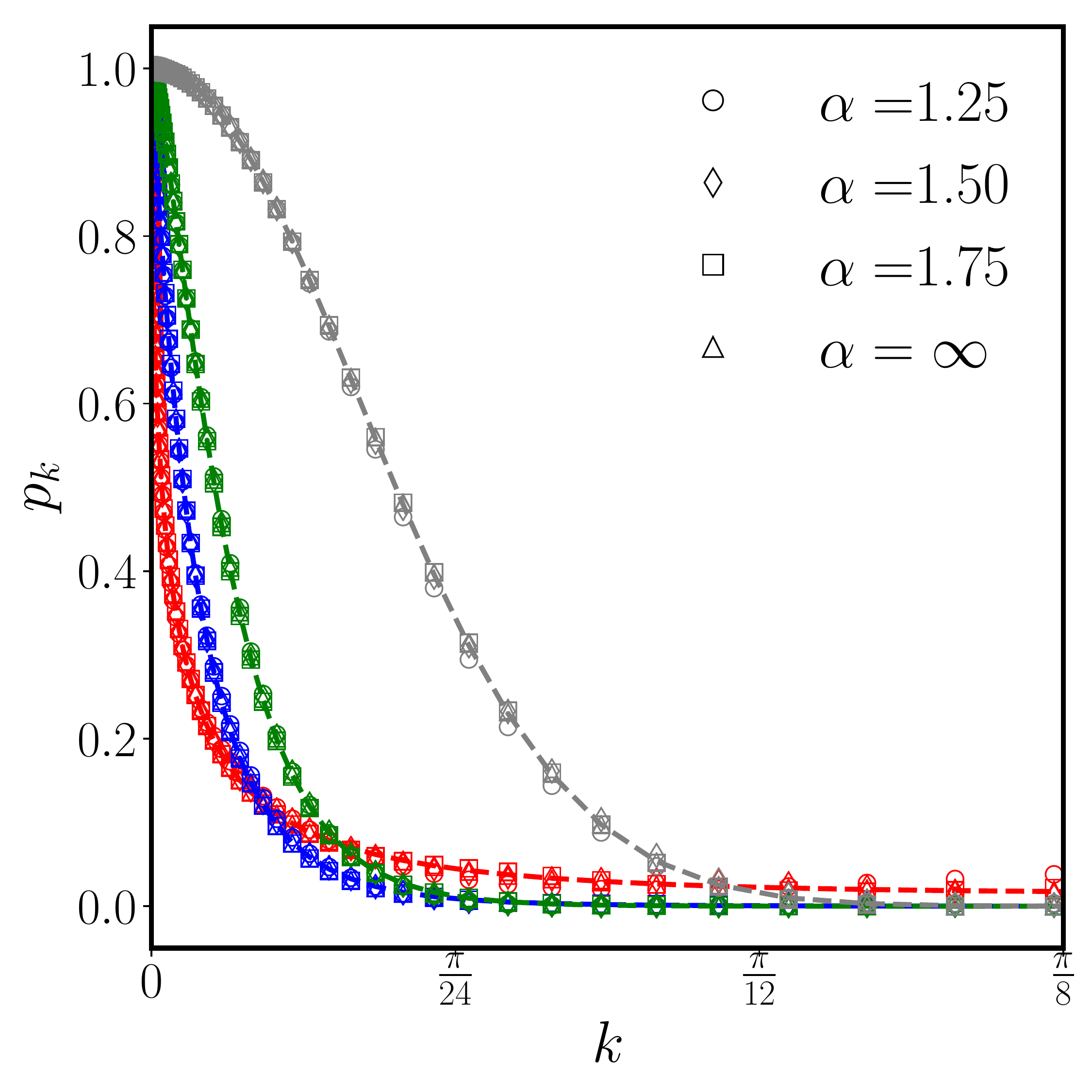}}
\subfigure[Excitations probability $\delta = 0.05 1$]{\label{FigSecV_3c}\includegraphics[width=.32\linewidth]{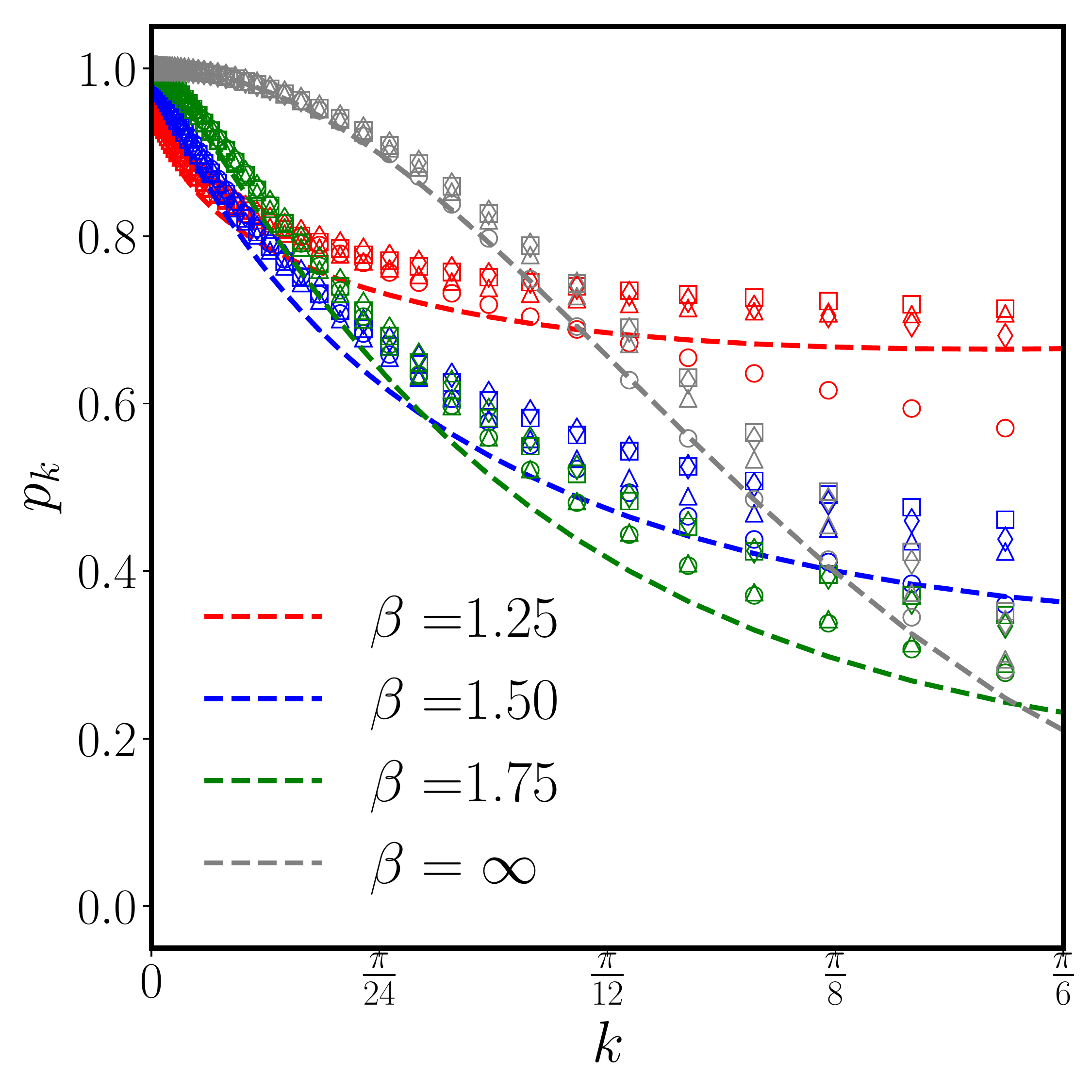}}
\caption{ \textbf{Kibble-Zurek mechanism in long-range Kitaev chains.} \textbf{(a)}: the dynamical phase diagram reporting the universal slow-dynamics exponents of the anisotropic Kitaev chain in the $(\alpha,\beta)$ plane. \textbf{(b,c)}: numerical analysis of Eq.\,\eqref{dyn_sys} compared with the analytic formula in Eq.\,\eqref{exc_prob_lr_kit} for intermediate and small dynamical rates $\delta=0.5,0.05$.\label{FigSecV_3}}
\end{figure*}

At variance, the anisotropic Kitaev model with weak long-range couplings in the $\alpha<\beta$ regime displays the aforementioned kinetic dominated scaling already at a second-order quantum critical point\,\cite{defenu2019universal}. In particular, its dynamical phase diagram, depicted in Fig.\,\ref{FigSecV_3a} contains four different regions, two of them (green and white in Fig.\,\ref{FigSecV_3a}) fulfil the Kibble-Zurek prediction both with the nearest neighbours universal exponents ($\theta=1/2$ in the white region) or with pairing dominated critical exponents ($\theta=(2\beta-2)^{-1}$ green region in Fig.\,\ref{FigSecV_3a}). The conventional prediction $\theta=z\nu/(1+z\nu)$ cannot be applied to the two red regions in Fig.\,\ref{FigSecV_3a}, where $\alpha<\beta$, to the point that in the upper portion of the red region the nearest-neighbour prediction for the dynamics $\theta=1/2$ remains valid deep in the regime where the equilibrium universal behaviour is dominated by long-range interactions.

Clearly, the absence of kinetic contributions to the critical dynamics only holds in the strict $\delta\to 0$ limit. So that non universal corrections still carry a sizeable contribution to the defect density from the power-law $\alpha$ as long $\delta\lesssim 1$ as it is shown in Fig.\,\ref{FigSecV_3b}, where a full numerical computation of the defect density for various points in the $(\alpha,\beta)$ plane (reported in different colours and shapes, see the legends in Fig.\,\ref{FigSecV_3}) is compared with the analytical prediction in Eq.\,\eqref{exc_prob_lr_kit} (dashed lines). Such non-universal corrections are rapidly washed out in the slow drive limit, see Fig.\,\ref{FigSecV_3c}, where the excitation probability at different $\alpha$ but with the same $\beta$ collapse on each other.

It is worth noting that the agreement between the analytic prediction in Eq.\,\eqref{dyn_sys} and the numerical result shown in Fig.\,\ref{dyn_sys} is limited by the $\delta^{2}$ contributions to the excitation probability, which, in turns, are generated by the finite edge-points of the present dynamical protocol. Actually, for a slow linear quench in the infinite interval $t\in [-\infty,\infty]$ the result in Eq.\,\eqref{dyn_sys} will remain valid independently on the $\delta$ value. Yet, in the present problem a variation of $h\in[-\infty,+\infty]$ will lead to the crossing of two critical points and it will naturally lead to more complications.

In summary, several diverse predictions exist for the defect scaling after slow quenches  in quantum many-body systems. In particular, the  finite size scaling argument in Eq.\,\eqref{fss_kz_arg} and the traditional Kibble-Zurek result in Eq.\,\eqref{def_sc_kz} remain consistent with each other and with the exact solution for quadratic Fermions, as long as $z\nu=1$. This last condition always hold for the fermionic system described in Sec.\,\ref{subsec_kit_chain}, but this is not the case for the interacting field theories described in Sec.\,\ref{q_rot_mod}, where the dynamical critical exponent $z\nu$ actually depends on the decay exponent, see Fig.\,\ref{Fig7}. In particular, the mean-field approximation produces the result $z\nu=1/2$ for rotor models, in agreement with the result observed in the Lipkin-Meshkov-Glick  model, which represents the $\alpha=0$ limit of such theories. In the following, we are going to examine such extreme case in details and show how the Kibble-Zurek mechanism is modified by interactions in the strong long-range regime.

\subsubsection{ Lipkin-Meshkov-Glick model}
\label{n_exc_lmg}

In the following, the difficulty to reconcile the finite size scaling prediction in Eq.\,\eqref{fss_kz_arg} with the perturbative result $\theta=d\nu/(1+z\nu)$\,\cite{polkovnikov2005universal}
is exemplified by the study of the flat interactions case $\alpha=0$ such as the Lipkin-Meshkov-Glick  model, whose equilibrium behaviour has been described in Sec.\,\ref{lmg_model}. Apart from its prototypical role, the interest in the Lipkin-Meshkov-Glick model is motivated by the possibility to experimentally study slow dynamics in this system thanks to cold atoms into cavity experiments\,\cite{brennecke2013real}, as well as to its relation with the BCS model\,\cite{dusuel2005finite}.
 
First numerical results on the scaling of the defect density after an adiabatic ramp crossing the quantum critical point of the Lipkin-Meshkov-Glick model could not be reproduced by the Kibble-Zurek formula in Eq.\,\eqref{def_sc_kz}, but they displayed qualitative agreement with the finite size scaling prediction in Eq.\,\eqref{fss_kz_arg}\,\cite{caneva2008adiabatic}. Yet, more intensive numerical studies unveiled a more complicated landscape where the adiabatic crossing of the equilibrium quantum critical point does not display any actual Kibble-Zurek scaling, but rather a universal behaviour as a function of the scaled variable $\Lambda=N\,\delta $\,\cite{acevedo2014new}; while non-analytic corrections for the defect scaling was found for quenches up to the critical point\,\cite{hwang2015quantum}.

This scenario can be safely reconstructed by the study of the effective critical theory depicted in Sec.\,\ref{lmg_eq}. However, since the effective harmonic theory, which describes the fully-connected problem at order $1/N$, was obtained at equilibrium, it is first convenient to generalise the treatment to the dynamical case. Our goal is to consider the Lipkin-Meshkov-Glick problem with time-dependent coupling $h(t)$, with the system initially prepared at equilibrium at any initial time $t_{i}$ and, then, manipulated across the quantum critical point. Thus, during the time-evolution the average expectation value of the global spin will change as the order parameter is modified by the dynamics as soon as $h(t)<h_{c}$. As a consequence, the assumption of small quantum depletions of the classical equilibrium expectation $\langle
\boldsymbol{S}\rangle$, which is at the basis of the Holstein-Primakov expansion in Eqs.\,\eqref{hp_z},\,\eqref{hp_+} and\,\eqref{hp_-}, is dynamically disrupted by the macroscopic change in the order parameter.

A simple solution to this difficulty is obtained by considering a time-dependent classical expectation for the Holstein-Primakov expansion via the time-dependent spin-wave approximation introduced in Ref.\,\cite{ruckriegel2011time}. This solution strategy for the time-dependent fully-connected problem has already been employed to characterise the chaotic dynamical phase which emerges upon the inclusion of additional nearest neighbour couplings on top of the Lipkin-Meshkov-Glick Hamiltonian\,\cite{lerose2017chaotic,lerose2019impact}. 

At leading order $1/N$ this procedure effectively decouples the classical evolution of the order parameter from the quantum fluctuations. Ramping the magnetic field slowly across the critical point $h(t)=h_{c}-\delta\,t$ for $t\in [-1/\delta,1/\delta]$ is equivalent to dynamically modify the frequency of both classical field and the quantum fluctuations according to the equilibrium formulas\,\eqref{dyn_gap} and\,\eqref{cl_gap}. In principle, an accurate description of the ramp dynamics at finite $\delta$ would need the description of the back-action of the displacement of the classical observable from its equilibrium configuration into the dynamics of the quantum mode.

However, in the adiabatic limit $\delta\to 0$ we can employ the classical adiabatic theorem\,\cite{landau1976mechanics} to conclude that the classical trajectory will remain close to the instantaneous solution $\theta(t)-\theta_{\rm eq}\approx \delta^{2}$ and $\varphi(t)-\varphi_{\rm eq}\approx \delta^{2}$, where the equilibrium contributions are $\varphi_{\rm eq}=0$ and $\theta_{\rm eq}$ is given in Eq.\,\eqref{mf_angle}. Yet, based on previous discussion, the classical $\delta^{2}$ correction is going to be superseded by the one arising from quantum fluctuations. Indeed, quantum fluctuations in the Lipkin-Meshkov-Glick problem are effectively described by a single harmonic mode adiabatically ramped across its fully degenerate quantum critical point. 

Interestingly, none of the results on defect scaling, presented at the beginning of Sec.\,\ref{kzm_gen}, apply to the present problem, since the general result derived by dynamical perturbation theory does not apply to Bose quasi-particles\,\cite{degrandi2009adiabatic}.
In fact, it was first noticed by asymptotic expansion that a quasi-static transformation of an Harmonic oscillator with linear time scaling of its frequency across the fully degenerate point $\omega(t)^{2}\approx (\delta t)^{2}$ produces non-adiabatic corrections which do not vanish in the $\delta\to 0$ limit\,\cite{bachmann2017dynamical}. Clearly, this result does not directly apply to the Lipkin-Meshkov-Glick case, since for a linear scaling of the control parameter $\lambda(t)=h(t)-h_{c}=\delta t$ the dynamical frequency for the spin wave model reads
\begin{align}
\label{lmg_kz_sc}
\omega(t)^{2}\approx \delta |t|
\end{align}
at leading order in the small-time $\delta$ expansion. Based on the conventional  adiabatic argument $\dot{\omega}(t)\ll \omega(t)^{2}$ the faster the drive vanishes across the fully-degenerate point, the stronger non-adiabatic effects shall be. Then, one may in principle expect the linear drive in Eq.\,\eqref{lmg_kz_sc} to be more adiabatic than the $\sim t^{2}$ case studied in Ref.\,\cite{bachmann2017dynamical} and to present a different non-adiabatic scaling.

In general, the characterisation of slow dynamics for different kind of excitations and dynamical scaling is very relevant to the problem of long-range interactions. Indeed, we have already shown that the quantum long-range Ising model in Eq.\,\eqref{h_lri} varies as a function of $\alpha$ from a critical point with Fermi quasi-particles ($\alpha>\alpha_{*}$) to a purely bosonic effective field theory ($\alpha<\frac{5}{3}d$). In the first case ($\alpha>\alpha_{*}$), the validity of the Kibble-Zurek argument follows from the derivation in Ref.\,\cite{dziarmaga2005dynamics}, which generally applies to critical systems with Fermi quasi-particles. In the intermediate case ($\alpha_{*}>\alpha>\frac{5}{3}d$) non-analytic scaling $\sim\delta^{\theta}$ follow from the dynamical perturbation theory result in Ref\,\cite{polkovnikov2005universal}, which qualitatively describes the Fermi quasi-particle case. However, this picture cannot be applied to Bose quasi-particles, whose large occupation numbers hinder the applicability of adiabatic perturbation theory\,\cite{degrandi2009adiabatic}.

Given this picture, it is not surprising that the problem of Kibble-Zurek scaling in fully-connected models such as the Lipkin-Meshkov-Glick has longly remained open. 
It has been recently shown that the universal slow dynamics across a quantum critical point with Bose quasi-particles always lies in the fully non-adiabatic regime and it is, therefore, not encompassed in the traditional Kibble-Zurek picture\,\cite{defenu2021quantum}.

Here, we are going to outline this current picture in the peculiar Lipkin-Meshkov-Glick case, where the quasi-particle excitation energy displays square-root scaling, see Eq.\,\eqref{lmg_kz_sc}. Therefore, we consider a single dynamically driven Harmonic mode with Hamiltonian
\begin{align}
\label{hoh}
H(t)=\frac{1}{2}\left(p^{2}+\omega(t)^{2}x^{2}\right).
\end{align}
which faithfully describes the dynamics in Eq.\,\eqref{h_lmg}, when adibatic $\delta^{2}$ corrections coming from the classical dynamics of the order parameter are neglected\,\cite{defenu2018dynamical}, see also Eq.\,\eqref{h_lmg}.

Fon any time-dependent frequency a complete set of time-dependent states $\psi_{n}(x,t)$ can be constructed, whose occupation is conserved by the dynamics\,\cite{lewis1967classical, lewis1968class,lewis1969exact}. Then, any dynamical state can be expanded into such basis with constant coefficients $\psi(x,t)=\sum \alpha_{n}\psi_{n}(x,t)$. We focus on a cyclic transformation where the system is initially in the ground state of the equilibrium Hamiltonian, thus, our focus is on the lowest energy time dependent state
\begin{align}
\label{dyn_eigen}
\psi_{0}(x,t)=\left(\frac{1}{2\pi\xi^{2}(t)}\right)^{\frac{1}{4}}e^{-\Omega(t)\frac{x^{2}}{2}}e^{-i\frac{\varphi(t)}{2}}.
\end{align}
with the effective time dependent frequency 
$\Omega(t)=-i\frac{\dot{\xi}(t)}{\xi(t)}+\frac{1}{2\xi^{2}(t)}$,
and the phase $\phi(t)=\int^{t}\frac{dt'}{2\xi^{2}(t')}$. According to the above equations the entire dynamics is determined in terms of the effective width $\xi(t)$, which satisfies the Ermakov-Milne equation
\begin{align}
\label{ermakov_eq}
\ddot{\xi}(t)+\omega(t)^{2}\xi(t)=\frac{1}{4\xi^{3}(t)}.
\end{align}
In general, also when excited dynamical states are considered, the solution of the quantum dynamical problem described by Hamiltonian\,\eqref{hoh} is fully determined by the classical trajectory described by Eq.\,\eqref{ermakov_eq}.

In order to determine the excitation density and the ground state fidelity with respect to the instantaneous equilibrium solution of the problem, we define the adiabatic basis $\psi_{n}^{\rm ad}(x,t)$, which is obtained taking the conventional time-independent Harmonic oscillator eigenstates and replacing the constant frequency with the time-dependent one\,\cite{dabrowski2016time}. Accordingly, one can expand the exact time-dependent state in terms of the adiabatic basis $\psi(x,t)=\sum c_{n}(t)\psi^{\rm ad}_{n}(x,t)$, leading to the following results for the excitation density
\begin{align}
\label{exc_den}
n_{\rm exc}(t)=\sum_{n\in 2\mathbb{N}}n|c_{n}|^{2}=\frac{\xi^{2}}{2\omega(t)}\left[\left(\frac{1}{2\xi^{2}}-\omega(t)\right)^{2}+\left(\frac{\dot{\xi}}{\xi}\right)^{2}\right],
\end{align} 
and the adiabatic ground-state fidelity
\begin{align}
\label{exp_fidelity}
f(t)=|c_{0}|^{2}=\frac{1}{\xi(t)}\sqrt{\frac{2\omega(t)}{\left(\frac{1}{2\xi^{2}}+\omega(t)\right)^{2}+\left(\frac{\dot{\xi}}{\xi}\right)^{2}}}.
\end{align}

In general, the solution of Eq.\,\eqref{ermakov_eq} cannot be found explicitly, but an analytic solution is possible for a cycle across the critical point $\omega(t)=0$ with the scaling form
\begin{align}
\label{freq_scal}
\omega(t)^{2}=\omega_{0}+\delta |t|
\end{align}
where $\delta>0$ is the drive rate. The linear scaling of $\omega(t)^{2}$ is the consequence of the gap scaling $z\nu=1/2$ of the equilibrium problem, while the parameter $\omega_{0}>0$ has been introduced in order to model a finite minimal gap arising due to finite size effects $\omega_{0}=N^{1/\nu_{*}}$, according to the result in Sec.\,\ref{classcal_slr_criticality}. 

It is convenient to explicitly write the Sch\"odinger equation of the proposed model for a quantum adiabatic cycle
\begin{align}
\label{full_mod}
\partial_{t}\psi(t)=\left[\frac{p^{2}}{2}+\left(\omega_{0}+\delta |t|\right)\frac{x^{2}}{2}\right]\psi(t),
\end{align}
The model in Eq.\,\eqref{full_mod} describes a cyclic transformation for the system and, in the limit $\delta\to 0$, it can be used to describe an adiabatic cycle in quantum systems with infinitely degenerate spectrum. According to the behaviour of the observables in the adiabatic limit $\delta\to 0$ the dynamical evolution described by Eq.\,\eqref{full_mod} presents three stages
\begin{enumerate}
\item Perturbative regime ($\omega_{0}>0$).
\item Kibble-Zurek regime ($\omega_{0}=0$, half-ramp to the quantum critical point).
\item Non-adiabatic regime ($\omega_{0}=0$, full ramp).
\end{enumerate}
Regime (1) occurs for a finite minimal frequency $\omega_{0}\neq 0$: there the adiabatic perturbation theory  result produces the analytic $\delta^{2}$ corrections predicted by dynamcal perturbation theory. Regime (2) is realised for a thermodynamic system ($\omega_{0}\to 0$) whose dynamics terminates at the quantum critical point $t=0$, where non-analytic corrections appear, which are encompassed by the Kibble-Zurek argument. The actual crossing of the quantum critical point only occurs in regime (3) and the actual non-adiabatic regime,  is realised, leading to rate independent corrections to the adiabatic observables, as it will be seen in the following.

The latter result can be easily shown rephrasing Eq.\,\eqref{full_mod} in a rate independent form via the transformations 
\begin{align}
\label{dim_trans}
t = \delta^{-\frac{1}{3}}s,\quad x= \delta^{-\frac{1}{6}}\tilde{x}
\end{align}
which reduce Eq.\,\eqref{full_mod} to the $\delta=1$ case. The expressions in Eqs.\,\eqref{exc_den} and\,\eqref{exp_fidelity} are invariant under the transformations in Eq.\,\eqref{dim_trans} in such a way that the fidelity and excitation density at real times can be obtained by $\tilde{\xi}(s)=\lim_{\delta\to 1}\xi(t)$ and $\tilde{\Omega}(s)^{2}=s$, provided that the endpoint of the dynamics is rescaled accordingly.

The crucial condition of adiabatic dynamics is for the system to start in the ground-state at the beginning of the  dynamics, i.e. $\displaystyle{\lim_{t \to -\infty}}\psi(t)=\psi_{0}^{\rm ad}(t)$, leading to the boundary conditions
\begin{align}
\label{bound_cond}
\lim_{t\to-\infty}\xi(t)^{2}=\frac{1}{2\omega(t)}; \quad \lim_{t\to-\infty}\dot{\xi}(t)^{2}=0.
\end{align}

As long as $\omega_{0}>0$ the instantaneous spectrum of the model remains gapped at $t=0$ and the scaled width $\tilde{\xi}(s)$ has to be evaluated at $s_{0}=\delta^{-2/3}\omega_{0}$. Inserting $\tilde{\xi}(s_{0})$ into the defect density and fidelity Eqs.\,\eqref{exc_den} and\,\eqref{exp_fidelity} yields the result for the two quantities for finite $\omega_{0}$. In the limit $\delta\to 0$ the scaled final time $s_{0}$ diverges and according to the conditions in Eq.\,\eqref{bound_cond} the adiabatic result is recovered apart from the expected perturbative corrections
\begin{align}
\label{ad_corr}
\lim_{\delta \to 0}n_{\rm exc}(t_{0})= o\left(\delta^{2}\right);\qquad \lim_{\delta \to 0}f(t_{0})=1-o\left(\delta^{2}\right).
\end{align}

More interestingly, when the dynamics terminates exactly at the critical point $s_{0}=\delta^{-2/3}\omega_{0}=0$ the  scaled width and its derivative remain finite
\begin{align}
\label{xi_0}
\lim_{s\to 0^{-}}&\tilde{\xi}^{2}(s)=\frac{\Gamma(p)\Gamma(p+1)}{2\pi p^{2p}},\\
\label{xi_dot_0}
\lim_{s\to 0^{-}}&2\dot{\tilde{\xi}}(s)\tilde{\xi}(s)=\frac{1}{\sqrt{3}}
\end{align}
where $p=1/3$.
The finiteness of the results in Eqs.\,\eqref{xi_0} and\,\eqref{xi_dot_0} corresponds to a vanishing fidelity in Eq.\,\eqref{exp_fidelity}. Consequently, the defect density diverges, see Eq.\,\eqref{exc_den}, but the heat (or excess energy) remains finite
 \begin{align}
 \label{heat}
 \lim_{t_{0}\to 0}Q(t_{0})\simeq \lim_{t_{0}\to 0}\omega(t_{0})n_{exc}(t_{0})\propto \delta^{\frac{1}{3}}.
\end{align}
The result in Eq.\,\eqref{heat} is consistent with the outcome of the impulse-adiabatic approximation at the basis of the KZM result\,\cite{degrandi2009adiabatic,dziarmaga2010dynamics} as well as with the result in Ref.\,\cite{hwang2015quantum}. 

\begin{figure*}[t!]
\centering
\subfigure[\,\, Effective width ]{\label{FigSecV_4a}\includegraphics[width=.3\textwidth]{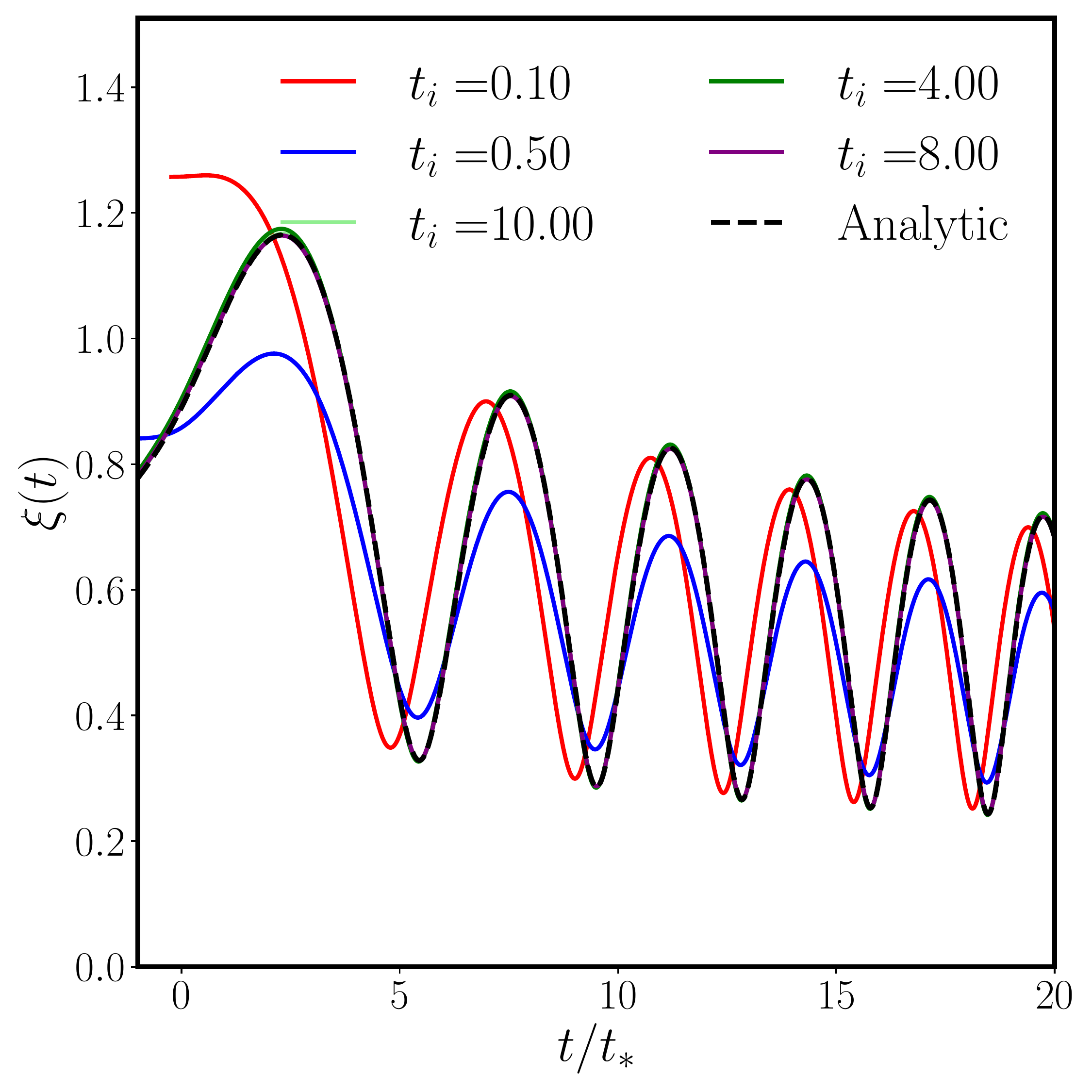}
}
\subfigure[\,\, Heat (analytics) ]{\label{FigSecV_4b}\includegraphics[width=.31\textwidth]{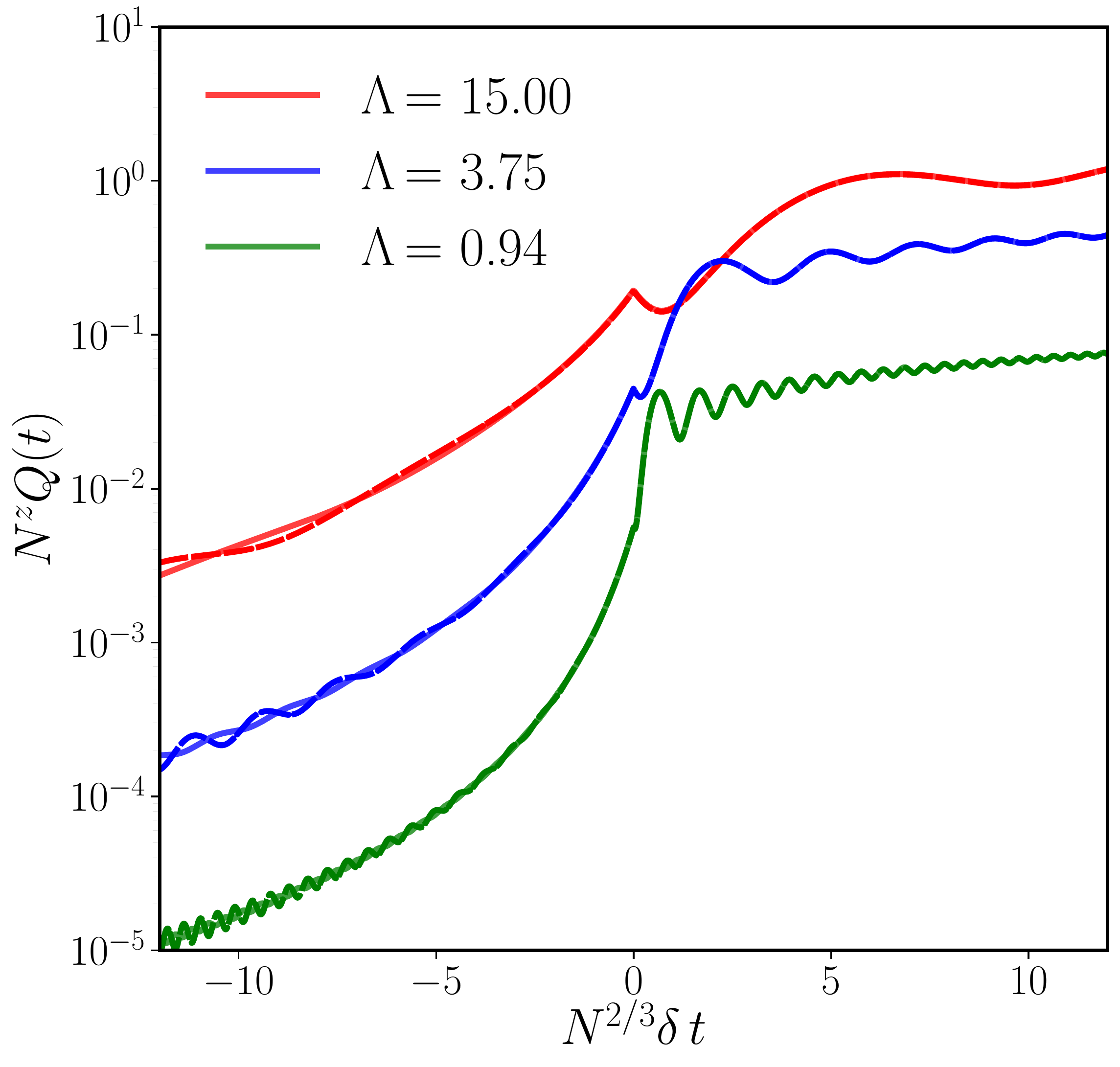}
}
\subfigure[\,\,Heat (numerics)]{\label{FigSecV_4c}\includegraphics[width=.31\textwidth]{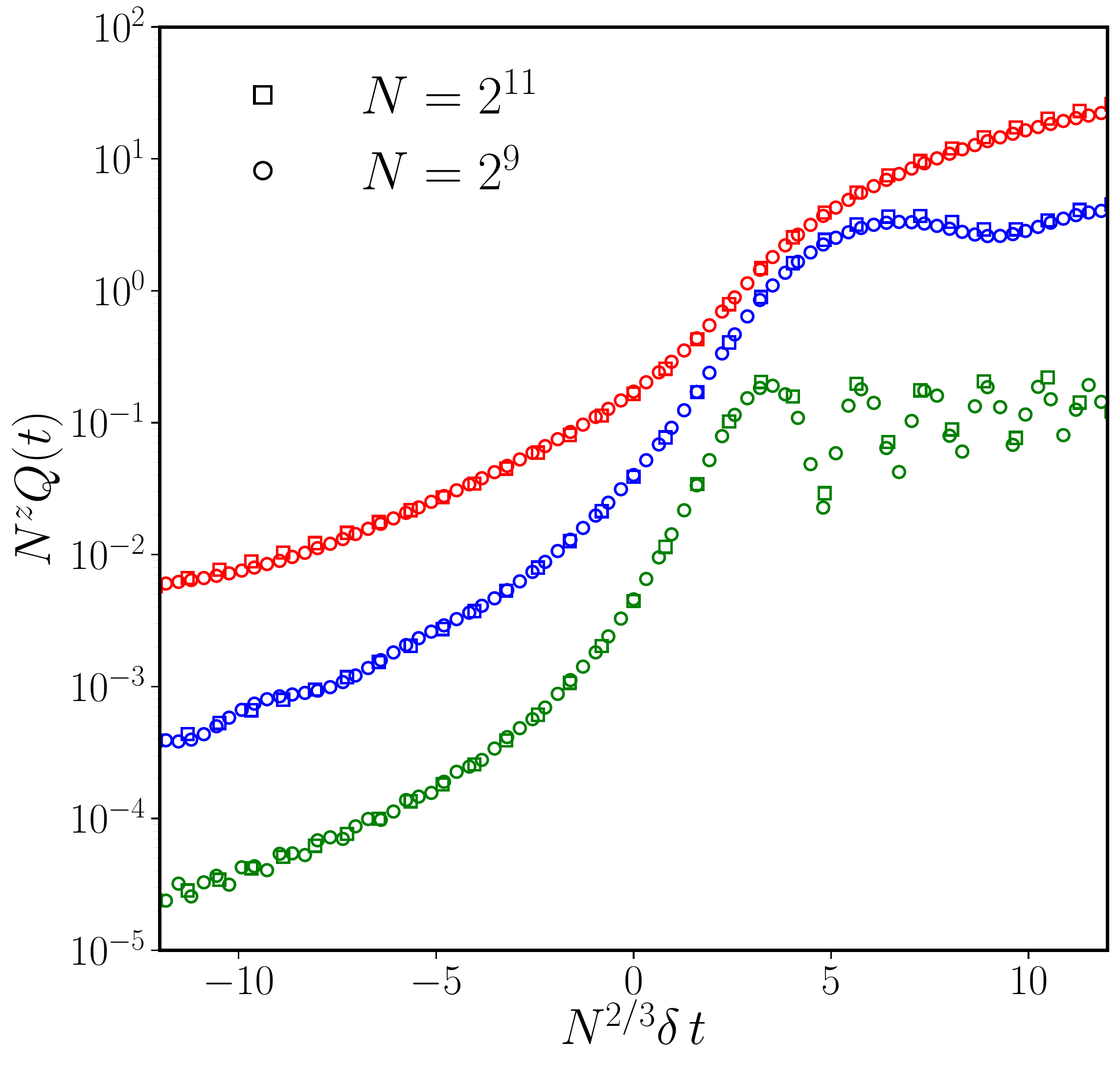}
}
\caption{\label{FigSecV_4} 
\textbf{Kibble-Zurek mechanism in the fully-connected model. (a)} The effective width $\xi(t)$ as a function of the effective time $t/\hat{t}$ for different values of the initial time $t_{i}$ the numerical solution of Eq.\,\eqref{ermakov_eq} is shown to collapse on the analytic result (black dashed line) reported in Refs.\,\cite{defenu2021quantum}. \textbf{(b,c)} display the heat curves obtained respectively via the effective model in Eq.\,\eqref{hoh} and via the full numerical solution of the time-dependent LMG model in Eq.\,\eqref{h_lmg} with time-dependent coupling $J=h_{c}-\delta\,t$ performed in Ref.\,\cite{acevedo2014new}. Each colour represents a different value of $\Lambda=N\delta$ with $N=2^{9}$ and $N=2^{11}$ (dashed and solid lines in panel b). Both the heat and the time variables have been rescaled following the notation of Ref.\,\cite{acevedo2014new}. As expected, the curves at different sizes collapse both in the theory and in the exact simulations. Despite the effective model in Eq.\,\eqref{hoh} is just an effective model, which does not account for the mean field energy shift, the similarity between the analytic (b) and numeric (c) curves is remarkable. Fig.\,\ref{FigSecV_4c} has been reproduced from Ref.\,\cite{acevedo2014new}.}
\end{figure*}

Regime (3) is obtained considering the case $\omega_{0}=0$ and taking the dynamics in $t\to\infty$ limit, which yields the $\delta$-independent results 
\begin{align}
\label{asy_exc_den}
\lim_{t\to\infty} n_{\rm exc}(t)&=\frac{1}{3}\\
\label{asy_fid}
\lim_{t\to\infty} f(t)&=\frac{\sqrt{3}}{2}
\end{align}
which characterise the non-adiabatic dynamics as they remain finite in the $\delta\to 0$ limit. The analytical results in Eqs.\,\eqref{asy_exc_den} and\,\eqref{asy_fid} are universal in the traditional of Kibble-Zurek mechanis result. So, they faithfully reproduce the slow drive limit $\delta\to 0$ of any dynamical protocol which crosses the critical point. The universality phenomenon is analysed in details in Ref.\,\cite{defenu2018dynamical}. On a less accurate, but perhaps more quantitative level, it can be numerically verified that the analytic solution accurately describes any drive $\omega'(t_{*})$ such that $|\omega'(t_{*})-\omega(t_{*})|^{2}\ll 1$, where $\omega(t)$ is given in Eq.\,\eqref{freq_scal}\,\cite{defenu2021quantum}.

For a finite thermodynamic system, we expect the dynamical gap not to completely vanish at the critical point, but to present a finite correction vanishing according to finite size scaling  $t_{0}\approx N^{-1/\nu_{*}}$, where $\nu_{*}=3/2$ according to Eq.\,\eqref{fss_flat_int}. Then, the residual scaled frequency only depends on the parameters combination $\Lambda=N\delta$ and, since the minimal scaled frequency reads $\tilde{\omega}^{2}(0)\approx\Lambda^{-2/3}$, it follows that the thermodynamic limit ($N\to\infty$) and the adiabatic one ($\delta\to 0$) do not commute. Rather, the same dynamical evolution for thermodynamical observables occurs for different sizes and drive rates as long as the combination $\Lambda$ remains fixed. The universal behaviour evidenced for the present Harmonic effective model faithfully reproduces exact numerical computations. Indeed, a comparison between the analytic and numerical analyses of the LMG model is shown in Fig.\,\ref{FigSecV_4} proving that the ``anomalous" scaling described in Ref.\,\cite{acevedo2014new} is perfectly justified by the effective model studied here and 
introduced in Ref.\,\cite{defenu2018dynamical}.

\subsubsection{Structural transitions}

Ion crystals and, in general, structural transitions occurring in non-local systems with competing interactions have first triggered the theoretical interest in the Kibble-Zurek scaling of non-homogeneous systems\,\cite{delcampo2010structural,chiara2010spontaneous,zurek2009causality}. In presence of inhomogeneity, the critical point occurs at different moments in the different regions of the system, restoring adiabaticity for dynamical transition where critical excitations propagate faster than the phase boundaries. A straightforward enough argument to justify the previous picture is found by generalising the scaling theory outlined in the beginning of Sec.\,\ref{kzm_gen} to the non-homogeneous case.

We consider a both spatial and time dependent control parameter $\lambda(x,t)$, such that the critical front occurs at $\lambda(x,t)\approx 0$, while in general one has
\begin{align}
\label{inhom_control_p}
\lambda(x,t)=\alpha(x-v_{\rm p}t)
\end{align}
where $v_{\rm p}>0$ is the velocity of the phase front. Locally, the inhomogeneous control parameter in Eq.\,\eqref{inhom_control_p} resembles the homogeneous case with ramp rate $\delta=\alpha\,v_{\rm p}$. Accordingly, all the locations of the systems where $\lambda(x,t)<0$ already lie in the symmetry broken phase and, then, they can communicate the orientation of the order parameter across the phase boundary at  $\lambda(x,t)\approx 0$ towards the symmetric regions of the system where $\lambda(x,t)>0$. The maximum velocity $\hat{v}_{\rm p}$ at which this communication occurs can be found via the relation $\hat{v}_{\rm p}=\hat{\xi}/\hat{t}$. As long as $v_{\rm p}\gg \hat{v}_{\rm p}$
inhomogeneity is not relevant, since the regions on the opposite side of the phase front are effectively decoupled. On the contrary,  defect formation is suppressed for $v_{\rm p}\ll \hat{v}_{\rm p}$ due to the symmetry broken regions of the system coordinating with the ones at $\lambda(x,t)>0$.

Following the discussion above one can use the conventional scaling relations for the homogeneous Kibble-Zurek mechanism to obtain $\hat{v}_{\rm p}\sim \delta^{\frac{(z-1)\nu}{z\nu+1}}\sim  \alpha^{\frac{(z-1)\nu}{\nu+1}}$, which, in turns, leads to the "critical" ramp rate
\begin{align}
\label{non_hom_kz}
\hat{\delta}\sim \alpha^{\frac{z\nu+1}{1+\nu}}.
\end{align}
At rates $\delta\gg \hat{\delta}$ the system effectively behaves as homegeneous and the traditional results for the excitations density are retrieved, conversely in the slow drive limit $\delta\ll \hat{\delta}$ inhomogeneity becomes relevant and can alter the universal Kibble-Zurek scaling. Accordingly, in the homogeneous limit the critical rate vanishes $\lim_{\alpha\to 0}\hat{\delta}=0$. Several examples of non-homogeneous Kibble-Zurek mechanism can be found in the literature\,\cite{zurek2008phase,schaller2008adiabatic,collura2010critical,dziarmaga2010dynamics_inhom}.

Thanks to their tuneability\cite{lemmer2015two}, trapped ion platforms played a crucial role both in the theoretical and experimental investigations of defects formation in the non-homogeneous realm\,\cite{schneider2012experimental,lemmer2015two}.  By adiabatically altering the trapping parameters,  it is possible to drive the system across the structural transition briefly outlined in Sec.\,\ref{sec_zig_zag}\,\cite{baltrusch2012quantum}. However, such procedure will naturally generate localized defects in agreement with the Kibble-Zurek theory\,\cite{schneider2012experimental}.  A similar phenomelogy is also expected for sudden quenches across the boundary of the structural transition\,\cite{landa2010quatum,delcampo2010structural}. Moreover, the dynamics of local defects in Coulomb crystals has been proposed to realise the Frenkel-Kontorova model\,\cite{pruttivarasin2011trapped,cormick2012structural}.

The experimental exploration of the quantum dynamics and formation of kinks in Coulomb crystals\,\cite{pyka2013topological,ulm2013observation} has shown good agreement with the theory expectation\,\cite{landa2010quatum}, providing a flexible tool to investigate defect formation according to the inhomogenous Kibble-Zurek mechanism\,\cite{delcampo2010structural,chiara2010spontaneous}.

\subsubsection{
  Cavity systems} \label{sec:CavityKibbleZurek}
\begin{figure}[]
\centering
\includegraphics[width=1\columnwidth]{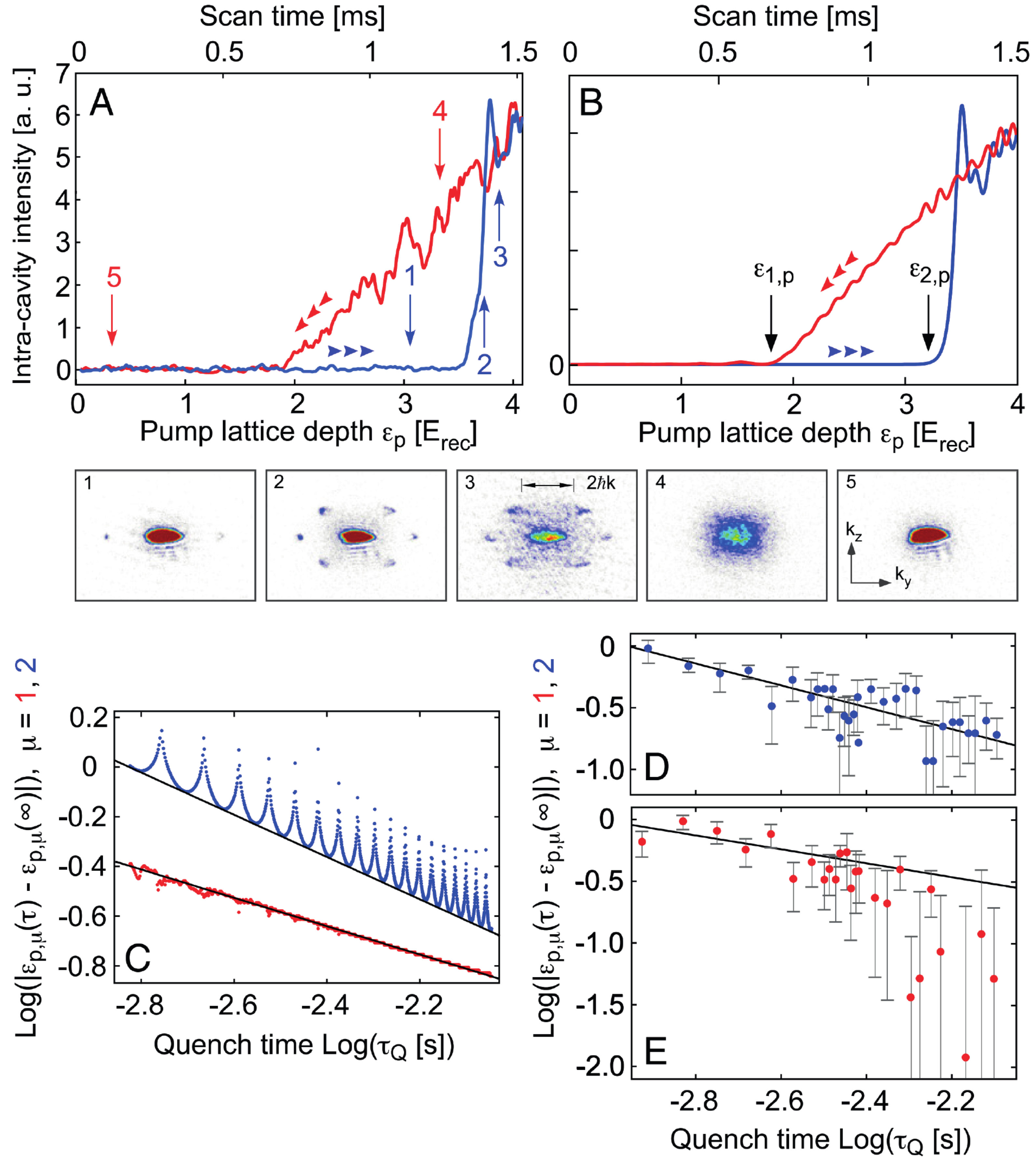}
\caption{{\bf Dynamical critical behavior at the self-organization phase transition.}  (\textbf{A}) Intracavity intensity while the transverse pump lattice depth $\epsilon_p$ is ramped up (blue) and down (red) in ramps of 1.5~ms, each. Below, momentum spectra (1--5) are shown, recorded at increasing times during the $\epsilon_p$-ramp, indicated by the numbered arrows in \textbf{A}. \textbf{B} Mean-field calculation according to \textbf{A} neglecting collisional interactions and assuming an infinite system. The points $\epsilon_{p,1}$ and $\epsilon_{p,2}$ indicate the upper and lower critical lattice depths.  \textbf{C} Mean-field calculations of $\epsilon_{p,1}$ and $\epsilon_{p,2}$ as a function of quench time, resulting in exponents of $(n_1,n_2)=(-0.57, 0.85)$ for power law fits. \textbf{D,E} Experimentally determined dependence of the upper and lower critical lattice depths on the quench time, together with solid lines reproducing the power law dependences of \textbf{C}. Figure reproduced from~\cite{klinder2015dynamical}.
}
\label{fig:KibbleZurekCavity}
\end{figure} 

Quench experiments based on quantum gases in optical cavities~\cite{baumann2011exploring,klinder2015dynamical} have also been interpreted within the framework of the Kibble Zurek mechanism~\cite{kibble1976topology,zurek1985cosmological,delcampo2014universality}. The global character of the cavity-mediated interaction inhibits the formation of domains and thus also of defects during the crossing of this second-order phase transition. However, remnants of the Kibble Zurek mechanism can be found in hysteretic behavior and in the symmetry breaking itself.

In the case of a retarded cavity-mediated interaction, i.e. where the cavity line width $\kappa$ is comparable to the recoil frequency $\omega_r$, pronounced dynamical hysteresis has been observed when crossing the self-organization phase transition \cite{klinder2015dynamical}, see Fig. \ref{fig:KibbleZurekCavity}. The intra-cavity light field, corresponding to the order parameter, shows a hysteresis loop that encloses an area exhibiting a power-law dependence upon the duration of the quench across the phase transition. Real-time observation of the intra-cavity field thus allows to identify at which coupling strength the system effectively freezes its dynamics, depending on the quench rate. A simple power-law model allows to extract dynamical exponents $z \nu$. However, a deeper interpretation would require a comprehensive extension of the concept of universality to driven-dissipative systems~\cite{sieberer2013dynamical,klinder2015dynamical}. In particular, it should be noted that these experimental observations appear not to follow the theoretical predictions outlined in Sec.\,\ref{n_exc_lmg} and in Refs.\,\cite{acevedo2014new,defenu2018dynamical} for isolated quantum systems.

In the limit of large cavity line width with respect to the atomic recoil frequency~\cite{baumann2011exploring}, the hysteresis loop is vanishing~\cite{klinder2015dynamical}, but the effect of the quench rate can be observed in the discrete symmetry breaking described in Sec.~\ref{sec:SymmetryBreaking}. The finite size of the system naturally leads to a small symmetry breaking field, completely dominating the symmetry breaking process in the limit of adiabatically crossing the phase transition. However, for a finite quench rate, the approach to the phase tranistion can be again divided into a quasi-adiabatic regime, where the system follows the control parameter, and an impulse regime, where the system is effectively frozen. For increasing quench rates of the transverse pump power, the coupling strength separating these two regimes is decreasing, as captured by Zurek's equation~\cite{zurek2005dynamics} $|\dot{\zeta}/\zeta|=\Delta/\hbar$, with $\zeta=(\Lambda_c-\Lambda)/\Lambda_c$ describing the distance to the critical point (see also Section~\ref{sec:SpinModelCavity}) and the energy gap between ground and first excited state $\Delta=\hbar \omega_0 \sqrt{1-\Lambda^2/\Lambda_c^2}$. Accordingly, in the experiments, the symmetry breaking for large quench rates becomes dominated by (quantum) fluctuations and increasingly independent of the symmetry breaking field. Quantitative agreement of the observations with the model was found~\cite{baumann2011exploring}.

\subsection{Dynamical Phase Transitions}
\begin{figure*}[t!]
\includegraphics*[width=2\columnwidth]{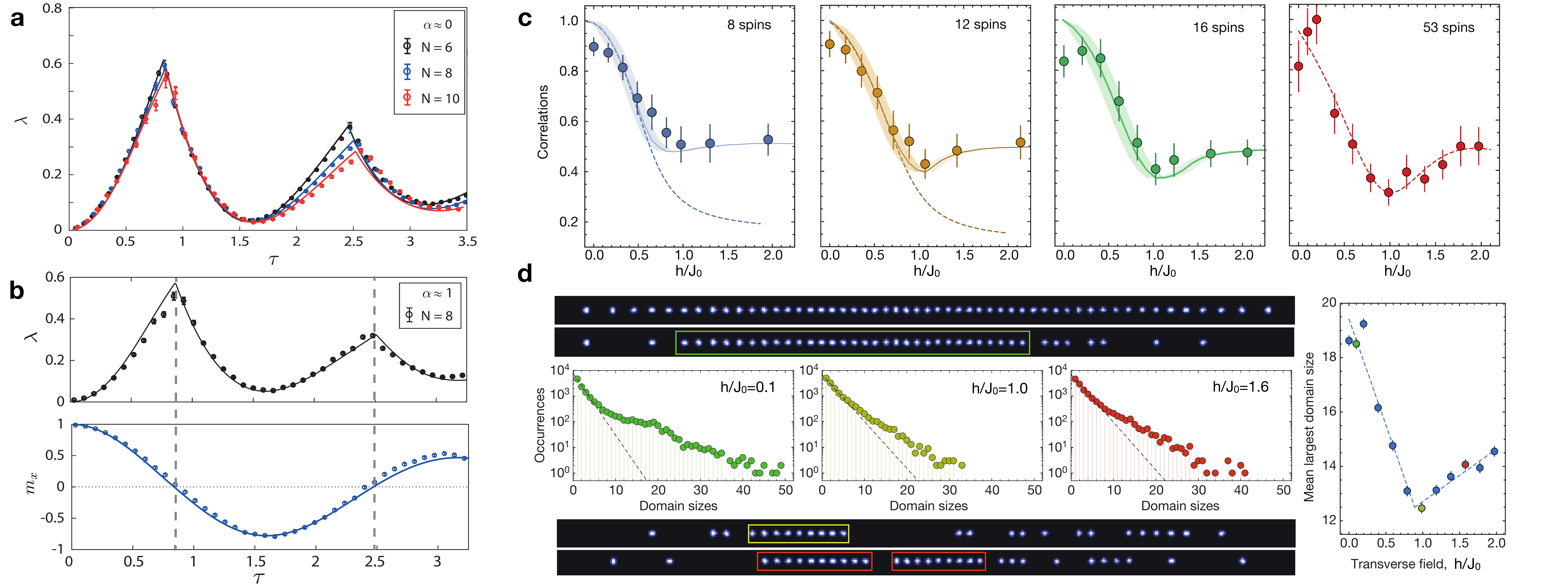}
\caption{{\bf Dynamical phase transitions:} Type I {\bf (a-b)} and type II {\bf (c-d)}. {\bf (a)} Measured rate function $\lambda$ for three different system sizes at $h/J_0\approx2.38$, with $\tau=t h$ being the dimensionless time. The kinks in the evolution become sharper for larger $N$. Here the rate function is defined based on the return probability to the ground state manifold, namely $\lambda(t) = N^{-1}\log(P_{\ket{\psi_0}}+P_{\ket{-\psi_0}})$, where $\ket{-\psi_0}=\ket{\uparrow\uparrow\uparrow\cdots\uparrow}_x$. {\bf(b)} Comparison between rate function $\lambda(t)$ and magnetization evolution $m_x(t)$. The inversion of the magnetization sign corresponds to the nonanalyticity of the rate function $\lambda(t)$. Solid lines are exact numerical predictions based on experimental parameters $(h/J_0=2)$. Adapted from Ref.\,\cite{jurcevic2017direct}. {\bf (c)} Long-time averaged values of the two-body correlations $C_2$, for different numbers of spins in the chain. Solid lines in are exact numerical solutions to the Schr\"odinger equation, and the shaded regions take into account uncertainties from experimental Stark shift calibration errors. Dashed lines in for $N=12,16$ are calculations using a canonical (thermal) ensemble with an effective temperature corresponding to the initial energy density. 
{\bf (d)} Domain statistics and reconstructed single shot images of 53 spins. Top and bottom: reconstructed images based on binary detection of spin state. The top image shows a chain of 53 ions in bright spin states. The other three images show 53 ions in combinations of bright and dark spin states. Center: statistics of the sizes of domains for three different values of $h/J_0$, plotted on a logarithmic scale. Dashed lines are fits to exponential functions, which could be expected for infinite-temperature thermal state. Long tails of deviations are clearly visible, and vary depending on $h/J_0$. Right: mean of the largest domain sizes in each single experimental shot. Dashed lines represent a piecewise linear fit, used to extract the transition point. The green, yellow, and red data points correspond to the transverse fields shown in the domain statistics data on the left center figure. Adapted from Ref. \cite{zhang2017observation}.}
\label{fig:DQPT}
\end{figure*}

One of the most relevant scaling phenomena in the far out-of-equilibrium realm is
provided by 
dynamical phase transitions\,\cite{zvyagin2016dynamical,mori2018thermalization}. In particular, after the sudden quench of a control parameter dynamical phase transitions may be classified in two main families. The first family displays a (possibly local) order parameter $A(t)$, whose long-time Cesaro's average $\bar{A}$, defined according to Eq.\,\eqref{ces_avg}, characterises different steady states\,\cite{moeckel2008interaction,eckstein2008nonthermal,eckstein2009thermalization,sciolla2010quantum,halimeh2017prethermalization,lang2018dynamics}. While this phenomenon is naturally observed for quenches across equilibrium symmetry breaking transitions, diverse dynamical phases may also arise in quantum systems, which do not posses any finite-temperature phase transition. There, following a sudden quench, the order parameter $A(t)$  always equilibrates to its normal phase expectation in the long time limit ($\bar{A}=0$ for ferromagnetic systems), but the dynamical phase transition can be observed in a sudden change in the scaling approach to equilibrium\,\cite{altman2002oscillating,barmettler2009relaxation,heyl2014dynamical,lang2018dynamics}.

Experimental evidences of this first kind of dynamical transitions have been found in a linear chain of trapped $^{171}$Yb$^+$ ion spins stored in a Paul trap\,\cite{zhang2017observation}. The system was initialised in the ferromagnetic product state $\ket{\psi_0}=\ket{\downarrow\downarrow\downarrow ...\downarrow}_x$ and, then, evolved according to the long-range Ising Hamiltonian in Eq.\,\eqref{h_lri}.  The dynamical quantum phase transition occurs when the ratio $h/J_0\sim1$, where $J_{0}$ is the strenght of long-range interactions ($V_{r}\propto J_{0}/r^{\alpha}$) and the order parameter changes abruptly from ferromagnetic to paramagnetic order. The observation of the dynamical transition has been obtained by measuring the late time average of the two-body correlator defined as:
\begin{equation}
    C_2=\frac{1}{N^2}\sum_{ij}\langle \sigma_i^x \sigma_j^x\rangle,
\end{equation}
after the quantum quench. 

The measured late time correlator $C_2$ features a "dip" at the critical point that sharpens scaling up the system size $N$ up to 53 $^{171}$Yb$^+$ qubits, as shown in Fig. \ref{fig:DQPT}c. Additional evidence of the occurrence of the dynamical phase transition can be also observed in higher-order correlations, such as the distribution of domain sizes throughout the entire chain, shown in Fig. \ref{fig:DQPT}d. The occurrence of the dynamical phase transition is observed in the decreased probabilities of observing long strings of aligned ions at the critical point $h/J_0\sim1$. This is shown measuring the mean largest domain size as a function of the transverse field strength, for late times and repeated experimental shots, which features a sharp transition at the critical point. 
Another recent experimental realization of dynamical phase transitions
within the LMG model was reported in \cite{muniz2020exploring}. 
The experiment was performed with large ensembles of $^{88}$Sr atoms in an optical cavity where magnetic interactions can be accurately tuned
\cite{norcia2018cavity} and reports the observation of distinct dynamical phases of matter in this system. A similar setup has been proposed also for the observation of dynamical phases of  the celebrated BCS model in superconductivity as a function of system parameters and the prepared initial states \cite{lewisswan2021cavity}.

The second family of dynamical phase transitions features periodic non analyticities in the Loschmidt return rate \cite{heyl2013dynamical}. It is convenient to define the return probability to the initial state $\ket{\psi_0}$ after a quantum quench under the Hamiltonian $H$ as $\mathcal{G}(t)=\bra{\psi_0} e^{-i H t}\ket{\psi_0}$. This quantity exhibits non-analycities that are formally analogous to the ones of the partition function of the corresponding equilibrium system, defined as $Z={\rm Tr}(e^{-H/k_B T})$  \cite{heyl2013dynamical}. Along this analogy, the complex counterpart of the thermodynamic free energy density $f=-N^{-1}k_B T \log(Z)$ is the rate function $\lambda(t)=-N^{-1} \log[\mathcal{G}(t)]$. This quantity, in the thermodynamic limit, exhibits dynamical real-time nonanalyticities that play an analogous role as the non-analytic behaviour of the free energy density of a thermodynamic system at equilibrium. 

As a consequence of the above statements, the non-analyticities in the return rate signal the occurrence of a dynamical quantum phase transitions at certain critical evolution times after the sudden quench. These phenomena recently generated a high degree of interest both from the theoretical\,\cite{heyl2018dynamical,mori2018thermalization} and experimental physics communities\,\cite{flaschner2018observation,jurcevic2017direct}. The first theoretical description of dynamical phase transitions in the return rates  have been showed in the case of the nearest-neighbor transverse-field Ising chain. There, non-analytic cusps in the return rate could be only observed after a sudden quench across the equilibrium critical point. It was shown by several subsequent examples that dynamical crossing an equilibrium phase boundary may not produces the aforementioned cusps in the return rates while sudden quenches within the same phase may produce type-II dynamical phase transitions\, \cite{andraschko2014dynamical,vajna2014disentangling}.

Therefore, the dynamical critical point for the appearance of type-II dynamical phase transitions does not need to coincide with the quantum critical point of the system at equilibrium. A further proof of this distinction comes from the strong dependence of the dynamical critical point on the initial state of the system\,\cite{halimeh2017prethermalization,lang2018dynamics}. In this perspective, long-range interactions have been shown to produce several additional dynamical phases with respect to the simple nearest neighbours case\,\cite{halimeh2017dynamical,homrighausen2017anomalous,defenu2019dynamical,uhrich2020out}. It is, thus, not surprising that first observation of type-II dynamical phase transitions have been detected in a trapped ion simulation of the long-range Ising Hamiltonian in Eq.\,\eqref{h_lri}.

The simulation was performed with a linear chain of trapped $^{40}$Ca$^+$ ion spins\,\cite{jurcevic2017direct}.  The system is prepared in the classical eigenstate which minimizes the ferromagnetic interactions $\ket{\psi_0}=\ket{\downarrow\downarrow\downarrow ...\downarrow}_x$, then a finite transverse field  is suddenly switched on (quenched), such that the Hamiltonian in Eq.\,\eqref{h_lri} lies in the $h>J_0$, with $J_0$ being the average nearest-neighbour spin-spin coupling. Fig. \ref{fig:DQPT}a displays the return rate $\lambda$, which exhibits clear non-analyticities at the critical times $t_c$. This behaviour can be related to the one of other global  observables, such as average magnetization $m_x=N^{-1}\sum_i \sigma_i^x$. Due to the final Hamiltonian having $h>J_0$, the $\mathbb{Z}_2$ symmetry of the Hamiltonian, which was broken in the initial state, is dynamically restored during the evolution at the critical times $t_{c}$, see Fig. \ref{fig:DQPT}b. This phenomenon is often referred as zero crossings and describes the oscillations performed by the order parameter within the slow decay, which appears at long times.

The correspondence between the zero crossings of the order parameter and the cusps of the return rate $\lambda(t)$ is not the only relation between the two families of dynamical phase transitions. Indeed, the dynamical critical points for type-I and type-II transitions were shown to coincide\,\cite{zunkovic2018dynamical,halimeh2017prethermalization}. More in general, the fundamental relations between thermodynamic equilibrium phases and their dynamical counterparts has been extensively explored not only in terms of order parameters\,\cite{heyl2018dynamical, titum2019probing, ajisaka2014nonequilibrium, zunkovic2018dynamical}, but also with respect to scaling and universality\,\cite{heyl2015scaling},
discrete or continuous symmetry breaking\,\cite{zunkovic2016dynamical,weidinger2017dynamical,huang2019dynamical} and nature of the quasi-particles\,\cite{syed2021dynamical}.

Free-fermionic systems, described by the Kitaev Hamiltonians studied in Sec.\,\ref{subsec_kit_chain}, played a prominent role both in the experimental and theoretical study of dynamical phase transitions. Indeed, despite the absence of a local order parameter in the equilibrium topological phase transition of the Kitaev chain, dynamical phase transitions also occur in these models\,\cite{vajna2015topological,budich2016dynamical,bhattacharya2017emergent,bhattacharya2017interconnections}, where they have been also experimentally observed\,\cite{flaeschner2018observation}.  The possibility of analytically solving free-fermionic models also in presence of long-range hopping or pairing produced a comprehensive understanding of how additional dynamical phases can be influenced by corrections to scaling in the spectrum as well as its relation with the results for the Ising model\,\cite{defenu2019dynamical}. Despite the absence of any local order parameter in free-fermi systems, a relation between the occurrence of cusps in the Loschmidt echo and the zero crossings of the (non-local) string order parameter\,\cite{uhrich2020out}

Despite its close relation to the Kitaev chain, see Sec.\,\ref{lrkc_i_rel}, the long-range Ising model presents a more complex phenomenology with respect to the Kitaev chain. This occurrence is related to the appearance of domain-wall confinement due to long-range interactions in the Ising model\,\cite{liu2019confined}, these
confined excitations behave like Stark-localized particles in an effective
confining potential\,\cite{lerose2020quasilocalized}, see also the next section. This domain-wall coupling was found to be the reason for the appearance of anomalous cusps in quantum quenches at sufficiently small transverse-field strengths\,\cite{halimeh2017dynamical,halimeh2020quasiparticle}, while the absence of quasiparticles coupling in the Kitaev chain disrupts the anomalous phase\,\cite{defenu2019dynamical}.

Critical quenches, where the post-quench Hamiltonian is 
critical are known to to yield long-time universal scaling behavior following the mechanism of 
aging \cite{chiocchetta2017dynamical}. 
These kind of phenomena are strongly influenced by long-range interactions  
have been studied in \cite{halimeh2021quantum}.
In particular, in the LMG model, depending on the type of quench, three behaviors where both the short-time dynamics and the stationary state at long times are effectively thermal, quantum, and genuinely non-equilibrium were identified. Each stationary state is characterized by distinct universality classes and static and dynamical critical exponents \cite{titum2020nonequilibrium}.

\subsection{Confinement}

\begin{figure*}[t!]
\centering
\includegraphics[width=2\columnwidth]{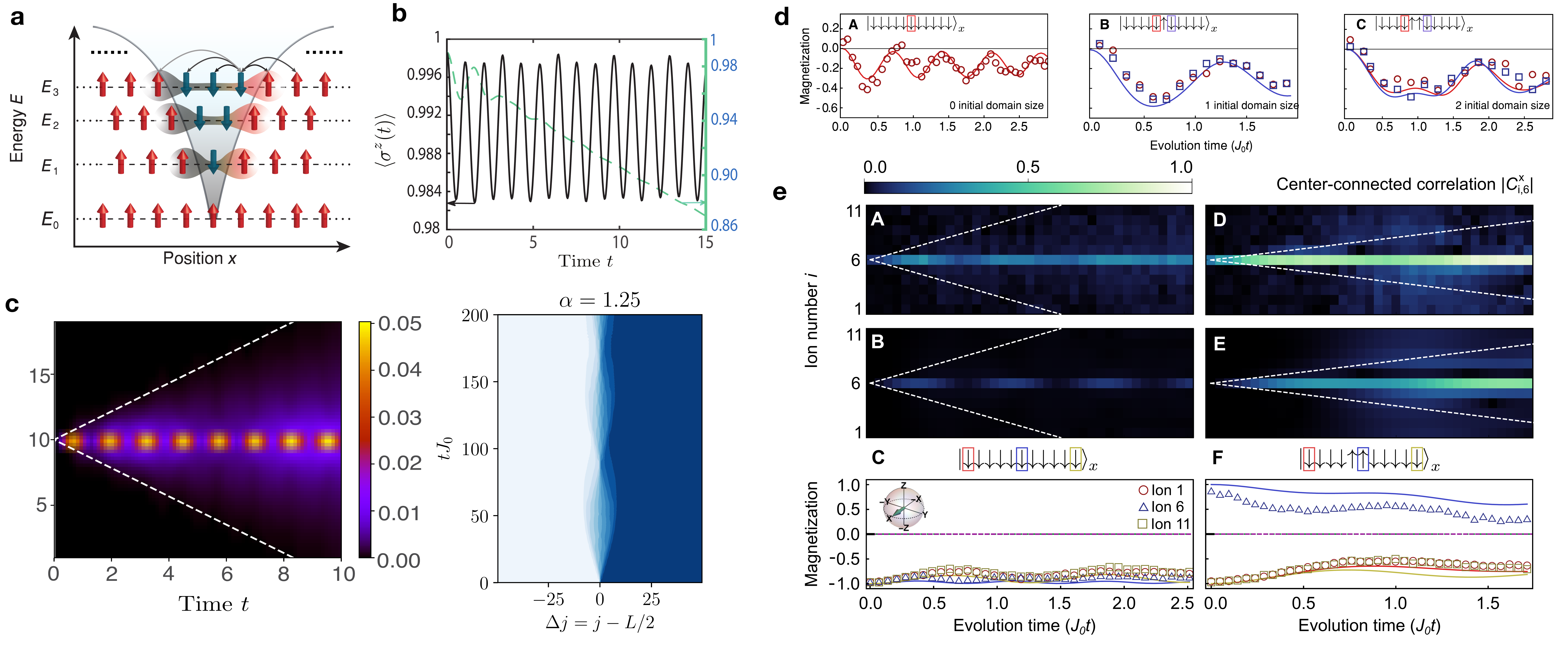}
\caption{{\bf Confinement in long-range spin systems}
  {\bf (a)} Magnetic domain walls in Ising spin chains can experience an effective confining potential that increases with distance, analogously to the strong nuclear force. This potential results in meson-like domain wall bound states (labeled $E_1$ to $E_3$) that influence the post-quench dynamics. Adapted from
  Ref. \cite{tan2021domain}.
{\bf (b)} Magnetization oscillation $\aver{\sigma_z(t)}$ (black line) versus time after quenching to $\alpha = 2.3$, $B_z=0.27 J_0$, for $N=20$. The dashed green lines show the magnetization for the TFIM with nearest neighbor interactions only. Numerical calculations adapted from Ref. \cite{liu2019confined}.
{\bf (c)} Left: Confinement of correlation in long-range system for $\alpha=2.3$ starting from the polarized state $\ket{\downarrow\downarrow\dots\downarrow} $, adapted from Ref. \cite{liu2019confined}. Right: Confinement of correlation in long-range system for $\alpha=1.25$ starting from the highly excited state $\ket{\downarrow\downarrow\dots\downarrow\uparrow\uparrow\dots\uparrow}$, adapted from Ref. \cite{lerose2019quasilocalized}.
{\bf (d)} Magnetization oscillations starting from low energy product states to probe the first three mesons' masses. Adapted from Ref. \cite{tan2021domain}. 
{\bf (e)} Confinement dynamics at $B_z/J_{0} \approx 0.75, L=11$. The top row shows the absolute value of experimental center-connected correlations $|C_{i,6}^x(t)|$ averaged over 2000 experiments. The middle row shows $|C_{i,6}^x(t)|$ calculated by solving the Schr\"{o}dinger equation. Dashed white lines show correlation propagation bounds (light cones) in the limit $\alpha \rightarrow \infty$  (nearest-neighbor interactions). The bottom row shows measured individual-spin magnetizations along their initialization axes, $\aver{\sigma^{z}_i(t)}$, averaged over 2000 experiments. Symbols represent magnetization data and solid colored curves represent theoretical magnetizations calculated by solving the Schr\"{o}dinger equation. Purple (green) dashed lines represent thermal expectation values calculated from a canonical (microcanonical) ensemble averaged over the three displayed spins. Adapted from Ref. \cite{tan2021domain}.
}
\label{fig_confinement}
\end{figure*}

As previously shown in section \ref{sec:Lieb-Rob}, long-range interactions can give rise to fast spreading of correlations. However, focusing on trapped ions systems in this section we will review a different regime in which long-range interactions allow the observation of the same phenomenology analogous to confinement. 

In general, spin models can be engineered to exhibit confinement of correlations and meson production. Ref. \cite{kormos2017real-time}, studied the case of a global quench with a nearest neighbor Ising Hamiltonian
\begin{equation}
\label{eq_confinement_H}
H=-J\sum_{i}\sigma_i^x \sigma_{i+1}^x + h_z\sum\sigma_i^z+ h_x\sum\sigma_i^x
\end{equation}
with both transverse $h_z$ and longitudinal field $h_x$. In this case the dynamics produces confinement of quasi-particles and magnetization oscillations with frequencies related to the mass/energy differences between the bound states most involved in the dynamics. In this setting, the quasiparticle excitation is mapped to domain walls whose separation is energetically suppressed by the longitudinal field, which causes the appearance of a ladder of discrete meson states in the low-energy spectrum of the system \cite{James2019Nonthermal}. Remarkably, after a quantum quench in these system, both correlation spreading and energy flow \cite{Mazza2019Suppression} are suppressed, even if the system is non-integrable and non-disordered.

A similar phenomenology can be also observed in long-range spin systems described by Hamiltonian(\ref{h_lri}), as theorized in Ref.\,\cite{liu2019confined} for low energy states and $\alpha<3$ (see Figs. \ref{fig_confinement}b-c) and in Ref.\,\cite{lerose2019quasilocalized} for highly excited states for $\alpha<2$ (see Fig. \ref{fig_confinement}d).
Interestingly, the confining potential induced by the long-range tail of the interaction on the domain walls acts, to a first approximation, as an effective longitudinal field that constrains the evolution of the spin excitations (see Fig. \ref{fig_confinement}a). Therefore, in the regime in which the transverse field $h_z$ is smaller than the spin-spin interaction $J_0$, long-range interactions cause a phenomenology analogous to the one found in the Hamiltonian in Eq.\,\eqref{eq_confinement_H}: the presence of bounds states results in slow spread of correlations and magnetization oscillations. 

The latter have been observed experimentally for a chain of up to 38 ions\,\cite{tan2021domain}, showing a mass scaling in agreement with theory in the low energy part of the spectrum. In the same work, a smaller chain of 11 ions was used to probe the first few bound states by preparing different initial product states and measuring the magnetization $\aver{\sigma^z_i(t)}$ at the center of the chain (for 0 initial domain walls) or next to the boundaries of the initial domain (for 2 initial domain walls). The initial states have been chosen to maximize the matrix elements of the magnetization between the prepared state $i$ and the adjacent higher-energy bound state $i+1$, allowing to extract the energy gap between these two states (see Fig. \ref{fig_confinement}d). Similarly, the slow spread of correlations have been observed by measuring two-body correlations of the central spin with the rest of the system, resulting in a much slower correlation spread compared a nearest neighbor Ising chain (see Fig.\,\ref{fig_confinement}e).

The possibility to engineer mesons in long-range interacting spin systems has sparked an increasing body of theoretical works on the existence of string breaking in a specific range of parameters  \cite{Verdel2020realtime} and mesons collisions \cite{karpov2020spatiotemporal,surace2021scattering}.

\subsection{Other dynamical phenomena}

Long-range interacting quantum systems have been explored also in different settings, including disordered fields or interactions or in a Floquet setting, where the system is subjected to a periodic drive.
In presence of disorder, long-range interacting quantum systems can exhibit many-body localization (MBL), where the system fails to thermalize at long times owing to the existence of an extensive set of quasi-local integrals of motion \cite{Nandkishore2015manybody,Abanin2019colloquium}. However, sufficiently long-range interactions can destroy many-body localization as shown in \cite{Yao2014many-body, Pino2014entanglement}. In this perspective, as it occurred for the XXZ model in Secs.\,\ref{sex_xxz} and\,\ref{sec_hd_bosons}, it is important to differentiate between the case of long-range exchange couplings, i.e. hopping terms in the Hubbard model representation, and long-range density-density interactions, i.e. Ising interactions in the spin formalism. In particular, for long-range hopping terms, analytical arguments have been used to predict the boundary $\alpha<3d/2$ \,\cite{Burin2015Localization} as a condition for delocalization in long-range spin systems governed by a XY Hamiltonian, while, in the case of long-range Ising interactions, the boundary value $\alpha_{*}=2d$ has been found\,\cite{Burin2015manybody}. Again in the case of long-range spin exchange, Ref. \cite{Safavi-Naini2019quantum} shows numerical evidence that a XY model is delocalized for $\alpha<1$ in one dimension, in contrast with the $\alpha_{*}=1.5$ result of Ref.\,\cite{Burin2015Localization}. This different prediction might be due to how dominant finite size effects are for system sizes that can be simulated exactly. In this respect, \cite{Maksimov2020many} studied the scaling with size of critical disorder for $\alpha<3/2d$. \cite{Nandkishore2017many} use bosonization arguments to show that MBL can arise in one-dimensional systems with $\sim r$ interactions and speculate that MBL can be observed in two-dimensional systems with $\log(r)$ interactions, and in three-dimensional systems with $1/r$ interactions. Interestingly, MBL has been predicted with mean field analysis \cite{Roy2019selfconsistent} on the disordered XXZ model with different power law exponents  for $\beta <1/2$ and $\beta<\alpha$, where $\alpha$ is the decay exponent long-range exchange couplings and $\beta$ the one of long-range Ising interactions. MBL has also been found numerically in all-to-all systems\,\cite{Sierant2019manybody} and fermionic system with long-range hopping\,\cite{Nag2019manybody}. 

An important feature of MBL in the presence of long-range density-density interactions is algebraic localization of the quasi-local integrals of motion (LIOMs) which characterize the MBL phase \cite{detomasi2019algebraic,Pino2014entanglement}. Conversely, in short range interacting systems, LIOMs are exponentially localized and entanglement entropy grows logarithmically.
However, since in MBL long-range systems LIOMs are algebraically localized one expects that entanglement entropy grows polynomially \cite{safavi2019quantum}.
In particular \cite{deng2020universal} showed that in a variety of models (XY, XXZ and Extended Hubbard Model) with power-law interactions there is a universal power-law growth of the entanglement entropy at the MBL transition. Experimental signatures of many-body localization in long-range systems, such as memory of the initial states \cite{Smith2016many} confirmed numerically by \cite{Wu2016understanding}, and slow growth of the second order Renyi entropy \cite{brydges2019probing}, have been observed in trapped ion chains up to 20 qubits. 

More recently, disorder-free, "stark" many-body localization\,\cite{Schulz2019stark,Nieuwenburg2019frombloch}, which has been predicted to be more resilient than standard MBL to long-range exchange couplings\,\cite{Bhakuni2020Stability}. Signatures of this type of disorder-free MBL has been observed in a trapped-ion chain of up to 25 qubits with long-range interactions decaying with $\alpha\sim1.3$ and a strong effective magnetic field gradient \cite{morong2021observation}. As mentioned in section \ref{sec:TI_techniques}, a large magnetic field makes the Ising model an effective XY model with long-range exchange couplings, and, in this case, the LIOMs are given by the Wannier-Stark states.
Conversely, in the case of long-range density-density interactions, one expects Hilbert-space fragmentation, which was also studied in short range interacting disordered spinless fermions\,\cite{lev2015absence, detomasi2019dynamics}.
In particular, in Hubbard models with polar interactions and nearest-neighbor hoppings\,\cite{li2021hilbert}
the power-law tail plays a crucial role because it induces Hilbert-space shattering and MBL-like localization in absence of any disorder, even for moderate ratios of the polar interactions versus hopping. 
This is not the case of models with both nearest-neighbor hopping and density-density interactions, where Hilbert-space fragmentation does not lead to disorder-free MBL \cite{detomasi2019dynamics}. 


Quantum systems with long-range interactions have been recently used to observe new phases of matter in periodically driven (Floquet) systems \cite{Else2016Floquet, Von2016Absolute,Khemani2016Phase, Yao2018Time} in which discrete time translational symmetry is spontaneously broken. The observation of time-crystalline behaviour has been achieved in a periodically driven 1D trapped ion chain with on-site static disorder \cite{zhang2017observationdiscrete} and a 3D disordered sample of NV-centers with dipolar interaction \cite{choi2017observation}. However, it has been shown numerically \cite{khemani2019brief} that both realizations did not realize a genuine discrete time crystal where MBL prevents the system to heat up to infinite temperature, but rather a pre-thermal (trapped ions) and critical (NV-centers) time crystal. Long-range interactions do play a special role in the case of pre-thermal discrete time crystals, where the temporal and spatial long-range order is exhibited only for low energy initial states \cite{Machado2020long-range}. Crucially, in one dimension, a long-range interacting system with power law $1<\alpha<2$ \cite{dyson1969existence} can exhibit a finite-temperature SSB phase, making an exception to the conventional Landau-Peierls argument that discrete symmetry breaking is forbidden for short-range interacting systems in one dimension. The pre-thermal discrete time crystal has been observed and characterized experimentally in a trapped ions chain of up to 25 spins \cite{kyprianidis2021observation}. Also in periodically driven many-body cavity QED systems, limit cycles and time crystalline behavior has been predicted and experimentally observed \cite{cosme2018dynamical,kessler2019emergent,kessler2020observation,georges2021dynamical}. In addition, even without providing a time-dependent external drive, many-body cavity QED systems can feature non-stationary periodically evolving states that emerge due to the competition between dissipative and coherent processes in long-range interacting systems, as has been recently experimentally observed \cite{dogra2019dissipation} and theoretically analyzed \cite{buca2019,chiacchio2019dissipation}.

Time crystals and, in general, Floquet dynamics has been also found to be a source of dynamical phase transition\,\cite{kosior2018dynamical,yang2019floquet}. Indeed, it has been found that novel dynamical transitions can be engineered by
periodic driving. In the particular case of the long-range Ising model, the periodic drive can stabilize phases, dubbed
Kapitza phases, with magnetic ordering without an equilibrium
counterpart\,\cite{lerose2019prethermal}.

\section{Conclusion and outlook}
\label{sec:conc}

The range of the effective interactions among the constituents of a system
is in general one of its main properties, and it can affect in many ways
the phase diagram, the critical properties and the dynamical
behaviour of physical observables. Therefore, the first natural question to be asked
both for classical and quantum systems is how the properties of the
system are modified by increasing the range of the interactions $V$, or
equivalently reducing the power exponent $\alpha$, where $V(r) \sim 1/r^{\alpha}$
for large inter-constituents distances $r$.

For classical systems, the effect of long-range interactions has been
systematically investigated both in the equilibrium and out-of-equilibrium
realms\,\cite{campa2014physics}. There, the range of interactions in most of the cases is given and
one studies its consequences on -- among others --
the ensemble equivalence, the thermodynamic
properties such as specific heat and the
occurrence of quasi-stationary states, i.e.
metastable configurations whose lifetime scales super-linearly
with the system size.

Of course, the same set of questions of how long-range interactions modify the
properties of models when the interactions are varied from the short-range
limit to the strong long-range regime is present also in the quantum realm.
In this paper we have reviewed the main atomic, molecular and optical (AMO)
systems in which long-range interactions are naturally present, but we have also
emphasized the fact that in many of such systems the range of interactions
can be controlled and varied giving raise to tunable values of
$\alpha$. This can be seen in the spirit of quantum simulations, where
one has a high degree of control on the system and on its crucial properties.

We have discussed in the main text most of the quantum models that it is possible to simulate,
focusing in particular on lattice and spin models. A variety of
spin models such as quantum Ising, XX and XXZ models (and their variants) with tunable
long-range interactions can be implemented. These spin models alongside
bosonic and fermionic models with long-range density-density interactions
provide an ample arena of models in which the long-range-ness of the
interactions plays a key role. If remarkable progress have been done in the
simulations of quantum long-range lattice models, many more models
have yet to find their way, such as bosonic and fermionic models with
long-range hopping (a task presently hard to be implemented) and long
multi-body and multi-spin terms \cite{andrade2021engineering}.

In particular, in experimental AMO systems the main challenges are centered on gaining more tunability of the spin-spin interactions through individual atom control. For example, trapped-ion systems are routinely used as quantum computing platforms \cite{Bruzewicz2019trapped, Wright2019benchmarking,Pino2021demonstration} where individual qubit control and detection are necessary ingredients to exploit the long-range connectivity of pairwise quantum logic gate operations. Leveraging on the same technological advances, trapped-ion simulators are posed to explore a wider range of physical models where long-range interactions and high connectivity play crucial roles, ranging from high energy physics \cite{martinez2016real, Muschik2017wilson},  spin-boson Dicke models \cite{Safavi-Naini2018verification,Gorman2018engineering}, to quantum spin glasses models \cite{rademaker2019bridging}.

Also many-body cavity QED systems have just demonstrated first results on tuning the interaction range. In a next step, the resulting many-body phases, phase transition, and associated phenomena including the Brazovskii transition, glasiness, or frustration have to be explored. Having these tunable range interactions compete with short-range collisional interactions will allow to enter strongly correlated regimes and to explore the rich universe of extended Hubbard models.

Several ubiquitous questions arising in the study of short-range (local) quantum systems find new life and impetus in the long-range domain. The fact that altering the range of the interactions qualitatively corresponds to varying the dimension of the system survives in the quantum realm, but associated to a modification of the dynamical exponent $z$ so that the universal behaviour of quantum long-range models effectively corresponds to the one of a classical model in the fractional $d+z$ dimension, with $z<1$. Therefore, the spatial dimensionality does not appear to play a crucial role in long-range systems as it does in the local case, since long-range couplings alter the spectral dimension of bare theory\,\cite{leuzzi2008dilute,millan2021complex}.
Moreover, the interplay between long-range interaction and unitary dynamics yields a plethora of novel dynamical phenomena absent in the classical limit, see Sec.\,\ref{sec:dyn}. Nevertheless, several workhorses of classical long-range physics such as ensemble in-equivalence and negative specific heats largely remain to be explored and exploited\,\cite{kastner2010nonequivalence}.

A point emerging in weak long-range systems
is that the long-range couplings induce a dispersion relation $\propto k^{\sigma}$ as opposed to the standard relation $\propto k^2$ in short-range systems. 
Given the nature of the dispersion relation in long-range systems, one can
expect -- and actually finds in some cases with microscopic calculations
-- that the effective low-energy model features fractional derivatives
(or fractional Laplacians). While several examples of this mechanism are found in classical dynamical systems, the corresponding studies of fractional quantum dynamics are still in their infancy\,\cite{helmrich2020signatures}.

The analysis of the different systems presented in this review ultimately
shows then that long-range interactions provide an "ingredient"
that we can control and use for different purposes. On the one hand, they can
be exploited to control the stationary states and the thermalization
properties. On the other hand they may
affect the phase diagram and the universality properties. Additionally, they can be a resource
in the quantum control of the system, providing an useful knob to control
the dynamics and the implementation of quantum information tasks, where they can
be used to improve the efficiency of control gates and the
unitary dynamics needed to modify in the desired way the quantum state of the
system.

Long-range properties can be also exploited also in typical quantum simulations
contexts, as highlighted in the simulation of dynamical gauge field theory
with AMO systems\,\cite{banuls2020simulating,davoudi2020towards,davoudi2021simulating}, where a suitably tailored
long-range interactions can be used to simulate the effect of dynamical
gauge fields. In a similar way, they can play a role in the study of quantum devices
and the thermodynamic aspects of quantum registers. 

The study of the possible uses of long-range interactions in quantum simulators
and devices is only at the beginning and it will benefit from 
(and motivate in turn) progress in systems in which the long-range nature
of the interactions can be controlled, as in the mode control of long-range
interactions with trapped ions. Several systems in which long-range interactions
may play a crucial role remain to be fully investigated, such as ultracold
fermionic gases with long-range interactions. We envision a significant
interplay between the study of new equilibrium phases and dynamical
regimes in quantum long-range systems and the focused embodiment of systems
with long-range coupling in quantum devices and simulators. We hope that
the present review may trigger such combined studies to fully exploit the
richness of quantum long-range systems.

\label{sec:concl}
\section*{Acknowledgements}

We thank all our colleagues within the long-range community that engaged us in many fruitful interactions and discussions along the years. In particular, we acknowledge useful correspondence on the content of this article with O. L. Acevedo, A. Lerose, G. Morigi and L. Quiroga.
We acknowledge A. Daley, D. O’Dell, 
L. Dell'Anna, Z.-X. Gong, J. A. S. Louren\c{c}o, M. Maghrebi, G. Morigi, D. Mukamel, 
G. Pupillo, A. M. Rey, L. Santos, J. Schachenmayer, P. Schauss for their valuable feedback and 
careful reading of the manuscript.
This work is supported by the Deutsche Forschungsgemeinschaft (DFG, German Research Foundation) under Germany’s Excellence Strategy EXC2181/1-390900948 (the Heidelberg STRUCTURES Excellence Cluster).
T.D. acknowledges funding from the Swiss National Science Foundation SNF: NCCR QSIT and the project “Cavity-assisted pattern recognition” (Project No. IZBRZ2 186312), and funding from EU Horizon 2020: ITN grant ColOpt (Project No. 721465). T.M.  acknowledges CNPq for support through Bolsa de produtividade em Pesquisa n.311079/2015-6 and the Serrapilheira Institute (grant number Serra-1812-27802). A.T. and S.R. ackownledge support by the MISTI GlobalSeed  Funds  MIT-FVG  Collaboration  Grant  "NV  centers  for  the  test  of  the  Quantum  Jarzynski  Equality (NVQJE)". G.P. acknowledges support by the DOE Office of Science, Office of Nuclear Physics, under Award no. DE-SC0021143, the Office of Naval Research (N00014-20-1-2695), the Army Research Lab (W911QX20P0063) and the Army Research office (W911NF21P0003).

\bibliographystyle{apsrmp4-1}
\bibliography{main}
\end{document}